\documentclass{aa}  
\usepackage{graphicx}       
\usepackage{natbib}
\usepackage{longtable}
\usepackage{supertabular}

\newcommand{\AGN}{{AGN}}

\newcommand{\BCDG}{{BCDG}}

\newcommand{\FIR}{\emph{FIR}}
\newcommand{\HI}{\textsc{H\,i}}
\newcommand{\Ha} {H$\alpha$}
\newcommand{\Hb} {H$\beta$}
\newcommand{\HII}{\textsc{H\,ii\ }}

\newcommand{\HIPASS}{{HIPASS}}

\newcommand{\IMF}{{IMF}}
\newcommand{\IRAS}{{IRAS}}

\newcommand{\LINER}{{LINER}}
\newcommand{\Lo} {$L_{\odot}$}
\newcommand{\Mo} {$M_{\odot}$}
\newcommand{\Myr}{Myr}          
\newcommand{\NED} {{NED}}
\newcommand{\NIR} {\emph{NIR}}

\newcommand{\SED}{{\sc Sed}}

\newcommand{\TDG} {{TDG}}

\newcommand{\UV}{\emph{UV}}

\newcommand{\WRBUMP} {WR bump}
\newcommand{\Zo} {Z$_{\odot}$}

\newcommand{\NOTe}{2.56m NOT}
\newcommand{\WHT}{4.2m WHT}
\newcommand{\INTe}{2.5m INT}

\newcommand{\WHa}{$W($H$\alpha)$}

\newcommand{\MHi}{$M_{\rm H\,I}$}
\newcommand{\Mdyn}{$M_{\rm Dyn}$}
\newcommand{\Mkep}{$M_{\rm Kep}$}
\newcommand{\Mdust}{$M_{\rm dust}$}
\newcommand{\Mkepl}{$M_{\rm Kep}/L_B$}

\newcommand{\MHil}{$M_{\rm H\, I}/L_B$}
\newcommand{\Mdynl}{$M_{\rm Dyn}/L_B$}
\newcommand{\Mdustl}{$M_{\rm dust}/L_B$}
\newcommand{\MHiMdyn}{$M_{\rm H\,I}/M_{\rm Dyn}$}

\newcommand{\kms}{km\,s$^{-1}$}

\newcommand{\valor}[3]{(#1$\pm$#2)$\times$10$^{#3}$}   

\newcommand{\tableline}{\hline}

\newcommand{\HeI}{He\,{\sc i}}
\newcommand{\HeII}{He\,{\sc ii}}
\newcommand{\Sii}{[\ion{S}{ii}] $\lambda\lambda$6716,6731}
\newcommand{\Oii}{[\ion{O}{ii}] $\lambda\lambda$3726,3729}

\newcommand{\Te}{$T_{\rm e}$}
\newcommand{\Ne}{$n_{\rm e}$}
\newcommand{\TeOiii}{$T_e$(\ion{O}{iii})}
\newcommand{\TeOii}{$T_e$(\ion{O}{ii})}
\newcommand{\TeNii}{$T_e$(\ion{N}{ii})}

\newcommand{\CHb}{$c$(H$\beta$)}
\newcommand{\Wabs}{$W_{abs}$}

\newcommand{\abox}{12+log(O/H)}
\newcommand{\lno}{log(N/O)}

\newcommand{\nodata}{...}
\begin{document}
   \title{Massive star formation in Wolf-Rayet galaxies\thanks{Based on observations made with NOT (Nordic Optical Telescope), INT (Isaac Newton 
Telescope) and WHT (William Herschel Telescope) operated on the island of La Palma jointly by Denmark, Finland, Iceland, Norway and Sweden (NOT) or 
the Isaac Newton Group (INT, WHT) in the Spanish Observatorio del Roque de Los Muchachos of the Instituto de Astrof\'\i sica de Canarias. 
}}

   \subtitle{II. Optical spectroscopy results}

   \author{\'Angel R. L\'opez-S\'anchez
          \inst{1,2}
		  \and
		  C\'esar Esteban\inst{2,3}
          }

   \offprints{\'Angel R. L\'opez-S\'anchez, \email{Angel.Lopez-Sanchez@csiro.au}}

\institute{CSIRO / Australia Telescope National Facility, PO-BOX 76, Epping, NSW 1710, Australia \and Instituto de Astrof{\'\i}sica de Canarias, C/ 
V\'{\i}a L\'actea S/N, E-38200, La Laguna, Tenerife, Spain \and Departamento de Astrof\'{\i}sica de la Universidad de La Laguna, E-38071, La Laguna, 
Tenerife, Spain}


   \date{Received March 12, 2009; Accepted September 26, 2009}

 
  \abstract
   {}
{We have performed a comprehensive multiwavelength analysis of a sample of 20 starburst galaxies that show the presence of a substantial
population of very young massive stars, most of them classified as Wolf-Rayet (WR) galaxies.
In this paper, the second of the series, we present the results
of the analysis of long-slit intermediate-resolution spectroscopy of star-formation bursts for 16 galaxies of our sample.}
  {We study the spatial localization of the WR stars in each galaxy. We analyze the excitation mechanism and derive the reddening coefficient, 
physical conditions and chemical abundances of the ionized gas. We study the kinematics of the ionized gas to check the
rotation/turbulence pattern of each system. When possible, tentative estimates of the Keplerian mass of the galaxies have been calculated.}
{Aperture effects and the exact positioning of the slit onto the WR-rich bursts seem to play a fundamental role in their detection. We checked that 
the ages of the last star-forming burst estimated using optical spectra agree with those derived from \Ha\ imagery.
Our analysis has revealed that a substantial fraction of the galaxies show evidences of perturbed kinematics. With respect to the results found in 
individual galaxies, we remark the detection of objects with different metallicity and decoupled
kinematics in Haro~15 and Mkn~1199, the finding of evidences of tidal streams in IRAS~08208+2816, Tol~9 and perhaps in SBS~1319+579, and the
development of a merging process in SBS~0926+606~A and in Tol~1457-262.}
  {All these results --in combination with those obtained in Paper I-- reinforce the hypothesis
that interactions with or between dwarf objects is a very important mechanism in the triggering of massive star formation in starburst
galaxies, specially in dwarf ones. It must be highlighted that only deep and very detailed observations --as these presented in this paper--
can provide clear evidences that these subtle interaction processes are taking place.}

\titlerunning{Massive star formation in Wolf-Rayet galaxies II: Spectroscopic results}

\authorrunning{L\'opez-S\'anchez \& Esteban}

   \maketitle
%

\section{Introduction}

Wolf-Rayet (WR) stars are the evolved descendants of the most massive, very hot 
and very luminous (10$^5$ to 
10$^6$ \Lo) O stars. In the so-called \citet{Conti76} and \citet{Maeder90,Maeder91} scenarios, WR stars are interpreted as central He-burning objects 
that have lost the main part of their H-rich envelope via strong winds. 
Hence, their surface chemical composition is dominated by He rather than H, along with elements 
produced by the nuclear nucleosynthesis. WN and WC stars show the products of
the CNO cycle (H-burning) and the triple-$\alpha$ (He-burning), respectively. The most massive O stars 
($M\geq$ \mbox{25 \Mo}\ for \Zo) became WR stars between 2 and 5 \Myr\ since their birth, spending only some few hundreds of thousands of years 
($t_{WR}\leq5\times10^5$~yr) in this phase \citep{MeynetMaeder05}. A review of the physical properties of WR stars was recently presented by 
\citet{Crowther07}.

The broad emission features that characterized the spectra of WR stars are often observed in extragalactic \HII regions. Actually, the so-called 
Wolf-Rayet galaxies make up a very inhomogeneous class of star-forming objects: giant \ion{H}{ii} regions in spiral arms, irregular galaxies, blue 
compact dwarf galaxies (BCDGs), luminous merging \IRAS\ galaxies, active galactic nuclei (AGNs), Seyfert 2 and low-ionization nuclear emission-line 
regions (LINERs). All objects have in common ongoing or recent star formation which has produced stars massive enough to evolve to the WR stage 
(Shaerer et al. 1999). 
There are two important broad features that reveal the presence of WR stars in the integrated spectra of an extragalactic \HII region:
\begin{enumerate}
\item A	blend of \ion{He}{ii} $\lambda$4686, \ion{C}{iii}/\ion{C}{iv} $\lambda$4650 and \ion{N}{iii} $\lambda$4640 emission lines originated in the 
expanding atmospheres of the most massive stars, the so-called {\bf blue WR Bump}. This feature is mainly due to the presence of WN stars. The broad, 
stellar, \ion{He}{ii} $\lambda$4686 is its main feature. Although rarely strong, the narrow, nebular \ion{He}{ii} $\lambda$4686  is usually 
associated with the presence of these massive stars, but its origin is still somewhat controversial \citep{Garnett91,G04,BKD08}.
\item The \ion{C}{iii} $\lambda$5698 and \ion{C}{iv} $\lambda$5808 broad emission lines, sometimes called the {\bf red or yellow WR Bump}. 
\ion{C}{iv} 
$\lambda$5808 is the strongest emission line in WC stars but it is barely seen in WN stars. The red WR bump is rarely detected and it is always 
weaker than the blue WR bump \citep{GIT00,FCCG04}.
\end{enumerate}

Making use of population synthesis models it is possible to determine the age of the bursts, the number of O and WR stars, the WN/WC ratio, the 
initial mass function (\IMF) or the mass of the burst. Therefore, the study of WR galaxies helps to widen our knowledge about both the massive star 
formation and the evolution of starbursts: they allow to study the early phases of starbursts and are the best direct measure of the upper end of the 
\IMF, a fundamental ingredient for studying unresolved stellar populations, \citep{SGIT00,GIT00,PSGD02,FCCG04,Buckalew05,Zhang07}, giving key 
constraints to 
stellar evolution models.


On the other hand, the knowledge of the chemical composition of galaxies, in particular in dwarf galaxies, is vital for understanding their 
evolution, star formation history, stellar nucleosynthesis, the importance of gas inflow and outflow and the enrichment of the intergalactic medium. 
Indeed, metallicity is a key ingredient for modelling galaxy properties, such it determines \UV, optical and \NIR\ \mbox{colors} at a given age 
(i.e., Leitherer et al. 1999), nucleosynthetic yields (e.g., Woosley \& Weaver 1995), the dust-to-gas ratio (e.g., Hirashita et al 2001), the shape 
of the interstellar extinction curve (e.g., Piovan et al. 2006), or even the WR properties \citep{Crowther07}.

The most robust method to derive the metallicity in star-forming and starburst galaxies is via the estimation of metal abundances and abundance 
ratios, in concrete through the determination of the gas-phase oxygen and nitrogen abundances and the nitrogen-to-oxygen ratio. The relationships 
between current metallicity and other galaxy parameters, such as colors, luminosity, neutral gas content, star formation rate, extinction or total 
mass, constraint galaxy evolution models and give clues about the current stage of a galaxy. For example, is still debated if massive star formation 
result in the instantaneous enrichment of the interstellar medium of a dwarf galaxy, or if the bulk of the newly synthesized heavy elements must cool 
before becoming part of the ISM that eventually will form the next generation of stars. Accurate oxygen abundance measurements of several \HII 
regions within a dwarf galaxy will increase the understanding of its chemical enrichment and mixing of enriched material. The analysis of the 
kinematics of the ionized gas will also help to understand the dynamic stage of galaxies and reveal recent interaction features. Furthermore, 
detailed analysis of starburst galaxies in the nearby Universe are fundamental to interpret the observations of high-z star forming galaxies, such as 
Lyman Break Galaxies \citep{EP03}, as well as quantify the importance of interactions in the triggering of the star-formation bursts, that seem to be 
very common at higher redshifts (i.e., Kauffmann \& White 1993; Springer et al. 2005). 


We have performed a detailed photometric and spectroscopic analysis of a sample of 20 WR galaxies. Our main aim is the study of the formation of 
massive stars in starburst galaxies, their gas-phase metal abundance and its relationships with other galaxy properties, and the role that the 
interactions with or between dwarf galaxies and/or low surface brightness objects have in 
the triggering mechanism of the star-formation bursts. In Paper~I \citep{LSE08} we exposed the motivation of this work, compiled the list of the 
analyzed WR galaxies (Table~1 of Paper~I) and presented the results of optical/\NIR\ broad-band and \Ha\ photometry. In this second paper we present 
the 
results of the analysis of intermediate-resolution long slit spectroscopy of 16 objects of our sample of WR galaxies --the results for the other 4 
objects have been published separately. In many cases, two or more slit positions have been used 
in order to analyze the most interesting zones, knots or morphological structures belonging to each galaxy or even surrounding objects.
In particular, these observations have the following aims:
\begin{enumerate} 
 \item Study the content and spatial location of the WR stars in each galaxy. We examine the spectra for the presence of the \ion{He}{ii} 
$\lambda$4686 emission line and/or the blue-\WRBUMP\ as well as for the red-\WRBUMP. The characteristics of the WR population can be derived by 
comparison with theoretical population synthesis models.
 \item Know the physical properties of the ionized gas: excitation mechanism, electron density, high and low ionization electron temperatures and 
reddening coefficient.
 \item Analyze the ionization structure and the chemistry of the gas (abundances of He, O, N, S, Ne, Ar, Fe and Cl) associated with different 
morphological zones in each galaxy, especially in those areas in which WR features are detected. This analysis is specially relevant in cases of 
interaction or merging processes because the regions may have different chemical composition and also would allow to discern between the \emph{tidal 
dwarf galaxy} (\TDG) or \emph{pre-existing dwarf galaxy} nature of nearby diffuse objects surrounding the main galaxy.   
 \item Determine the radial velocities of different star-formation bursts, galaxies in the same system and/or objects in possible interaction. The 
distance to the main galaxy is also calculated. 
 \item Study the velocity field via the analysis of position-velocity diagrams in order to understand the kinematics of the ionized gas associated to 
different members in the system in order to know their evolution (rotation, interactions features, fusion evidences, movements associated to 
superwinds...). The Keplerian mass has been estimated in objects showing solid-body rotation.  
 \item Obtain independent estimations of the age of the last star-forming burst via the comparison with stellar population synthesis models. 
 \item Study the stellar population underlying the bursts using the analysis of absorption lines (i.e. \ion{Ca}{ii} H,K, \ion{Mg}{i} 
$\lambda\lambda$5167,5184, \ion{Na}{i} $\lambda\lambda$5890,5896, \ion{Ca}{ii} triplet).
\item Finally, the spectral energy distribution (\SED) has been analyzed in some cases in order to constrain the properties of the underlying stellar 
population.
\end{enumerate}

This paper mainly presents the analysis of the ionized gas within our WR galaxy sample.
In \S2 we describe our observations, some details of the data reduction processes, and describe some useful relations. In \S3 we describe the 
physical properties ($T_e$, $n_e$, reddening coefficient, excitation mechanism), the chemical abundances and the kinematics of the ionized gas for 
each galaxy. Finally, the most important results derived from our spectroscopic study, including a comparison with the ages derived from the \Ha\ 
photometry and an estimation of the age of the underlying stellar component using the \SED, are summarized in \S4. 
A detailed analysis of the O and WR populations and the comparison with theoretical models is presented in Paper~III. The global analysis of our 
optical/NIR data will be shown in Paper IV. The final paper of the series (Paper~V) will compile the properties derived using data from other 
wavelengths (UV, FIR, radio and X-ray) 
and complete the global analysis combining all available multiwavelength data of our WR galaxy sample. It is, so far, the most complete and 
exhaustive data set of this kind of galaxies, involving multiwavelength results and analyzed following the same procedures. 

\section{Observations and data reduction and analysis}

\subsection{Spectroscopic observations}

We obtained intermediate-resolution long slit spectroscopy for all our sample WR galaxies
except for NGC~5253, for which high-resolution echelle spectroscopy was taken (see L\'opez-S\'anchez et al. 2007 for details). We used three 
telescopes to carry out these observations: 2.5m \emph{Isaac Newton Telescope} (INT), 2.56m \emph{Nordical Optical Telescope} (NOT), and 4.2m 
\emph{William Herschel Telescope} (WHT), all located at Roque de los Muchachos Observatory (ORM, La Palma, Spain). 
The details of these observations are the following: 
\begin{enumerate}
\item {\bf Observations at the 2.5m INT}. We used the IDS (\emph{Intermediate Dispersion Spectrograph}) instrument attached at the Cassegrain focus 
in December 1999. A EEV CCD 2148$\times$4200 pixel array, with a pixel size of 13.5 $\mu$m, was used, that corresponds to an spatial resolution of 
0.40$\arcsec$ pix$^{-1}$. The slit was 2.8$\arcmin$ long and 1$\arcsec$ wide. We used the R400V grating, that has a dispersion of 104.5 \AA\ 
mm$^{-1}$ (1.40 \AA\ pix$^{-1}$) and an effective spectral resolution of 3.5 \AA. The spectra cover the wavelength range from 3200 to 7700 \AA. The 
absolute 
flux calibration was achieved by observations of the standard stars Feige~56, Hiltner~600 and Feige~110 \citep{M88}. 

\item {\bf Observations at the 4.2m WHT}. We completed two observation runs in this telescope on December 2000 and December 2002. In both cases, the 
double-arm ISIS (\emph{Intermediate dispersion Spectrograph and Imaging System}) instrument located at the Cassegrain focus of the 
telescope was used. The dichroic used to separate the blue and red beams was set at 5400 \AA. The slit was 3.7\arcmin\ long and 1\arcsec\ wide. We 
used different configurations in each observing run:
  \begin{enumerate}
  		\item {\bf December 2000}: 
    		\begin{itemize}
    		\item \emph{Blue arm}: an EEV CCD with a 4096 $\times$ 2048 pixels array and 13 $\mu$m size was used. The spatial resolution was  
0.20$\arcsec$ pix$^{-1}$. The grating was R600B, giving a dispersion of 33~\AA~mm$^{-1}$ (0.45~\AA~pix$^{-1}$) and an effective spectral
resolution of 1.8~\AA. The observed spectral range was 3600 -- 5200 \AA. 
    		\item \emph{Red arm}: we used a TEX CCD with a configuration of 1024$\times$1024 pixels of 24 $\mu$m pixel size, having a spatial 
resolution of 0.36$\arcsec$ pix$^{-1}$. The grating R316R, that has a dispersion of 66~\AA~mm$^{-1}$ (0.93~\AA~pix$^{-1}$)  and an effective spectral 
resolution of 2.6~\AA, was used, covering the spectral range  5400 -- 6800 \AA.
  		\end{itemize}
  		\item {\bf December 2002}: 
  			\begin{itemize}
  	   		 \item \emph{Blue arm}: We used the same CCD that previously indicated but the R1200B grating, that gives a dispersion of 17 \AA\ 
mm$^{-1}$ (0.23~\AA~pix$^{-1}$) and an effective spectral resolution of 0.86 \AA. The spectral range was 4450 -- 5480 \AA.
    		\item \emph{Red arm}: A Marconi CCD with 4700$\times$2148 pixels array and 14.5 $\mu$m pixel size was used. The spatial resolution was 
0.20$\arcsec$ pix$^{-1}$, hence identical to that provided in the blue arm. We used the R316R grating covering the spectral range 5700 -- 8600 \AA.
		\end{itemize}
  \end{enumerate}
The absolute flux calibration was achieved by observations of the \citet{M88} standard stars {\bf G191B2B} and Feige 34 (December 2000) and Feige 15, 
Feige 110, Hiltner 600 and Hz44 (December 2002). 
 
\item {\bf Observations at the 2.56m NOT}. We completed three observation run at this telescope, always using the ALFOSC (\emph{Andaluc\'{\i}a Faint 
Object Spectrograph and Camera}) instrument and a Loral/Lesser CCD detector (2048 $\times$ 2048 pixels) 
with a pixel size of 13.5 $\mu$m and spatial resolution of 0.19$\arcsec$ pixel$^{-1}$. The slit was 6.4$\arcmin$ 
long and 1$\arcsec$ wide. 
We used several configurations:
  \begin{enumerate}
   \item {\bf 20 March 2004}. We used grism \#7 that has a dispersion of 111 \AA\ mm$^{-1}$ (1.5 \AA\
pix$^{-1}$) and a spectral resolution of 7.5 \AA, covering the spectral range \mbox{3200 -- 6800~\AA.}
   \item {\bf 4 April 2005 and 26--27 April 2006}. We used two different grisms to obtain the blue and the red ranges of the optical spectrum. Grism 
\#14, which has a dispersion of 104 \AA\ mm$^{-1}$ (1.4 \AA\ pix$^{-1}$) and a spectral resolution of 7.0 \AA, was used to cover the spectral range 
3300 -- 6100~\AA. This grism has a low efficiency for $\lambda \leq$4000 \AA. 
Spectra in the red range were obtained using the grism \#8, that has a dispersion of 96 \AA\ mm$^{-1}$ (1.3 \AA\
pix$^{-1}$), a spectral resolution of 6.5 \AA\ and covers the spectral range \mbox{5800 -- 8300 \AA.}
  \end{enumerate}
The spectrophotometric standard star Feige 56  \citep{M88} was used for flux calibrating all the spectra obtained with this telescope.
\end{enumerate}
In all observations, three or four exposures for each slit position were taken to get a good S/N ratio and to remove cosmic rays. 
Table~\ref{observaciones_espectros} compiles all the intermediate-resolution long-slit spectroscopy observations performed for the 16 WR galaxies 
included in this paper.

\begin{table*}[t!]
\centering
  \caption{\footnotesize{Journal of the intermediate-resolution long-slit spectroscopy observations, carried out using the \INTe, \NOTe\ and \WHT\ 
telescopes and with the instrumentation explained in text. In last column, $\sec z$ represents the average airmass in which the observations were 
made.}}
  \label{observaciones_espectros}
  \tiny
  \begin{tabular}{lcccc@{\hspace{3pt}}  ccc@{\hspace{3pt}}c c}
  \noalign{\smallskip}
    \tableline
	\noalign{\smallskip}
Galaxy & Tel. & Date & Exp. Time & Spatial R. & Grism & P.A. & Spectral R. & $\Delta\lambda$ & $\sec z$ \\
      &           &      &    [s]  & [$\arcsec$/pix]&  &[$\rm^{\circ}$] & [\AA ]& [\AA] \\ 
    \tableline
    \noalign{\smallskip} 
Haro 15        & INT & 99/12/27& 3$\times$1200 & 0.40 & R400V&  41 & 3.5 & 3500-7700 & 1.67  \\
               & INT & 99/12/27& 3$\times$1200 & 0.40 & R400V& 117 & 3.5 & 3500-7700 & 1.40 \\
                \noalign{\smallskip}
Mkn 1199       & INT & 99/12/28& 3$\times$1200 & 0.40 & R400V&  32 & 3.5 & 3500-7700 & 1.02 \\
               & INT & 99/12/28& 3$\times$1200 & 0.40 & R400V&  53 & 3.5 & 3500-7700 & 1.00 \\
			    \noalign{\smallskip} 
Mkn 5          & WHT & 02/12/27& 3$\times$700  & 0.20 & R1200&  90 & 0.86 & 4300-5100 & 1.90 \\
               & WHT & 02/12/27& 3$\times$700  & 0.20 & R136R&  90 & 2.6 & 5700-7800 & 1.90 \\
               & INT & 99/12/29& 3$\times$1200 & 0.40 & R400V&   0 & 3.5 & 3500-7700 & 1.54 \\
               & INT & 99/12/29& 3$\times$1200 & 0.40 & R400V& 349 & 3.5 & 3500-7700 & 1.47 \\   
                \noalign{\smallskip}
POX 4          & WHT & 00/12/30& 3$\times$1800 & 0.20 & R600B&  25 & 1.8 & 3650-5100 & 1.60 \\
               & WHT & 00/12/30& 3$\times$1800 & 0.36 & R136R&  25 & 2.6 & 5300-6650 & 1.60 \\
                \noalign{\smallskip}
UM 420         & WHT & 00/12/30& 4$\times$1800 & 0.20 & R600B&  90 & 1.8 & 3650-5100 & 1.14 \\
               & WHT & 00/12/30& 4$\times$1800 & 0.36 & R136R&  90 & 2.6 & 5300-6650 & 1.14 \\
               & INT & 99/12/28& 3$\times$1200 & 0.40 & R400V&  90 & 3.5 & 3500-7700 & 1.23 \\
                \noalign{\smallskip}				
IRAS 08208+2816& INT & 99/12/28& 3$\times$1200 & 0.40 & R400V&  10 & 3.5 & 3500-7700 & 1.11  \\
               & INT & 99/12/28& 3$\times$1200 & 0.40 & R400V& 345 & 3.5 & 3500-7700 & 1.01  \\
			   & INT & 99/12/28& 3$\times$1200 & 0.40 & R400V& 355 & 3.5 & 3500-7700 & 1.32  \\
                \noalign{\smallskip}
SBS 0926+606A  & WHT & 02/12/27& 3$\times$600  & 0.20 & R1200&  27 & 0.86 & 4300-5100 & 1.18 \\
               & WHT & 02/12/27& 3$\times$600  & 0.20 & R136R&  27 & 2.6 & 5700-7800 & 1.18 \\
                \noalign{\smallskip}
SBS 0948+532   & WHT & 00/12/31& 3$\times$1800 & 0.20 & R600B& 114 & 0.86 & 3650-5100 & 1.10 \\
               & WHT & 00/12/31& 3$\times$1800 & 0.36 & R136R& 114 & 2.6 & 5300-6650 & 1.10 \\
                \noalign{\smallskip}
SBS 1054+365   & INT & 99/12/29& 3$\times$1200 & 0.40 & R400V&  55 & 3.5 & 3500-7700 & 1.15  \\
                \noalign{\smallskip}
SBS 1211+540   & WHT & 00/12/31& 3$\times$1800 & 0.20 & R600B& 138 & 1.8 & 3650-5100 & 1.12 \\
               & WHT & 00/12/31& 3$\times$1800 & 0.36 & R136R& 138 & 2.6 & 5300-6650 & 1.12 \\ 
                \noalign{\smallskip}
SBS 1319+579   & WHT & 02/12/27& 3$\times$600  & 0.20 & R1200&  39 & 0.86 & 4300-5100 & 1.48 \\
               & WHT & 02/12/27& 3$\times$600  & 0.20 & R136R&  39 & 2.6 & 5700-7800 & 1.48 \\
                \noalign{\smallskip}
SBS 1415+437   & WHT & 02/12/27& 3$\times$600  & 0.20 & R1200&  20 & 0.86 & 4300-5100 & 1.54 \\
               & WHT & 02/12/27& 3$\times$600  & 0.20 & R136R&  20 & 2.6 & 5300-6650 & 1.54  \\
                \noalign{\smallskip}
III Zw 107     & INT & 99/12/28& 3$\times$1200 & 0.40 & R400V&   0 & 3.5 & 3500-7700 & 1.18 \\
			    \noalign{\smallskip}			
Tol 9          & INT & 99/12/27& 4$\times$1200 & 0.40 & R400V&  49 & 3.5 & 3500-7700 & 1.90  \\
               & NOT & 06/04/28& 4$\times$1200 & 0.19 & g14  & 109 & 7.0 & 3300-6100 & 1.85 \\
			   & NOT & 06/04/27& 3$\times$900  & 0.19 & g8   & 109 & 6.5 & 5800-8300 & 1.92  \\
                \noalign{\smallskip}
Tol 1457-262a  & NOT & 05/04/04& 3$\times$900  & 0.19 & g14  & 155 & 7.0 & 3300-6100 & 1.92 \\  
               & NOT & 05/04/04& 3$\times$900  & 0.19 & g8   & 155 & 6.5 & 5800-8300 & 1.75 \\
                \noalign{\smallskip}
Arp~252        & NOT & 05/04/04& 3$\times$900  & 0.19 & g14  & 342 & 7.0 & 3300-6100 & 1.67 \\  
               & NOT & 05/04/04& 3$\times$900  & 0.19 & g8   & 342 & 6.5 & 5800-8300 & 1.56  \\
	\noalign{\smallskip}    
  \tableline
  \end{tabular}
\end{table*}

\subsection{Reduction of the spectra}

All the data reduction were completed at the IAC. 
IRAF\footnote{IRAF is distributed by NOAO which is operated by AURA Inc.,
under cooperative agreement with NSF.} 
software was used to reduce the CCD frames (bias
correction, flat-fielding, cosmic-ray rejection, wavelength and flux
calibration, sky subtraction) and extract the one-dimensional spectra. 
The correction for atmospheric extinction was performed using an average curve for the continuous atmospheric extinction at Roque de los Muchachos 
Observatory.
For each two-dimensional spectra several apertures were defined along the spatial direction to extract the final one-dimensional spectra of each 
galaxy or emission knot. The apertures were centered at the brightest point  of each  aperture and the width was fixed to obtain a good 
signal-to-noise spectrum. In case of having the optical spectrum separated in two different wavelength intervals, identical apertures in both 
spectral ranges were used.

\subsection{Analysis of the spectra}

IRAF software was also used to analyze the one-dimensional spectra. 
Line intensities and equivalent widths were measured by integrating 
all the flux in the line between two given limits and over a local continuum
estimated by eye. In the cases of line blending, a multiple Gaussian profile 
fit procedure was applied to obtain the line flux of each individual line. 
We used the standard assumption $I$(\Hb)=100 to compute the line intensity ratios. 
The identification and adopted laboratory wavelength of the lines, as well
as their errors, were obtained following Garc\'{\i}a-Rojas et al. (2004); Esteban et al. (2004).

\subsubsection{Distance to the galaxies}

We computed the distance to the galaxies using the brightest emission lines (\Ha\ and [\ion{O}{iii}] $\lambda$5007) in our optical spectra. We 
assumed a Hubble flow with $H_0$=75 km s$^{-1}$ Mpc$^{-1}$, $q_0$=0.5, and corrected for Galactic Standard of Rest. The distance we derived for each 
galaxy is listed in Table~1 in Paper~1. All values agree well within the errors with the distances quoted by the \NED, except for Tol~9. For this 
galaxy, we measure a radial velocity of $v_r=3441$ \kms, while previous observations suggested $v_r=3190$ \kms\ \citep{Lauberts89}. We consider that 
our value is more appropriate because the maximum of the \HI\ emission detected in Tol~9 \citep{LS10} shows the same radial velocity than that our 
optical spectrum provides. 

\subsubsection{Correction for reddening}

The reddening coefficient, \CHb, was derived from the Balmer decrement. 
However, in extragalactic objects the fluxes of nebular Balmer lines may be affected by absorptions produced by the underlying stellar population 
(mainly B and A stars). We performed an iterative procedure to derive simultaneously the reddening coefficient and the equivalent 
widths of the absorption in the hydrogen lines, $W_{abs}$, to correct the observed line intensities for both effects. 
We assumed that $W_{abs}$ is the same for all the Balmer lines and used the
relation given by \citet{MB93} to the absorption correction,
\begin{eqnarray}
c(H\beta)=\frac{1}{f(\lambda)} \log\Bigg[\frac{\frac{I(\lambda)}{I(H\beta)}\times
\Big(1+\frac{W_{abs}}{W_{H\beta}}\Big)} {\frac{F(\lambda)}{F(H\beta)}\times 
\Big(1+\frac{W_{abs}}{W_{\lambda}}\Big)}\Bigg],
\end{eqnarray}
for each detected hydrogen Balmer line. In this equation, $F(\lambda)$ and $I(\lambda)$ are the observed and the theoretical fluxes (unaffected by 
reddening or absorption); $W_{abs},\ W_{\lambda}$ and $W_{H\beta}$ are the equivalent widths of the underlying stellar absorption, the considered 
Balmer line and H$\beta$, respectively, and $f(\lambda)$ is the reddening curve normalized to \Hb\ using the \citet{Cardelli89} law. We always  
considered the theoretical ratios between these pairs of \ion{H}{i} Balmer lines expected for case B recombination given by \citet{SH95} assuming the 
electron temperature and density computed independently for each region. Using three different Balmer lines (i.e., \Ha, \Hb\ and H$\gamma$) an unique 
value for \CHb\ and $W_{abs}$ is computed. However, in the case of using four or more Balmer lines, several solutions are derived, so we considered 
the values that provide the best match between the corrected and the theoretical Balmer line ratios as representative of the region. An 
example of this method using 5 Balmer line ratios is shown in Figure~\ref{um420reddening} when analyzing UM~420.

Tables A.1, A.3, A.5, A.7, A.9, A.11, A.13, A.15 and A.17 show the dereddened line intensity ratios and their associated errors for all the regions 
and galaxies, as well as the adopted $f(\lambda)$ for each emission line. In these tables, we also include other important quantities as: the size of 
the extracted aperture, its relative distance to the main region of the galaxy, the observed \Hb\ flux (uncorrected for extinction), the adopted
values of \CHb\ and \Wabs\ and the equivalent widths of \Ha, \Hb, H$\gamma$ and [\ion{O}{iii}] $\lambda$5007. Colons indicate errors of the order or 
larger than 40\%.

\subsubsection{Physical conditions of the ionized gas}

We studied the physical conditions and chemical abundances of the ionized gas from the 1-D spectra of each galaxy or knot.
We used a two-zone approximation to define the temperature structure of the nebulae, assuming the electron temperature, \Te,
provided by the [\ion{O}{iii}] ion, \TeOiii, as the representative temperature for high ionization
potential ions and $T_e$(\ion{N}{ii}) or $T_e$(\ion{O}{ii}) for the low ionization potential ones. \TeOiii is obtained from the [\ion{O}{iii}] 
($\lambda$4959+$\lambda$5507)/$\lambda$4363 ratio, $T_e$(\ion{N}{ii}) from [\ion{N}{ii}] ($\lambda$6548+$\lambda$6583)/$\lambda$5755 and 
$T_e$(\ion{N}{ii}) from 
[\ion{O}{ii}] ($\lambda$3727+$\lambda$3729)/($\lambda$7319+$\lambda$7330). 
$T_e$s are calculated making use of the five-level program for the analysis of emission-line nebulae that is included
in IRAF NEBULAR task \citep{SD95}. 
Notice that we used an updated atomic dataset for O$^+$, S$^+$, and S$^{++}$ for NEBULAR. The references are indicated in Table~4 of \citet{GRE04}.

When one of the high/low ionization electron temperatures can not be computed, we used the linear relation between \TeOiii\ and \TeOii\ provided by 
\citet{G92},
\begin{eqnarray}
T_e(\textsc{O~ii}) = T_e(\textsc{N~ii}) =0.7 \times T_e(\textsc{O~iii}) + 3000, 
\end{eqnarray}
to estimate the unknown electron temperature.
In the case that no direct estimate of the electron temperature can be obtained, we considered the \TeOiii\ and 
\TeOii\ pairs that reproduce the total oxygen 
abundance obtained by applying the \citet{P01a,P01b} empirical method (see below), also assuming the \citet{G92} relation, that is the same equation 
that Pilyugin uses in his empirical calibrations.

The electron density of the ionized gas, \Ne, was usually computed via the [\ion{S}{ii}] $\lambda\lambda$6716,6731 doublet, although sometimes the 
[\ion{O}{ii}] $\lambda\lambda$3726,3729 doublet was also used. Regions showing \Ne$<$100 cm$^{-3}$ are below the low-density limit and hence we 
considered \Ne=100 cm$^{-3}$ in those cases.

\citet{VO87} 
proposed diagnostic diagrams plotting two different excitation 
line ratios, such as [\ion{O}{iii}]/\Hb\ versus [\ion{S}{ii}]/\Ha, for classifying the 
excitation mechanism of ionized nebulae. 
\HII\ regions (or \HII or starburst galaxies) lie into a narrow band within these diagrams, but
when the gas is ionized by shocks, accretion disks or cooling flows (in the case of 
AGNs or LINERs)
its position is away from the locus of \HII regions. 
We used the analytic relations given by \citet{Do00} and \citet{KD01} between different line ratios to check the nature of the excitation 
mechanism of the ionized gas within the bursts. Figure~\ref{mkn1199diag} shows an example of these diagrams applied to  the regions analyzed in the 
galaxy Mkn~1199.

\subsubsection{Chemical abundances}

Once the electron density and temperature are estimated, the ionic abundances 
can be derived for each region. All the ionic abundances except He$^+$ and Fe$^{++}$ were calculated using the
IRAF NEBULAR task \citep{SD95} from the intensity of collisionally excited lines.
We assumed a two-zone scheme for deriving the ionic abundances,
adopting \TeOiii\ for the high ionization potential ions O$^{++}$,
Ne$^{++}$, S$^{++}$, Ar$^{++}$, Ar$^{3+}$ and Cl$^{++}$; and $T_e$(\ion{N}{ii}) or $T_e$(\ion{O}{ii}) for the
low ionization potential ions O$^+$, N$^+$, S$^+$ and Fe$^{++}$.

The He$^+$/H$^+$ ratio was computed from the \ion{He}{i} lines intensities and  
using the predicted line emissivities calculated by \citet{SSM96} for the \TeOiii\ and \Ne\ assumed for each region.
We also corrected for collisional contribution following the calculations by \citet{B02}. Self-absorption effects were 
not considered. 

Fe$^{++}$ abundances were derived via the [\ion{Fe}{iii}] $\lambda$4658 emission line. We used a 34 level model-atom that includes the collision 
strengths of \citet{Z96} and the transition probabilities of \citet{Q96}

We computed the total abundances of O, N, S, Ne, Ar and Fe. We always adopt O/H = 
O$^+$/H$^+$ +  O$^{++}$/H$^+$ to determine the total oxygen abundance. We detect 
the \ion{He}{ii} $\lambda$4686 line in several objects, but the relative contribution of He$^{++}$ to the 
total amount of helium is negligible, implying that O$^{3+}$ has also a very low abundance in the nebula, thus we did not 
consider its contribution to the total O/H ratio. For deriving the nitrogen abundance we assumed the standard ionization
correction factor (icf) by \citet{PC69}: N/O = N$^+$/O$^+$, which is a reasonably good approximation for the 
excitation degree of the observed 
galaxies. We used the icf provided by \citet{PC69} to derive the total neon abundance. We computed the total sulphur abundance when both S$^+$/H$^+$ 
and S$^{++}$/H$^+$ ratios are available using the icf given by the photoionization models by \citet{S78}. The total argon abundance was 
calculated considering the icfs proposed by \citet{ITL94}. The total iron abundances were obtained from the Fe$^{++}$/H$^+$ ratio and the icf given 
by \citet{RR05}.

As we said before, when direct estimations of the electron temperature were not available, we resorted to empirical calibrations.  \citet{P01a,P01b} 
performed a detailed analysis of observational data combined with photoionization models to obtain the oxygen abundance from the relative intensities 
of strong optical lines. 
\citet{P01a} gives the empirical calibration between the $R_{23}$ and $P$ (an indicator of the hardness of the ionizing radiation) parameters and the 
oxygen abundance in moderately high-metallicity \ion{H}{ii} regions, \abox$\geq$8.3. \citet{P01b} provides the empirical calibration for the 
low-metallicity branch. Unless indicated, we always used Pilyugin empirical calibrations to derive the electron temperatures and the 
chemical abundances of the ionized gas in those regions where no direct determination of \Te\ was possible. Sometimes, we estimated the total oxygen 
abundance making use of the \citet{D02} or \citet{PP04} empirical calibrations, which involve the [\ion{N}{ii}] $\lambda$6583/\Ha\ ratio.

\subsubsection{Estimation of the Keplerian and Dynamical masses}

The bidimensional spectra have been used to perform a position-velocity diagram via the analysis of the brightest emission line profiles (\Ha\ 
and/or  [\ion{O}{iii}] $\lambda$5007) for all the objects. Although the main objective is the analysis of the kinematics of the ionized gas, in some 
cases we estimated the Keplerian mass (\Mkep) of the galaxies assuming that the kinematics are representative of circular rotation. 
We usually considered the half of the maximum velocity difference, $\Delta v$, the half of the spatial separation corresponding to these maxima, $r$, 
and applied the equation:   
\begin{eqnarray}
M_{kep}\, [ M_\odot] \sim 233 \times r~[{\rm pc}] \bigg(\frac{\Delta v~[{\rm km~s^{-1}}]}{\sin i}\bigg)^2,
\end{eqnarray}
assuming circular orbits and Keplerian dynamics \citep{LSER04b}. We remark that the result of this equation \emph{is not} the total dynamical mass of 
the galaxy but only the total mass contained within a circle of radius $r$. The inclination angle, $i$, is defined as that found between the plane of 
the sky and the plane of the galaxy (hence, $i$=90$^{\circ}$ in an edge-on galaxy and $i$=0$^{\circ}$ in a face-on galaxy). We usually estimated this 
angle assuming that the elliptical shape of the galaxy is just a consequence of its orientation.  
 
When 21-cm \HI\ data were available in the literature, we computed both the neutral gas mass, \MHi, and the dynamical mass, \Mdyn, using the typical 
relations (i.e., Dahlem et al. 2005). We estimated the rotation velocity of the neutral gas considering $\Delta v = \frac{W_{\rm H\, I}}{2 \sin i}$ 
and adopted the maximum radius observed in our deep optical images. Therefore, as the extension of the neutral gas is usually larger than the 
extension of the stellar component, our estimations of \Mdyn\ may be underestimated. The gas depletion timescale defined by \citet{SCM03} was 
computed using \MHi\ and the total star-formation rate (SFR) derived for each galaxy in Paper~I. When FIR data are available, we estimated the mass 
of the warm dust, \Mdust, using the equations given by \citet{HSH95}. Although \Mkep, \Mdyn\ and \Mdust\ should be considered only tentative values, 
their comparison gives important clues about the galaxy type, its dynamic and the fate of the neutral gas.

\section{Results}

\subsection{NGC 1741 - HCG 31 AC}

The spectroscopic analysis of \object{NGC~1741} (member AC in the galaxy group HCG~31) was presented in detail in \citet{LSER04a}. Our spectra show 
an evident broad blue \WRBUMP\ and the \HeII\ $\lambda$4686 emission line in this object (Figure~\ref{wrfig1}). A careful re-analysis of the data 
reveals an evident and broad red \WRBUMP\ in the same region (Figure~\ref{wrfig2}), in agreement with previous observations of the same object 
\citep{GIT00}. The spectrum of members F1 and F2 (TDG s candidates) show the \HeII\ $\lambda$4686 emission line; member F2 also seems to show a blue 
\WRBUMP. The analysis of the kinematics of the group reveals almost simultaneous interaction processes involving several objects. 

\subsection{Mkn 1087} 
 
\object{Mkn~1087} is a luminous blue compact galaxy and the main galaxy in a group in interaction. Although some authors did observe WR features in 
this galaxy \citet{KJ85,Va97}, others did not \citep{VC92}, and therefore it was classified as \emph{suspected} WR galaxy by \citet{SCP99}. Our 
analysis of Mkn~1087 was presented in \citet{LSER04b}; we did not detected any WR feature (Figure~\ref{wrfig1}) in any important star-forming region 
in or surrounding Mkn~1087.


\subsection{Haro 15}

\object{Haro~15} is a blue compact galaxy well studied in all frequencies, including optical spectroscopy  \citep{HG85,MBB91,Kong02,Shi05}. 
\citet*{SCP99} listed Haro 15 as a WR galaxy because of the detection of the \ion{He}{ii} $\lambda$4686 emission line by \citet{KovoContini99}. Our 
analysis confirms the presence of WR stars (nebular and broad \HeII\ $\lambda$4686) in the bright star-forming region A (Figure~\ref{wrfig1}) .  

\begin{figure}[t!]
\centering
\includegraphics[angle=270,width=\linewidth]{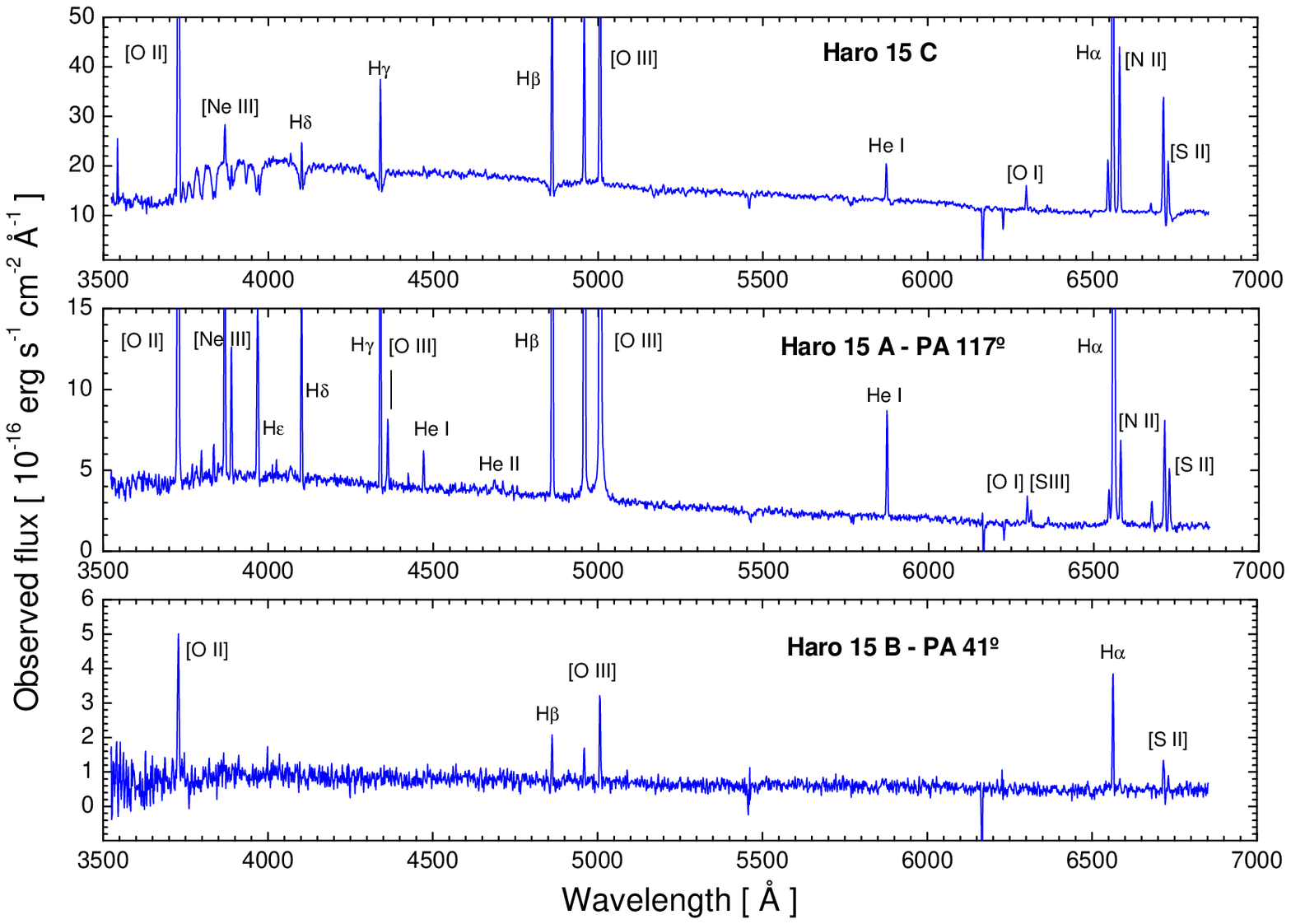}
\caption{\footnotesize{IDS INT spectra for regions C, A and B of Haro 15. Fluxes are not corrected for reddening. The most important emission lines 
have been labeled. See Figure~3 in Paper~I for the identification of the regions.}}
\label{haro15espectros}
\end{figure}

Our long-slit spectroscopy covers the four main regions observed in Haro~15 (see Figure~3 of Paper~I): the center (C), the bright region
A at the ESE, the relatively bright \HII region D at the WNW (both observed with the slit 
with PA~117$^\circ$) and the knot B (at the NE, observed with the slit with PA~41$^\circ$). Figure~\ref{haro15espectros} shows the spectra of the 
three brightest objects, whereas Table~\ref{haro15lineas} compiles the emission line ratios and other properties of the spectra of each region.

The spectrum of the center of Haro~15 shows both nebular emission lines and stellar absorptions; these  absorptions are observed mainly in 
the \HI\ Balmer lines. However, the spectrum of region A is entirely dominated by nebular emission lines, where we detect  [\ion{O}{iii}] 
$\lambda$4363 and \ion{He}{ii} $\lambda$4686. Because of the faintness of B and D, few emission lines are observed in these regions. 

\subsubsection{Physical conditions of the ionized gas}

A direct estimation of $T_e$(\ion{O}{iii}) was computed in knot A because of the detection of [\ion{O}{iii}] $\lambda$4363. We used empirical 
calibrations to estimate the electron temperatures for the rest of the regions; the results are shown in Table~\ref{haro15abun}. The spectrum of the 
central region of Haro~15 has a marginal detection of [\ion{O}{iii}] $\lambda$4363. Using this value we derived $T_e$[\ion{O}{iii}]$\sim$9700~K, 
which is very similar to that obtained using empirical calibrations.  
[\ion{S}{ii}] $\lambda$6717 is blended with a skyline in all spectra, and hence we can not compute \Ne. 
We assumed a value of 100 cm$^{-3}$ for all regions. Comparing the line ratios with diagnostic diagrams, we 
found that all knots can be classified as typical \HII regions. 

\subsubsection{Chemical abundances}

\begin{figure}[t!]
\centering
\includegraphics[angle=270,width=\linewidth]{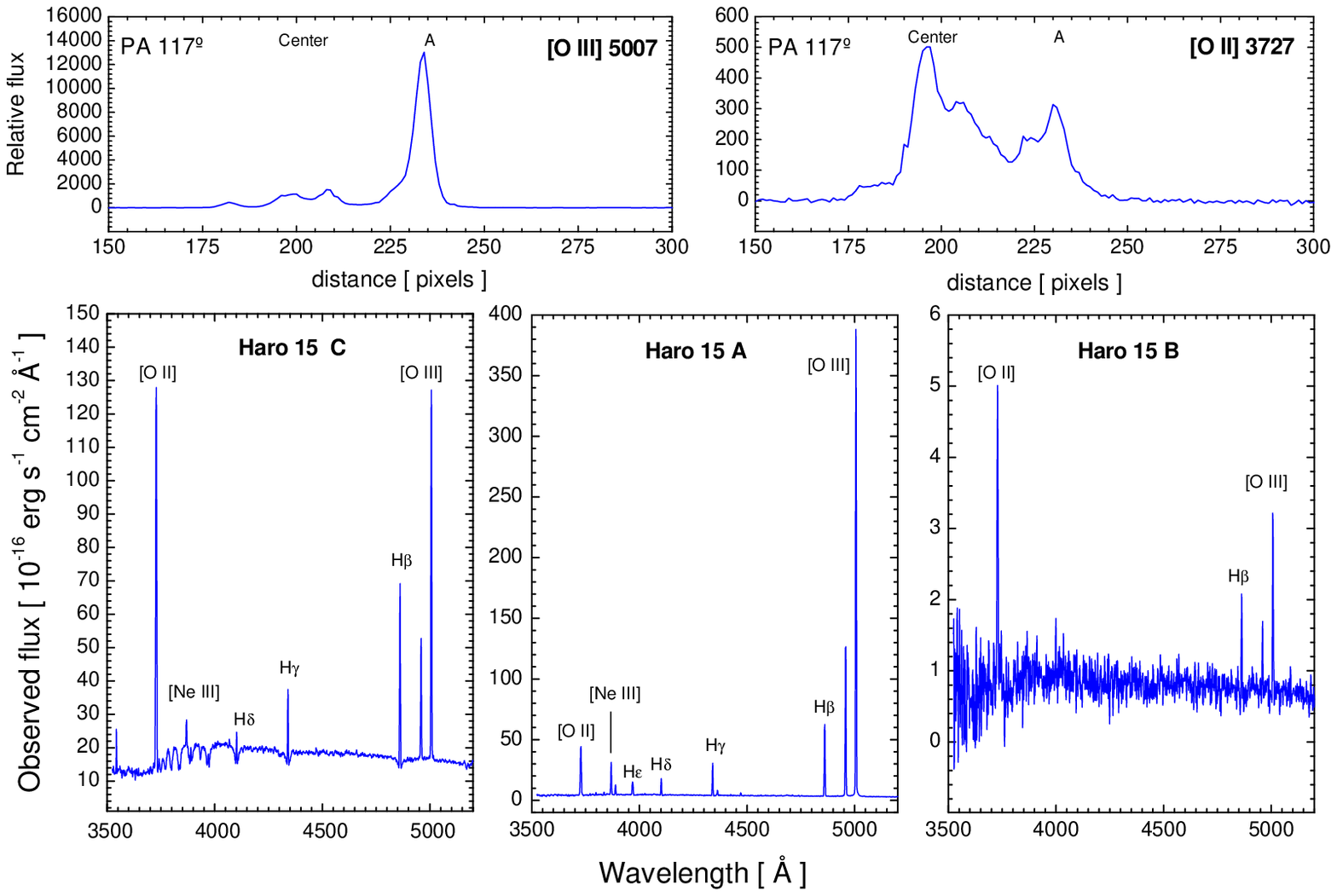}
\caption{\footnotesize{Lower panels: zoom of the spectra of regions C (center), A  and B of Haro 15 (bottom). Fluxes are not corrected for reddening. 
Upper panels: spatial distribution of the relative flux of [\ion{O}{ii}] $\lambda$3727 (upper, 
right) and [\ion{O}{iii}] $\lambda$5007 (upper, left) emission lines along the slit of PA 117$^{\circ}$.}}
\label{haro15espectros2}
\end{figure}

Table~\ref{haro15abun} compiles all the chemical abundances computed for each region of Haro 15. We found a significant difference between the oxygen 
abundance at the center of Haro~15, \abox=8.37, and that found in the ESE region (knot A), \abox=8.10, which is larger than the uncertainties. This 
fact suggests that, although their projected distance is very small (5.5 kpc), both objects seem to have experienced a different chemical evolution. 
The values for the oxygen abundance in knots B and D --at larger distances from the center of the galaxy-- are slightly lower than that at the center 
of Haro 15, perhaps indicating a possible radial abundance gradient in the disk of the galaxy. However, this may be no applicable to knot B because 
it could be an independent object due to its decoupled kinematics (see below). Regions 
A, B and D show a much lower N/O ratio than the center of the galaxy. 


Figure~\ref{haro15espectros2} shows the relative flux of the 
[\ion{O}{ii}] $\lambda$3727 and [\ion{O}{iii}] $\lambda$5007 emission lines along the slit. As we can see, knot A has 
a much higher ionization degree compared to the other regions. This should be a consequence of the extreme youth of the 
star formation in this zone an its lower metallicity. Both things suggest the different nature of this object. 

\subsubsection{Kinematics of the ionized gas}

\begin{figure}[t!]
\centering
\includegraphics[angle=270,width=\linewidth]{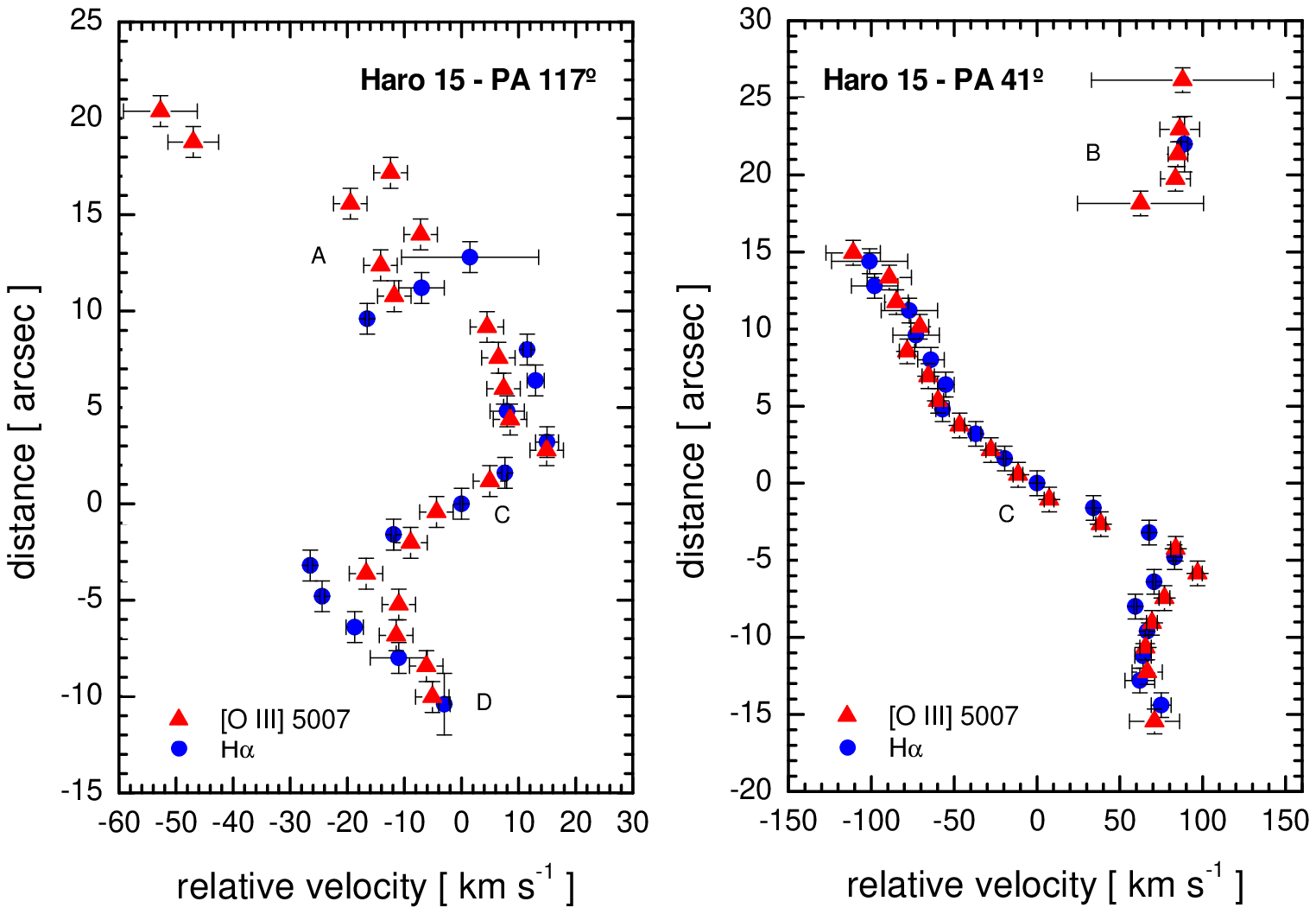}
\caption{\footnotesize{Position-velocity diagrams for the two slit positions observed in Haro 15. Both the \Ha\ and the [\ion{O}{iii}] $\lambda$5007 
profiles have been analyzed. East is on top in both diagrams. See Figure~3 in Paper~I for the identification of the regions.}}
\label{haro15curvas}
\end{figure}

Figure~\ref{haro15curvas} shows the position-velocity diagrams for the two slit positions observed in Haro~15. We analyzed both the \Ha\ and the 
[\ion{O}{iii}] $\lambda$5007 profiles, extracting 4 pixels bins (1.6 arcsec) and considering the velocity of the center of the galaxy as reference. 
Both emission lines give almost identical results. The diagram of PA 41$^{\circ}$ shows an apparent rotation pattern, although some divergences 
are found at the SW. Knot B is clearly decoupled from the rotation of the disk, suggesting that it is an external object. The interaction between 
knot B and Haro 15 could be the responsible of the distortion observed at the SW of the diagram of PA 41$^{\circ}$ because the object and this zone 
of the galaxy disk show similar radial velocities. Another possibility is that knot B is a \TDG\ but, in this case, 
the material that formed B should come from the external parts of the disk of Haro 15 because its chemical abundances are 
more similar to those of knot D than to those of the central region. 

The diagram with PA~117$^{\circ}$ shows a clear sinusoidal pattern with differences of around 40 km s$^{-1}$. This feature is common in 
processes involving galaxy interaction or merging. Although it is not completely clear, region A seems to be kinetically coupled with the rotation of 
the galaxy. This fact and the different chemical composition indicate that knot A is probably an external object suffering a merging process with 
Haro~15.     

Assuming that the position-velocity diagram of PA~41$^{\circ}$ from $-$5 to 15 arcsec from the center is consequence of circular rotation, we can 
estimate the Keplerian mass of Haro~15. Considering a radial velocity of $\sim$80~km~s$^{-1}$ within a distance of $\sim$13 arcsec (5.46~kpc) 
and assuming 
$i$=55$^{\circ}$ from our optical images
[\citet{GG81} found $i$=57$^{\circ}$], we derive $M_{kep}$=1.21$\times$10$^{10}$ \Mo\ and \Mkepl$=0.35$. 
The neutral and dynamical masses of Haro~15 are 
\MHi=5.54$\times$10$^9$ \Mo\ and \Mdyn=3.65$\times$10$^{10}$ \Mo\ \citep{GG81}. Thus, our Keplerian mass is $\sim$33\% of the the 
total mass. The  $M_{\rm H\, I}/L_{\odot}$  and the $M_{dust}/L_{\odot}$ ratios (0.16 and 0.6$\times$10$^{-4}$, respectively) suggest that Haro~15 
is a Sb or Sc spiral \citep{Bettoni03}. The gas depletion timescale is 2.3~Gyr, showing that the system still possesses a huge amount of material 
available to form new stars.


\subsection{Mkn 1199}

\citet{IT98} observed \object{Mkn~1199} but they did not detect the [\ion{O}{iii}] $\lambda$4363 emission line. However, 
they reported the blue and red \WRBUMP s and nebular \ion{He}{ii} $\lambda$4686 emission. Hence Mkn 1199 was classified as a WR galaxy 
\citep{SCP99}. \citet*{GIT00} revisited its WR properties indicating the existence of WNL, WNE and WCE stellar populations.

\begin{figure}[t!]
\centering
\includegraphics[angle=270,width=\linewidth]{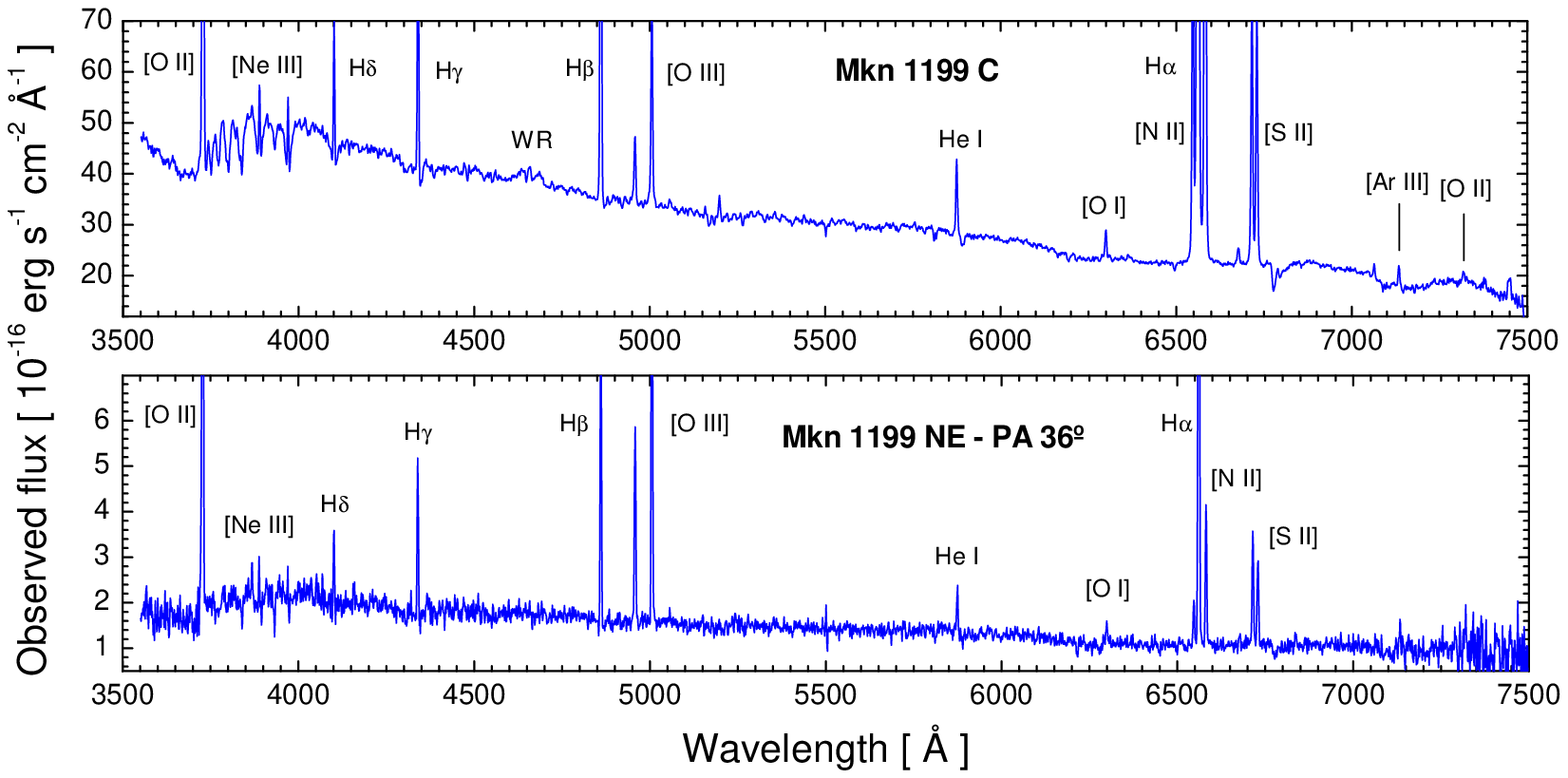}
\caption{\footnotesize{IDS INT spectra for the center of Mkn 1199 and the dwarf companion galaxy at its NE. Fluxes are not corrected for reddening. 
The most important emission lines have been labeled. See Figure~5 in Paper~I for the identification of the regions.}}
\label{mkn1199espectros}
\end{figure}

We obtained intermediate-resolution long slit spectroscopy of Mkn~1199 using two different slit positions (see Figure~5 of Paper~I). We analyzed five 
different regions: the center of Mkn 1199 (C), the companion dwarf object (NE) and knots A, B and D, all extracted with a slit with PA 36$^{\circ}$ 
but region A, for which the slit with PA 53$^{\circ}$ was used. The spectra of the two brightest regions (C and NE) are shown in 
Figure~\ref{mkn1199espectros}, while the emission line intensities and other important properties of the spectra are compiled in 
Table~\ref{mkn1199lineas}. The spectrum of the center of Mkn~1199 shows stellar absorptions in both the \HI\ Balmer and the \ion{He}{i} lines, as it 
was previously noticed by \citet{IT98}. The blue \WRBUMP\ and a probable \HeII\ $\lambda$4686 are detected in C (Figure~\ref{wrfig1}). Our spectrum 
also shows a tentative detection of the red \WRBUMP\ in this region (Figure~\ref{wrfig2}). The spectrum of the dwarf companion galaxy at the NE has a 
lower S/N ratio than the obtained for Mkn 1199, but all relevant emission lines are clearly identified. Stellar absorptions are also found in this 
region but are fainter than those observed in C. The spectra of the rest of the knots only show the brightest emission lines.

\subsubsection{Physical conditions of the ionized gas}

\begin{figure}[t!]
\centering
\includegraphics[angle=270,width=\linewidth]{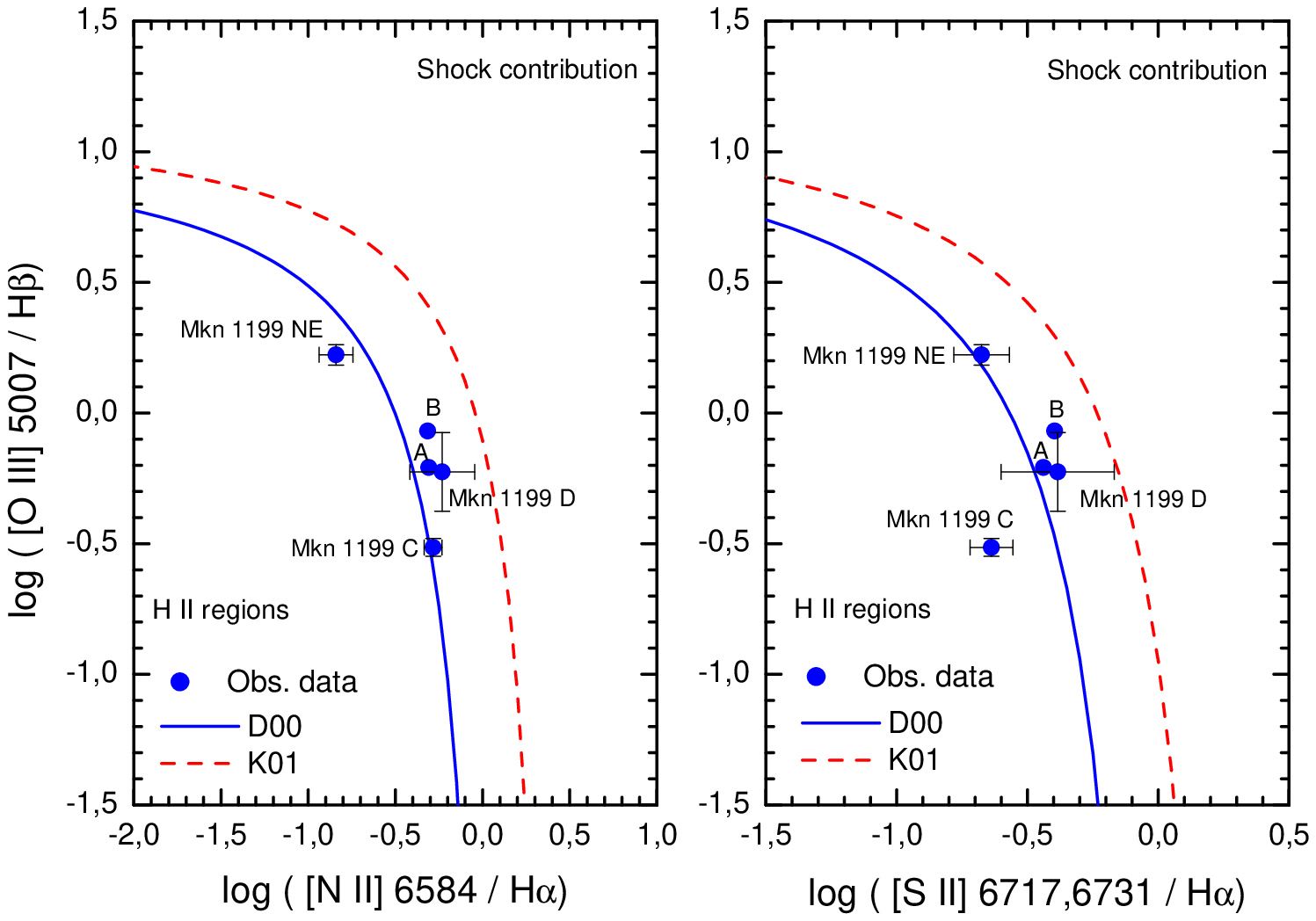}
\caption{\footnotesize{Comparison of some line intensity ratios in several regions of Mkn 1199 with the diagnostic diagrams proposed by Dopita et al. 
(2000) and Kewley et al. (2001).}}
\label{mkn1199diag}
\end{figure}

The [\ion{O}{iii}] $\lambda$4363 line was not detected in any region, hence we computed the electron temperatures and chemical abundances using 
empirical calibrations (see Table~\ref{mkn1199abun}). The electron temperatures found in C are rather low. In this zone, we detected [\ion{N}{ii}] 
$\lambda$5755, which gives a $T_e$[\ion{N}{ii}]$\sim$6740~K, a value very similar to that estimated using the empirical method. We also detect the 
[\ion{O}{ii}] $\lambda\lambda$7319,7330 doublet in the spectrum of C; we obtain $T_e$(\ion{O}{ii})$\sim$6910 K. Hence, we consider that the electron 
temperature values obtained using the empirical calibrations are entirely reliable for this object, assuming $T_e(low)$=6800~K. 
The high ionization electron temperature was computed using Garnett's relation, $T_e(low)$=5400~K. Except for region C, the electron densities 
derived from the [\ion{S}{ii}] $\lambda\lambda$6717,6730 doublet were in the low-density limit, $n_e\sim$100 cm$^{-3}$. The values for the reddening 
coeficient are relatively high for C and B, suggesting the presence of an important amount of dust in those regions. The comparison of some line 
intensity ratios with the diagnostic diagrams proposed by \citet{Do00} and \citet{KD01} indicates that all regions can be classified as \HII regions 
(see Figure~\ref{mkn1199diag}).  

\subsubsection{Chemical abundances}

Table~\ref{mkn1199abun} compiles all the chemical abundances computed for the knots analyzed in Mkn~1199. The oxygen abundance found at the center of 
the system is very high, \abox=8.75$\pm$ (indeed, it highest metallicity region found in this work). However, the oxygen 
abundance of the dwarf companion galaxy at the NE is almost 0.3 dex lower, \abox=8.46. The N/O ratios derived for 
them are also very different. Hence, this result reinforces the hypothesis that they are independent galaxies in the first stages of a minor merging. 

It is worthy to notice that, although the line intensities observed by \citet{IT98} in Mkn~1199 are quite similar to those compiled in 
Table~\ref{mkn1199lineas}, the oxygen abundance derived by these authors is \abox=8.19$\pm$0.18, extremely low when compared with our values. 
However, \citet{GIT00} using the same data that \citet{IT98} and the empirical calibration based on  the [\ion{N}{ii}]$\lambda$6583/\Ha\ ratio 
proposed by \citet{vZee98} obtain \abox=9.13. We consider that \abox=8.75 is a more appropriate value for the oxygen abundance at the center of 
Mkn~1199, that also agrees with our tentative estimations of electron temperatures.

Knots A, B and D show oxygen abundances slightly lower than C, between 8.6 and 8.7, but not as low as that reported for the companion galaxy. 
Interestingly, A, B and D show a N/O ratio very similar to that observed at the center of the galaxy, but much higher than in the dwarf companion. 
This confirms the different chemical evolution of the disk of Mkn~1199 and the companion galaxy. We consider that knots A, B and D are giant \HII 
regions located in the spiral arms of Mkn~1199 and that this galaxy could have a slight radial metallicity gradient along its disk. Finally, we think 
that the triggering mechanism of the intense star-formation activity found in both, Mkn~1199 and its dwarf companion, is a likely consequence of the 
strong interaction they are experiencing. 

\subsubsection{Kinematics of the ionized gas}

\begin{figure}[t!]
\centering
\includegraphics[angle=270,width=\linewidth]{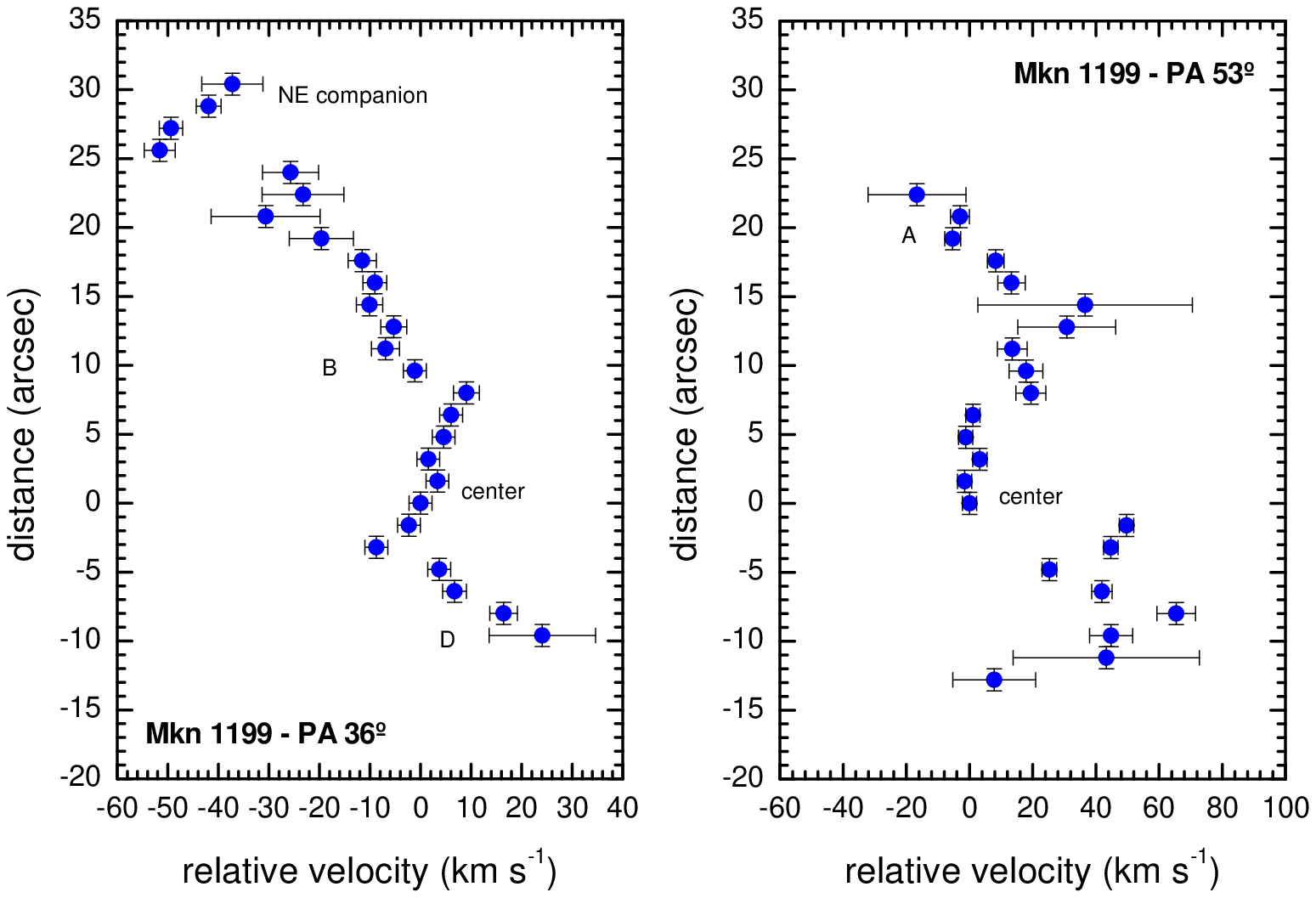}
\caption{\footnotesize{Position-velocity diagrams for the two slit positions observed in Mkn 1199 using the \Ha\ profiles. NE is up in both diagrams. 
See Figure~5 in Paper~I for the identification of the regions.}}
\label{mkn1199curvas}
\end{figure}

Figure~\ref{mkn1199curvas} shows the position-velocity diagrams for the two slit positions in Mkn~1199. They were obtained extracting 3 pixels bins 
(1.2 arcsec) across the \Ha\ profile and considering the center of Mkn~1199 as reference. The diagram with PA 36$^{\circ}$, that crosses the center 
of Mkn~1199 and the companion galaxy, may be explained by the rotation of the disk of Mkn~1199, because it shows a velocity gradient between 30 \kms\ 
at SW (knot D) and $-$30 \kms\ at NE (around knot B). However, a sinusoidal pattern is also seen in the brightest region of Mkn~1199, within the 
central 10$\arcsec$. This feature may be consequence of an entity kinetically decoupled from the disk, such a bar or the bulge of the galaxy, that 
seems to be counter-rotating, although it may be also an interaction feature. On the other hand, although there are only four points in the diagram 
in that zone, the dwarf galaxy at the NE seems to be rotating. This object, that has an elliptical shape in the optical images, may be observed 
edge-on, and hence it would explain both its kinematics and morphology. The diagram with PA 53$^{\circ}$ also seems to show a velocity gradient from 
the SW to the NE (where knot A is located) but this gradient is broken at the central 10$\arcsec$. The maximum velocity difference is around 80 \kms.

It is difficult to estimate the Keplerian mass of Mkn~1199 because it is seen almost face-on. However, considering that the velocity difference 
between its center and the external regions within a radius of 10$\arcsec$ (2.6 kpc) is around 30~\kms\ and adopting an inclination angle of 
15$^{\circ}$, we derive $M_{kep}\sim$8.2$\times$10$^9$ \Mo. 
For the companion dwarf object, assuming that it is edge-on  ($i$=90$^{\circ}$) and considering a velocity difference of 10 km s$^{-1}$ within a 
radius of 2.5$\arcsec$ (1.3 kpc), we compute \mbox{$M_{kep}\sim$2.9$\times$10$^7$ \Mo.} 

Using the \HI\ data given by \citet{DC04}, we compute  $M_{\rm H\, I}$=1.22$\times$10$^9$ \Mo\ and $\Delta v_{\rm H\, I}$=85 km s$^{-1}$. 
Assuming a radius of 25$\arcsec$ (6.55 kpc) and $i$=15$^{\circ}$, the dynamical mass of Mkn~1199 is $M_{dyn}=$1.7$\times$10$^{11}$ \Mo. 
The warm dust mass using the \FIR\ data is  $M_{dust}$=3.1$\times$10$^6$ \Mo. Following the classification provided by \citet*{Bettoni03}, the 
$M_{\rm H\, I}$/\Lo=0.042 and $M_{dust}$/\Lo=1.1$\times$10$^{-4}$ ratios are not compatible: while the the first corresponds to the typical values 
for S0 galaxies, the second ratio indicates that Mkn~1199 should be an Sc or Sd, more similar to the actual morphological classification of Sb. 
Furthermore, less than 1\% of the total mass of the system is neutral hydrogen and the gas depletion timescale is very low (0.4~Gyr). All these 
facts suggest that a substantial fraction of the neutral \HI\ gas has been expelled to the intergalactic medium, perhaps as a consequence of the 
interaction with the dwarf companion galaxy. An \HI\ map obtained using a radio-interferometer would be needed to confirm such hypothesis.


\subsection{Mkn 5}

\citet{C91} included \object{Mkn~5} in his catalogue of WR galaxies because of the detection of the nebular \ion{He}{ii} $\lambda$4686 line 
by \citet{French80}. However, \citet{IT98} only observed the blue \WRBUMP, without any trace of the nebular \ion{He}{ii} 
emission \citep*{SCP99}. \citet{GIT00} detected \ion{N}{iii} $\lambda$4640, implying the presence of WNL stars within the starburst.

\begin{figure}[t!]
\centering
\includegraphics[angle=270,width=\linewidth]{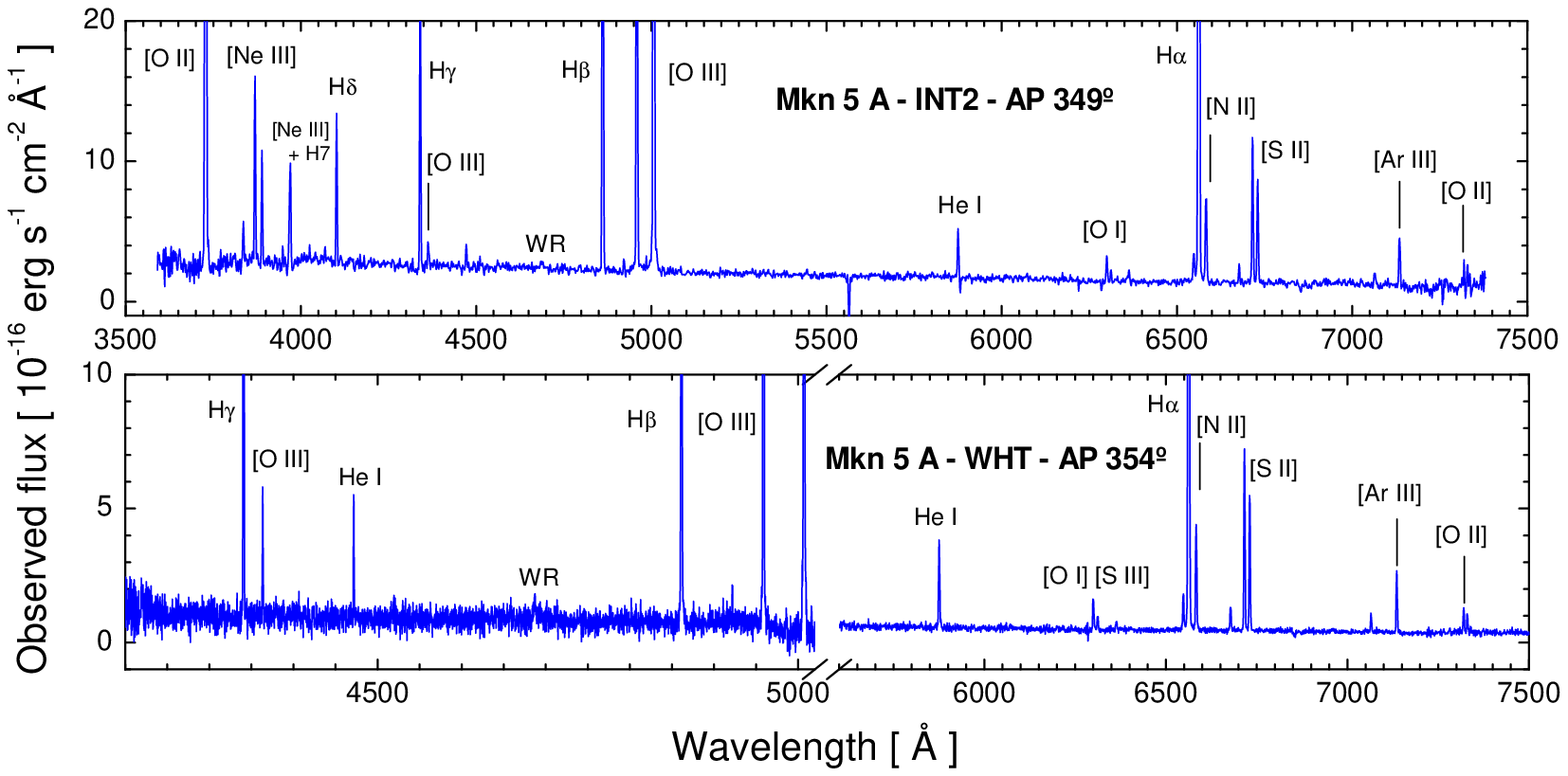}
\caption{\footnotesize{Spectra for the region A of Mkn~5 obtained using IDS at the INT (PA $349^{\circ}$) and ISIS at the WHT (PA $354^{\circ}$). 
Fluxes are not corrected for reddening. The most important emission lines have been labeled. See Figure~7 in Paper~I for the identification of the 
regions.}}
\label{mkn5espectros}
\end{figure}

We used three slit positions to get the spectroscopic data of Mkn~5 (see Figure~7 of Paper~I), two of them using the \INTe\ and an additional 
position using \WHT. All have a very similar PA, 0$^{\circ}$ (INT-1), 354$^{\circ}$
(WHT) and 349$^{\circ}$
(INT-2). 
All slit positions cover region A but only two cross knot B. We analyzed the three spectra extracted for region A independently to check the quality 
of the results. Figure~\ref{mkn5espectros} shows the spectra of the region~A obtained with the slit positions with PA $349^{\circ}$ and PA 
$354^{\circ}$. For region~B we only analyzed the spectrum extracted using the slit position with PA~0$^{\circ}$ (INT-1) because of its higher 
signal-to-noise. Table~\ref{mkn5lineas} compiles all the emission line fluxes and other characteristics of each spectrum. Our spectra confirm the 
presence of a nebular \ion{He}{ii} $\lambda$4686 line on top of a broad emission line in region A (Figure~\ref{wrfig1}). Although the WHT spectrum 
has high S/N and spectral resolution, we do not detect the red \WRBUMP\ in this region (Figure~\ref{wrfig2}).
All spectra are dominated by nebular emission but some absorptions in the \HI\ Balmer lines are also detected; these are more evident in knot B.

\subsubsection{Physical conditions of the ionized gas}

The three spectra obtained for region A show the [\ion{O}{iii}] $\lambda$4363 emission line and hence we computed the electron temperature using the 
direct method. As it is seen in Table~\ref{mkn5abun} all \Te\ values are in agreement within the errors, being the average value 
$T_e$[\ion{O}{iii}]$\sim$12500~K. The low ionization temperature was derived using Garnett's relation, $T_e$[\ion{O}{ii}]$\sim$11700 K. The spectrum 
INT-2 shows a tentative detection of [\ion{O}{ii}] $\lambda\lambda$7318,7330, that gives 
$T_e$[\ion{O}{ii}]$\sim$11950~K, in agreement with the electron temperature derived using Garnett's relation. We used \citet{P01b} empirical 
calibration to derive \Te\ in knot B, but this determination is very much uncertain and perhaps overestimated due to the faintness of the spectrum. 
Electron densities are always below the low-density limit (100 cm$^{-3}$) except for knot B (although it also has a large error).

The values of the reddening coefficient derived for region A are somewhat different in the different spectra. Perhaps, this apparent inconsistency is 
a consequence of an irregular distribution of dust within Mkn~5, as we suggested in our analysis of the optical/NIR colors (see \S~3.5.1 of Paper~I). 
For knot B, we assumed  $W_{abs}$=1.5 \AA\ and the \CHb\ computed via the \Ha/\Hb\ ratio. The diagnostic diagrams for knots A and B agree with the 
loci of typical \HII regions. 

\subsubsection{Chemical abundances}

The WHT spectrum does not cover the [\ion{O}{ii}] $\lambda\lambda$3726,29 doublet, hence we used [\ion{O}{ii}] $\lambda\lambda$7318,7330 to compute 
the O$^+$/H$^+$ ratio. As we see in Table~\ref{mkn5lineas}, the agreement between the emission line ratios for all the three spectra extracted for 
region A is very good. Table~\ref{mkn5abun} compiles the chemical abundances obtained for Mkn~5; for region A all values are quite similar and in 
agreement with previous results found in the literature. Averaging all data and minimizing errors, we derive for A the following chemical abundances:
12+log(O/H)=8.07$\pm$0.04, log(N/O)=$-$1.38$\pm$0.07, log (S/O)=$-$1.62$\pm$0.11, log(Ne/O)=$-$0.80$\pm$0.13 and log(Ar/O)=$-$2.31$\pm$0.12. 
These values are very similar to the results provided by \citet{IT99}.
On the other hand, the oxygen abundance estimated for knot B, 12+log(O/H)=7.89$\pm$0.17, lower than that derived for the main star-forming region but 
consistent within the errors. If real, this difference may suggests different chemical evolutions between the two regions.

\subsubsection{Kinematics of the ionized gas}

\begin{figure}[t!]
\centering
\includegraphics[angle=270,width=\linewidth]{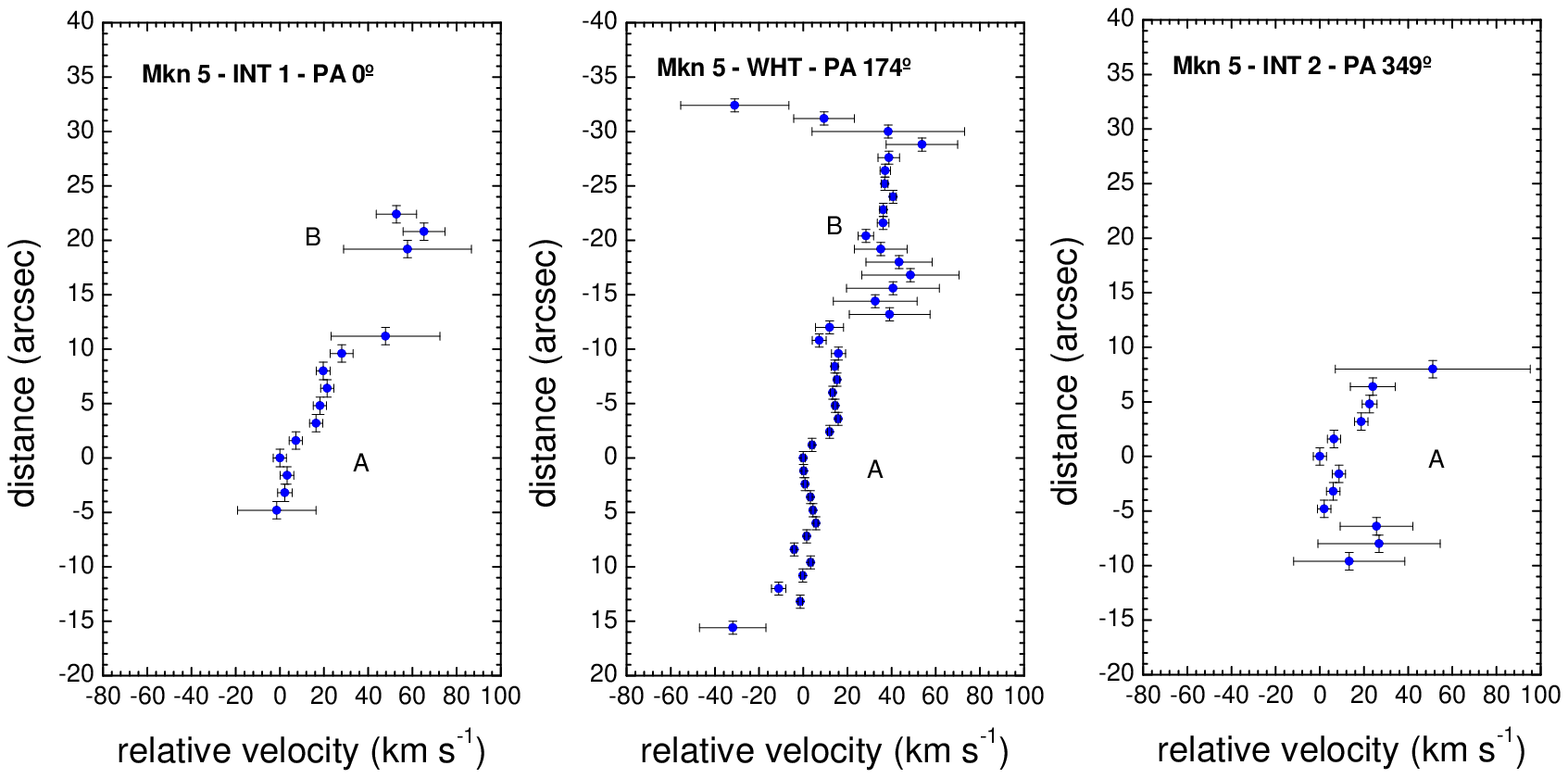}
\caption{\footnotesize{Position-velocity diagrams for the slit positions observed in Mkn 5 using the [\ion{O}{iii}] $\lambda$5007 profile. N is up in 
all diagrams. See Figure~7 in Paper~I for the identification of the regions.}}
\label{mkn5curvas}
\end{figure}


Figure~\ref{mkn5curvas} shows the position-velocity diagrams obtained for the three slit positions observed in Mkn~5. For the \WHT\ spectrum we 
analyzed the \Ha\ profile considering 6 pixels bins (1.2 arcsec), while we used the [\ion{O}{iii}] $\lambda$5007 profile (brighter than the \Ha\ 
profile) extracting 4 pixels bins (1.6 arcsec) from the \INTe\ spectra. As we see, the agreement between the three diagrams is very good. The best 
diagram is that provided by the analysis of the \WHT\ spectrum, that shows a velocity gradient of around $\sim$50~\kms. The velocity of knot B with 
respect to that found in region A is $\sim$40~\kms. Although the uncertainties are important, we detect a slight reverse in the velocity of region 
A, with an amplitude of $\sim$20 \kms\ (it is more evident in the \WHT\ diagram), that seems to show a sinusoidal pattern in that area.  

Assuming that the global velocity gradient is mainly a consequence of the rotation of the galaxy, we may estimate the Keplerian mass of the system. 
We found $M_{Kep}\sim~2.1\times~10^9$~\Mo\ assuming $i$=90$^{\circ}$, $\Delta v\sim$27~\kms\ and $r\sim21\arcsec$ (1.22 kpc). Using \HI\ data 
\citep*{Paturel03}, 
we derive $M_{\rm H\, I}$ = (7.2$\pm$0.9)$\times$10$^7$ \Mo\ and $M_{Dyn}\sim$3.6$\times$10$^9$~\Mo. Although both \Mkep\ and \Mdyn\ are similar, 
notice that they are low limits because we are assuming that Mkn~5 is an edge-on galaxy.
The mass-to-luminosity ratios are $M_{Kep}/L_{\odot}$=7.98, $M_{Dyn}/L_{\odot}$=13.7 and $M_{\rm H\, I}/L_{\odot}$ =0.27. 
The \HI\ mass is quite low for a dwarf or irregular galaxy, being only 2\% of the total mass. The gas depletion timescale is 
$\sim$1.8~Gyr, high for a starburst galaxy. All these facts suggest that something has happened with the atomic gas of Mkn~5: or it has been consumed 
forming stars at a high rate until some few hundred Myr ago (nowadays it has decreased) or it has been expelled to the intergalactic medium. Indeed, 
\citet{TM81} found indications of \HI\ gas at slightly different radial velocities ($\sim300$ \kms) using single-dish data. 
An interferometric \HI\ map will be crucial to understand the fate of the neutral gas in this blue compact dwarf galaxy. 


\subsection{IRAS 08208+2816}

The first spectroscopic data of \object{IRAS~08208+2816} were obtained by \citet{HGJL99}, who reported the detection of both the nebular and broad 
\ion{He}{ii} $\lambda$4686 emission line. They also detected the red \WRBUMP, \ion{C}{iv} $\lambda$5808, suggesting the presence of both WNL and WCE 
populations in the starburst. \citet*{SCP99} included IRAS~08208+2816 in the latest catalogue of WR galaxies .

\begin{figure}[t!]
\centering
\includegraphics[angle=270,width=\linewidth]{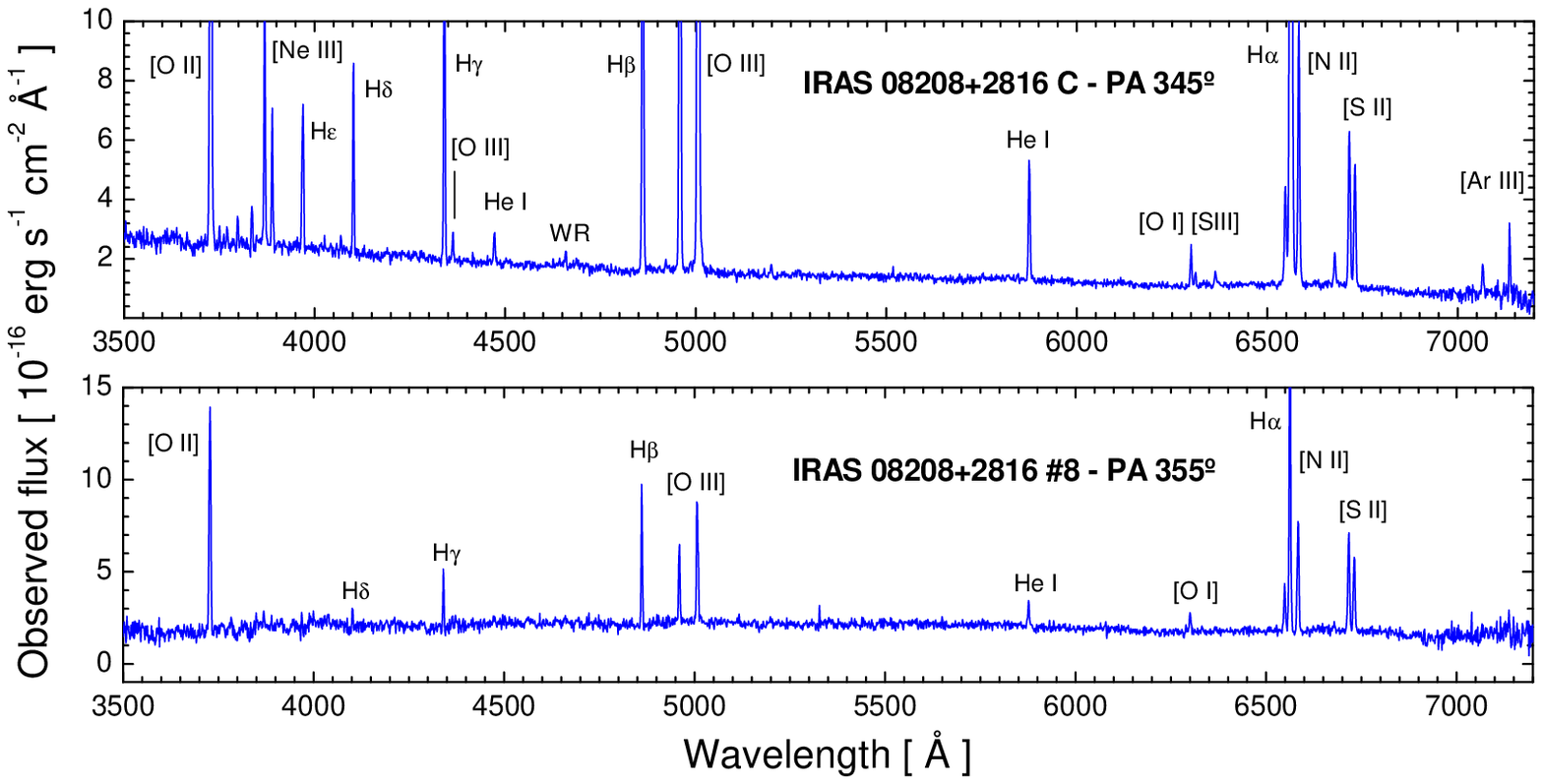}
\caption{\footnotesize{IDS INT spectra for the center of IRAS 0828+2816 and knot \#8. Fluxes are not corrected for reddening. The most important 
emission lines have been labeled. See Figure~9 in Paper~I for the identification of the regions.}}
\label{iras08208espectros}
\end{figure}

We used three slit positions with the IDS spectrograph at the \INTe\ to cover all bright knots within the galaxy (see Figure~9 of Paper~I). 
The slit position with PA 345$^{\circ}$ crosses a bright star and the center of IRAS~0828+2816, the slit position with PA 355$^{\circ}$ covers the 
center and knots \#8, \#10 and \#1 (that is very affected by the light of the bright star)
and the slit position with PA 10$^{\circ}$ crosses \#3, \#4 (very weak) and \#5, although it also may be contaminated by some emission from the 
center and knot \#8. Table~\ref{iras08208lineas} compiles the properties of the five regions that we analyze spectroscopically. The center of the 
galaxy, C, corresponds to the brightest region extracted using the slit position with PA 345$^{\circ}$. This spectrum and that obtained for knot \#8 
are shown in  Figure~\ref{iras08208espectros}. Although all spectra are dominated by nebular emission and do not show any evidence of absorptions in 
the \HI\ Balmer lines, we detect a slight decrease of the continuum level on the blue range of the spectra of the faintest objects . This fact may be 
explained by both an important extinction in these objects and by the possible presence of an evolved underlying stellar population. The broad 
\ion{He}{ii} $\lambda$4686 line is weakly detected in the central region of the galaxy but the nebular \HeII\ $\lambda$4686 line is not identified in 
this spectrum (Figure~\ref{wrfig1}). A faint red \WRBUMP\ at around $\sim$5800 \AA\ seems also to be observed in this region (Figure~\ref{wrfig2}). 
Spectra with higher S/N ratio and spectral resolution are needed to get a proper value of the \WRBUMP\ fluxes. 

\subsubsection{Physical conditions of the ionized gas}

We detect the weak auroral [\ion{O}{iii}] $\lambda$4363 emission line at the center of the galaxy, its flux value was used to estimate the electron 
temperature following the direct method. For the rest of the objects we used empirical calibrations to compute \Te, all results are compiled in 
Table~\ref{iras08208abun}. [\ion{O}{iii}] $\lambda$4363 is barely detected in knots \#3 and \#5, but they were not considered in the analysis because 
of their large errors. All electron temperatures derived using empirical methods, remarking those found in knot \#8, are systematically lower than 
those computed in the central region following the direct method. All objects can be classified as \HII regions  following the typical diagnostic 
diagrams.  

The reddening coefficient was computed using all available \HI\ Balmer lines in each spectrum. We obtained very different values: while the central 
region and knots in the northern tail have a low reddening coefficient, \CHb$\sim$0.12, knots located in the southern tail show a higher value, 
\CHb$\sim$0.43. This fact seems to indicate an inhomogeneous distribution of the dust within the galaxy, being the southern regions dustier than the 
rest of the system. \citet*{HGJL99} also derived low extinction values in the center of IRAS~08208+2816; they explained this considering the 
presence of a galactic wind in the galaxy. But we do not detect such structures in our deep \Ha\ images (see Figure~10 of Paper~I).

\subsubsection{Chemical abundances}

Table~\ref{iras08208abun} compiles all chemical abundances computed for the different knots analyzed in IRAS~08208+2816. The oxygen abundance of the 
central region, derived using the direct method, is \abox=8.33$\pm$0.08, and its nitrogen-to-oxygen ratio is \lno$=-0.89\pm0.11$. This value is 
higher than the N/O ratio expected for a galaxy with an oxygen abundance of \abox$\sim$8.3, which should be \lno$\sim-$1.2. If this effect is real, 
it may be due to nitrogen pollution by the winds of the WR stars, as we confirm to occur in the case of NGC~5253 \citep{LSEGRPR07}. 

For the rest of the objects, the oxygen abundances were calculated using \citet{P01a,P01b} empirical calibrations. Although all estimations are 
slightly higher than the value found in the central region, we notice a significant difference in the case of knot \#8, that has \abox$\sim8.64$ 
(i.e., almost the solar value). Knots \#3 and \#5 show a tentative detection of [\ion{O}{iii}] $\lambda$4363 in their spectra, for which we derive an 
oxygen abundance $\sim$0.12 -- 0.15 dex lower than that estimated using empirical calibrations (see Table~\ref{iras08208abun}). This trend is 
also found in the central region, for which we derive \abox$\sim$8.41 following the empirical calibrations. 
Hence, the values obtained using the Pilyugin method may be somewhat overestimated for this galaxy. 

In any case, the chemical abundance differences seem to be real in knot \#8, first because its oxygen abundance is $\sim$0.3 dex higher than that 
found in the central region, and second because its N/O ratio is also the highest, $\log({\rm N/O})\sim-0.84$, and consistent with the value expected 
for a galaxy with almost solar metallicity.
This result indicates that knot \#8 could be an object more chemically evolved that the others. 
Because of this and its position within the system, knot \#8 may even correspond to the center of an independent galaxy that is in a process of 
merging with another galaxy which nucleus coincides with knot C of IRAS~08208+2816.

\subsubsection{Kinematics of the ionized gas}

\begin{figure}[t!]
\centering
\includegraphics[angle=270,width=\linewidth]{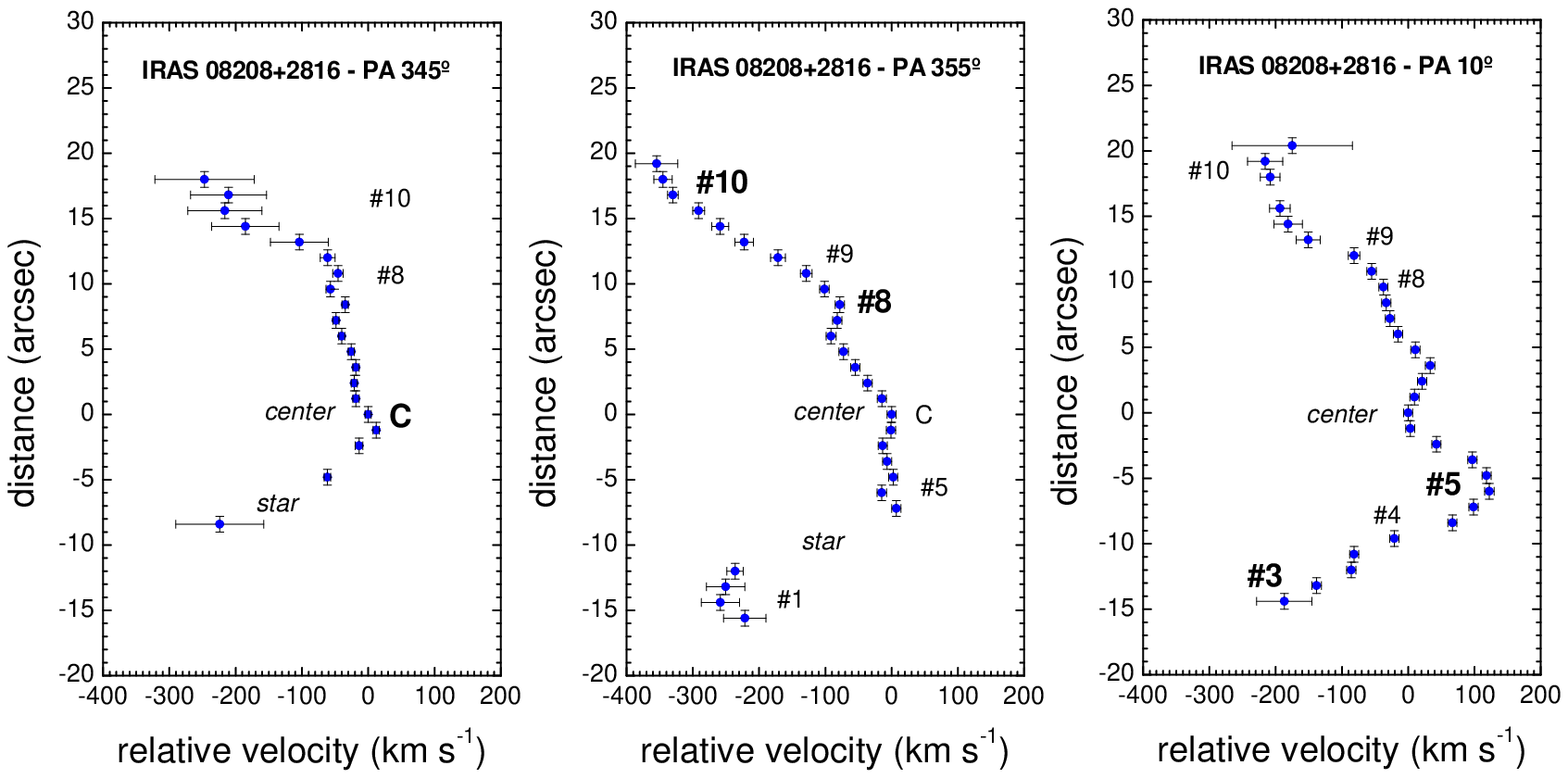}
\caption{\footnotesize{Position-velocity diagrams for the three slit positions observed of IRAS 08208+2816 using the [\ion{O}{iii}] $\lambda$5007 
profile. N is up in all diagrams. We have included the position of the regions (see Figure~9 in Paper~I for their identification), emphasizing those 
analyzed by spectroscopy. Notice that the lacking of data at the southern edge of the southern tail in the diagrams with PA 345$^{\circ}$ and 
355$^{\circ}$ is because of the contamination by a bright star.}}
\label{iras08208curvas}
\end{figure}

Figure~\ref{iras08208curvas} shows the position-velocity diagrams obtained for the three slit positions. The [\ion{O}{iii}] $\lambda$5007 profile 
was analyzed considering 3 pixel bins (1.2$\arcsec$) and taking the center of IRAS~08208+2816 as reference. As it was clearly seen in the 
bidimensional spectra, this object possesses very interesting kinematics, remarking a probable tidal stream in the northern tail. In these areas we 
observe a velocity gradient larger than 300~\kms\ within 12$\arcsec$ \mbox{(11~kpc)} in the slit position crossing the northern region of the system 
(PA=355$^{\circ}$). This velocity difference of 300~\kms\ is the same that  \citet{Perryman82} reported between these objects. As we commented 
before, the spectra crossing the southern tail is affected by the contamination from the nearby bright star, but the slit position with 
PA=10$^{\circ}$ is free of such contamination allowing the kinematic analysis of the southern 
zone. Again, we found an important velocity gradient towards negative values, that cannot be explained by a rotating disk. Furthermore, this diagram 
shows 
an evident sinusoidal pattern with an amplitude larger than 50~\kms\ at the center of the galaxy. This is an 
additional evidence that we are observing a process of merging. In general, the agreement between the three diagrams 
is very good, for example, the velocity found for knot \#1 using the slit 
position with PA~355$^{\circ}$ is $\sim-$250~\kms, that corresponds quite well with the velocity observed at the end of the southern tail (knot 
\#3, with $\sim-$200 \kms) using the slit position with PA~10$^{\circ}$.

Although the kinematics of the system is not supported by rotation, we have computed a tentative estimation of the Keplerian mass of the 
system. Assuming $i$=90$^{\circ}$, $\Delta v\sim$30 \kms\ (using the diagram with PA 345$^{\circ}$ that seems to be less affected by the tidal tails) 
and a radius of $\sim$20$\arcsec$ (18.4 kpc), we derive  \Mkep$\sim$3.9$\times$10$^9$ \Mo. The mass-to-luminosity ratio is quite low, 
\Mkepl$\sim$0.08, hence we are very probably underestimating the mass of the system. There are no \HI\ data available for this galaxy in the 
literature, but it should be really interesting to compare the kinematics of the neutral gas with that found here for the ionized gas. 
The warm dust mass is high, \Mdust=8.84$\times$10$^6$~\Mo, giving \Mdustl$\sim$1.73$\times$10$^{-4}$.

\subsection{IRAS 08339+6517}

Our complete analysis of the physical properties, chemical abundances and kinematics of the ionized gas in \object{IRAS~08339+6517} and its companion 
dwarf 
galaxy was presented in \citet{LSEGR06}. We reported weak spectral features that could be attributed to the blue \WRBUMP\  at the center of the 
galaxy. The kinematics of the ionized gas showed an interaction pattern that indicates that the \HI\ tidal tail detected by \citet{CSK04} in 
the direction of the dwarf companion galaxy has been mainly formed from material stripped from the main galaxy. A star-forming region in the 
outskirts of the galactic disk may be a \TDG\ candidate.


\subsection{POX 4}

The first indications of WR features in \object{POX~4} were noticed by \citet{KJ85} and  \citet{CTM86}\footnote{These authors named POX~4 as 
C~1148-203.}, 
because both detected the broad \ion{He}{ii} $\lambda$4686 emission line. Therefore, \citet{C91} included POX~4 in his catalogue of WR galaxies. 
\citet{Masegosa91} also suggested the presence of WR stars in one or two regions of the galaxy\footnote{These authors named POX~4 as C~1148-2020 in 
their Table~1 and as Tol~1148-202 in their Table~2. Following the \NED, the appropriate name of this galaxy is IRAS~11485-2018 = POX~4.}. 
\citet{VC92} confirmed the presence of a high number of O and WN stars in the brightest region of POX~4 and detected the \ion{He}{ii} $\lambda$4686 
emission line in other knot \citep*{SCP99}. 
We do detect the nebular \HeII\ $\lambda$4686 line with a very good S/N ratio in the brightest knot of POX~4. Both the broad \ion{He}{ii} 
$\lambda$4686 and \ion{C}{iv} $\lambda$5808 lines are  clearly identified in this region too (Figures~\ref{wrfig1} and \ref{wrfig2}).

\begin{figure}[t!]
\centering
\includegraphics[angle=270,width=\linewidth]{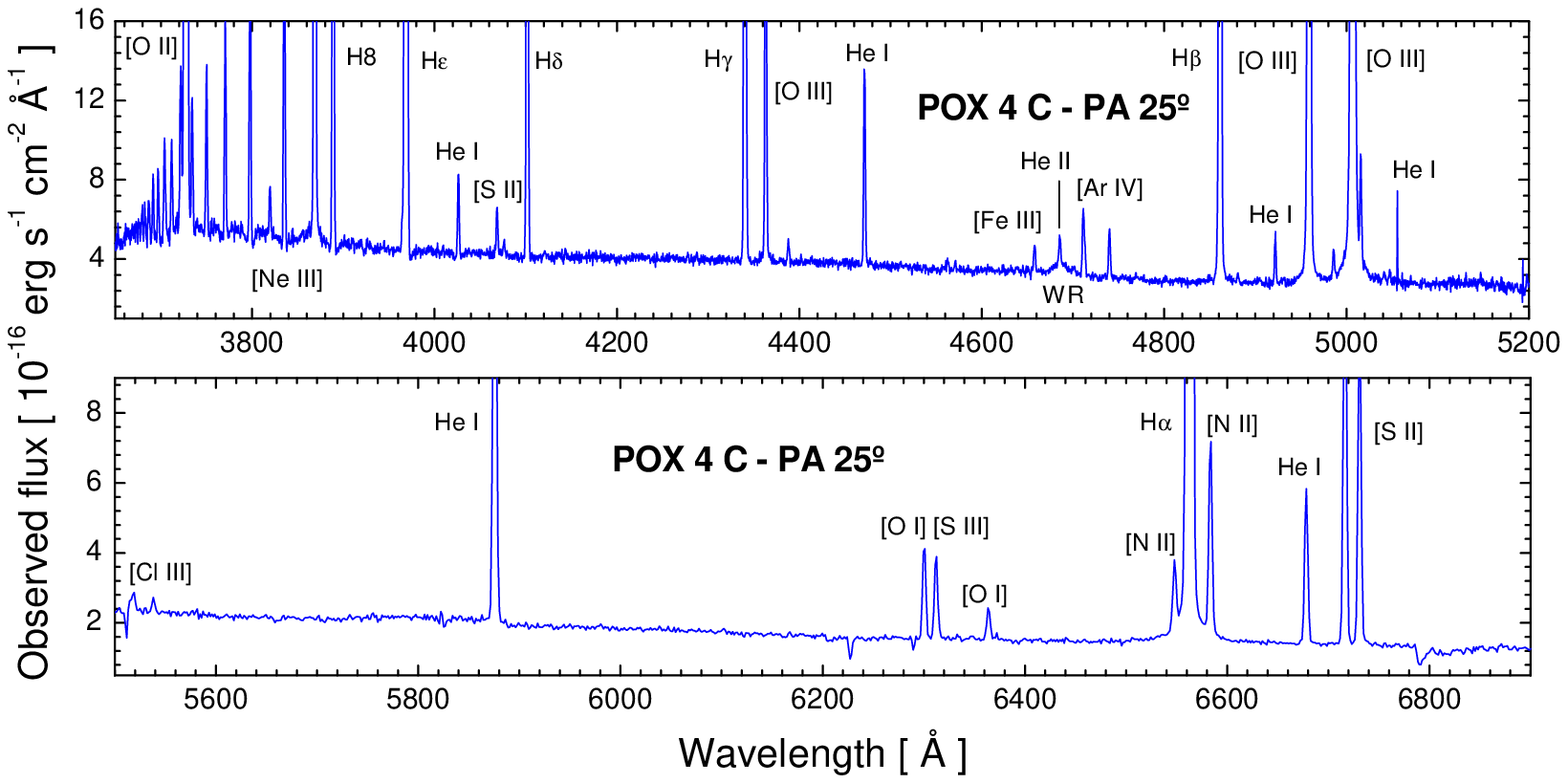}
\caption{\footnotesize{ISIS \WHT\ spectrum for the center of POX~4. Fluxes are not corrected for reddening. The most important emission lines have 
been labeled. See Figure~11 in Paper~I for the identification of the regions.}}
\label{pox4espectros}
\end{figure}

We used ISIS at the \WHT\ to obtain the spectroscopic data of POX~4 and its dwarf companion galaxy (knot \#18 following Figure~11 in Paper~I). The 
position angle of the slit was set to 25$^{\circ}$. The spectrum of the center of POX~4 (Figure~\ref{pox4espectros}) is dominated by intense emission 
lines and does not show any evidence of stellar absorption in the \HI\ or \HeI\ lines. As it was previously noticed by \citet{ME97}, broad 
low-intensity assymetric wings are detected in the profiles of the brightest emission lines (\Ha\ and [\ion{O}{iii}]).  The clear detection of 
the nebular \ion{He}{ii} $\lambda$4686 emission on the top of a broad feature indicates the presence of WR stars at the center of 
the galaxy. However, we do not see the red \WRBUMP\ despite the high S/N and spectral resolution of our spectrum.

The spectrum of the dwarf companion object has a rather low S/N ratio and only the brightest emission lines are detected. However, it shows clear 
absorption features in the \HI\ Balmer lines, indicating the presence of evolved stellar populations underlying the starburst. Table~\ref{pox4lineas} 
compiles all emission lines detected in the center of POX~4 and in the dwarf companion object, as well as other important properties of their 
spectra.

\subsubsection{Physical conditions of the ionized gas}

The physical conditions and chemical abundances of the ionized gas in POX~4 are compiled in Table~\ref{pox4abun}. The 
electron temperature calculated for the center of POX~4 using [\ion{O}{iii}] $\lambda$4363 is $T_e$(\ion{O}{iii}) = 14000$\pm$600 K, suggesting that 
it is a low metallicity object. Although [\ion{N}{ii}] $\lambda$5755 is barely detected, 
we preferred to use Garnett's relation to determine $T_e$(\ion{O}{ii}) from $T_e$(\ion{O}{iii}). 
The electron density was computed using the [\ion{S}{ii}] $\lambda$6717/$\lambda$6731 ratio, yielding \Ne$\sim$250~cm$^{-3}$. Notice that the 
estimation of \Ne\ using the [\ion{Ar}{iv}] $\lambda$4711/$\lambda$4740 ratio gives a very similar value, $\sim$270~cm$^{-3}$, although it has a 
higher uncertainty. 
For the dwarf companion object we used empirical calibrations to estimate \Te, being its electron density in the low-density limit.  The comparison 
of the observed  [\ion{O}{iii}]$\lambda$5007/\Hb\ and [\ion{N}{ii}]$\lambda$6584/\Ha\ ratios with the diagnostic diagrams 
let to classify all regions as starbursts.

The reddening coefficient at the center of POX~4 was determined using 7 \HI\ Balmer lines, obtaining \CHb=0.08$\pm$0.01. 
A similar low value of the reddening has been found for the companion object.

\subsubsection{Chemical abundances}

The oxygen abundance derived for the center of POX~4, \abox=8.03$\pm$0.04, agrees with that found in the literature 
(e.g. Kobulnicky \& Skillmann 1996 reported 7.97$\pm$0.02). The nebular \ion{He}{ii} $\lambda$4686 emission line is clearly detected and, therefore, 
some O$^{+3}$ contribution is expected in the nebular gas, however this contribution is found to be marginal, $\sim$0.01--0.02 dex. The oxygen 
abundance derived for the companion object using empirical calibrations, \abox=8.03$\pm$0.14, is the same than that found in POX~4. The values of the 
N/O ratio are also similar, \lno$=-1.54\pm0.06$ and $-1.60$. Despite the uncertainties, the resemblance of the chemical abundances may suggest that 
the dwarf companion is not an independent object, as \citet{ME99} concluded, but a \TDG\ candidate.

\subsubsection{Kinematics of the ionized gas}

\begin{figure}[t!]
\centering
\includegraphics[angle=270,width=\linewidth]{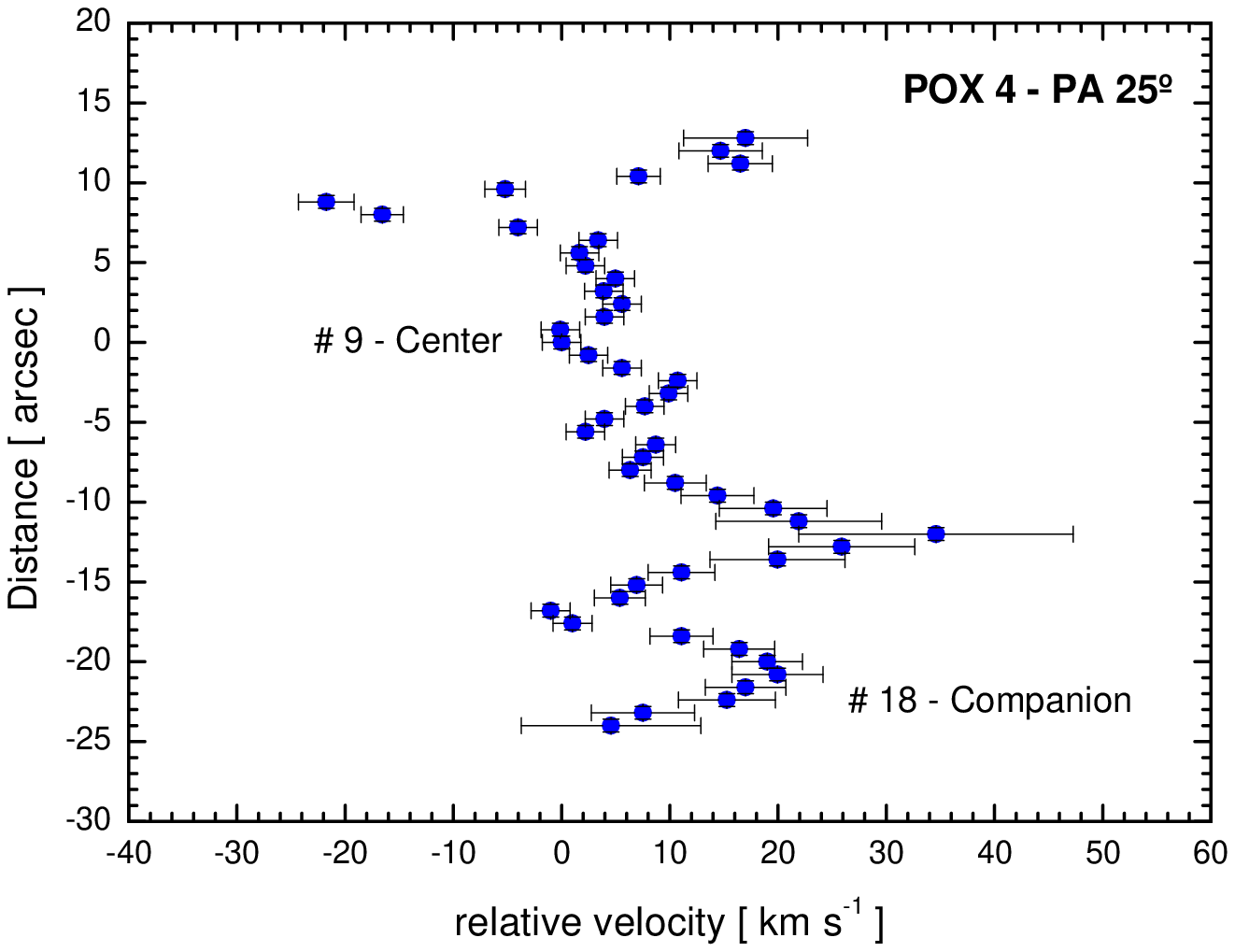}
\caption{\footnotesize{Position-velocity diagram for the slit position observed in POX~4 using the [\ion{O}{iii}] $\lambda$5007 profile. NE is up. 
See Figure~11 in Paper~I for the identification of the regions.}}
\label{pox4curvas}
\end{figure}

The position-velocity diagram obtained for the slit-position taken for POX~4 is shown in Figure~\ref{pox4curvas}. 
Because of its higher intensity, we used the [\ion{O}{iii}] $\lambda$5007 profile instead of the \Ha\ profile, extracting 4 pixel bins (0.8 arcsec) 
and taking as reference the velocity found in the center of the galaxy (knot \#9, see Figure 11 in Paper~I). The diagram shows a clear irregular 
pattern without any rotation evidence, indicating that the movement of the ionized gas of the system is rather chaotic 
but of small amplitude. \citet{ME99} proposed that the companion (\#18) has actually gone through 
POX~4, being the origin of the peculiar ring-morphology of the galaxy and the strong star-forming bursts observed 
throughout all the system. The diffuse dwarf companion, that possesses a 
radial velocity similar to that observed at the center of POX~4, also 
shows peculiar kinematics, being in some sense coupled with the movement of the ionized gas in the main galaxy. Our optical data do not suggest that 
the dwarf companion object is a \TDG\ because we do not detect any other dwarf galaxy that interacted with POX~4 and stripped some material from it.
\citet{ME99} indicated a velocity difference of $\sim$130 \kms\ between POX~4 and the companion galaxy, but this is not confirmed in our deeper 
spectroscopic data. 

Taking into account the complex kinematic structure shown in Figure~\ref{pox4curvas}, it is clear that 
we can not obtain a confident estimate of the Keplerian mass of POX~4. However, assuming a ratio of 
$M_{Kep}/L_B\sim$0.1, it would be of the order of $M_{Kep}\sim5\times10^8$ \Mo. The \emph{\HI\ Parkes All-Sky Survey} (\HIPASS; Barnes et al. 2001) 
shows a tentative detection of \HI\ emission. 
Our group has performed 21-cm observations of POX~4 using the interferometer \emph{Australia Telescope Compact
Array} (ATCA). Although a detailed description and analysis of such observations will be presented elsewhere \citep{LS10}, the
very preliminary analysis suggests that the system possesses a lot of neutral gas. The \HI\ kinematic are perturbed in the position of the dwarf 
companion object but it shows the same radial velocity we found using optical spectroscopy. 
An independent \HI\ cloud, that has the same radial velocity that POX~4, is found at $\sim$4.5$\arcmin$ ($\sim$60~kpc) at the south. It shows a clear 
alignment with both the bright center of POX~4 and the dwarf companion object, suggesting a very probable interaction in the past. A detailed 
analysis of the \HI\ observations will confirm o discard the \TDG\ nature of the dwarf companion object surrounding POX~4.



\subsection{UM 420}

\begin{figure}[t!]
\centering
\includegraphics[angle=270,width=\linewidth]{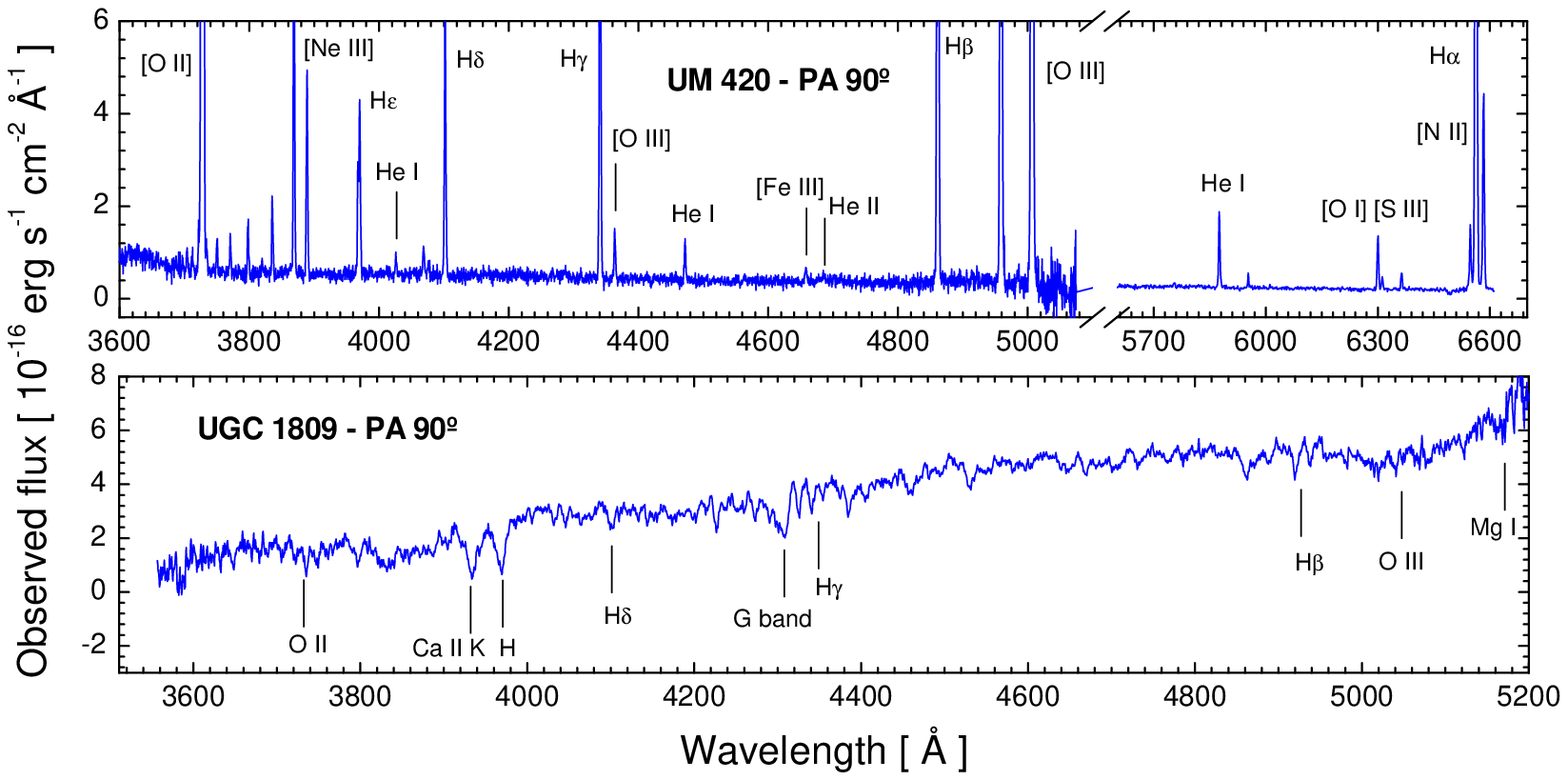}
\caption{\footnotesize{ISIS \WHT\ spectrum for the center of UM~420 (\emph{top}) and the galaxy UGC~1809 (\emph{bottom}). Fluxes are not 
corrected for reddening. The most important emission lines have been labeled. }}
\label{um420espectros}
\end{figure}

\citet{IT98} reported the detection of the broad \ion{He}{ii} $\lambda$4686 emission line in \object{UM~420}, being therefore included in the latest 
catalogue of WR galaxies \citep*{SCP99}. The reanalysis of their spectra performed by \citet{GIP01} also suggests the presence of the \ion{C}{iv} 
$\lambda$4658 and \ion{C}{iv} $\lambda$5808 emission lines, indicating the possible existence of WCE stars in the starburst.

Figure~\ref{um420espectros} shows the spectra of UM~420 and UGC~1809 for our slit position (see Figure~13 of Paper~I). Notice the huge difference 
between both spectra: while the spectrum of UM~420 is dominated by emission lines, the spectrum of UGC~1809 only shows stellar absorption features as 
\ion{Ca}{ii} H,K, G-band and \ion{Mg}{i} $\lambda$5167. Hence, we may classify this object as a S0 spiral galaxy with redshift of 
$z$=0.0243. The radial velocity of UGC~1809 is $v_r$=7290~\kms, in excellent agreement with the value given by the \NED\ ($v_r$=7306~\kms) but much 
lower than the radial velocity of UM~420 ($v_r$=17507~\kms). That confirms that both galaxies are not physically related.

The spectra of UM~420 do not show absorption features. We observe, although with large error, the nebular \HeII\ $\lambda$4686
line on top of a very faint broad feature (Figure~\ref{wrfig1}). 
We do not detect the red \WRBUMP\ despite the good S/N ratio and the clear detection of the weak [\ion{N}{ii}] $\lambda$5755 auroral line 
(Figure~\ref{wrfig2}).
The list of the emission lines observed in UM~420 is compiled in Table~\ref{pox4lineas}. 

\subsubsection{Physical conditions of the ionized gas}

\begin{figure}[t!]
\centering
\includegraphics[angle=270,width=\linewidth]{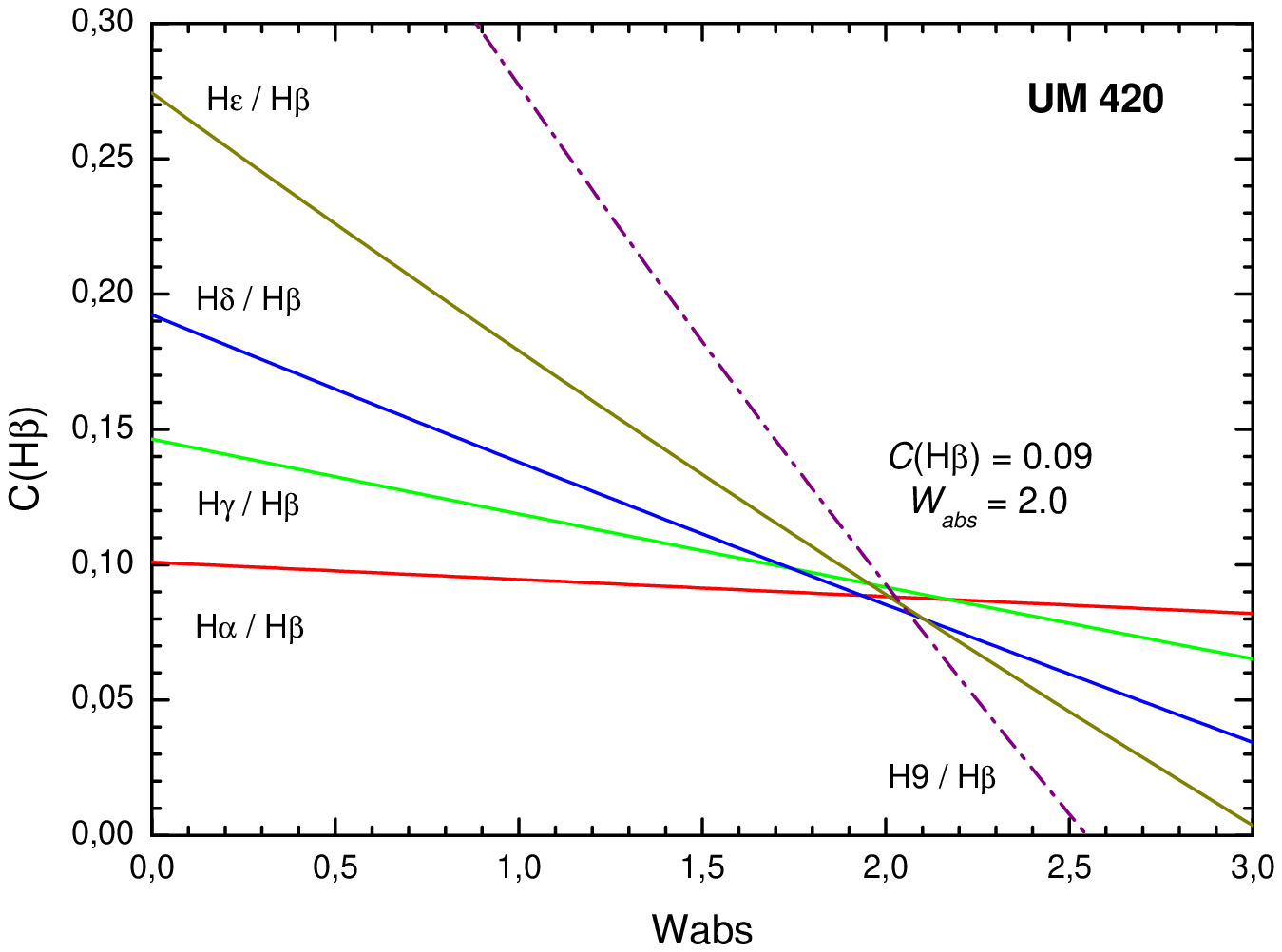}
\caption{\footnotesize{Interactive estimation of \CHb\ and $W_{abs}$ using the six brightest \HI\ Balmer lines detected in the spectrum of 
UM~420. Note the excellent agreement in the behaviour of all lines.}}
\label{um420reddening}
\end{figure}

Using the [\ion{O}{iii}] $\lambda$4363 line intensity we compute  \TeOiii=13200$\pm$600~K in UM~420. Although it has a 
large error, the detection of the auroral 
[\ion{N}{ii}] $\lambda$5755 emission line indicates \TeNii$\sim$11800 K, that is similar to the low ionization temperature given by Garnett's 
relation between  \TeOiii\ and \TeOii. The electron density computed using the \Sii\ doublet is in the low-density limit, but that 
estimated using the \Oii\ lines gives \Ne$\sim$140 cm$^{-3}$. The reddening coefficient and the underlying stellar absorption in the \HI\ Balmer 
lines were computed using 5 ratios between the \HI\ Balmer lines and give very consistent results, which mean values are   \CHb=0.09$\pm$0.01 and 
$W_{abs}$=2.0$\pm$0.1 
(see Figure~\ref{um420reddening}). The comparison of the observed line flux ratios with the diagnostic diagrams clearly identify UM~420 as a 
starburst galaxy.

\subsubsection{Chemical abundances}

Table~\ref{pox4abun} lists all the chemical abundances computed for UM~420. The derived oxygen abundance is \abox=7.95$\pm$0.05, in excellent 
agreement with the value given by \citet{IT98}. 
Our estimation does not consider the small contribution of O$^{+3}$ 
(that should exist because of the detection of the \HeII) but it is smaller than 0.01 dex. The N/O ratio, \lno=$-1.11\pm0.08$, is also 
similar to the value reported by these authors. It is interesting to notice that its N/O ratio is 
higher than that expected for its oxygen abundance, which should be around 0.4 dex lower. This fact was previously reported by \citet{PKP04}, who 
suggested that the overabundance of nitrogen may be produced by pollution by the large amount of WR stars present in the violent starburst triggered 
by galaxy merging. 
The value of the neon abundance in UM~420, log(Ne/O)=$-$0.71$\pm$0.13, is also similar to that given by  \citet{IT98}. We derive 
log(S/O)=$-$1.66$\pm$0.15, that is the typical value for \BCDG s with the same oxygen abundance \citep{IT99}.

\subsubsection{Kinematics of the ionized gas}

\begin{figure}[t!]
\centering
\includegraphics[angle=270,width=\linewidth]{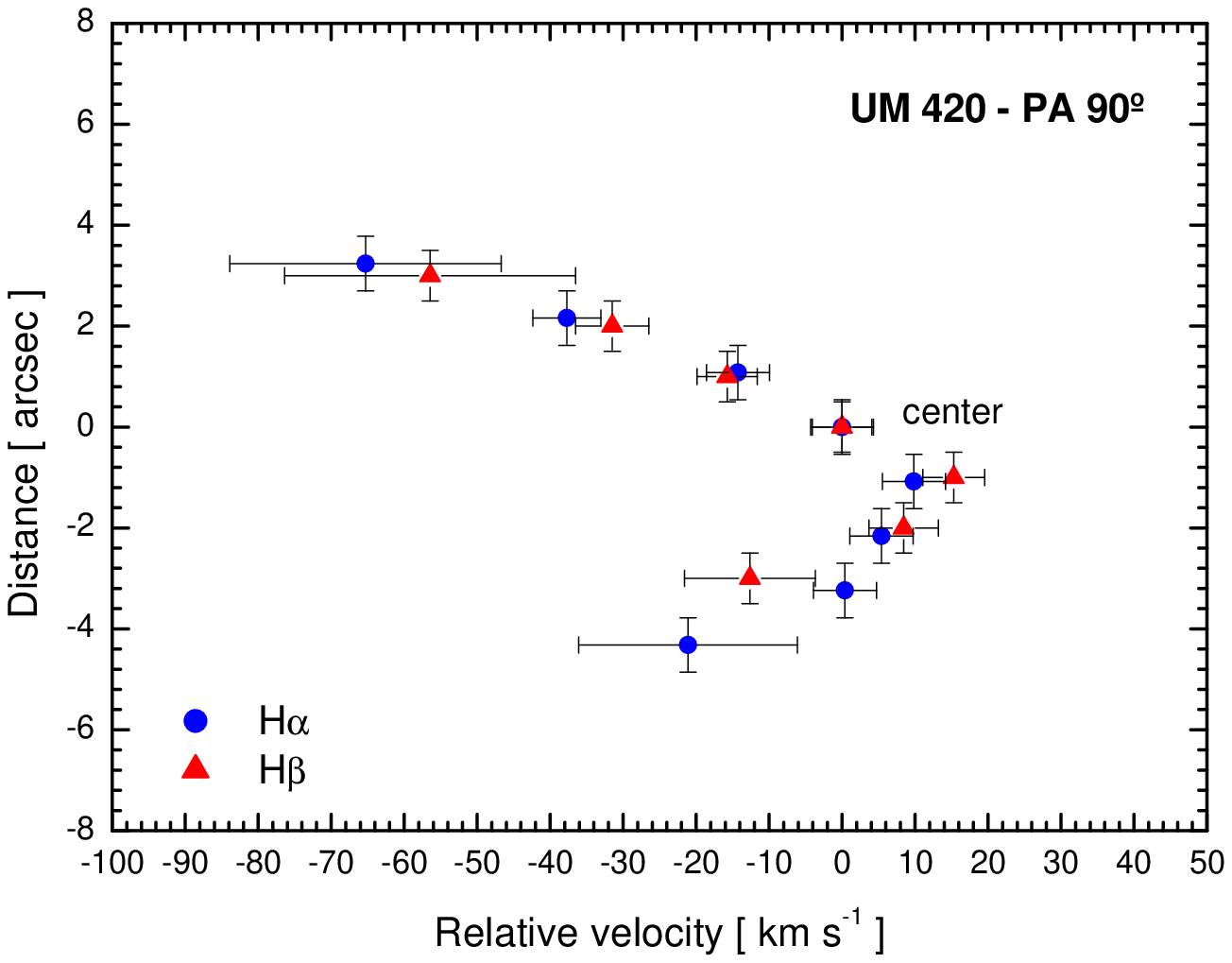}
\caption{\footnotesize{Position-velocity diagram for the slit position observed in UM 420 using both the \Ha\ and \Hb\ profiles. W is up.}}
\label{um420curvas}
\end{figure}

Figure~\ref{um420curvas}  shows the position-velocity diagram obtained for the slit position 
with PA~90$^{\circ}$ observed in UM~420. Both the \Ha\ and \Hb\ profiles were 
analyzed, extracting 3 pixel bins (1.08$\arcsec$) for \Ha\ and 4 pixel bins (0.8$\arcsec$) for \Hb. The diagram is identical in both cases. Although 
the number of points is small, we notice a velocity gradient of around 30 \kms\ from the eastern region to the center of the galaxy, but this 
tendency is reversed in the western areas of UM~420. Indeed, in this region a negative velocity gradient of  $\sim$70~\kms\ is found within only 
4$\arcsec$ ($\sim$4.6 kpc). 
The diagram does not allow to get any determination of the Keplerian mass. There are not \HI\ data of the galaxy to derive its neutral or dynamical 
masses.


\subsection{SBS 0926+606}

\object{SBS~0926+606} actually is a pair of compact nearby objects (see Figure~15 of Paper~I). SBS~0926+606~A has been spectroscopically studied by 
Izotov and collaborators with the aim of determining the primordial helium abundance \citep{ITL97,IT98} and the chemical 
abundances of heavy elements in \BCDG s \citep{IT99}. Subsequent spectroscopic studies were performed by \citet{PerezMontero03} and 
\citet{Kniazev04}.
\citet*{ITL97} indicated the presence of broad low-intensity components in both the \Ha\ and [\ion{O}{iii}] $\lambda$5007 
profiles. These authors also detected the blue \WRBUMP, strongly contaminated by nebular emission, observing both the nebular and broad \ion{He}{ii} 
$\lambda$4686 emission lines. 
Hence, \citet*{SCP99}  included SBS~0926+606~A in their catalogue of WR galaxies. \citet{GIT00} revisited the properties of the massive 
stars in this galaxy.  Until now, there were no spectroscopic data for SBS~0926+606~B.

\begin{figure}[t!]
\centering
\includegraphics[angle=270,width=\linewidth]{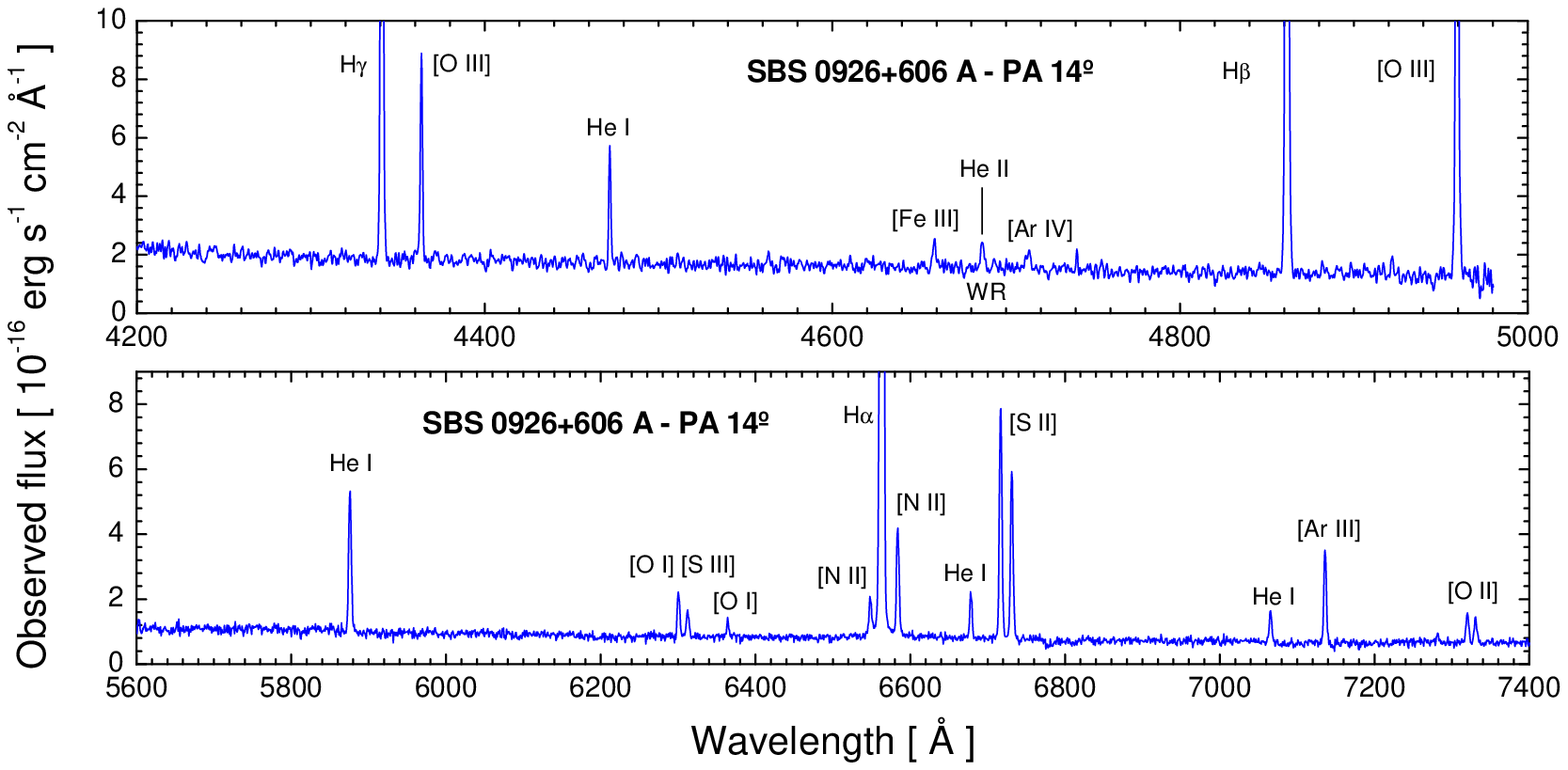}
\caption{\footnotesize{ISIS \WHT\ spectrum for SBS 0926+606 A. Fluxes are not corrected for reddening. The most important emission lines have been 
labeled. See Figure~15 in Paper~I for the identification of this region.}}
\label{sbs0926espectros}
\end{figure}

Figure~\ref{sbs0926espectros} shows our ISIS \WHT\ spectrum obtained for SBS~0926+606~A using a long-slit with a PA of 14$^{\circ}$. It mainly 
crosses the subregion A2 defined using our optical data (see Figure~15 in Paper~I).  As we see in Figure~\ref{sbs0926espectros}, the spectrum has a 
good spectral resolution, but it only covers between 4200 and 5000 \AA\ in the blue range and between 5600 and 7400 \AA\ in the red range. 
Therefore, we did not observe the [\ion{O}{ii}] $\lambda$3726,29 doublet and the bright [\ion{O}{iii}] $\lambda$5007 emission line. 
The spectrum obtained for the galaxy B is very noisy and only shows the brightest emission lines.

Table~\ref{pox4lineas} compiles all the emission line fluxes and the main properties of each spectrum. The nebular \HeII\ $\lambda$4686 emission line 
is clearly detected in SBS~0926+606~A (Figure~\ref{wrfig1}). Both the blue and red (Figure~\ref{wrfig2}) \WRBUMP s are not observed despite the high 
S/N and spectral resolution. There are not substantial stellar absorptions in the spectrum of SBS~0926+606~A, but we detect broad low-intensity 
assymetric wings in the \Ha\ profile. 

\subsubsection{Physical conditions of the ionized gas}

Table~\ref{pox4abun} compiles the values found for the electron temperatures of the ionized gas within these objects. For member~A, 
$T_e$(\ion{O}{ii}) was computed using the direct method because of the detection of [\ion{O}{iii}] $\lambda$4363 line. 
$T_e$(\ion{O}{ii}) was estimated using Garnett's relation. The 
spectrum obtained for member B does not allow to calculate $R_{23}$, so  
we used the $N_2$ ratio and the empirical calibrations given by \citet*{D02} and \citet{PP04} to estimate \Te. As empirical 
calibrations involving the $N_2$ ratio seem to overestimate the actual abundance (we will discuss this in Paper~III), the electron temperatures 
derived for galaxy~B may be underestimated. 

The electron density, computed using the [\ion{S}{ii}]~$\lambda\lambda$6717,31 doublet, was always at the low-density limit. For both galaxies, the 
reddening coefficient was derived using the 3 brightest \HI\ Balmer lines. 
The comparison of the emission lines ratios with the diagnostic diagrams indicates that both galaxies are starbursts.

\subsubsection{Chemical abundances}

Table~\ref{pox4abun} compiles the chemical abundances derived in this galaxy pair. Because the [\ion{O}{ii}] $\lambda\lambda$3726,29 doublet is not 
covered in our spectra, we used [\ion{O}{ii}] $\lambda\lambda$7318,7330 fluxes to compute the O$^+$/H$^+$ ratio. The chemical abundances derived for 
SBS~0926+606 A are 12+log(O/H) = 7.94$\pm$0.08, 
log(N/O) = $-$1.45$\pm$0.09, log (S/O) = $-$1.60$\pm$0.13 and log(Ar/O) = $-$2.34$\pm$0.13; very similar to those 
obtained by \citet{IT99}. 
The oxygen abundance found for member~B using the $N_2$ empirical calibrations is
\abox$\sim$8.15, somewhat higher than that found in galaxy~A, but as we commented above, it may be overestimated. 
The N/O ratios of both galaxies are rather similar, so both object should have suffered a similar chemical evolution. In 
fact, they are dwarf galaxies with a very similar $M_B$ (see Paper~I).

\subsubsection{Kinematics of the ionized gas}

\begin{figure}[t!]
\centering
\includegraphics[angle=270,width=\linewidth]{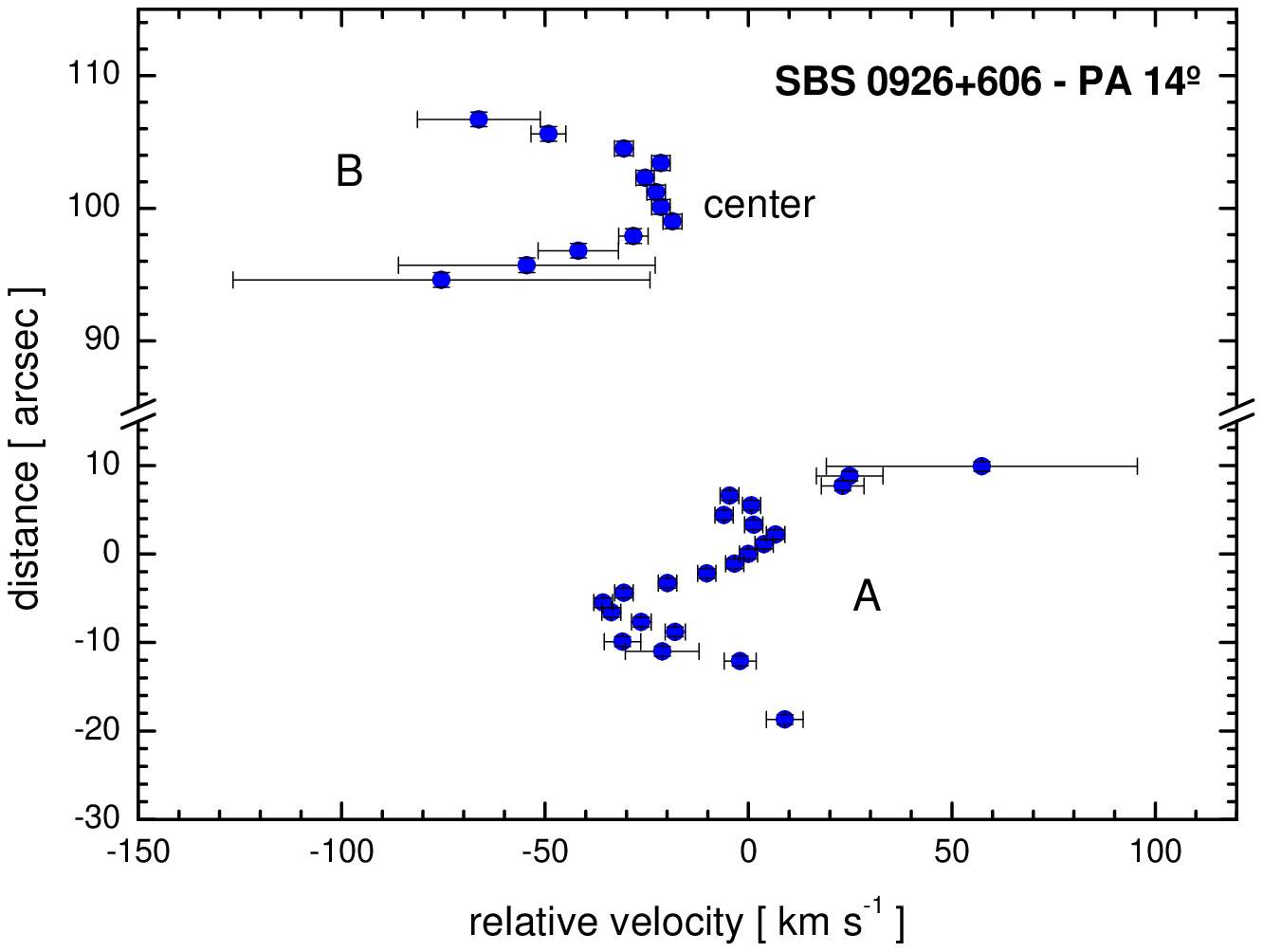}
\caption{\footnotesize{Position-velocity diagram for the slit position with PA 14$^{\circ}$ observed in SBS~0926+606 using the \Ha\ profile. 
Notice that the y-axis is broken. NE is up. See Figure~15 in Paper~I for the identification of the regions.}}
\label{sbs0926curvas}
\end{figure}

Figure~\ref{sbs0926curvas} shows the position-velocity diagram derived from the slit position observed in SBS~0926+606. We extracted 4 pixel bins 
(0.8 arcsec) across the \Ha\ profile. SBS~0926+606~A shows a clear sinusoidal pattern, with and amplitude of around 50~\kms, suggesting that the 
double nucleus we found in this galaxy (see Figures~15 and 16 of Paper~I) may be a consequence of an advanced merging 
process between two objects. The northern outskirts of SBS 0926+606~A seems to 
be partially decoupled from the sinusoidal pattern (there is a difference of around 60 km s$^{-1}$ with respect to the central velocity). On the 
other hand, SBS~0926+606~B also shows a perturbed kinematics, because both its northern and southern edges have similar radial velocities. The 
elongated shape of SBS~0926+606~B, the two tails towards the west detected in our deep images and the disturbed kinematics suggest that the 
interaction that this galaxy is experiencing --most probably with SBS~0926+606~A-- is very close to the plane perpendicular to the line of sight, and 
therefore SBS~0926+606~B is observed almost edge-on. In any case, we do not detect any morphological feature, such as the debris of a tidal tail or a 
diffuse non-stellar object, between both galaxies thus their possible interaction is nowadays no very strong. 

Finally, the complexity of the position-velocity diagram shown in Figure~\ref{sbs0926curvas} does not allow to determine the Keplerian mass of the 
galaxy. Using the \HI\ data given by \citet{Pustilnik02}, we derive  $M_{\rm H\, I}$=(9.6$\pm$3.6)$\times$10$^8$ \Mo\ and $M_{\rm H\, 
I}$=(8.1$\pm$3.6)$\times$10$^8$~\Mo\ for A and B, respectively, that indicate mass-to-luminosities ratios of $M_{\rm H\, I}/L_B$=0.75 and 0.59. The 
gas depletion timescales are 1.7~Gyr for A and 5.5~Gyr for B. Assuming half of the amplitude of the \HI\ velocity ($\sim$60 km s$^{-1}$ for both 
galaxies) and effective radii of  $\sim$10$\arcsec$ (2.71~kpc) for A and $\sim$20$\arcsec$ (5.42~kpc) for B, we estimate dynamical masses of  
$M_{Dyn}\sim$2.3$\times$10$^9$~\Mo\ and $M_{Dyn}\sim$4.5$\times$10$^9$~\Mo\ for A and B, respectively. The mass-to-luminosity ratios, 
$M_{Dyn}/L_B$=1.8 and 3.3 for A and B, respectively, which are values typical for \BCDG s \citep{HKP05}. However, the $M_{\rm H\, I}/M_{Dyn}$ ratios 
are high, 0.42 and 0.18 for A and B, respectively, indicating that a considerable amount of the total mass of the galaxies ($\sim$42\% for A) 
is neutral hydrogen. All these values indicate that the system still possesses a huge amount of fresh material from which new stars may be born. 
Indeed, the \HI\ profile obtained by \citet{Thuan99} shows two peaks, that coincide with the optical velocities of the galaxies, embedded in an 
common \HI\ envelope. This fact strongly suggests that a lot of neutral gas should be found between both galaxies. An HI\ map obtained using a 
radio-interferometer would be necessary to study the distribution and kinematics of the neutral gas, giving key clues 
about the evolution of the system. 


\subsection{SBS 0948+532}

\object{SBS~0948+532} was studied by Izotov and collaborators \citep*{ITL94,TIL95,IT98,GIT00,IT04}. \citet*{SCP99} included this \BCDG\ in their 
catalogue of WR galaxies because of the detection of both the broad and nebular \ion{He}{ii} $\lambda$4686 emission lines in the spectra presented by 
\citet*{ITL94}. 
The reanalysis performed by \citet*{GIT00} indicated 
the presence of WNL stars and a rather noisy red \WRBUMP. 

\begin{figure}[t!]
\centering
\includegraphics[angle=270,width=\linewidth]{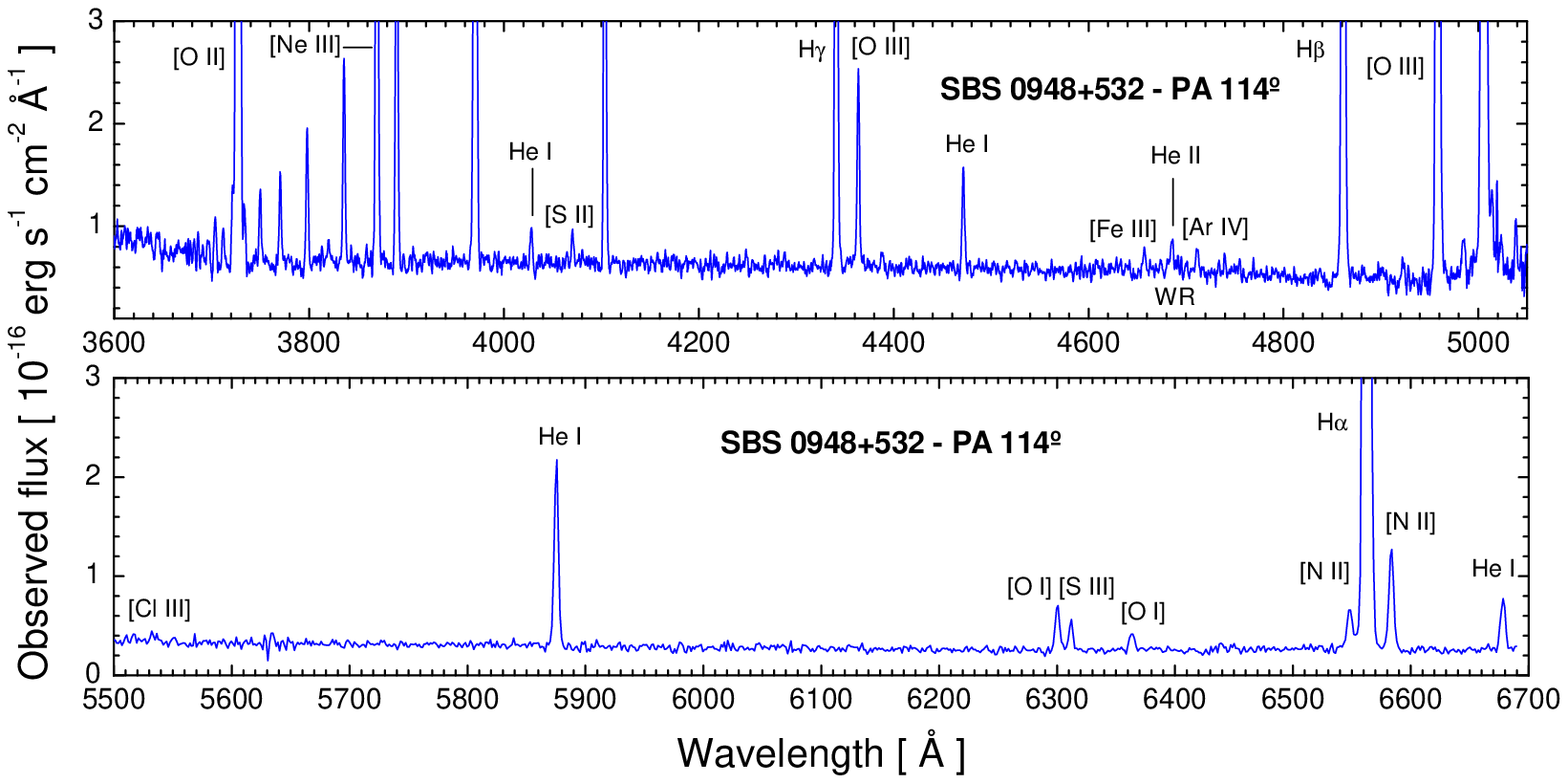}
\caption{\footnotesize{ISIS \WHT\ spectrum for SBS 0948+532 using a slit with PA 114$^{\circ}$. Fluxes are not corrected for reddening. The most 
important emission lines have been 
labeled.}}
\label{sbs0948espectro}
\end{figure}

Figure~\ref{sbs0948espectro} shows our ISIS \WHT\ spectrum of SBS~0948+532 using a slit position with PA 114$^{\circ}$. The emission line fluxes of 
the detected lines and other properties of the spectrum are compiled in Table~\ref{sbs1054lineas}. No stellar absorptions are observed in this 
spectrum. We detect the broad and nebular \ion{He}{ii} $\lambda$4686 emission lines (Figure~\ref{wrfig1}), but the red \WRBUMP\ (Figure~\ref{wrfig2}) 
is not seen besides the good S/N ratio of our spectrum.

\subsubsection{Physical conditions of the ionized gas}

The intensity of [\ion{O}{iii}] $\lambda$4363 was used to compute the high ionization electron temperature; the low ionization electron temperature 
was estimated using Garnett's relation. These values are compiled in Table~\ref{sbs1054abun}. The electron density, $n_e\sim$250 cm$^{-3}$, was 
derived using the [\ion{O}{ii}] $\lambda\lambda$3726,3729 doublet. Despite of its high error, the value of the \Ne\ estimated from the [\ion{Ar}{iv}] 
$\lambda$4711/$\lambda$4740 ratio is similar, $\sim$260~cm$^{-3}$. 
The reddening coefficient was estimated with a good precision because of the detection of many \HI\ Balmer lines.
The equivalent width of the \HI\ Balmer stellar absorption lines, \Wabs, is very small, suggesting that the underlying population of evolved stars is 
not important. The comparison of the observed emission line fluxes with the diagnostic diagrams confirms that the gas is ionized by the strong $UV$ 
emission of the massive stars. 

\subsubsection{Chemical abundances}

Table~\ref{sbs1054abun} lists all the chemical abundances computed for SBS~0948+532. The value of the oxygen abundance is  \abox=8.03$\pm$0.05. 
The N/O ratio is log(N/O)=$-1.42\pm0.08$, the typical found for objects with the metallicity of this galaxy. In general, all our chemical abundances 
for this BCDG agree with those estimated by \citet{IT99} within the errors.

\subsubsection{Kinematics of the ionized gas}

\begin{figure}[t!]
\centering
\includegraphics[angle=270,width=\linewidth]{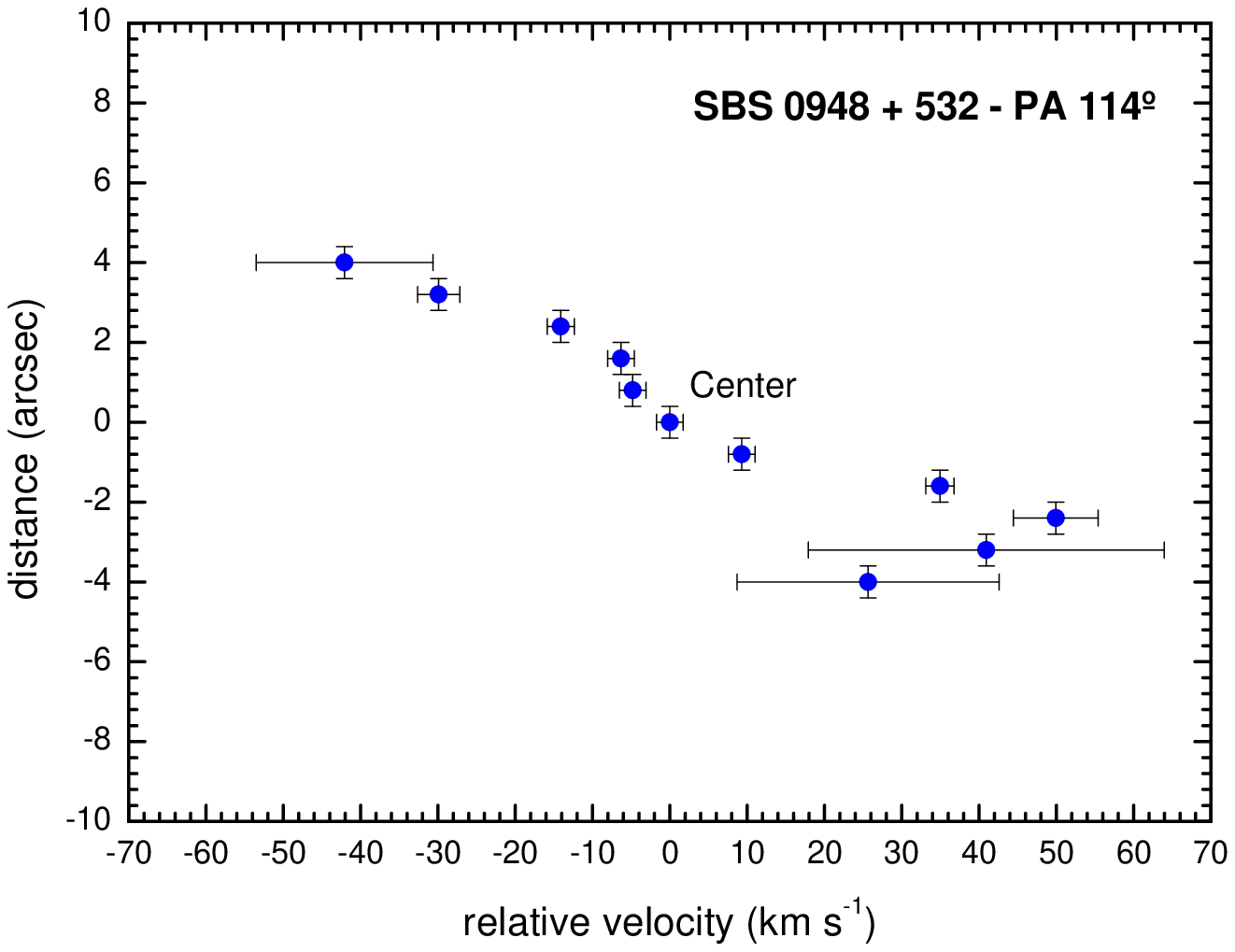}
\caption{\footnotesize{Position-velocity diagram for the slit position with PA 114$^{\circ}$ observed in SBS 0948+532 using the [\ion{O}{iii}] 
$\lambda$5007 profile. NW is up (see Figure~17 in Paper~I).}}
\label{sbs0948curvas}
\end{figure}

Figure~\ref{sbs0948curvas} shows the position-velocity diagram obtained for the slit position observed in SBS~0948+532 (see Figure~17 in Paper I). 
We analyzed the 
[\ion{O}{iii}] $\lambda$5007 profile extracting 4 pixel bins (0.8 arcsec). Although the diagram only shows 11 points because of the compact size of 
this \BCDG, a kind of rotation pattern, with an velocity gradient of around 100 \kms, is observed. This feature seems to be disturbed in the SE 
areas, that is precisely the region where the faint arc is seen in our optical images (see Figure~17 of Paper~I). 
Assuming that the kinematics of the galaxy is due to rotation,  
$i$=90$^{\circ}$ and $\Delta v\sim$50~\kms\ within a radius of $\sim 4 \arcsec$ (3.63 kpc), we derive a Keplerian mass of  
$M_{Kep}\sim$2.1$\times$10$^9$ \Mo\ and 
$M_{Kep}/L_{\odot}\sim$0.57. Unfortunately, no additional \HI\ or \FIR\ data are available for this galaxy.


\subsection{SBS 1054+365}

The first spectroscopic study of \object{SBS~1054+365} was performed by \citet{ITL97}, who detected the broad \ion{He}{ii} $\lambda$4686 emission 
line. \citet*{GIT00}, \citet{IT04} and \citet*{Buckalew05} confirmed the WR feature of this low-metallicity galaxy. This \BCDG\ is included in the 
latest WR galaxies catalogue \citep*{SCP99}.

\begin{figure}[t!]
\centering
\includegraphics[angle=270,width=\linewidth]{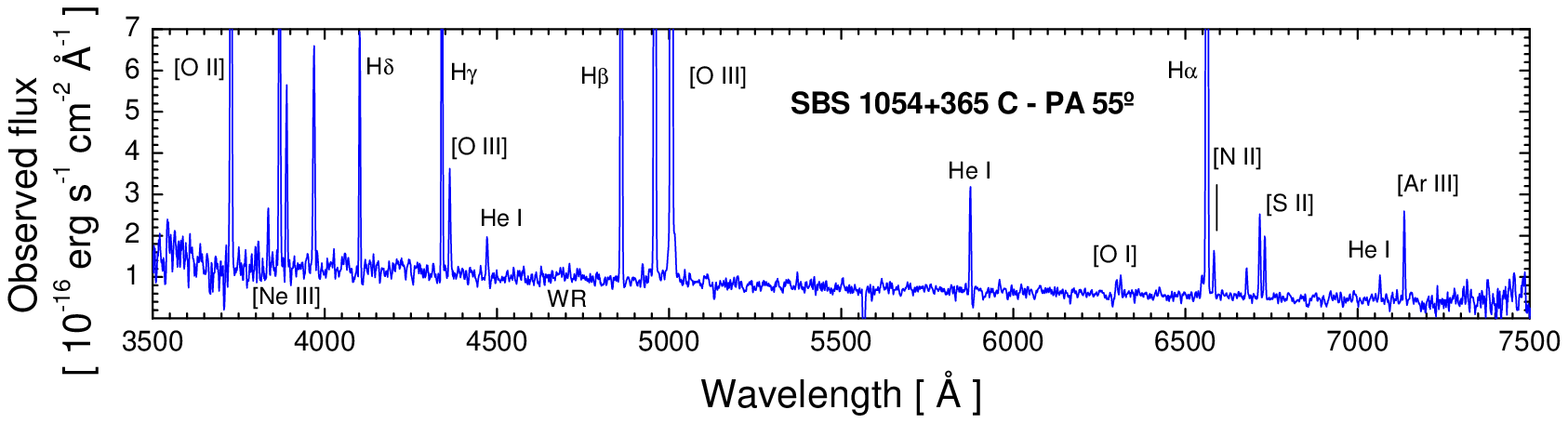}
\caption{\footnotesize{IDS \INTe\ spectrum for SBS 1054+364 using a slit with PA 55$^{\circ}$. Fluxes are not corrected for reddening. The most 
important emission lines have been labeled. See Figure~19 in Paper~I for identification of the region.}}
\label{sbs1054espectro}
\end{figure}

We used the IDS spectrograph attached at the \INTe\ to get the spectroscopy of SBS~1054+365. The slit position was set at 55$^{\circ}$, crossing the 
main body of the galaxy along its major axis (see Figure~19 of Paper~I). Hence, we observed knot \emph{b}, the central bright region C and part of 
the star-forming semi-ring located at the west (knot \emph{a}). We only analyzed the physical conditions and the chemical abundances of the ionized 
gas in region C and knot \emph{b}. Figure~\ref{sbs1054espectro} shows the spectrum of the center of SBS~1054+365. Table~\ref{sbs1054lineas} compiles 
the dereddened line intensity ratios and other properties of the spectrum. As we see, the spectrum is dominated by nebular emission without any 
features of stellar 
absorptions. We do detect the nebular \ion{He}{ii} $\lambda$4686 emission line onto a broad stellar \ion{He}{ii} line (Figure~\ref{wrfig1}). The red 
\WRBUMP\ is not detected (Figure~\ref{wrfig2}) perhaps because of the lacking of enough S/N in our spectrum. 


\subsubsection{Physical conditions of the ionized gas}

The electron temperatures in the central region were computed using the [\ion{O}{iii}] $\lambda$4363 emission line intensity and Garnett's 
relation between \TeOiii\ and \TeOii. For knot \emph{b} we used the \citet{P01a,P01b} empirical calibrations. The electron density was estimated 
using the [\ion{S}{ii}] $\lambda\lambda$6716,6731 doublet, being in the low-density limit in the central region. The values of the reddening 
coefficient are rather low. 
The comparison of the [\ion{O}{iii}]$\lambda$5007/\Hb,  [\ion{N}{ii}]$\lambda$6584/\Ha\ and [\ion{S}{ii}]$\lambda\lambda$6716,6730/\Ha\ ratios with 
the diagnostic diagrams allows to classify all knots as typical \HII regions.

\subsubsection{Chemical abundances}

Table~\ref{sbs1054abun} compiles the chemical abundances derived for SBS~1054+364. The oxygen abundance computed for the center of the galaxy is  
\abox=8.00$\pm$0.07, and its N/O ratio is log(N/O)=$-1.41\pm0.09$, 
in excellent agreement with the values obtained by \citet{IT99}. 
Despite of its higher error, the oxygen abundance  and N/O ratio estimated 
for knot \emph{b} are very similar to those found at the center of the galaxy.

\subsubsection{Kinematics of the ionized gas}

\begin{figure}[t!]
\centering
\includegraphics[angle=270,width=\linewidth]{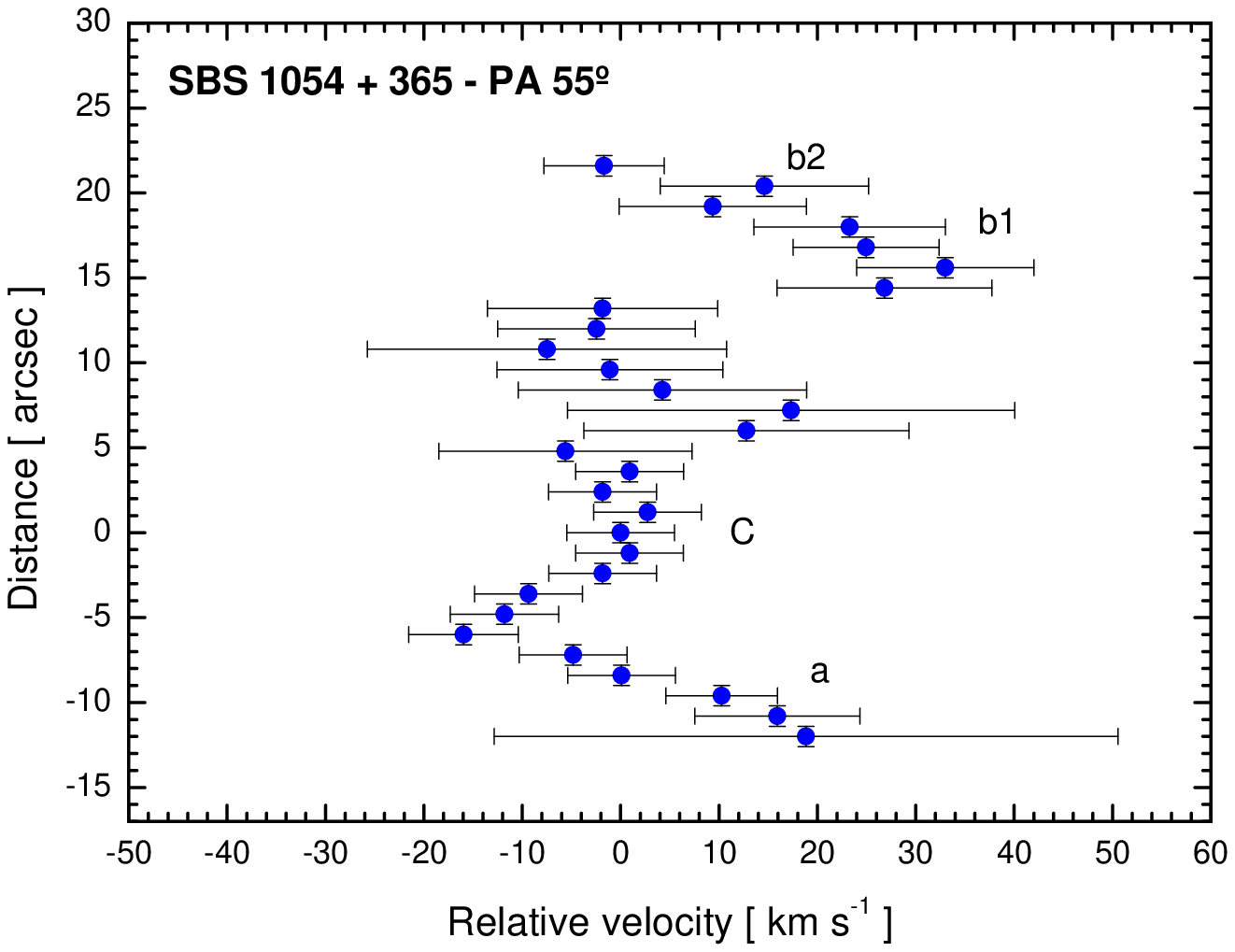}
\caption{\footnotesize{Position-velocity diagram for the slit position observed in SBS~1054+365 using the \Ha\ profile. NE is up. See Figure~19 in 
Paper~I for identification of the regions.}}
\label{sbs1054curvas}
\end{figure}

We used our bidimensional spectrum for the slit position with PA 55$^{\circ}$ 
to build the position-velocity diagram shown in Figure~\ref{sbs1054curvas}. We extracted 3 pixel bins (1.2 arcsec) across the \Ha\ profile and took 
as reference the brightest region of the galaxy. The diagram does not show a clear rotation pattern but several changes in 
the velocity distribution. The central region seems to show a velocity gradient of around 40 \kms\ between  
$-$5$\arcsec$ and 10$\arcsec$. This feature was previously noticed by \citet{Zasov00} but their lower spatial 
resolution did not permit to see the small amplitude velocity variations 
(see their Figure~3b). These authors suggested that this velocity gradient is consequence of the rotation of the galaxy. Our diagram also indicates 
that the SW region, the partial ring where knot \emph{a} is located, does not follow 
the kinematics of the center of the galaxy, showing a velocity variation of 40~\kms\ in 7.2$\arcsec$. On the other hand, the kinematics of the region 
\emph{b}, that has an inverted velocity gradient of around 40~\kms, also seem to be decoupled from the movement of the gas in the central region, 
that shows a positive velocity gradient of $\sim$25~\kms\ in its NE area. However, the amplitude of all the velocity variations seen 
in Figure~\ref{sbs1054curvas} are rather small and cover spatial extensions of the order of several hundred pc. Therefore, it is possible that they 
could be due to local movements of the bulk of the ionized gas due to the combined action of winds or supernova explosions.

In any case, assuming that rotation is present at the center of the galaxy and considering $\Delta v\sim$20~km s$^{-1}$ within a radius of 
$\sim$16$\arcsec$ (624 pc) and an inclination of $i$$\sim$60$^{\circ}$ (determined using 
its optical sizes) we have computed a tentative value for the Keplerian mass of $M_{Kep}\sim$7.8$\times$10$^7$~\Mo, that indicates 
$M_{Kep}/L_{\odot}\sim$1.19. Using the \HI\ data provided by \citet{Zasov00}, we estimated  \MHi=(6.08$\pm$0.59)$\times$10$^7$~\Mo\ and 
\MHil$\sim$0.93. Assuming a radius of 35$\arcsec$ (1.37~kpc) and the same inclination angle, we estimated a dynamical mass of 
\Mdyn$\sim$1.5$\times$10$^9$~\Mo. The neutral gas to total mass ratio, \MHiMdyn$\sim$0.04, is typical for \BCDG s but its total mass-to-light ratio, 
\Mdynl$\sim$22.9, is quite high \citep*{Salzer02,HKP05}. That may suggest that the dynamic of the system is perturbed, but only an interferometric 
\HI\ map can confirm this issue. The gas depletion timescale is higher than 2.6~Gyr, indicating that the galaxy still possesses a huge 
amount of fresh material available to create new generations of stars.


\subsection{SBS 1211+540}

\begin{figure}[t!]
\centering
\includegraphics[angle=270,width=\linewidth]{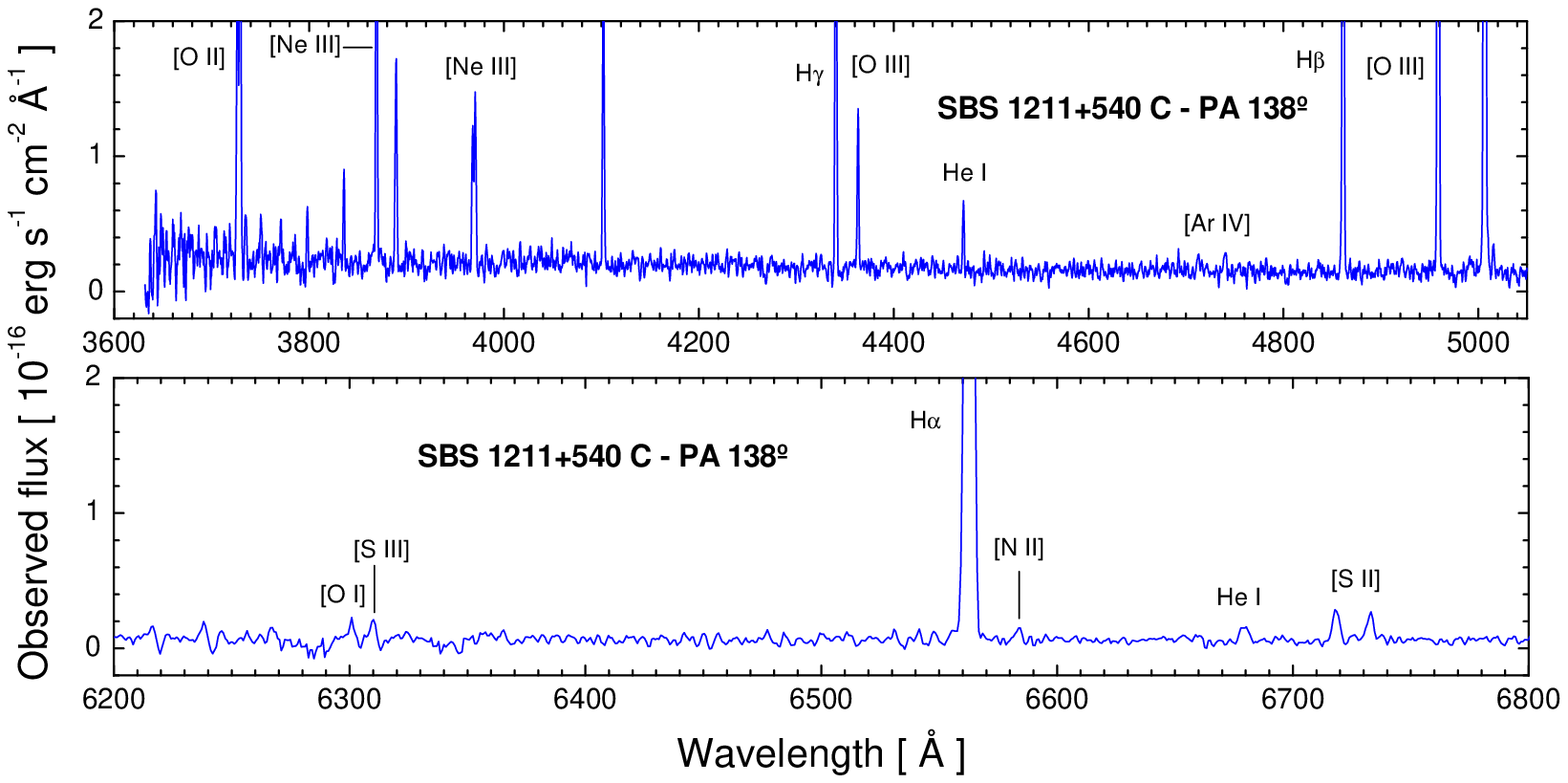}
\caption{\footnotesize{ISIS \WHT\ spectrum for SBS 1211+540 using a slit with PA 138$^{\circ}$. Fluxes are not corrected for reddening. The most 
important emission lines have been labeled. See Figure~21 in Paper~1 for identification of the region.}}
\label{sbs1211espectros}
\end{figure}

\object{SBS~1211+540} was included in the study of chemical abundances in  \BCDG s performed by Izotov and collaborators 
\citep*{Izotov91,TIL95,IT98,GIT00,IT04}. 
WR features was firstly reported by \citet*{ITL94}, who detected the nebular \ion{He}{ii} $\lambda$4686 emission line \citep*{SCP99}. However, the 
re-analysis performed by \citet*{GIT00} only indicates the presence of the broad emission line. 

Figure~\ref{sbs1211espectros} shows our ISIS \WHT\ spectrum of the center of SBS~1211+540; Table~\ref{sbs1054lineas} compiles all its derived 
properties. The spectrum is dominated by the nebular emission showing no traces of stellar absorptions. We do 
not detect the blue \WRBUMP\ or the nebular \ion{He}{ii} $\lambda$4686 emission, although the spectrum has a low S/N ratio (Figure~\ref{wrfig1}). The 
spectral range where the red \WRBUMP\ is located was not observed, but this feature should not be expected because of the very low metallicity of 
SBS~1211+540.

\subsubsection{Physical conditions of the ionized gas}

We derive a very high electron temperature, $T$(\ion{O}{iii})=17100$\pm$600 K, using the direct method (see 
Table~\ref{sbs1054abun}). The low ionization temperature was estimated considering Garnett's relation. Both the [\ion{O}{ii}] 
$\lambda\lambda$3726,3729 and [\ion{S}{ii}] $\lambda\lambda$6717,6731 doublets were used to compute the electron 
density, that is \mbox{$n_e=320\pm50$~cm$^{-3}$.} The reddening coefficient was determined using all available \HI\ Balmer lines with errors lower 
than 20\%. The comparison of the emission line ratios with the diagnostic diagrams confirms the starbursting nature of this \BCDG.

\subsubsection{Chemical abundances}

Table~\ref{sbs1054abun} compiles all the chemical abundances derived for SBS~1211+540. The oxygen abundance, \abox=7.65$\pm$0.04, and the N/O ratio, 
log(N/O)=$-1.62\pm0.10$, are in excellent agreement with the values given by \citet{IT99}. Hence, it is the lowest metallicity object analyzed in 
this work. 
The rest of chemical abundances, log(S/O)$\sim-$1.47 and log(Ne/O)$\sim-$0.75, are also similar to those determined by these authors.

\subsubsection{Kinematics of the ionized gas}

\begin{figure}[t!]
\centering
\includegraphics[angle=270,width=\linewidth]{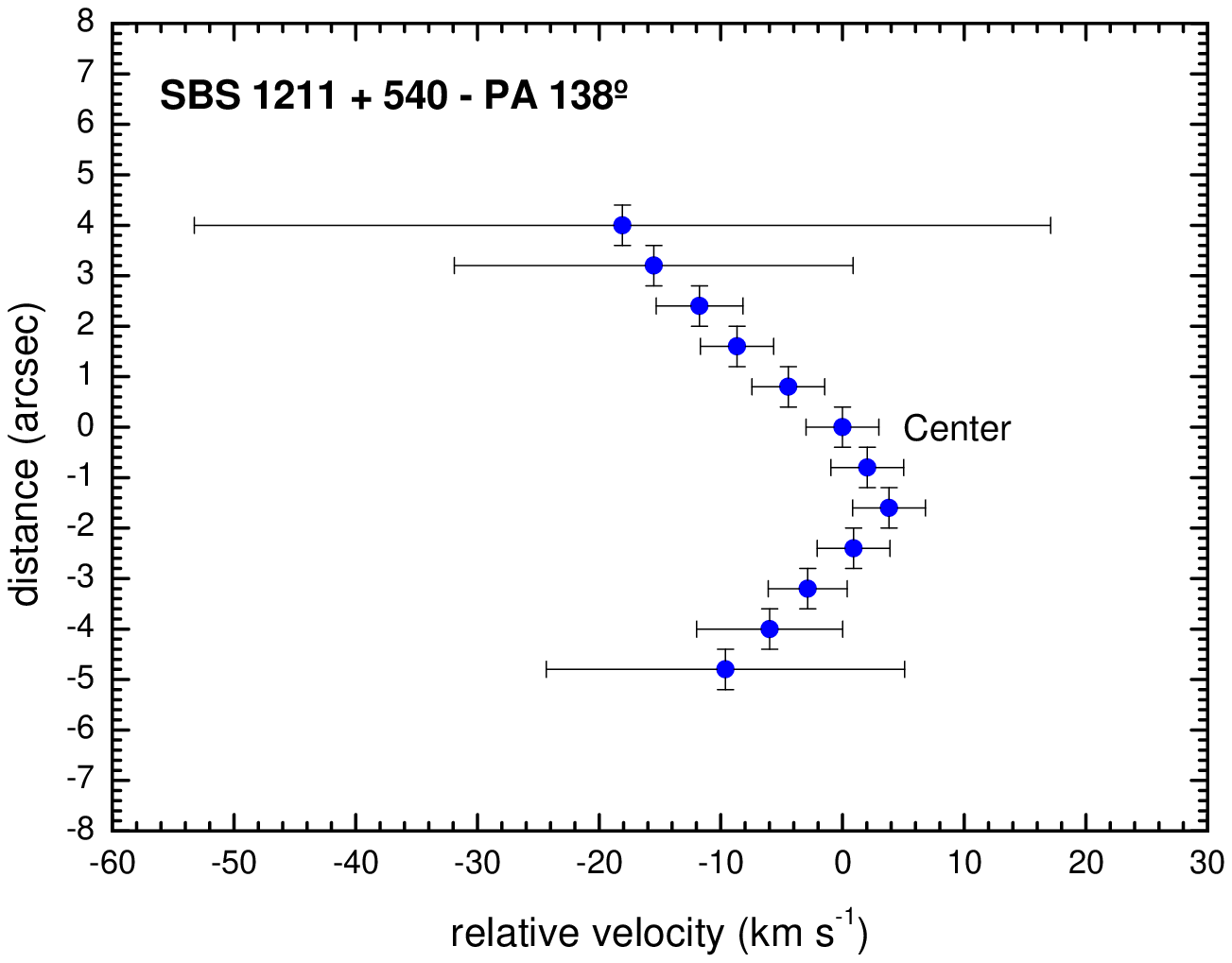}
\caption{\footnotesize{Position-velocity diagram for the slit position with PA~138$^{\circ}$ observed in SBS~1211+532 using the [\ion{O}{iii}] 
$\lambda$5007 profile. NW is up. See Figure~21 in Paper~I for identification of the regions.}}
\label{sbs1211curvas}
\end{figure}

Figure~\ref{sbs1211curvas} shows the position-velocity diagram obtained using the bidimensional spectrum of SBS~1211+540. The slit 
position we used, with a PA of 138$^{\circ}$, 
crosses the center of the galaxy but not knot \emph{a} that, as it was explained in \S3.13.2 of Paper~I, also shows nebular emission. We 
extracted 4 pixel bins (0.8 arcsec) across the [\ion{O}{iii}] $\lambda$5007 profile (the brightest line) and considered the velocity of the center as 
reference. The position-velocity diagram does not show a clear rotation pattern, only a reverse of the velocity gradient at 
the center of the galaxy. In any case, the amplitude of the velocity variations are very small. If we do not consider the 
four lowest points at the SE of Figure~\ref{sbs1211curvas} where we detected two faint plumes (see Figure~21 of Paper~I), we may assume that the 
kinematics is explained by rotation with a velocity gradient of $\sim$20--30 \kms.  

Considering that the gas is rotating with the parameters described above, we may derive the Keplerian mass of the galaxy. Assuming  $\Delta 
v\sim$30~\kms\ within a radius of $\sim$5$\arcsec$ (315~pc) and an inclination angle of $i\sim$50$^{\circ}$ (value derived from the optical shape of 
the galaxy), we find  \Mkep$\sim$1.13$\times$10$^8$~\Mo\ and \Mkepl$\sim$3.6. Using the \HI\ data provided by \citet*{HKP05}, we derive 
\MHi=(2.4$\pm$0.4)$\times$10$^7$~\Mo\ and \Mdyn$\sim$1.14$\times$10$^8$~\Mo, and hence \MHiMdyn$\sim$0.21, \MHil$\sim$0.76 and \Mdynl$\sim$3.6. The 
fact that the dynamical mass (determined using radio data) and the Keplerian mass (estimated using our optical data) completely agree indicates that 
we have probably overestimated the rotation velocity of the ionized gas and/or the actual extension of the neutral gas is much larger that the 
optical extent. If the first assumption is true, it would indicate that, besides the rotation, there is an additional velocity component, that may be 
connected with the detection of a very faint plume at the NW in our deep optical images. In any case, the high proportion of neutral mass estimated 
for this \BCDG\ (21\%) and the high value for the gas depletion timescale, 2.5~Gyr, indicate that SBS~1211+540 possesses a huge amount of fresh 
material available for the birth of new stars.    


\subsection{SBS 1319+579}

\begin{figure}[t!]
\centering
\includegraphics[angle=270,width=\linewidth]{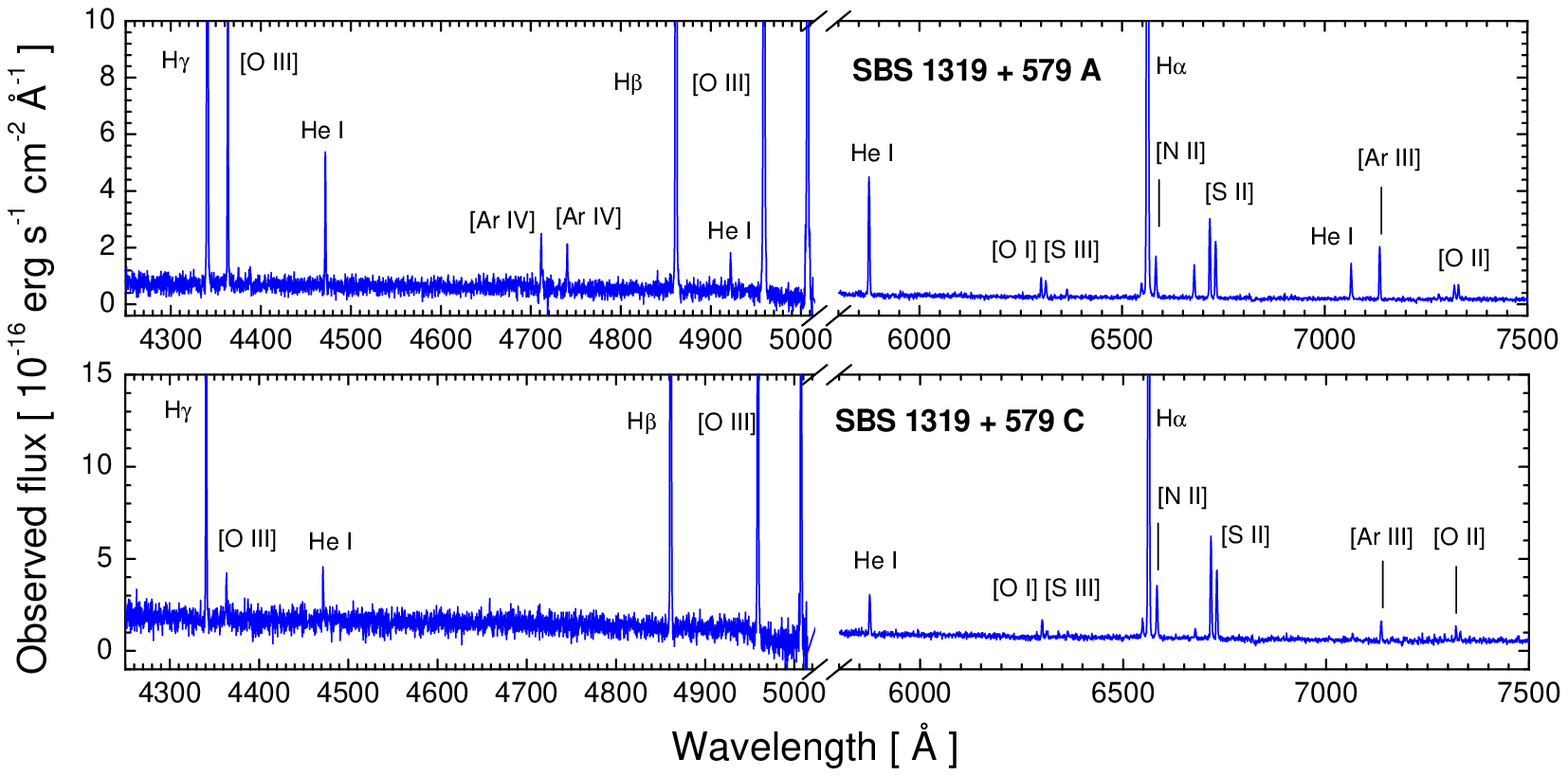}
\caption{\footnotesize{ISIS \WHT\ spectrum for regions A (\emph{top}) and C (\emph{bottom}) of SBS 1319+579 obtained with a slit with 
PA~49$^{\circ}$. Fluxes are not corrected for reddening. The most important emission lines have been labeled. See Figure~23 in Paper~I for 
identification of the regions.}}
\label{sbs1319espectros}
\end{figure}

The only bibliographic references of \object{SBS~1319+579} are from Izotov and collaborators 
\citep*{ITL97,IT98,IT99,GIT00,IT04}. \citet*{SCP99} included SBS~1319+579 in their WR galaxies catalogue because \citet*{ITL97} 
reported the detection of the broad and nebular \ion{He}{ii} $\lambda$4686 emission lines. \citet*{GIT00} 
indicated the presence of WNL and WCE populations in the galaxy. 

We used the ISIS spectrograph at the \WHT\ with a slit position with a PA of 39$^{\circ}$ to analyze the ionized gas along the main axis of the 
galaxy (see Figure~23 of Paper~I). We got spectroscopic data of regions  A, B, C, \emph{d} and \emph{e} but we only analyzed A, B and C because of 
their higher S/N ratio. The spectra of the two brightest regions A and C are shown in Figure~\ref{sbs1319espectros}, and Table~\ref{sbs1319lineas} 
compiles the dereddened flux ratios for all knots. Region B is the only one that shows some stellar absorptions in its spectra. We do not have a 
clear detection of the blue \WRBUMP\ or the nebular \HeII\ $\lambda$4686 in any region (Figure~\ref{wrfig1}).
A careful analysis of the spectrum indicates a tentative detection of both the broad and the nebular \ion{He}{ii} in knot~A (see Paper~III).
We do not detect the red \WRBUMP\ in that region (Figure~\ref{wrfig2}).

\subsubsection{Physical conditions of the ionized gas}

The [\ion{O}{iii}] $\lambda$4363 line is measured in all the regions and therefore we could determine $T_e$(\ion{O}{iii}) via the direct method. The 
low ionization electron temperatures were estimated using Garnett's relation. We found a significant difference in the electron temperatures found 
for the brightest regions A and C, $T_e$(\ion{O}{iii})$\sim$13400 and 11500 K, respectively.  The electron density, computed using the 
[\ion{S}{ii}]~$\lambda\lambda$6317,31 doublet, was always in the low-density limit. 
The reddening coefficient derived for regions A and C is low and similar to the 
Galactic reddening. All regions can be classified as starbursts following the results given by the analysis of the diagnostic diagrams. 

\subsubsection{Chemical abundances}

Because of the lacking of [\ion{O}{ii}] $\lambda\lambda$3726,29 flux values, we used [\ion{O}{ii}] $\lambda\lambda$7318,30 to compute the O$^+$ 
abundance. 
All the results for the chemical abundances derived in SBS~1319+579 are compiled in Table~\ref{sbs1319abun}. The oxygen abundance found in all 
regions are similar within the errors, \abox$\sim8.10$, although that
computed in region A, \abox=8.05$\pm$0.06, 
is slightly lower to that found in region C, \abox=8.15$\pm$0.07. 
Knot A shows a high excitation degree, log(O$^{++}$/O$^+$)=0.77, something that 
is not observed in the other regions. 
The N/O ratios are very similar in A and B [\lno=$-$1.53$\pm$0.10 in A] but also slightly different than in region C, which has 
\lno=$-$1.38$\pm$0.10. 
All the chemical abundances are consistent with those reported by \citet{IT99}. 

\subsubsection{Kinematics of the ionized gas}

\begin{figure}[t!]
\centering
\includegraphics[angle=270,width=\linewidth]{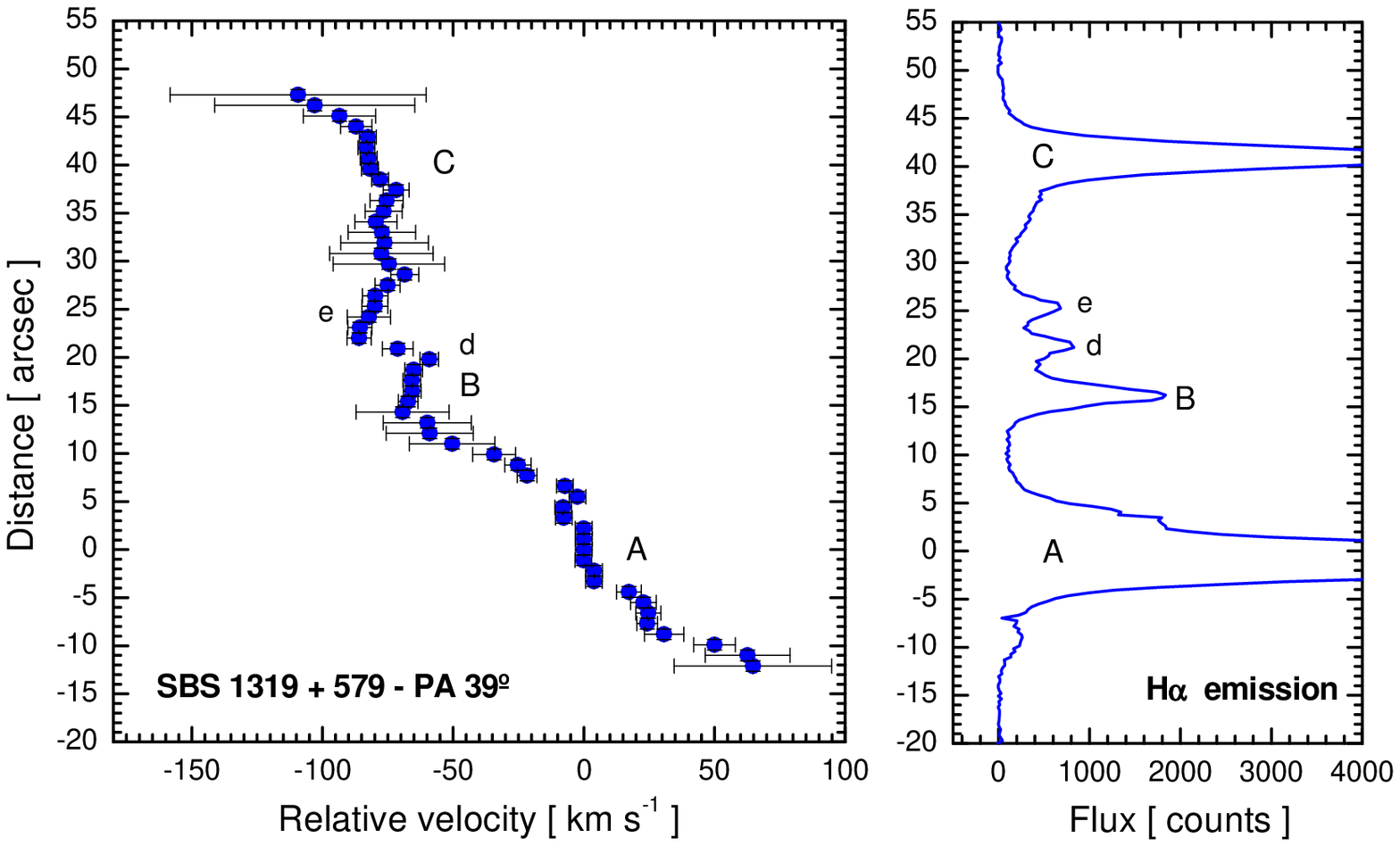}
\caption{\footnotesize{Position-velocity diagram for the slit position observed in SBS 1319+579 using the \Ha\ profile. The relative intensity of the 
\Ha\ emission along the spatial direction is also shown, identifying all observed regions. NE is up. See Figure~23 in Paper~I for identification of 
the regions.}}
\label{sbs1319curvas}
\end{figure}

Figure~\ref{sbs1319curvas} shows the position-velocity diagram obtained from our bidimensional spectrum using a slit position with PA 39$^{\circ}$. 
We extracted 4 pixel bins (0.8 arcsec) across the \Ha\ profile. The relative intensity of the \Ha\ emission along the spatial direction in also shown 
in this figure. Although the velocity continuously decreases from the eastern regions ($v\sim-$105~\kms) to the western areas ($v\sim$65~\kms) of the 
galaxy, the velocity gradient is not the same across the system. We observe two tendencies: from region C to region B (velocity difference of 
$\sim$40~\kms\ in 30$\arcsec$) and from region B to region A (velocity difference of $\sim$130~\kms\ in 28$\arcsec$). This behavior may suggest that 
there is a tidal stream moving from B to A in the direction away from the observer, but our deep images do not show such tail or any morphological 
feature that support this hypothesis. Another explanation to this feature may be the assumption that they are two systems, as we suggested from the 
morphology of the \Ha\ images and the chemical abundances may indicate, with different kinematics and in interaction. If this idea is correct, we 
should expect to observe distortions in the 
kinematics of the gas with higher amplitudes that those we see. However, because of the high inclination angle that the galaxy seems to have, 
$i\sim$70$^{\circ}$, we cannot discard any of both hypothesis. 

Considering that the kinematic pattern is consequence of the rotation of the galaxy and assuming $i\sim$70$^{\circ}$ and $\Delta v\sim$88~\kms\ 
within a radius of $\sim$30$\arcsec$ (4.2~kpc), we derive a Keplerian mass of \Mkep$\sim$8.6$\times$10$^9$~\Mo. The corresponding mass-to-luminosity 
ratio, \Mkepl$\sim$2.14, is high for that expected for a dwarf galaxy with the properties observed in SBS~1319+579. However, if we consider that only 
the NE region (from C to B) is rotating with a $\Delta v\sim$45~\kms, we now find \Mkep$\sim$1.7$\times$10$^9$~\Mo\ and \Mkepl$\sim$0.42, similar to 
the values found in other \BCDG s \citep*{HKP05}. This fact seems to confirm that the kinematics surrounding region A are disturbed and not produced 
by rotation. 
Using the \HI\ data provided by \citet{HPKK07}, we derive \MHi=$1.64\times10^9$~\Mo\ and \Mdyn$\sim 1.4\times10^{10}$~\Mo, assuming a 
rotation velocity of 109~\kms\ within 4.5 kpc and the same inclination angle. The neutral gas accounts for only the 12\% of all the mass of the 
system. Its \Mdynl\ value, $\sim$3.5, is the expected in \BCDG s. However, the gas depletion timescale is huge for a starburst galaxy, $\tau\sim$10.8 
Gyr. This may suggest that the star formation is not very efficient in the system, perhaps because the \HI\ gas has been expelled from the galaxy.    
An \HI\ map obtained using a radio-interferometer that includes both SBS~1319+579 and the nearby spiral NGC~5113 would be fundamental to understand 
the dynamics and evolution of this system.


\subsection{SBS 1415+437}

\begin{figure}[t!]
\centering
\includegraphics[angle=270,width=\linewidth]{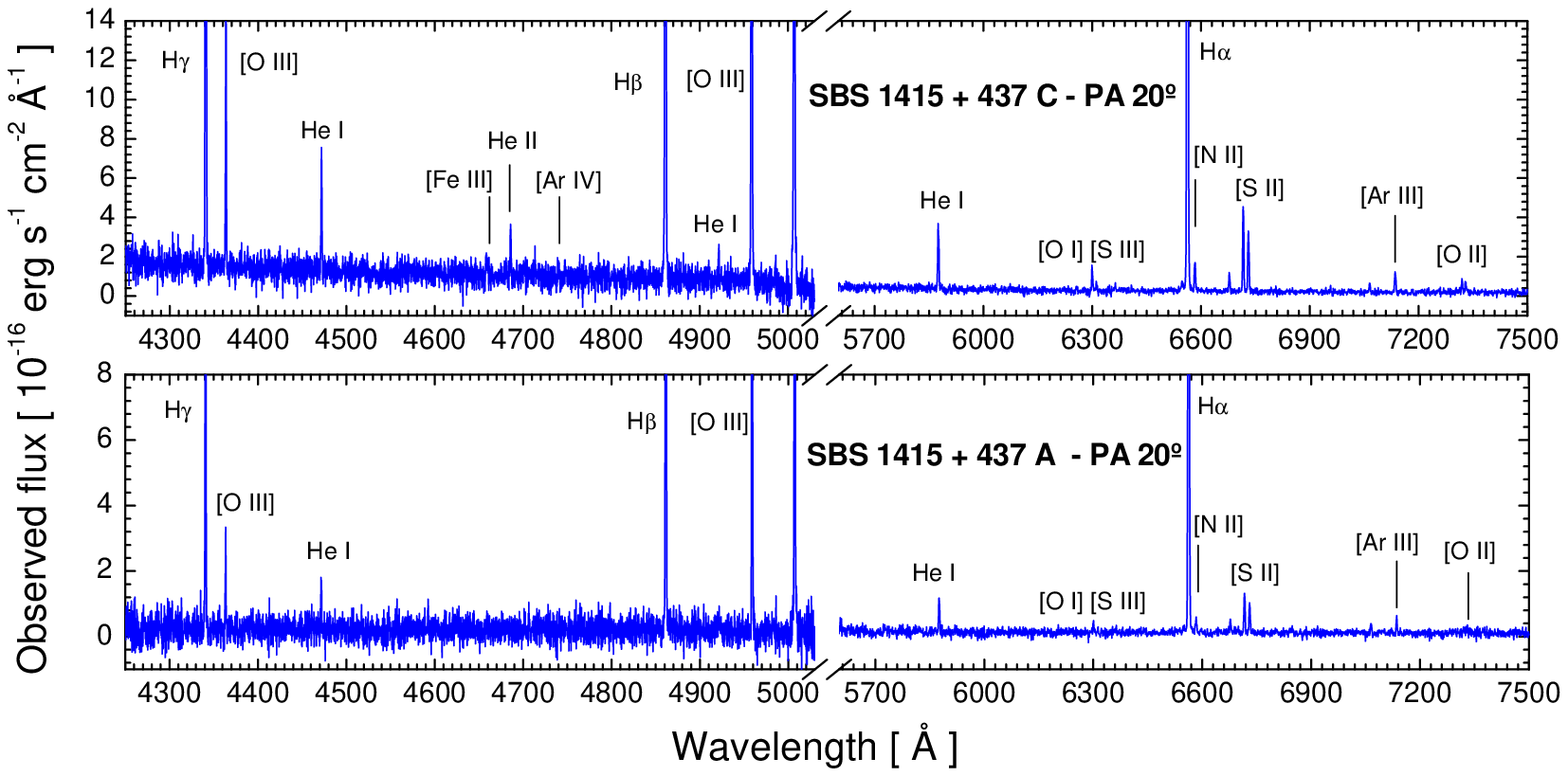}
\caption{\footnotesize{ISIS \WHT\ spectrum for regions C (\emph{top}) and A (\emph{bottom}) of SBS~1415+437 using a slit with PA 20$^{\circ}$. Fluxes 
are not corrected for reddening. The most important emission lines have been labeled. See Figure~25 in Paper~I for identification of the regions.}}
\label{sbs1415espectros}
\end{figure}

The first spectroscopic data of \object{SBS~1415+437} were reported by \citet*{TIL95}, who determined an oxygen abundance of 12+log(O/H)=7.51, being 
one of the less-metallicity galaxies known. A later reanalysis of the same spectrum raised this value to 7.59 
\citep{IT98,IT99,TIF99}.
Their spectrum shows the broad and nebular \HeII\ emission lines, and therefore SBS~1415+437 was included in the latest WR galaxies catalogue 
\citep*{SCP99}. 
Subsequent spectroscopic analysis were published by 
\citet{Melbourne02,Melbourne04,Guseva03,IT04} and \citet{Lee04}.

Figure~\ref{sbs1415espectros} shows the spectra of regions A and C obtained using the instrument ISIS at the \WHT\ and a slit with PA 20$^{\circ}$ 
that crosses the main body of the galaxy (see Figure~25 of Paper~I). Although we detected some emission lines in knot~B, we have not 
analyzed its properties because of the low S/N ratio of its spectrum. All spectra are dominated by nebular emission; no stellar absorptions are 
detected. Table~\ref{sbs1319lineas} compiles all the line intensities ratios and other properties of the spectra. 
Although we do not see the broad blue \WRBUMP, the nebular \HeII\ $\lambda$4686 emission line is well detected in the spectrum of region~C 
(Figure~\ref{wrfig1}). 
We do not observe the red \WRBUMP\ (Figure~\ref{wrfig2}) besides the good S/N ratio and spectral resolution.  

\subsubsection{Physical conditions of the ionized gas}

The electron temperatures were computed using the direct method and are very high, $T$(\ion{O}{iii})=16400 and 15500~K for C and A, respectively. 
\TeOii\ was estimated using Garnett's relation. The electron density was derived using the [\ion{S}{ii}] $\lambda$6716,31 doublet and was below the 
low-density limit. The reddening coefficient found in region C is extremely low, \CHb$\sim$0.01, and identical to that determined by 
\citet{Guseva03}. However, the higher value of \CHb\ in region A, \CHb=0.16, suggests an inhomogeneous distribution of dust in the galaxy. The 
comparison of the observed [\ion{O}{iii}]$\lambda$5007/\Hb\ and [\ion{N}{ii}]$\lambda$6584/\Ha\ ratios with the predictions given by the diagnostic 
diagrams confirm their starbursting nature.

\subsubsection{Chemical abundances}

Table~\ref{sbs1319abun} compiles our results of the chemical abundances derived in SBS~1415+437. These data confirm the very low-metallicity of the 
galaxy, being the oxygen abundances \abox=7.58$\pm$0.05 (for C) and 7.61$\pm$0.06 (for A). The N/O ratio for both objects, log(N/O)=$-1.57\pm0.08$, 
is the expected for such low metallicities. Our results are in very good 
agreement with the abundances obtained by \citet{Guseva03}.

\subsubsection{Kinematics of the ionized gas}

\begin{figure}[t!]
\centering
\includegraphics[angle=270,width=\linewidth]{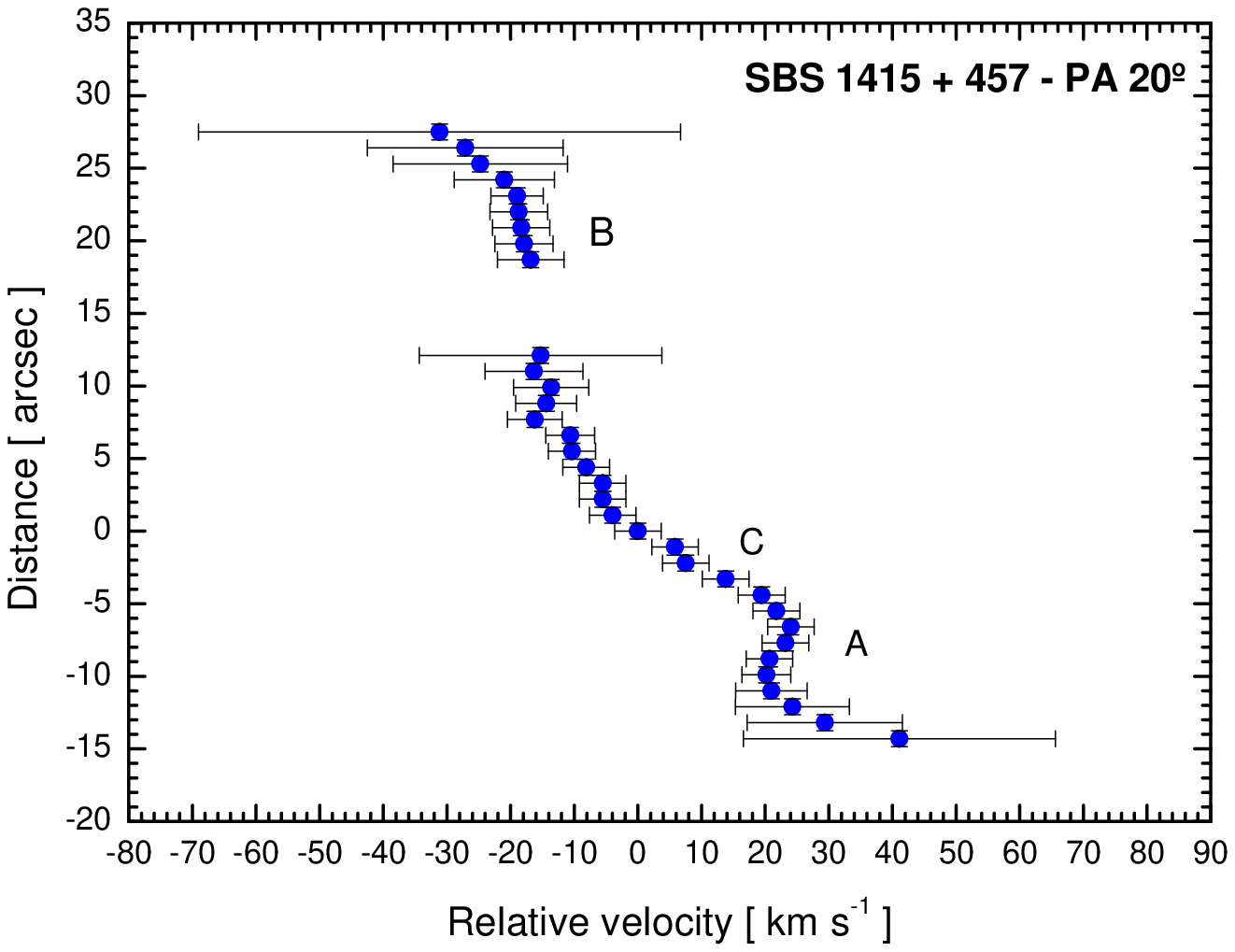}
\caption{\footnotesize{Position-velocity diagram for the slit position with PA~20$^{\circ}$ observed in SBS~1415+437 using the \Ha\ profile. NE is 
up. See Figure~25 in Paper~I for identification of the regions.}}
\label{sbs1415curvas}
\end{figure}

The position-velocity diagram shown in Figure~\ref{sbs1415curvas} was obtained extracting 4 pixel bins (0.8$\arcsec$) along the \Ha\ profile of our 
bidimensional spectrum. We took the velocity of the brightest object C as reference. We observe that the velocity continuously decreases from the SW 
regions ($v\sim$30~\kms, where region A is located) to the NE areas ($v\sim-$30~\kms, where region B is found), that may be attributed to the 
rotation of the galaxy. Some kinematic divergences are detected between regions A and C. However, because of the low amplitude of such variations 
(less than 15~\kms), they may just be a consequence of local movements in the ionized gas. Our position-velocity diagram is similar in both shape and 
values to that obtained by \citet*{TIF99} using a slit with a PA of 22$^{\circ}$ (see their Figure~10). They also reported the 
peculiar kinematic behavior we observe between A and C.   

The Keplerian mass we estimate for this galaxy, assuming a rotation velocity of $\sim$30~\kms\ within a radius of  $\sim$25$\arcsec$ (1.13~kpc) and 
an inclination angle of $i\sim$75$^{\circ}$ (determined using the shape of the galaxy we see in our optical images), is 
\Mkep$\sim$2.5$\times$10$^8$~\Mo, and its mass-to-luminosity ratio  \Mkepl=2.5. The  \HI\ mass estimated by \citet{HKP05} is 
\MHi=(9.64$\pm$0.65)$\times$10$^7$ \Mo. 
Using their data of $W_{\rm H\,I}$ and considering a radius of 40$\arcsec$ (1.8~kpc), we estimate a dynamical mass of \Mdyn=4.9$\times$10$^8$~\Mo. 
With these data, we derive \MHil=0.96, \Mdynl=4.9 and \MHiMdyn=0.20, that are the typical values found for \BCDG s \citep*{Salzer02,HKP05}. These 
estimations are more reliable that those given by \citet*{TIF99} because we are using recent data with a higher 
sensibility. Both the gas depletion timescale ($\sim$3.2~Gyr) and the fact that 1/5 of the mass of the system is neutral gas indicate 
that SBS~1415+437 possesses a huge reservoir of fresh material available for new star-forming phenomena. 


\subsection{III Zw 107}

\begin{figure}[t!]
\centering
\includegraphics[angle=270,width=\linewidth]{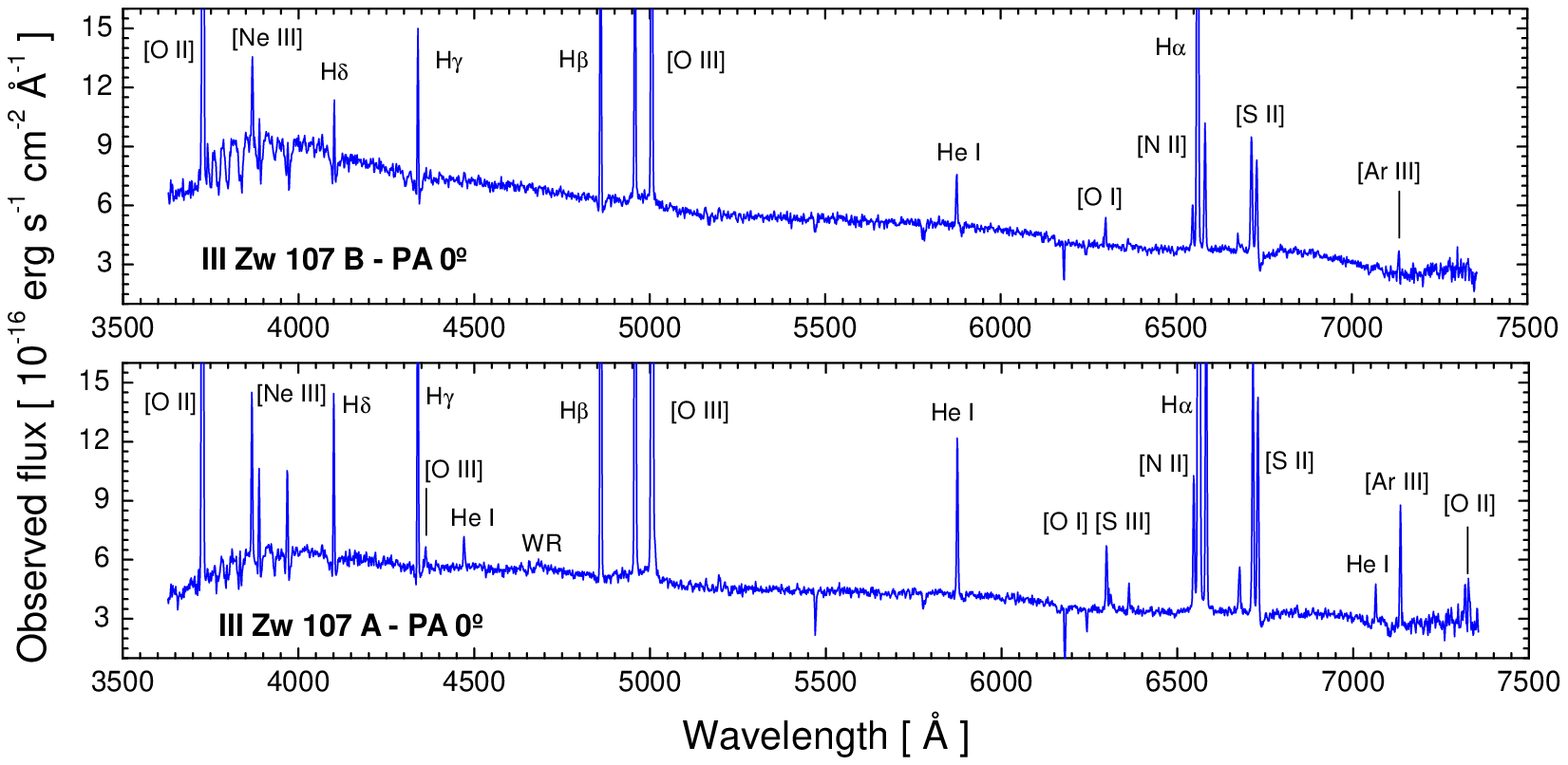}
\caption{\footnotesize{IDS INT spectra for the regions A (\emph{bottom}) and B (\emph{top}) of IRAS III Zw 107. Fluxes are not corrected for 
reddening. The most important emission lines have been labeled. See Figure~27 in Paper~I for identification of the regions.}}
\label{iiizw107espectros}
\end{figure}

\object{III~Zw~107}  was analyzed using spectroscopy by \citet{Sargent70,Gallego97} and \citet{KJ85}. The last authors detected a continuum excess in 
the spectral region of the blue \WRBUMP\ in the southern object, and hence \citet*{SCP99} included this \BCDG\ in their catalogue of WR galaxies. 

A slit with a PA of 0$^{\circ}$ was used in the IDS spectrograph at the \INTe\ to observe III~Zw~107 (see Figure~27 of Paper~I). Three different 
regions, A, B and C, were extracted. 
The optical spectra of regions A and B are shown in Figure~\ref{iiizw107espectros}. Region A possesses important underlying stellar absorption 
features. 
The faint region C is not very evident from our optical images, but it is clearly identified at the north of region B in our bidimensional spectrum.
Table~\ref{iiizw107lineas} compiles all the emission intensity ratios and other properties  
of the spectra of III~Zw~107. We clearly detect the broad \HeII\ $\lambda$4686 feature in the spectrum of knot A (Figure~\ref{wrfig1}), the same 
region where \citet{KJ85} indicated as WR-rich. However, we do not see the red \WRBUMP\ (Figure~\ref{wrfig2}) perhaps because of the relatively low 
spectral resolution of these \INTe\ spectra.
[\ion{O}{iii}] $\lambda\lambda$4959,5007 and \Ha\ show broad wings in their profiles, more evident in the spectrum of region A.

\subsubsection{Physical conditions of the ionized gas}

In region~A we computed both \TeOiii\ and \TeOii\ using the direct method because of the detection of [\ion{O}{iii}] $\lambda$4363 and the 
[\ion{O}{ii}] $\lambda\lambda$7319,7330 doublet. The values are $T_e$(\ion{O}{iii})=10900$\pm$900~K and $T_e$(\ion{O}{ii})=10500$\pm$800~K. The 
electron temperatures for knots B and C were estimated using empirical calibrations. All 
results are compiled in Table~\ref{iiizw107abun}. The electron density was computed using the  [\ion{S}{ii}] $\lambda\lambda$6716,31 doublet and it 
was in the low-density limit for regions B and C.

The comparison of the spectra shown in Figure~\ref{iiizw107espectros} indicates that the spectral energy distribution of the continuum in region A is 
less steeper than that observed in region B. This effect may be explained by the contribution of 
a more evolved stellar population in A. However, the analysis of the reddening coefficient using the \HI\ Balmer lines
of this regions gives a much higher value in region A, \CHb$\sim$0.68, than in region B, \CHb$\sim$0.15, and hence it seems that a difference in 
extinction is the explanation to the different slope of the continuum. The diagnostic diagrams for all regions agree with the loci of typical \HII 
regions. 


\subsubsection{Chemical abundances}

Table~\ref{iiizw107abun} lists all the chemical abundances computed for the bursts analyzed in~III~Zw 107. The oxygen abundance of the region A, 
derived using the direct method, is \abox=8.23$\pm$0.09. This value is similar to that reported by  \citet{KJ85}, 
but more than 0.3 dex higher than that reported by \citet{Gallego97}.
The oxygen abundances of regions B and C were estimated using the \citet{P01a,P01b}
empirical calibrations, yielding values of \abox$\sim$8.31, similar to the abundance of region A within the errors. 
The rest of the chemical abundances are also similar in all regions, just slightly lower in A. The results of the chemical abundances found between B 
and C are essentially identical besides all the uncertainties involved in their determination.

\subsubsection{Kinematics of the ionized gas}

\begin{figure}[t!]
\centering
\includegraphics[angle=270,width=\linewidth]{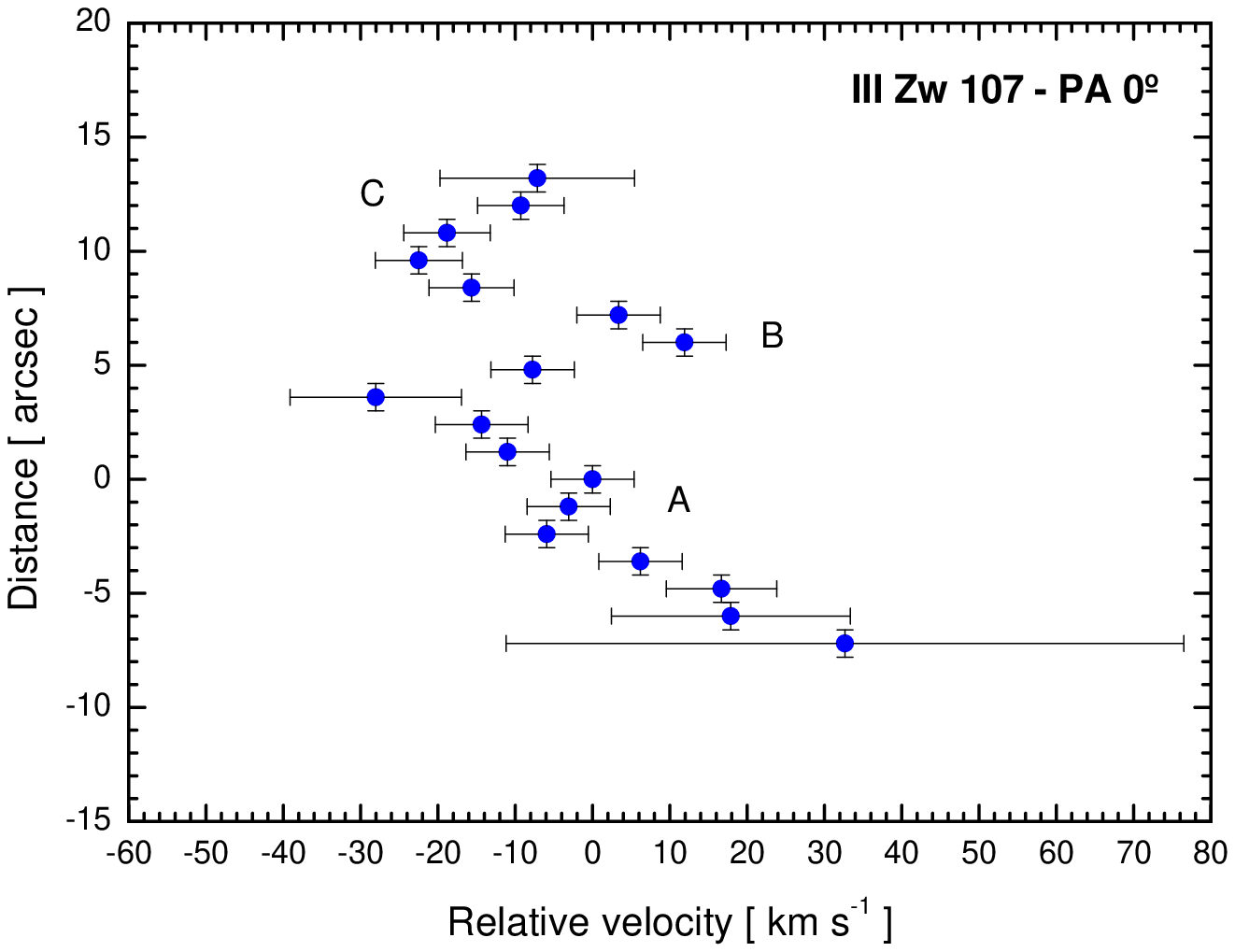}
\caption{\footnotesize{Position-velocity diagram for the slit position with PA~$^{\circ}$ observed in III~Zw~107 using the \Ha\ profile. N is up. See 
Figure~27 in Paper~I for identification of the regions.}}
\label{iiizw107curvas}
\end{figure}

The position-velocity diagram obtained using our bidimensional spectrum 
is shown in Figure~\ref{iiizw107curvas}. We extracted 3 pixel bins (1.2~arcsec) along the \Ha\ profile, taking as reference the velocity of region A, 
the brightest knot. We observe a negative velocity gradient between the southern regions of the galaxy ($\sim$30~\kms) and region A ($\sim-$20~\kms), 
but between this knot and region B a reverse of the velocity of $\sim$40~\kms\ is found within 4$\arcsec$. The velocity difference between region B 
and C is $\sim-$40~\kms. Hence, although the velocity amplitudes are not large and the spatial resolution is not very high, the position-velocity 
diagram seems to show a sinusoidal pattern  This feature may suggest interaction or merging phenomena between the two brightest knots seen in 
III~Zw~107. This hypothesis would explain the existence of the tail found in the deep optical images (see Figure~27 of Paper~I). It is also possible 
that the velocity gradient observed at the south of the galaxy is a consequence of the movement of the ionized gas within/towards that tail. 

We have performed a tentative estimation of the Keplerian mass of III~Zw~107 using the position-velocity diagram shown in 
Figure~\ref{iiizw107curvas}. Considering that we observe the galaxy edge-on ($i\sim$90$^{\circ}$) and assuming  $\Delta v\sim$30~\kms\ within a 
radius of $\sim$10$\arcsec$ (3.9~kpc), we estimate \Mkep$\sim$8.2$\times$10$^8$~\Mo\ and \Mkepl$\sim$0.05. These values are very low compared with 
the neutral gas mass and the dynamical mass derived using the radio data \citep{Paturel03}, \MHi=\valor{6.7}{1.2}{9}~\Mo, 
\Mdyn$\sim$1.8$\times$10$^{10}$~\Mo\ (the dynamical mass was estimated using a radius of 20$\arcsec$=7.8~kpc and the half of the \HI\ width, 
$\sim$100~\kms), and the mass-to-luminosity ratios derived from them, \MHil=0.38 and \Mdynl$\sim$1. Indeed, if all these values are right, around 
37\% of the mass of the system is neutral gas. 
We consider that, because of the detection of an important population of old stars within the galaxy and its relatively high metallicity, the 
dynamical mass of III~Zw~107 has been probably underestimated. Hence, the \HI\ distribution should be several times larger than the optical extent. 
The comparison of the velocity amplitudes between the optical ($\sim$30~\kms) and the radio ($\sim$100~\kms) data strongly supports this idea. 
Perhaps, the neutral gas has been expelled and/or dispersed as a consequence of the possible interaction or merging 
between the two main objects observed in III~Zw~107. An interferometric \HI\ map of this galaxy is needed to answer to all these issues. In any case, 
the gas depletion timescale is high, $\tau\sim$3.5 Gyr, indicating that there is still a lot of neutral gas available to form new stars. 



\subsection{Tol 9}

\begin{figure}[t!]
\centering
\includegraphics[angle=270,width=\linewidth]{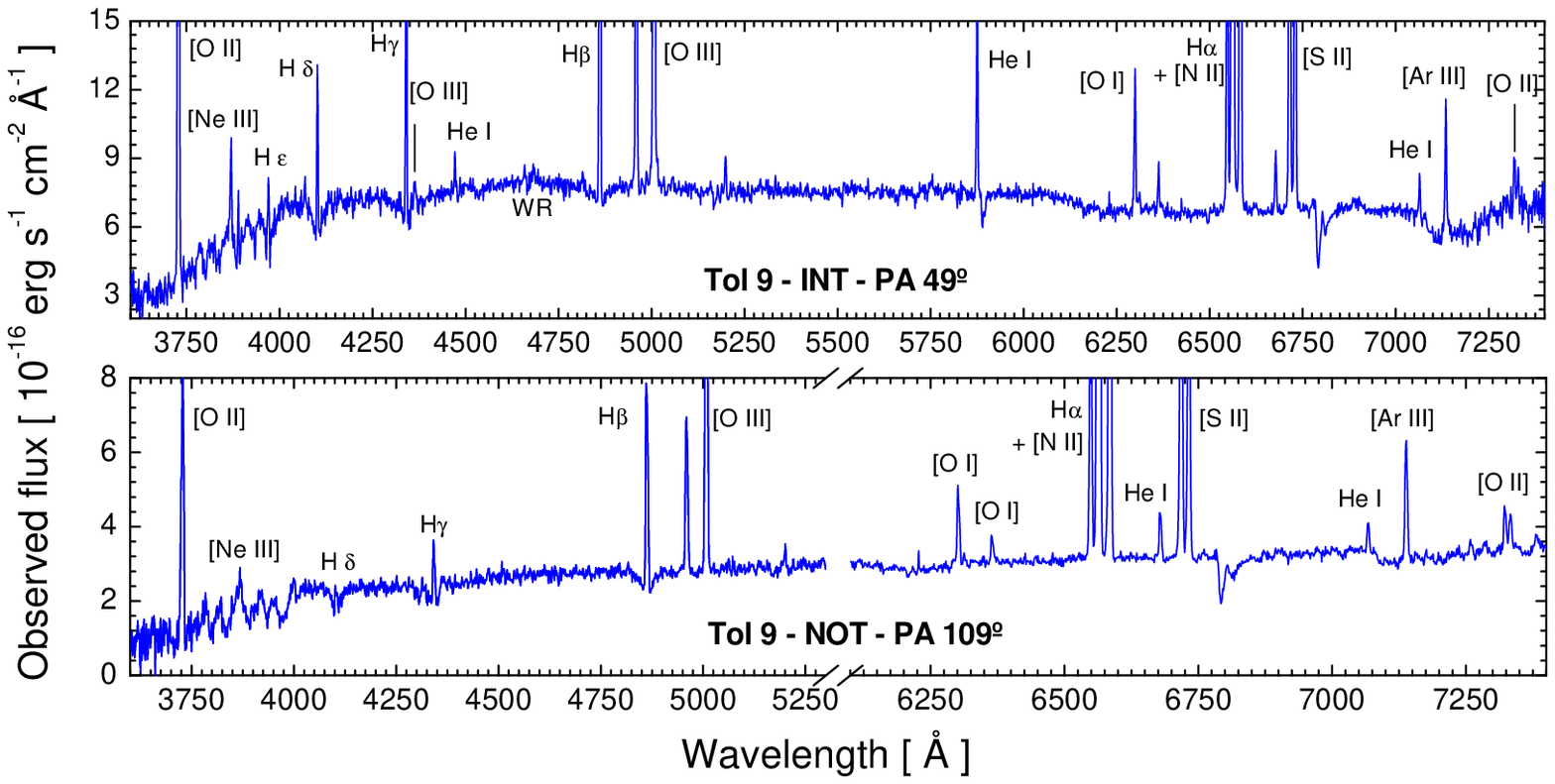}
\caption{\footnotesize{IDS INT spectrum  (\emph{bottom}) using a slit with PA 49$^{\circ}$ and ALFOSC \NOTe\ spectrum (\emph{top}) using a slit of PA 
109$^{\circ}$ of Tol 9. Fluxes are not been corrected for reddening. The most important emission lines have been labeled. See Figures~29 and 30 in 
Paper~I for identification of the regions.}}
\label{tol9espectros}
\end{figure}

The WR nature of \object{Tol~9} has been controversial. \citet{Penston97} indicated a probable detection of a faint emission line around 
$\lambda$4686. \citet{KS86} did not find the blue \WRBUMP\ or the \HeII\ emission line but suggested the detection of the red \WRBUMP.  Both 
\citet{C91} and \citet*{SCP99} included Tol~9 in their lists of candidate WR galaxies.

We observed two slit positions of Tol~9 (see Figure~29 of Paper~I). For the first one, we used the IDS instrument at the \INTe, 
using a slit with a PA of 49$^{\circ}$ that crosses the center of Tol~9 and the dwarf companion galaxy located at the SW. The second slit position 
was taken with the ALFOSC instrument at the \NOTe, choosing a slit position almost perpendicular to that used at the \INTe. This last slit position 
was centered at the SW of the center of the galaxy and the PA was set to 109$^{\circ}$,  our main objective was to analyze the kinematics and 
properties of the filamentary structure of ionized gas found 
in our deep \Ha\ images (see Figure~30 of Paper~I). The list of all the emission lines observed using both slits, as well as other important 
properties of the spectra, 
are shown in Table~\ref{iiizw107lineas}. As we see, both spectra show very similar line intensities. Notice that the radial 
velocity obtained for Tol~9 using our optical spectra is $\sim$200~\kms\ more positive than the value previously reported and listed in the \NED\ 
\citep{Lauberts89}.

The spectra obtained for Tol~9 show nebular emission and a continuum dominated by stellar absorptions in the \HI\ Balmer lines. 
We also observe an important decrement in the continuum at the blue range of the spectra. 
This feature can be explained by both the contribution of the older stars and an high extinction. We detect both the blue \WRBUMP\ and the 
nebular \HeII\ $\lambda$4686 emission line in the \INTe\ spectrum (Figure~\ref{wrfig1}), indicating the presence of WNL stars at the center of Tol 9. 
However, we do not have a clear detection of the red \WRBUMP\ (Figure~\ref{wrfig2}) although there are some evidences that it is there. If WNL stars 
are seen in Tol 9, WCE stars should also exist because of the high metallicity of the ionized gas. Deep spectroscopy with higher S/N ratio and 
spectral resolution is therefore needed to confirm this feature.   


\subsubsection{Physical conditions of the ionized gas}

The \INTe\ spectrum of Tol~9 shows [\ion{O}{iii}] $\lambda$4363, [\ion{N}{ii}] $\lambda$5755 and the [\ion{O}{ii}] $\lambda\lambda$7319,7330 doublet, 
and hence we derived the electron temperatures using the direct method. The results (that are compiled in Table~\ref{iiizw107abun}) are  
\TeOiii$\sim$7600~K and \Te(low)$\sim$8300~K. For the spectrum obtained using the 
\NOTe\ data we computed \TeOii\ using the [\ion{O}{ii}] lines and estimate the \TeOiii\ using Garnett's relation. In all cases, the electron 
densities were in the low-density limit. As we said, the values for the reddening coefficients are high, about  \CHb$\sim$0.45. The 
absorption equivalent widths in the \HI\ Balmer lines, computed interactively with \CHb, are large and give values of $W_{abs}\sim$6--8 \AA. The 
comparison of the data with the diagnostic diagrams classifies all regions as starbursts, but it seems that there is a small shock contribution to 
the ionization of the gas in the region analyzed with the \NOTe\ spectrum.

\subsubsection{Chemical abundances}

Table~\ref{iiizw107abun} compiles all the chemical abundances computed in Tol~9, showing almost identical results for both spectra. The average value 
of the oxygen abundance derived using the direct method is \abox=8.57$\pm$0.10.
This value is around 0.8 dex higher than that provided by \citet{KS86}. 
The average N/O ratio, \lno=$-$0.81$\pm$0.11, is the expected for a galaxy with the oxygen abundance found in Tol~9. 
The rest of the chemical abundances computed for this galaxy averaging both set of data are log(S/O)=$-1.62\pm0.12$, log(Ne/O)=$-0.72\pm0.14$ and 
log(Ar/O)=$-2.45\pm0.15$.

\subsubsection{Kinematics of the ionized gas}

\begin{figure}[t!]
\centering
\includegraphics[angle=270,width=\linewidth]{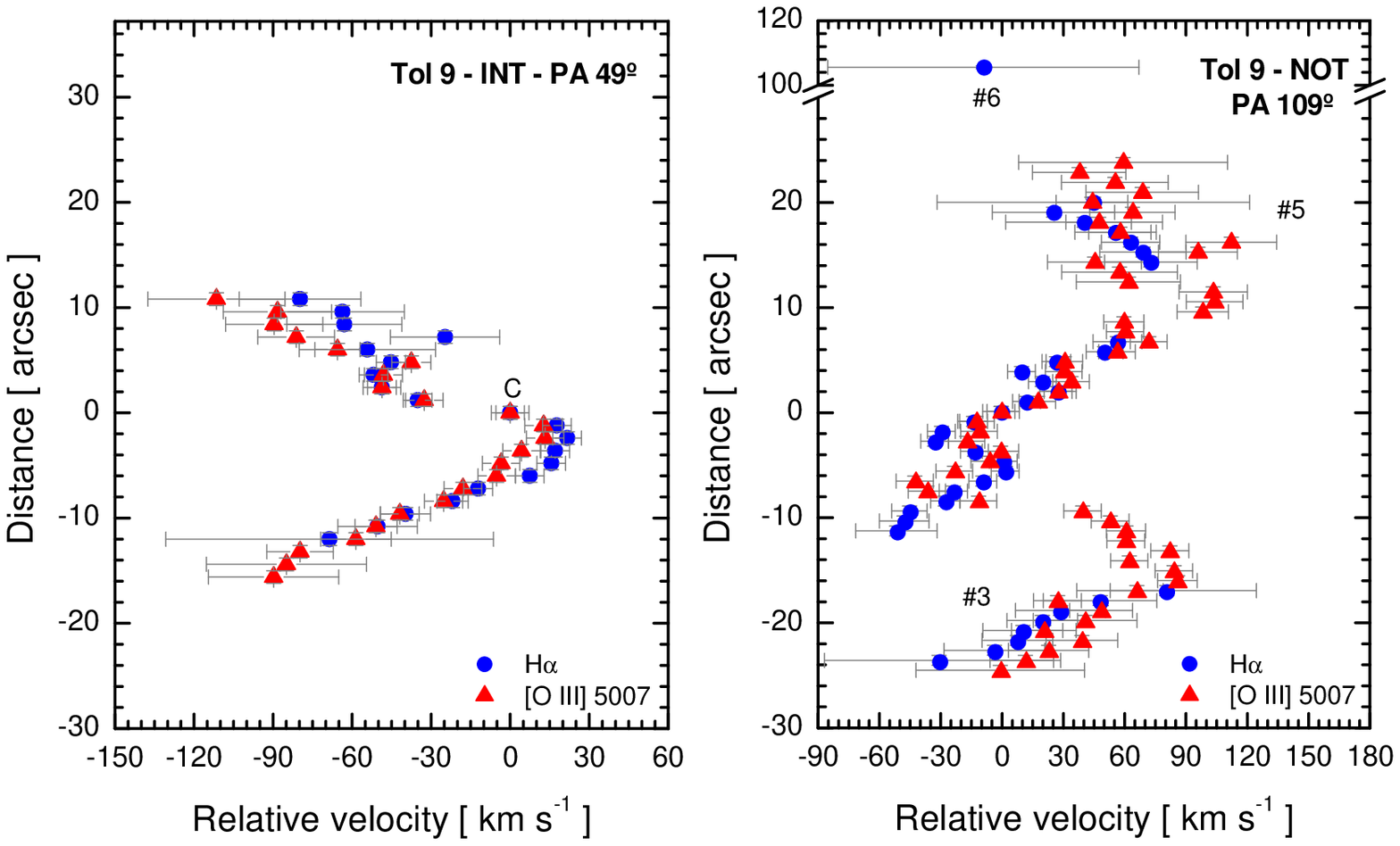}
\caption{\footnotesize{Position-velocity diagrams for the slit positions observed in Tol 9: PA 49$^{\circ}$ (\emph{left}, using \INTe\ data) and PA 
109$^{\circ}$ (\emph{right}, using \NOTe\ data). Both the \Ha\ (circles) and the [\ion{O}{iii}] $\lambda$5007 (triangles) profiles were analyzed. 
Notice that the y-axis is broken in two parts in the right diagram. NE is up in the left diagram and NW is up in the right diagram. See Figures~29 
and 30 in Paper~I for identification of the regions.}}
\label{tol9curvas}
\end{figure}

The kinematics of the ionized gas in Tol~9 were analyzed using our bidimensional spectra. We extracted 3 pixel bins (1.2$\arcsec$) and 5 pixel bins 
(0.95$\arcsec$) along the \Ha\ and the [\ion{O}{iii}] $\lambda$5007 profiles for the \INTe\ and the \NOTe\ spectra, respectively. The reference 
velocity was always chosen in the brightest region. The position-velocity diagrams are shown in Figure~\ref{tol9curvas}. Although the diagrams have 
an excellent agreement between the \Ha\ and the [\ion{O}{iii}] $\lambda$5007 results, their interpretation is not easy. 

First, the diagram with AP 49$^{\circ}$, that crosses the center of Tol~9, does not show a rotation pattern. Indeed, we observe two velocity 
gradients: while the velocity changes $\sim$120~\kms\ from the NE regions to the center, this tendency is completely reversed in the SW region, that 
shows a velocity variation of $\sim-$120 \kms. Notice that the center of Tol~9 is not located in the (0,0) position of this diagram because the 
maximum of the \Ha\ emission is displaced $\sim$10$\arcsec$ towards the SW of the center. Perhaps the velocity pattern we seen at this PA is a 
combination of rotation in the center and NE areas and the velocity gradient produced by the optical tail we detected in our deep optical images that 
connects Tol~9 with a dwarf companion galaxy located at the SW (see 
Figure~29 of Paper~I). Assuming $\Delta v\sim$70~\kms\ 
within a radius of $\sim12\arcsec$ (2.5~kpc) and an inclination angle of $i\sim$50$^{\circ}$, we estimate a Keplerian mass of 
$\sim$1.2$\times$10$^9$~\Mo\ and a mass-to-luminosity radio of \Mkepl$\sim$0.16. This last value is very low compared with that found in similar 
objects, and therefore we 
think that the mass of the galaxy has been underestimated. Interferometric \HI\ measurements are needed to get a reliable estimation of the dynamical 
mass of this galaxy because single-dish observations (such those provided by \HIPASS) would blend the \HI\ gas in Tol~9 and the nearby spiral 
ESO~436-46. This analysis will be performed using the new \HI\ ATCA data for the galaxy group where Tol~9 resides obtained by our group \citep{LS10}.  
Using the \FIR\ fluxes, the warm dust mass is \Mdust$\sim$2.41$\times$10$^6$~\Mo. Following the analysis performed by \citet*{Bettoni03}, the 
derived mass-to-luminosity ratio, \Mdustl$\sim$3.1$\times$10$^{-4}$, agree with the typical value found in spiral galaxies 
(\Mdustl$\sim$2$\times$10$^{-4}$).

On the other hand, the position-velocity diagram obtained using the slit with PA~109$^{\circ}$, shows several reverses in the velocity of the ionized 
gas. Regions \#3 and \#5, that we identified as two filamentary structures using our deep \Ha\ map (see Figure~30 of Paper~I), are labeled in this 
diagram. They show a similar kinematic behavior (a variation of $\sim-$90 \kms\ in their velocities within the same distance, $\sim12\arcsec$ 
(2.5~kpc), with respect to the 
center of the system. This kinematic structure, which is not coincident with any stellar distribution, reminds an expanding bipolar bubble, 
reinforcing the hypothesis that the \Ha\ envelope surrounding Tol~9 is consequence of some kind of galactic wind. Knot \#6 is barely detected at 
around 100$\arcsec$ (21~kpc) from the maximum of \Ha\ emission, but it 
seems to show a radial velocity similar to that observed at the ending of the filament \#5. 
This fact suggests that knot \#6 has been kinematically coupled to the main 
filamentary structure. Maybe, knot \#6 has been expelled from the expanding bubble. 3D optical spectroscopy is needed to clarify all these issues, as 
well as to compare the kinematics of the stellar and ionized gas components.  



\subsection{Tol 1457-262}

\object{Tol~1457-262} was studying using spectroscopy by \citet*{Winkler88,Terlevich91,KD01,Westera04} and \citet{Buckalew05}.  
\citet*{SCP99} included Tol~1457-262 in their WR galaxies catalogue because \citet{Contini96} detected the broad \ion{He}{ii} $\lambda$4686 emission 
line in the brightest region of the western object. This feature was also reported by \citet{Pindao99}. Both authors also detected WR features in 
another region of the same object, although while \citet{Contini96} observed the nebular \ion{He}{ii} $\lambda$4686 emission line, \citet{Pindao99} 
only reported the detection of the blue \WRBUMP. 

\begin{figure}[t!]
\centering
\includegraphics[angle=270,width=\linewidth]{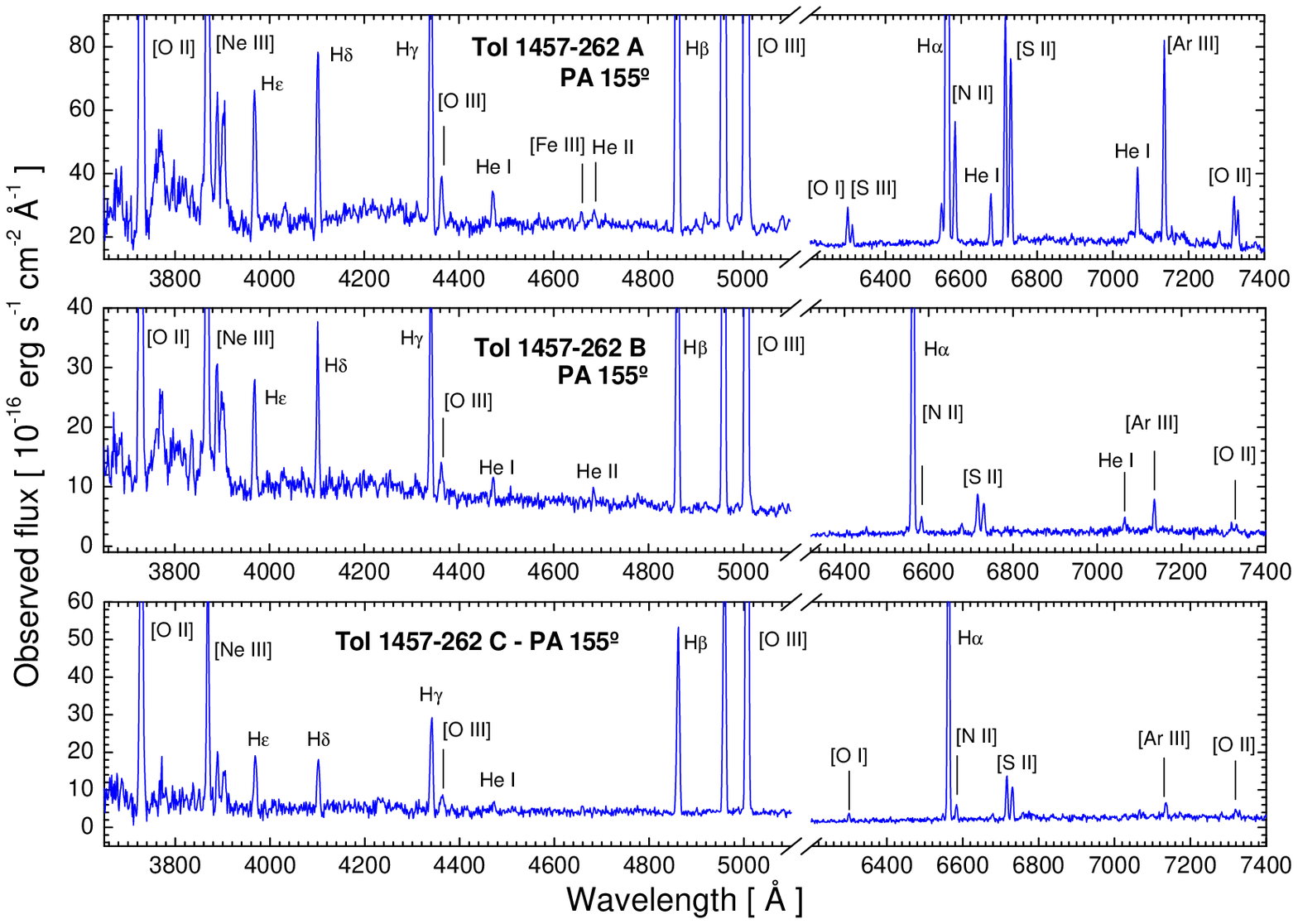}
\caption{\footnotesize{ALFOSC \NOTe\ spectra of the regions analyzed in Tol 1457-262. Fluxes are not corrected for reddening. The most important 
emission lines have been labeled. See Figure~31 in Paper~I for identification of the regions.}}
\label{tol1457espectros}
\end{figure}

Figure~\ref{tol1457espectros} shows the spectra extracted for regions A, B and C of the western object in Tol~1457-262 (\emph{Object 1}, see 
Figure~31 of Paper~I) using the instrument ALFOSC at the \NOTe\ and a slit with a PA of 155$^{\circ}$.
Table~\ref{tol1457lineas} compiles all the line intensities ratios and other properties of all the spectra 
analyzed in this galaxy. Stellar absorptions are barely detected, indicating that the nebular 
emission strongly dominates their spectra. Broad low-intensity wings are detected in the \Ha\ profile in region A.
We observe the nebular \ion{He}{ii} $\lambda$4686 emission line in the spectrum of region A; 
this feature is also observed in the spectrum of region~B (Figure~\ref{wrfig1}). 
We do not see the blue or the red (Figure~\ref{wrfig2}) \WRBUMP s in any region besides the good S/N of the spectra.

\subsubsection{Physical conditions of the ionized gas}

The electron temperatures were computed using the direct method because of the detection of [\ion{O}{iii}] $\lambda$4363 and [\ion{O}{ii}] 
$\lambda\lambda$7318,7330 lines in all the spectra. The derived values, that are compiled in Table~\ref{tol1457abun}, agree well with the empirical 
relation between  \TeOii\ and \TeOiii\ provided by \citet{G92}.
Despite their similar ionization degree, region B has an electron temperature, $T_e$(\ion{O}{iii})$\sim$15200~K, that is higher than that found in 
regions A and C, $T_e$(\ion{O}{iii})$\sim$14000~K. The electron densities found in 
regions A and C are similar, \Ne$\sim$200~cm$^{-3}$, but \Ne\ was in the low-density limit in region B. The determination of the reddening 
coefficient for region A was done using 5 \HI\ Balmer ratios, yielding consistently a very 
high value, \CHb=0.83$\pm$0.03. However, the \CHb\ found in region B gave a negative value. This result may not attributed to a bad flux calibration 
because adjacent regions A and C do not have this problem. Hence, we assumed \CHb$\sim$0 in region B\footnote{However, the Galactic value using 
\citet{SFD98} is \CHb=0.23$\pm$0.02.} and scaled the blue and red spectra considering the theoretical ratio between the \Ha\ and \Hb\ fluxes 
for the electron temperature estimated for this region. The comparison of the emission line ratios with the diagnostic diagrams indicates that all 
regions can be classified as starbursts. 

\subsubsection{Chemical abundances}

The chemical abundances computed for the regions analyzed in Tol~1457-262 are listed in Table~\ref{tol1457abun}. The oxygen abundance derived for the 
brightest knot (region A) is \abox=8.05$\pm$0.07, similar to that found in adjacent region~C, \abox=8.06$\pm$0.11, but higher than the oxygen 
abundance computed in region B, \abox=7.88$\pm$0.07. This would suggest that regions A and B have experienced different chemical evolution. However, 
the rest of the chemical abundances are relatively similar in all regions and only slightly higher in A and C. Remarkably weak are the [\ion{N}{ii}] 
emission lines, that give \lno$=-$1.57$\pm$0.11 and $-$1.61$\pm$0.12 for A and B, respectively. Our results do not agree with those reported by 
\citet*{Masegosa94} in region A, for which they computed \abox=8.23. These are the first abundance determinations for the rest of the objects. 


\subsubsection{Kinematics of the ionized gas}

\begin{figure}[t!]
\centering
\includegraphics[angle=270,width=\linewidth]{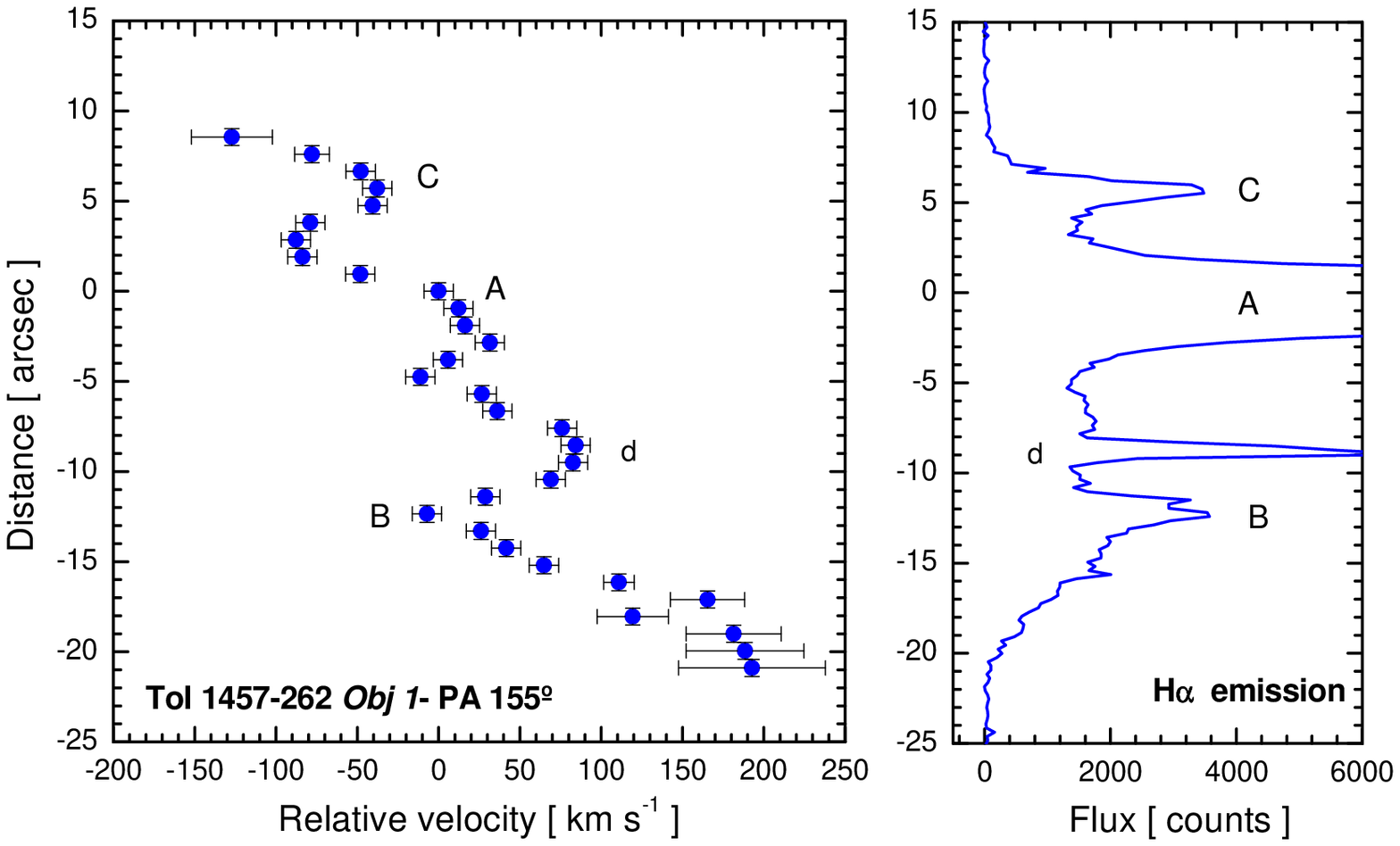}
\caption{\footnotesize{Position-velocity diagram for the slit position observed in \emph{Object~1} of Tol 1457-262 using the \Ha\ profile. The 
relative intensity of the \Ha\ emission along the spatial direction is also shown, identifying all observed regions. NW is up. See Figure~31 in 
Paper~I for identification of the regions.}}
\label{tol1457curvas}
\end{figure}

Figure~\ref{tol1457curvas} shows the position-velocity diagram of the slit position 
with PA~155$^{\circ}$ analyzed in \emph{Object~1} in Tol~1457-262. We extracted 5 pixel 
bins (0.95$\arcsec$) along the \Ha\ profile and took as reference the center of region A (the maximum of the \Ha\ emission). 
Figure~\ref{tol1457curvas} includes the relative intensity of the \Ha\ emission along the spatial direction. Although a velocity gradient between the 
northern regions ($v\sim-$120~\kms) and the southern areas  ($v\sim$190~\kms) is seen, we notice several velocity reverses along the system, being 
the most important that found between regions A and C, that has an amplitude larger than 100~\kms. These features indicate that the kinematics of the 
different star-forming regions found within \emph{Object~1} of Tol~1457-262 are decoupled. 
Indeed, the observed sinusoidal pattern suggests that the system is experiencing a merging process.
The fast increasing of the velocity between knot~B 
and the southernmost regions may be related to the movement of the material within the faint tail we detected in our deep optical images (see 
Figure~31 of Paper~I). We consider that knot C is a \TDG\ candidate because of its kinematics and chemical abundances are similar to those derived in 
bright region A.

Assuming that the general kinematics pattern is consequence of rotation, we derived a tentative Keplerian mass of  
\Mkep$\sim$6.2$\times$10$^9$~\Mo\ and a mass-to-luminosity ratio of \Mkepl$\sim$0.34 for \emph{Object 1} in Tol~1457-262. We considered $\Delta 
v\sim$60~\kms\  within a radius of $r\sim$15$\arcsec$ (4.95~kpc) and an inclination angle of $i\sim$55$^{\circ}$ (from our optical images). 
\HIPASS\ provides a detection of the \HI\ 
gas in this galaxy, for which we derive \MHi=4.7$\times$10$^9$~\Mo. However, this estimation for the neutral gas mass is for all the objects that 
compose Tol~1457-262 and therefore an interferometric \HI\ map is needed to quantify the amount of \HI\ and the dynamical mass of each member. The 
total luminosity of the system, computed from our optical data, is $L_B$=1.82$\times$10$^{10}$~\Lo, and hence the neutral hydrogen mass-to-luminosity 
radio of Tol~1457-262 is \MHil$\sim$0.26. This value is higher than the typical \MHil\ found in similar galaxies, indicating 
the large amount of neutral gas in Tol~1457-262. 



\subsection{Arp 252}

\begin{figure}[t!]
\centering
\includegraphics[angle=270,width=\linewidth]{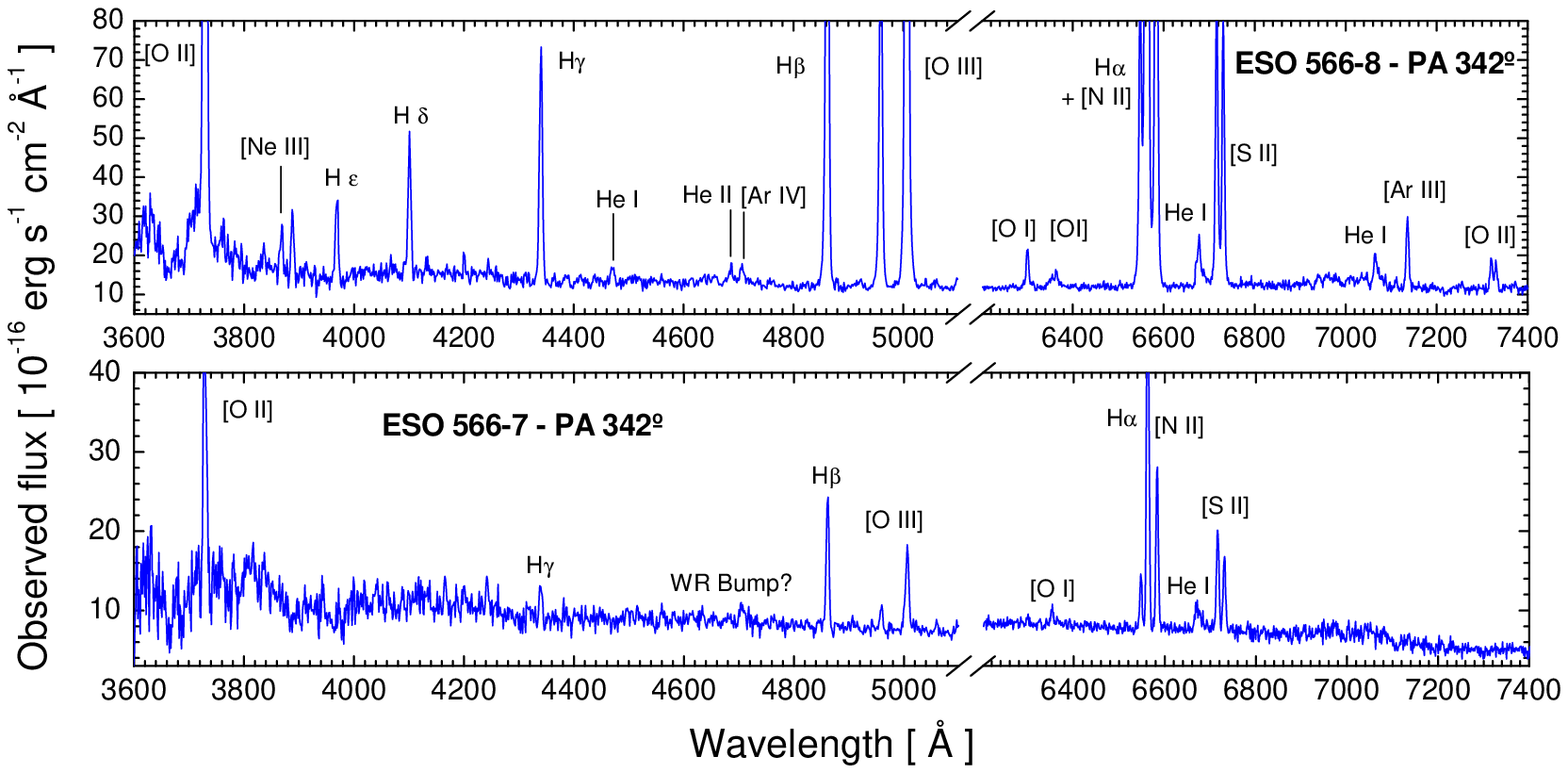}
\caption{\footnotesize{ALFOSC \NOTe\ spectra for the galaxy pair ESO 566-8 (\emph{top}) and ESO 566-7 (\emph{bottom}) that constitute Arp 256. Fluxes 
are not been corrected for reddening. The most important emission lines have been labeled. See Figure~34 in Paper~I for identification of the 
galaxies.}}
\label{eso566espectros}
\end{figure}

\object{Arp~252} is a pair of interacting galaxies designed \mbox{ESO 566-8 (galaxy A)} and ESO 566-7 (galaxy B).
Their spectroscopic properties were analyzed by \citet{PenaRM91} and \citet{Masegosa91}. These authors detected the blue \WRBUMP\ in 
\mbox{ESO~566-7}\footnote{\citet*{Masegosa91} named this object C~0942-1929A, but it is incorrect following \citet*{SCP99}.}. The WR feature was 
confirmed by \citet{Pindao99}, and hence ESO~566-7 was included in the latest catalogue of WR galaxies \citep*{SCP99}. \citet{Contini96} reported a 
tentative detection of the nebular \ion{He}{ii} $\lambda$4686 emission line in ESO~566-8, and therefore it was listed as a suspected WR galaxy 
\citep*{SCP99}.

Figure~\ref{eso566espectros} shows the optical spectra of ESO~566-8 and ESO~566-7 obtained using the instrument ALFOSC at the \NOTe\ using a slit 
with PA 342$^{\circ}$ (see Figure~34\footnote{Notice that the real slit position in Figure~34 of Paper~I is $342^{\circ}=-18^{\circ}$ and not 
$18^{\circ}$.} in Paper~I). The spectrum of 
ESO 566-8 shows many emission lines, but the spectrum of ESO~566-7 only shows a few. All line 
intensity ratios are listed in Table~\ref{tol1457lineas}. Although the spectra are dominated by the emission of the ionized gas, they also show some 
stellar absorptions, that are more evident in the weakest \HI\ Balmer lines observed in \mbox{ESO~566-7.} We detect the nebular \HeII\ $\lambda$4686 
emission line onto a faint broader feature in the spectrum of ESO~566-8 (Figure~\ref{wrfig1}). The red \WRBUMP\ is also observed in this object 
(Figure~\ref{wrfig2}), confirming the presence of both WNL and WCE stars in this galaxy.
Besides both  \citet*{Masegosa91} and \citet{Pindao99} reported a detection of the blue \WRBUMP\ in ESO 566-7, our spectrum does not show this 
feature. This may be consequence of the slit position chosen to get the spectrum of this galaxy, that does not cross along the major body of  ESO 
566-7 but almost perpendicular to it (see Figure~34 in Paper~I).

\subsubsection{Physical conditions of the ionized gas}

The detection of the [\ion{O}{ii}] $\lambda\lambda$7318,7329 doublet and the [\ion{N}{ii}] $\lambda$5755 emission line in the spectrum of 
\mbox{ESO~566-8} allowed the direct determination of the low ionization electron temperature, that gave values of \TeOii$\sim$9300~K and 
\TeNii$\sim$9000~K. Averaging 
these numbers, we estimate \Te(low)=9100$\pm$800 K. The high ionization electron temperature was computed using  Garnett's relation. 
The electron temperatures in ESO~566-7 were estimated via empirical calibrations. The electron densities were derived using the [\ion{S}{ii}] 
$\lambda\lambda$6617,6730 doublet. All results are tabulated in Table~\ref{tol1457abun}. The reddening coefficient are  
\CHb$\sim$0.49 in ESO~566-8 and \CHb$\sim$0.27 in ESO~566-7. 
The comparison of the observed emission line ratios with the diagnostic diagrams 
indicates that the nature of the ionization of the gas in ESO~566-7 is photoionized, but some shocks contribution seems to be present in ESO~566-8.

\subsubsection{Chemical abundances}

Table~\ref{tol1457abun} compiles all the chemical abundances derived for the galaxy members of Arp~252. The oxygen abundance in ESO~566-8 (galaxy A) 
is \abox=8.46$\pm$0.11 and was computed using the direct method. The empirical calibration provided by \citet{P01a} suggests that the oxygen 
abundance in ESO~566-7 (galaxy B), \mbox{\abox=8.50$\pm$0.16,} similar to that found in ESO~566-8. The N/O ratio is also similar in both galaxies, 
\lno$=-$0.76$\pm$0.12 in ESO~566-8. The value of the oxygen abundance in ESO 566-7 
reported by \citet*{Masegosa91}, \abox=8.54, agrees with our estimations.  

\subsubsection{Kinematics of the ionized gas}

\begin{figure}[t!]
\centering
\includegraphics[angle=270,width=\linewidth]{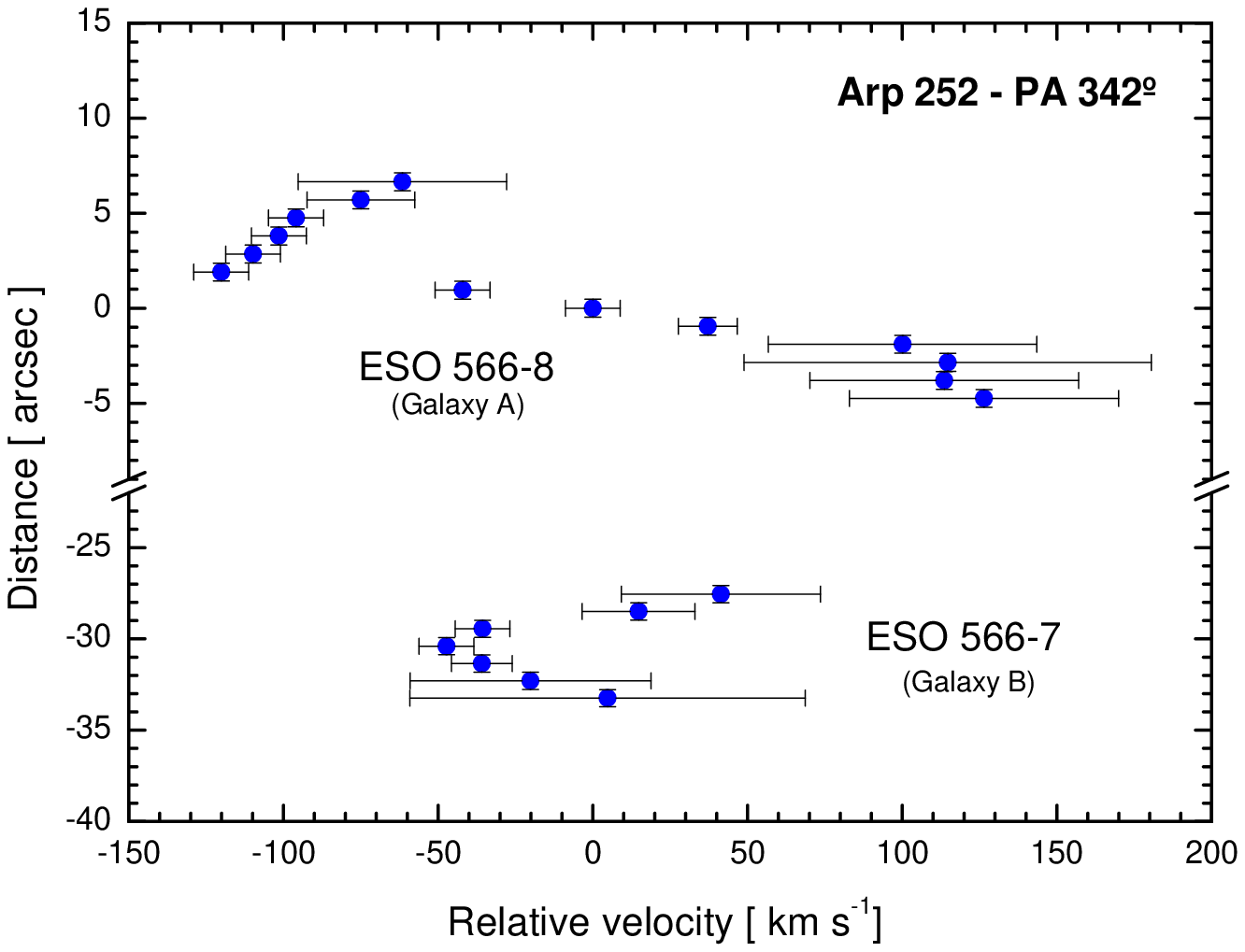}
\caption{\footnotesize{Position-velocity diagram for the slit position observed in Arp 252 using the \Ha\ profile.  Notice that the y-axis is broken 
in two parts. N is up. See Figure~34 in Paper~I for identification of the galaxies.}}
\label{eso566curvas}
\end{figure}

Figure~\ref{eso566curvas} shows the position-velocity diagram obtained for Arp~252 analyzing the bidimensional spectrum using a slit with 
PA~342$^{\circ}$. The \Ha\ profile was analyzed extracting 5 pixel bins 
(0.95$\arcsec$) and taking as reference the maximum of the emission in ESO~566-8.  As we see, both galaxies have a slight velocity difference of 
50~\kms\ between their centers. However, we appreciate important differences in the kinematics of the 
galaxies: while A seems to be rotating in its central regions (there is a velocity variation of $-$130~\kms at the north and 100~\kms\ at the south), 
B shows a distorted kinematic pattern. The velocity gradients observed in the upper and lower regions of galaxy A may correspond to the 
tidal streams induced in the long tails we observe in the optical images (see Figure~34 of Paper~I). 

We performed a tentative determination of the Keplerian mass of the galaxies. We assumed a velocity of  $\Delta v\sim$100~\kms\ within a radius of 
$r\sim$5$\arcsec$ (3.15~kpc) in ESO~566-8 and a velocity of $\Delta v\sim$30~\kms\ within a radius of $r\sim$3$\arcsec$ (1.89~kpc) in ESO~566-7. 
For both we 
considered an inclination angle of $i=90^\circ$, hence our Keplerian mass determinations are low limits to the real ones. We consider, however, that 
this assumption is not bad because in ESO~566-8 the northern tidal tail shows a high inclination angle with respect to the plane of the sky and 
ESO~566-7 shows its 
long southern tail almost in the plane of the sky and its disk seems to be edge-on. We estimated Keplerian masses of 
\Mkep$\sim$7.3$\times$10$^9$~\Mo\ for ESO~566-8 and \Mkep$\sim$4.0$\times$10$^8$~\Mo\ for ESO~566-7, 
that indicate a mass-to-luminosity ratios of  \Mkepl$\sim$0.21 and 0.05 for ESO~566-8 and ESO~566-7, respectively.
We consider that the Keplerian mass in ESO~566-7 has been highly underestimated.  
The warm dust mass of Arp~252, computed using the \FIR\ fluxes, is \Mdust$\sim$6.3$\times$10$^6$ \Mo.

Arp~252 is not detected in \HIPASS, and hence we can not derive the neutral gas mass and the dynamical mass. However, considering the absolute 
magnitude of the main galaxy,  $M_B=-20.9$, and despite the distance to the system ($D\sim$130~Mpc) we should expect some \HI\ emission. Hence, or 
Arp~252 does not have too much neutral gas or it has been lost in the intergalactic medium because of tidal effects. Finally, it would be very 
interesting to analyze the kinematics of knots \emph{c} and \emph{d} to check their probable \TDG\ nature (see Figure~34 of Paper~I).



\subsection{NGC 5253}

The echelle spectrophotometric analysis of the BCDG \object{NGC~5253} was presented in \citet{LSEGRPR07}. We measured the intensities of a large 
number of permitted and forbidden emission lines in four zones of the central part of the galaxy. The physical conditions of the ionized gas were 
derived using a large number of different line intensity ratios. Chemical abundances of He, N, O, Ne, S, Cl, Ar, and Fe were determined following the 
standard methods. We detected, for the first time in a dwarf starburst galaxy, faint \ion{C}{ii} and \ion{O}{ii} recombination lines. We confirmed 
the presence of a localized N enrichment in certain 
zones of the center of the galaxy and suggested a possible slight He overabundance in the same areas. We shown that the enrichment pattern agrees 
with that expected for the pollution by the eyecta of WR stars. The amount of enriched material needed to produce the observed overabundance is 
consistent with the mass lost by the number of WR stars estimated in the starbursts. 

The analysis of the \HI\ data provided by the \emph{Local Volumen \HI\ Survey} project \citep{Koribalski08} reveals that the neutral gas kinematics 
within NGC~5253 has a velocity gradient along the optical minor axis of the galaxy; it does not show any sign of regular rotation \citep{LS08a}. Some 
authors suggested that this feature is an outflow, but most likely its origin is the disruption/accretion of a dwarf gas-rich companion \citep{KS08} 
or the interaction with another galaxy in the M\,83 subgroup. The finding of a distorted \HI\ morphology in the external parts of the galaxy supports 
this hypothesis. A comprehensive analysis of the neutral gas within NGC~5253 will be presented elsewhere \citep{LS10}.


\section{Summary}

We have presented a detailed analysis of the ionized gas within 16 Wolf-Rayet galaxies using long-slit intermediate-resolution optical spectroscopy. 
In many cases, more than two star-forming regions have been studied per galaxy. We have analyzed the physical properties of the ionized gas, deriving 
their electron temperatures, electron density, the reddening coefficient and the stellar absorption underlying the \HI\ Balmer lines. We have 
confirmed that the excitation mechanism of the ionized gas in all bursts is mainly photoionization and not due to shock excitation as it happens 
in \AGN s and \LINER s. In the majority of the cases, we have computed the chemical abundances of O, N, S, Ne, Ar and Fe using the direct 
determination of the electron temperature (see second column in Table~2). When these data were not available, we used the empirical calibration of  
\citet{P01a,P01b} to get an estimation of the metallicity of the ionized gas. We have estimated the oxygen abundance of many new regions within the 
sample galaxies and refined the chemical properties of some of them, remarking regions in HCG~31, Mkn 1087, Mkn~1199, III~Zw~107, Tol~9, Tol~1457-262 
and NGC~5253.

\begin{table}[h!]
\centering
  \caption{\footnotesize{Comparison between the oxygen abundance of the regions analyzed here and their previous estimations compiled from the 
literature. The second column indicates if \Te\ was computed using the direct method (D) or via empirical calibrations (EC) in this work.}}
  \label{compara}
  \tiny
  \begin{tabular}{l@{\hspace{8pt}} c@{\hspace{6pt}}       c@{\hspace{12pt}}  c@{\hspace{12pt}}  c@{\hspace{12pt}}}   
  \tableline
   \noalign{\smallskip}
 & & \multicolumn{2}{c@{\hspace{4pt}}}{12+$\log$(O/H)}   &    \\
\cline{3-4}
\noalign{\smallskip}
Galaxy & \Te       &   This work     & Previous W. &   Ref  \\
\noalign{\smallskip}
\tableline
\noalign{\smallskip}

HCG 31 AC	    & D  &8.22$\pm$0.05 & 8.31        & VC92 \\
                &    &              & 8.3$\pm$0.2 &  R03 \\
HCG 31 A1       & EC &8.22$\pm$0.10 & \nodata   &  \nodata \\
HCG 31 B	    & D  &8.14$\pm$0.08 & 8.34$\pm$0.20 & R03 \\
HCG 31 E	    & D  &8.13$\pm$0.09 & \nodata & \nodata \\  
HCG 31 F1	    & D  &8.07$\pm$0.06 & 8.1$\pm$0.2 & R03 \\	
HCG 31 F2	    & D  &8.03$\pm$0.10 & \nodata & \nodata \\	
HCG 31 G	    & D  &8.15$\pm$0.07 & \nodata & \nodata	\\
HCG 31 H        & EC &8.3$\pm$0.2   & \nodata & \nodata \\
Mkn 1087	    & EC &8.57$\pm$0.10 & 8.55    &  VC92 \\
Mkn 1087 N	    & EC &8.23$\pm$0.10 & \nodata & \nodata	\\
Mkn 1087 \#7    & EC &8.54$\pm$0.10 & \nodata & \nodata \\
Haro 15 C	    & EC &8.37$\pm$0.10 &  8.33   & S05	 \\
Haro 15 A	    & D  &8.10$\pm$0.06 & \nodata & \nodata \\
Haro 15 B       & EC &8.21$\pm$0.14 & \nodata & \nodata \\
Mkn 1199	    & D  &8.75$\pm$0.12 &  8.19$\pm$0.18  & IT98 \\ 
                &    &              &  9.13 &  GIT00 \\
Mkn 1199 NE     & EC &8.46$\pm$0.13 &  \nodata & \nodata   \\
Mkn 5 A	        & D  &8.07$\pm$0.04 &  8.04$\pm$0.04 & IT99	 \\
Mkn 5 B         & EC &7.89$\pm$0.17 & \nodata  &\nodata \\
IRAS 08208+2816 C & D  &8.33$\pm$0.08 & 	\nodata & \nodata \\
IRAS 08208+2816 \#8& EC &8.64$\pm$0.15 & \nodata & \nodata \\
IRAS 08339+6517	 & EC &8.45$\pm$0.10 &	\nodata & \nodata   \\
IRAS 08339+6517 c.& EC &8.38$\pm$0.10 &	\nodata & \nodata \\
POX 4	        & D  &8.03$\pm$0.04 &	7.97$\pm$0.02  & KS96\\
POX 4 comp.   	& EC &8.03$\pm$0.14 &  \nodata & \nodata   \\
UM 420	        & D  &7.95$\pm$0.05 &	7.93$\pm$0.05  & IT98 \\
SBS 0926+606 A 	& D  &7.94$\pm$0.08 &	7.95$\pm$0.01  & IT98 \\
SBS 0926+606 B  & EC &8.15$\pm$0.16 &	\nodata & \nodata \\
SBS 0948+532    & D  &8.03$\pm$0.05 &	8.00$\pm$0.01  & IT98 \\
SBS 1054+365    & D  &8.00$\pm$0.07 &	7.97$\pm$0.02  & IT98    \\
SBS 1054+365 b	& EC &8.13$\pm$0.16 &   \nodata & \nodata \\
SBS 1211+540    & D  &7.65$\pm$0.04 &	7.64$\pm$0.01 & IT98 \\
SBS 1319+579 A$^a$	& D  &8.05$\pm$0.06 &	8.13$\pm$0.01 & ITL97 \\
                    &    &              &   8.09$\pm$0.03 & IT99   \\
SBS 1319+579 B$^a$	& D  &8.12$\pm$0.10 &	7.95$\pm$0.10 & ITL97   \\
SBS 1319+579 C$^a$	& D  &8.15$\pm$0.07 &	8.15$\pm$0.03 & ITL97 \\
                    &    &              &   8.11$\pm$0.01 & IT99	  \\
SBS 1415+437 C	& D  &7.58$\pm$0.05 &	7.61$\pm$0.01 & G03\\
SBS 1415+437 A	& D  &7.61$\pm$0.06 &	7.62$\pm$0.03 & G03	 \\
III Zw 107 A	& D  &8.23$\pm$0.09 &	8.20  & KJ85 \\ 
                &    &              &   7.90  & G97	 \\
III Zw 107 B,C     & EC &8.31$\pm$0.12 & \nodata & \nodata \\
Tol 9	        & D  &8.57$\pm$0.10 &	7.73 & KS86	 \\
Tol 1457-262 A	& D  &8.05$\pm$0.07 &   8.23  & M94 \\
Tol 1457-262 B	& D  &7.88$\pm$0.07 &  \nodata & \nodata \\
Tol 1457-262 C	& D  &8.06$\pm$0.11 &	\nodata & \nodata \\
ESO 566-8	    & D  &8.46$\pm$0.11 &  \nodata & \nodata \\
ESO 566-7	    & EC &8.50$\pm$0.16 &  8.54 &  M91\\
NGC 5253 A	    & D  &8.18$\pm$0.04 &  8.12$\pm$0.06 & K97 \\
NGC 5253 B	    & D  &8.19$\pm$0.04 &  8.19$\pm$0.07 & K97	 \\
NGC 5253 C	    & D  &8.28$\pm$0.04 &  8.16$\pm$0.12 & K97	 \\
NGC 5253 D	    & D  &8.31$\pm$0.07 & \nodata & \nodata	 \\
\noalign{\smallskip}
\tableline
  \end{tabular}
  \begin{flushleft}
  $^a$ We follow ITL97 names in SBS~1319+579. Notice, however, than IT99 named region~A to SBS~1319+579~C and region~B to SBS~1319+579~A; they did 
not consider SBS 1319+579~B because of its higher uncertainties. 
  REFERENCES: G97: \citet{Gallego97}; GIT00: \citet*{GIT00}; G03: \citet{Guseva03}; IT98: \citet{IT98}; IT99: \citet{IT99}; KS96: \citet{KS96}; K97:  
\citet{Kobulnicky97}; KJ85: \citet{KJ85}; KS86: \citet{KS86}; M91; \citet*{Masegosa91}; M94: \citet*{Masegosa94}; R03: \citet{R03}; S05: 
\citet{Shi05};  VC92: \citet{VC92}.
  \end{flushleft}
  \end{table}

 \begin{figure*}[t!]
\centering
\includegraphics[angle=90,width=0.725\linewidth]{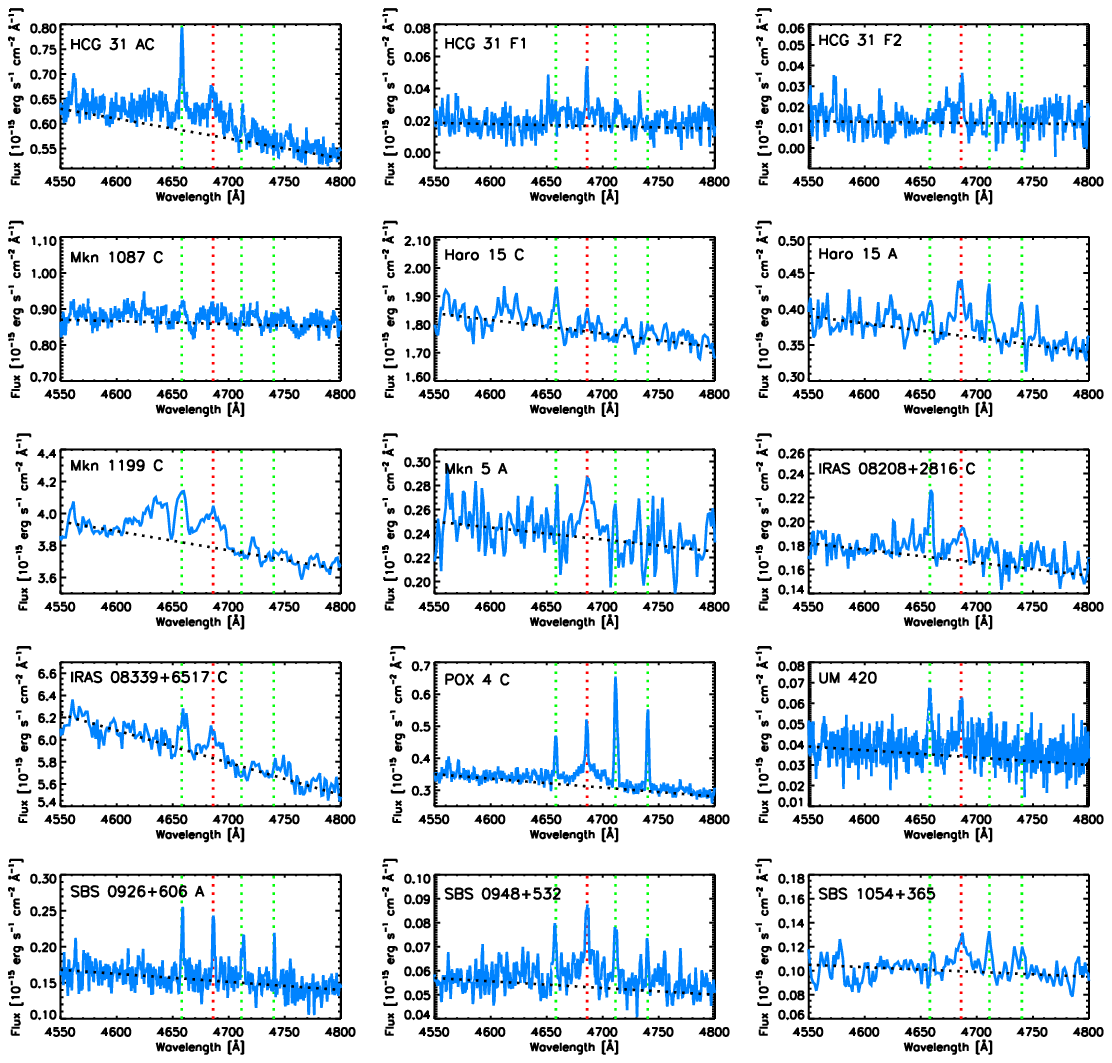} \\
\includegraphics[angle=90,width=0.725\linewidth]{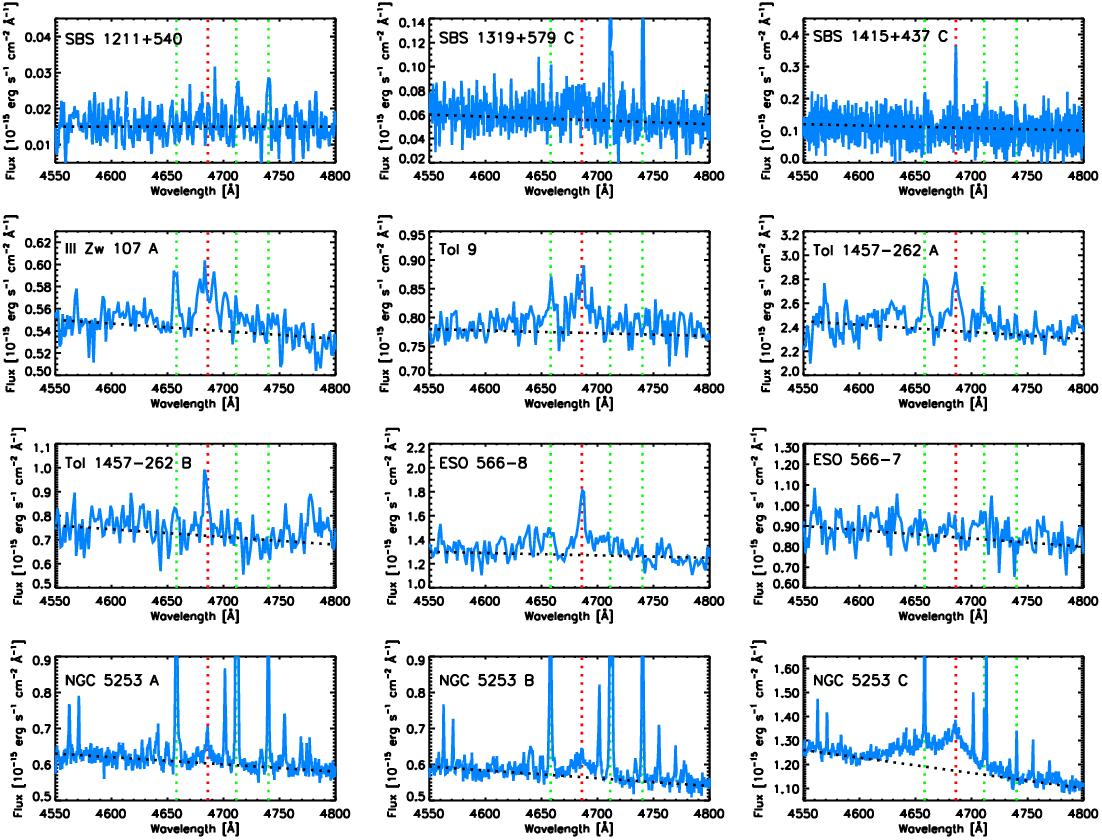} 
\protect\caption[ ]{\footnotesize{
Detail of the spectra of the main regions within our galaxy sample showing the zones around the blue \WRBUMP. The red dotted line represents the 
position of the \ion{He}{ii} $\lambda$4686 emission line, the blue dotted lines indicate the position of [\ion{Fe}{iii}] $\lambda$4658 and 
[\ion{Ar}{iv}] $\lambda\lambda$4711,4740 emission lines. The black dotted line represents the continuum level fitted by eye.}}
\label{wrfig1}
\end{figure*}

\begin{figure*}[t!]
\centering
\includegraphics[angle=90,width=0.75\linewidth]{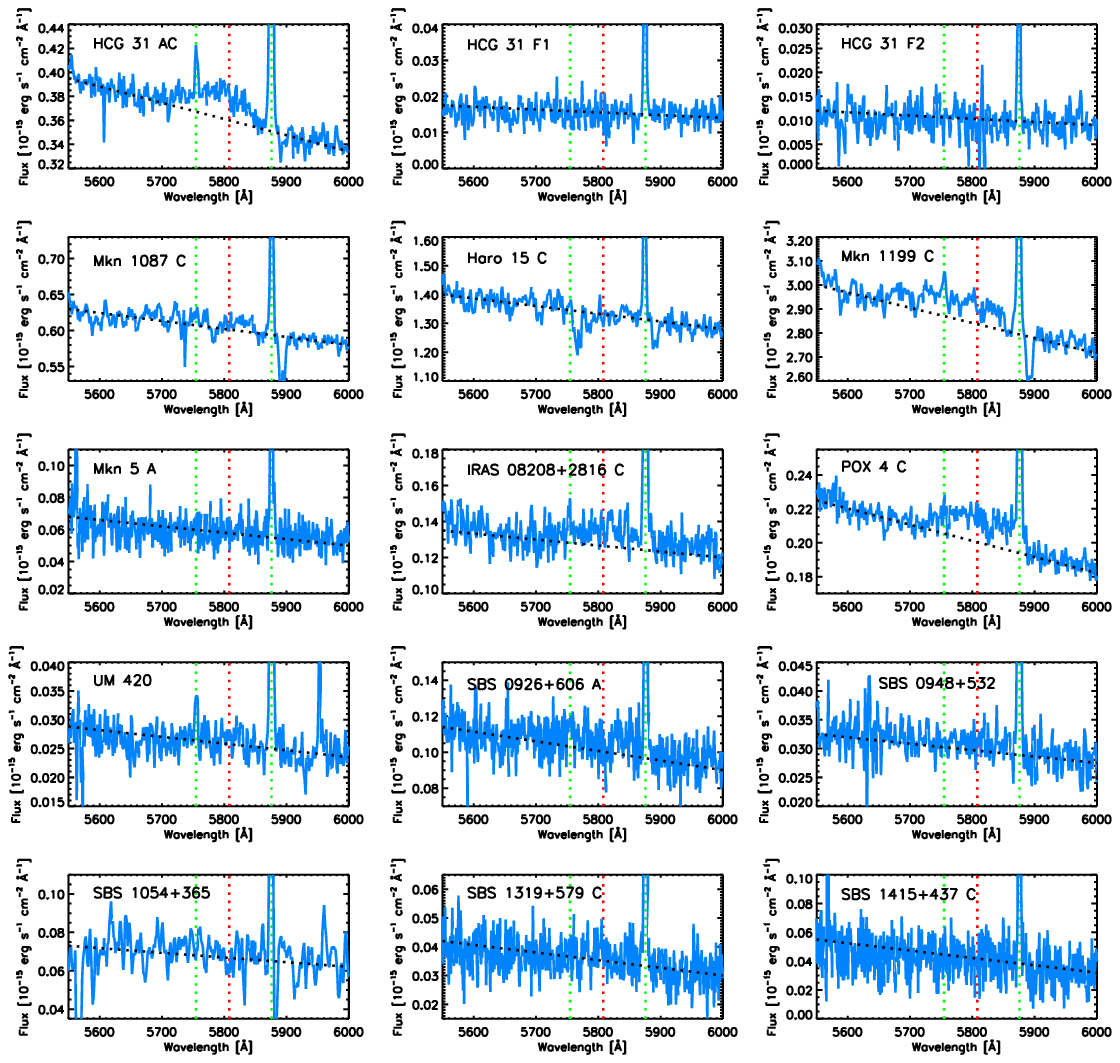} \\
\includegraphics[angle=90,width=0.75\linewidth]{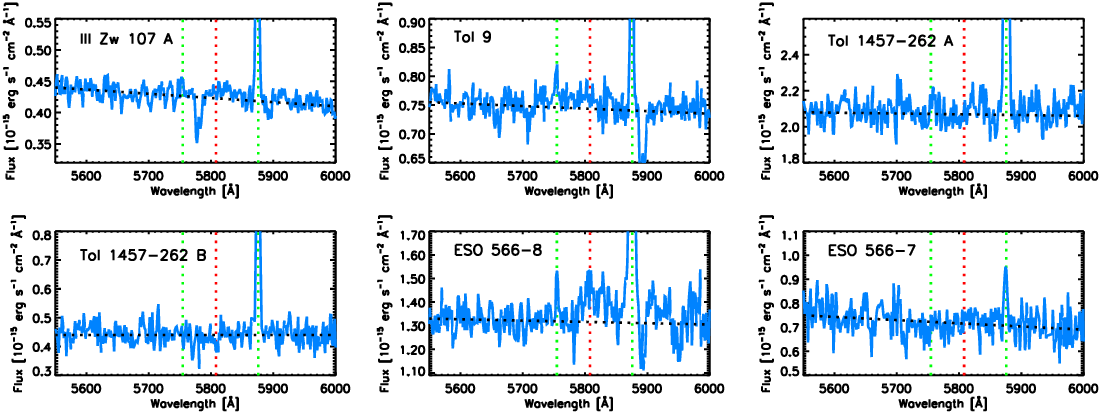} 
\protect\caption[ ]{\footnotesize{
Detail of the spectra of the main regions within our galaxy sample showing the zones around the red \WRBUMP. The red dotted line represents the 
position of the \ion{C}{iv} $\lambda$5808 emission line, the blue dotted lines indicate the position of [\ion{N}{ii}] $\lambda$5755 and \ion{He}{i} 
$\lambda$5875 emission lines. The black dotted line represents the continuum level fitted by eye. The red \WRBUMP\ is clearly identified in HCG~31~AC 
and POX~4, detected Mkn~1199 and ESO 566-8 and it also seems to be observed in IRAS~08208+2816 and Tol~9. We do not have data for SBS~1211+540 and 
NGC~5253 in this spectral range.}}
\label{wrfig2}
\end{figure*}

The derived physical and chemical properties were usually in agreement with previous 
observations reported in the literature. Table~\ref{compara} compares our oxygen abundance determinations with those previously reported in the 
literature. As we see, the majority of the results agree well with previous estimations, but there are important differences in Mkn~1199, III~Zw~107, 
Tol~9 and Tol~1457-262.

Including the data of the four systems analyzed in previous papers, a very useful database of objects with 
oxygen abundances between 7.58 and 8.75 in units of \abox\ is provided. In Papers~IV and V we will explore this database comparing their properties 
with other data derived from both our deep optical/NIR images and other multiwavelength observations available in the literature.  

We have confirmed the detection of Wolf-Rayet features in the majority of the galaxies, as we should expect because our sample was extracted from the 
latest WR galaxies catalogue \citep*{SCP99}. 
We have reported the detection of broad WR features in 20 regions within 16 systems. Figure~\ref{wrfig1} shows a detail of the optical spectrum in 
the 4600--4750 \AA\ range (blue \WRBUMP) of all important objects; faint regions with very low S/N have been excluded. We have indicated the spatial 
localization of the massive stars in each system. WR features are sometimes found in different knots within a same galaxy (HCG~31, Haro~15, 
Tol~1457-262, NGC~5253). The \ion{He}{ii} $\lambda$4686 emission line is unambiguously detected in 14 regions (HCG~31~AC and F1, Haro~15~A, Mkn~5, 
POX~4, UM~420, SBS~0926+606~A, SBS~0948+532, SBS~1415+437~C, Tol~1457-262~A and B, ESO~566-8 and NGC~5253~A and D), being specially strong in POX~4. 
Only in three objects previously listed as WR galaxies (Mkn~1087, SBS~1211+540 and ESO~566-7) we do not detect any feature that can be attributed to 
the presence of this sort of massive stars. 
We consider that aperture effects and the exact positioning of the slit onto the WR-rich bursts seem to play a fundamental role in the detection of 
the WR features. 

As it was expected, the red \WRBUMP\ (the broad \ion{C}{iii} $\lambda$5696 and \ion{C}{iv} $\lambda$5808 emission lines) is much more difficult to 
observe. Figure~\ref{wrfig2} shows a detail of the optical spectrum in the 5550--6000 \AA\ range (red \WRBUMP) of all important objects for which 
data is this spectral range is available. The broad \ion{C}{iv} $\lambda$5808 emission line is clearly identified in 2 galaxies (HCG~31~AC and 
POX~4), detected in 2 galaxies (Mkn~1199 and ESO~566-8) and it also seems to be observed in other 2 galaxies (IRAS~08208+2816 and Tol~9). However, we 
do not see this feature in galaxies for which previous detection have been reported [Mkn~5, UM~420, SBS~0926+606, SBS~0948+532, SBS~1054+365 and 
SBS~1319+579A, \citet{GIT00}]. We will discuss this result in Paper~III. A detailed quantitative analysis of the WR/(WR+O) and WC/WN ratios and the 
comparison with theoretical evolution models will be also performed in Paper~III.

\begin{figure}[t!]
\centering
\includegraphics[angle=270,width=\linewidth]{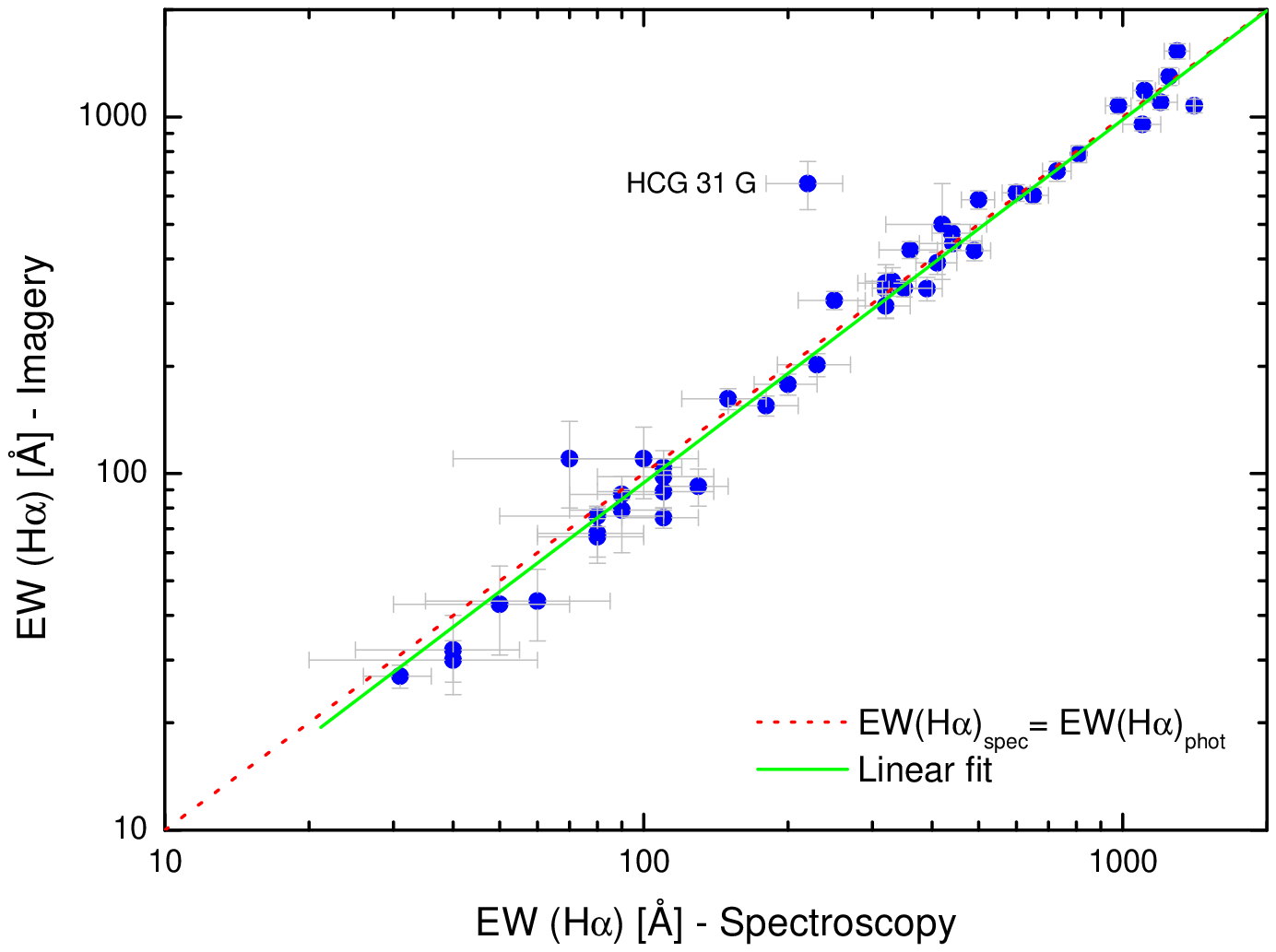}
\caption{\footnotesize{Comparison between the \Ha\ equivalent widths derived from our \Ha\ images and those obtained from the analysis of the optical 
spectrum for every particular star-forming region analyzed in this work.}}
\label{whawha}
\end{figure}

As it was commented in Paper~I, one of the best methods of determining the age of the last star-forming burst is through the \Ha\ equivalent width 
since it decreases with time. Our spectroscopic observations provide an independent estimation of \WHa\ within every knot. We have checked the 
correspondence between the values we obtained in our deep \Ha\ images (see Table~7 in Paper~I) with those derived from the spectroscopic data. 
Figure~\ref{whawha} plots such correlation, showing the excellent agreement of both kinds of data in almost all objects. Indeed, the linear fit to 
the data practically coincides with a $x=y$ function. Small divergences are found in very few objects, but they can be explained because of 
considerable differences in the relative sizes for which the photometric/spectroscopic values were extracted. The most evident case is member G in 
HCG~31, for which we only extracted the spectrum of a small knot at its NW, but the \WHa\ value derived from the images considers the flux of all the 
galaxy, that possesses a global star-formation activity lower than that observed in the NW knot.

\begin{figure}[t!]
\centering
\includegraphics[angle=270,width=\linewidth]{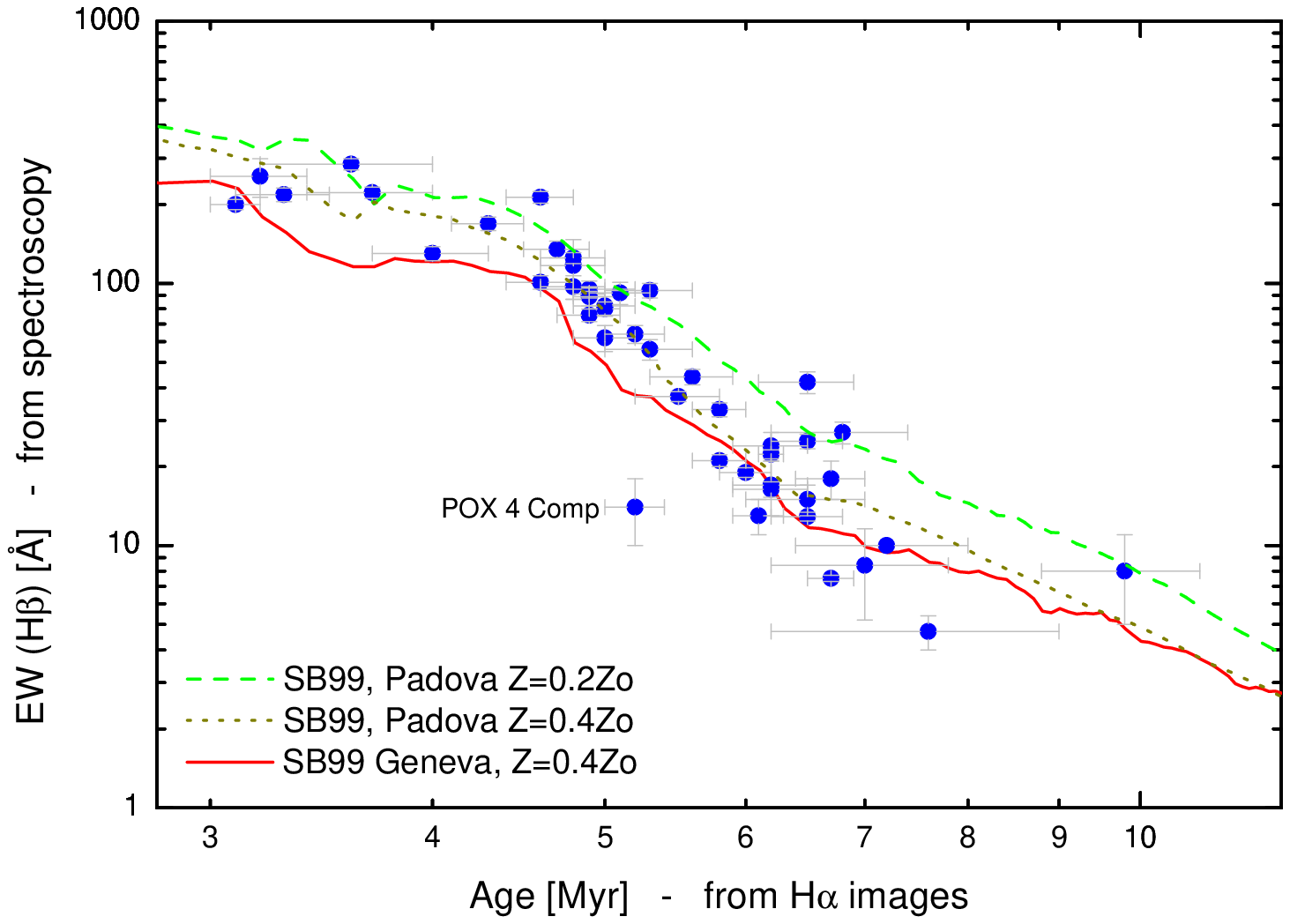}
\caption{\footnotesize{\Hb\ equivalent width vs. age of the most recent star-forming burst diagram comparing the predictions given by the 
evolutionary synthesis models provided by STARBURST~99 \citep{L99}. We include the $Z/Z_{\odot}$=0.4 model originally included in STARBURST~99 using 
Geneva tracks and two new models with $Z/Z_{\odot}$ = 0.2 and 0.4 than consider Padova tracks (Vazquez \& Leitherer, 2005). The age was computed from 
the \WHa\ given by our images; $W$(\Hb) was determined from our optical spectra.}}
\label{agewhb}
\end{figure}

Figure~\ref{agewhb} shows the comparison of our data with the theoretical predictions provided by the last release of the STARBURST~99 \citep{L99} 
models which uses Padova tracks
\citep{VL05}. We assumed an instantaneous burst 
with a Salpeter IMF, a total mass of $10^6$ \Mo, and a metallicity of $Z/Z_{\odot}$ = 0.2 and 0.4, the most common values according to the oxygen 
abundance of the majority of the knots. For comparison, we have also included the predictions of the original STARBURST~99 models, that consider 
Geneva tracks, for $Z/Z_{\odot}$=0.4. The ages of the last star-formation event are those estimated from the \WHa\ determined from our deep images, 
while the \Hb\ 
equivalent widths are those directly measured from our spectra. Thus, both set of data come from independent observations. As we see, 
the agreement is excellent, and therefore we are quite confident in the determination of the age of the star-forming regions. As it should be 
expected, the predictions given by the new models are better than those obtained from the old models. The only data point that does not follow the 
models is the dwarf companion object surrounding POX~4. However, as we explained in Paper~I, the values of the \WHa\ have been taken from images 
obtained by \citet{ME99} and seem to be somewhat overestimated.

\begin{figure*}[t!]
\centering
\begin{tabular}{cc}
\includegraphics[angle=90,width=0.47\linewidth]{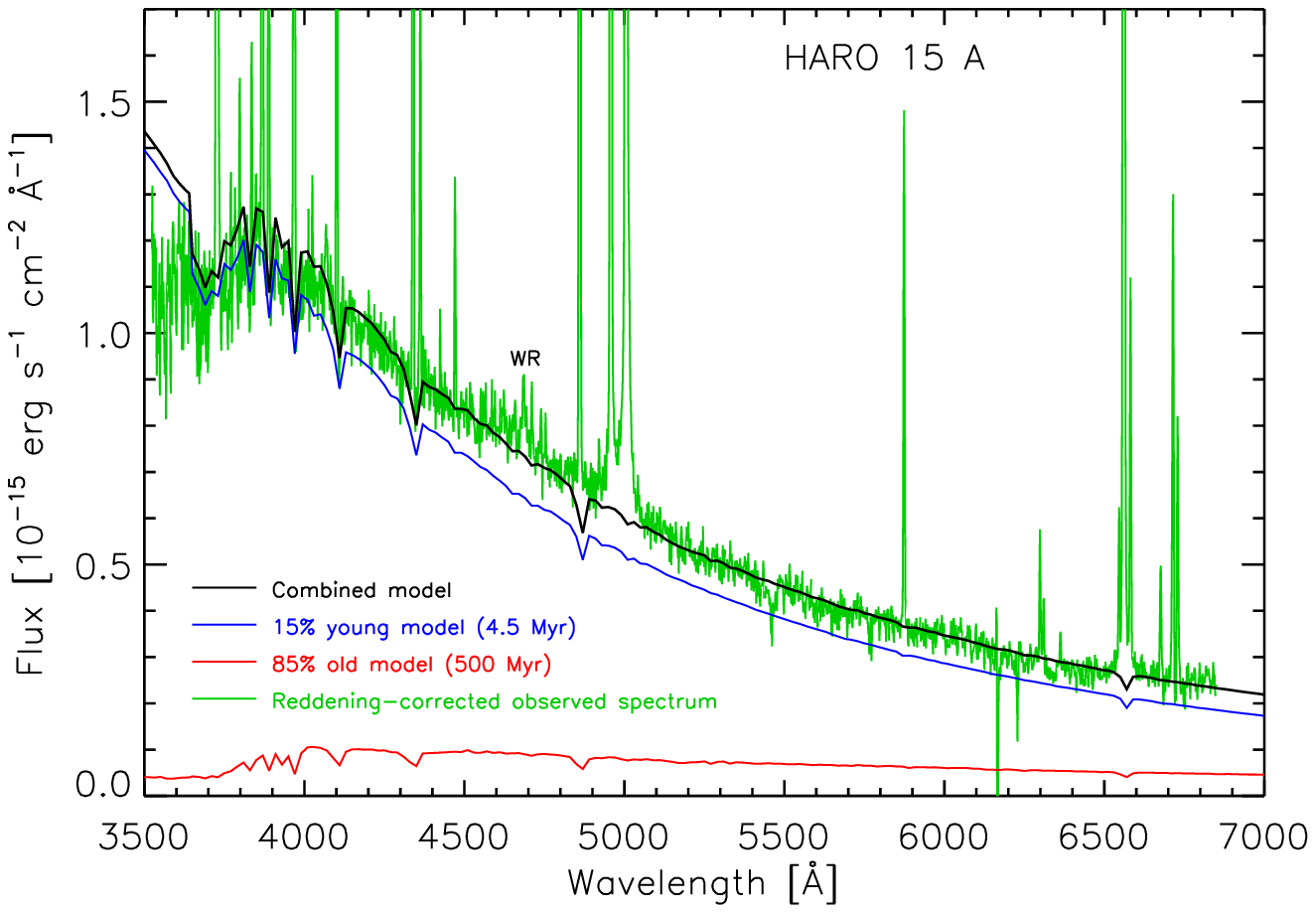} & 
\includegraphics[angle=90,width=0.47\linewidth]{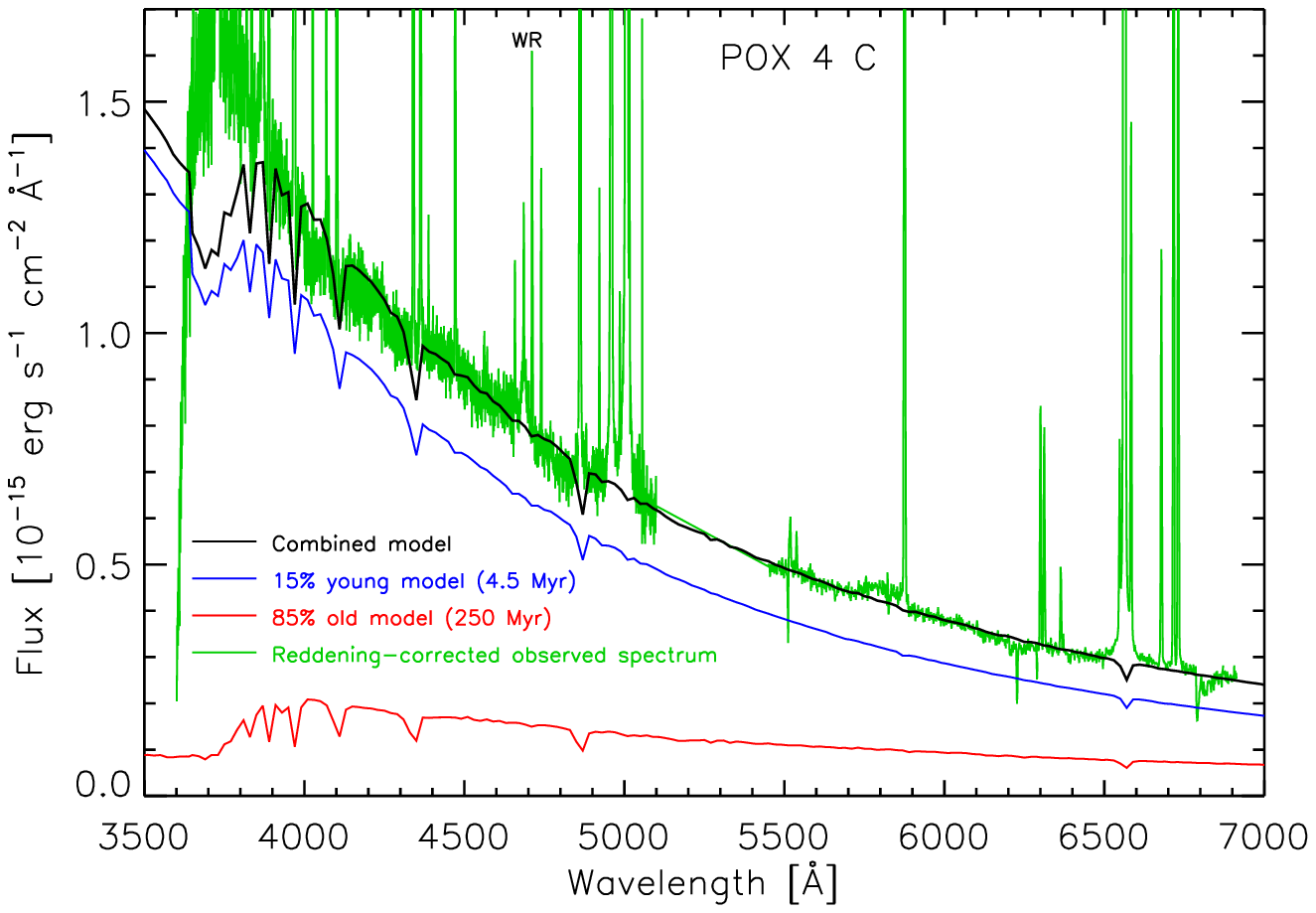}\\
\end{tabular}
\caption{\footnotesize{Spectra of region A in Haro~15 (\emph{left}) and the center of POX~4 (\emph{right}) compared with synthetic continuum 
spectral energy distributions obtained using the PEGASE.2 \citep{PEGASE97} code. The gray/green continuous line is the extinction-corrected spectrum, 
the upper continuous line corresponds to a model with an age of 4.5 Myr (young population model), whereas the lower continuous line is a 500 Myr (for 
Haro~15 A) or a 250 Myr (for POX~4 C) model (old population model). The shape of our observed derredened spectra fit in both cases with a model with 
a contribution of 15\% for the young population and 85\% for the old population (continuous black line over the galaxy spectrum).}}
\label{sed}
\end{figure*}

Although absorption features have been only detected in some of the object, 
all of them possess an old stellar population underlying the bursts, as we 
commented in the analysis of the optical/\NIR\ colors in Paper~I. This fact is evident from the values of the \Wabs\ derived from our spectra using 
the \HI\ Balmer lines, that seems to increase with increasing metallicity. A very powerful method to constraint the ages of the stellar populations 
within a starburst galaxy is the analysis of its spectral energy distributions (SED), although have the degeneracy problem between the interstellar 
extinction and the age of the old stellar population. 
In some objects, we have checked the results given by this method considering our estimation of the reddening contribution derived from the Balmer 
decrement, as we previously did in our analysis of the stellar populations in IRAS~08339+6517 \citep{LSEGR06}. We have made use of the PEGASE.2 code 
\citep{PEGASE97} to produce a grid of theo\-retical SEDs for an instantaneous burst of star formation and ages between 0 and 10~Gyr, assuming a 
$Z_{\odot}$ metalli\-city and a Salpeter IMF with lower and upper mass limits of 0.1 \Mo\ and 120 \Mo. Although the grid include the ionized gas 
emission, we have neglected it because its contribution to the continuum is almost irrelevant. In Figure~\ref{sed} we show the extinction-corrected 
spectra of the region A in Haro~15 (\emph{left}) and the center of POX~4 (\emph{right}) and two synthetic continuum spectral energy distributions 
assuming young (\emph{blue line}) and old (\emph{red line}) populations. We considered the ages derived from \WHa\ representative for the young 
population ages (4.5~Myr in both cases) and the ages of the underlying component estimated from the optical/\NIR\ colors  (500~Myr in Haro~15, 
250~Myr in POX~4). As we expected, none of the individual synthetic spectra fitted our observed SED. We then constructed a model than combines both 
young and old models. For both cases, the best fits are found when 15\% of the 4.5~Myr model and 85\% of the old model are considered. As we see, 
this combined model is in very good agreement with the shape of our derreddened spectrum. We conclude that, although the star formation activity is 
very intense in these starbursts, an important underlying old stellar population is usually found in the galaxies, indicating previous star-forming 
phenomena and ruling out the hypothesis that some of them are pristine dwarf galaxies.


Our study of the kinematics of the ionized gas and the morphology and environment included in this and previous papers 
of our group has revealed that 14 up to 20 of the analyzed galaxies show rather clear kinematical and/or morphological evidences of interaction or 
merging. The morphological evidences were presented and discussed in Paper I. The 
kinematical evidences presented here are of diverse nature: presence of objects with velocities decoupled from the main rotation pattern (Mkn~1087, 
Haro~15), sinusoidal velocity patterns that suggest a merging process (HCG~31~AC, Mkn~1199, IRAS~08208+2816, SBS~0926+606~A, III~Zw~107, 
\emph{Object~1} in Tol 1457-262), reverses in the velocity distribution
(Tol~9, Arp~252), indications of tidal streaming (HCG~31, IRAS~08208+2816, SBS~1319+579, Tol~9) or the presence of TDG candidates 
(HCG~31~F1 and F2, Mkn~1087, IRAS~08339+6517, POX~4, Tol 1457-262). The interaction could be between a spiral --or, in general, a non-dwarf-- galaxy 
(HCG~31, IRAS~08208+2816, Tol 1457-262, III~Zw~107 and Arp~252), between and spiral or non-dwarf galaxy and a dwarf one (Mkn~1087, Haro~15, Mkn~1199, 
IRAS~08339+6517, Tol~9), between two dwarf galaxies (POX~4, SBS~0926+606, SBS~1319+579). 
In the case of NGC~5253, we have a dwarf starburst galaxy that has suffered a possible interaction with a galaxy in the M~83 subgroup or with the 
spiral galaxy M~83 itself \citep{LS08a}.
These results reinforce the hypothesis that interaction with or between dwarf 
objects is an important mechanism to trigger the massive star 
formation in this kind of starbursts. These neighboring interacting dwarf or low-luminosity objects are only detected 
when a systematic and detailed analysis of the morphology, environment, chemical composition and kinematics of the 
objects are carried out. 

Finally, our detailed spectroscopical analysis has provided with some clear evidences of chemical differences within the 
objects or of interacting objects of different metallicities. For example, Mkn~1087, Haro~15 and Mkn~1199 are in clear interaction with dwarf 
galaxies with lower O/H and N/O ratios. NGC~5253, IRAS~08208+2816, and Tol~1457-262 show zones with different chemical compositions. In the case of 
NGC~5253 this is produced by localized pollution by massive stars, but in the cases of IRAS~08208+2816 and Tol~1457-262 the different chemical 
composition seem to be because the regions correspond to different galaxies in interaction. Apart from NGC~5253, two more galaxies: IRAS~08208+2816 
and UM~420, show a localized high N/O that could be a signature of contamination by WR winds.


\begin{acknowledgements}

\'A.R. L-S thanks C.E. (his PhD supervisor) for all the help and very valuable explanations, talks and discussions along these years. He also 
acknow\-ledges Jorge Garc\'{\i}a-Rojas, Sergio Sim\'on-D\'{\i}az and Jos\'e Caballero for their help and friendship during his PhD, extending this 
acknowledge to all people at Instituto de Astrof\'{\i}sica de Canarias (Spain). The authors thank B\"arbel Koribalski (CSIRO/ATNF) for her help 
analyzing HIPASS data.
\'A.R. L-S. \emph{deeply} thanks to Universidad de La Laguna (Tenerife, Spain) for force him to translate his PhD thesis from English to Spanish; he 
had to translate it from Spanish to English to complete this publication. 
This work has been partially funded by the Spanish Ministerio de Ciencia y Tecnolog\'{\i}a (MCyT) under project AYA2004-07466. 
This research has made use of the NASA/IPAC Extragalactic Database (NED) which is operated by the Jet Propulsion Laboratory, California Institute of 
Technology, under contract with the National Aeronautics and Space Administration. 

\end{acknowledgements}

\listofobjects

\appendix
\normalsize

\section{Tables}

\begin{table*}[t!]
\centering
  \caption{\footnotesize{Dereddened line intensity ratios with respect to $I$(\Hb)=100 for knots analyzed in Haro 15.}}
  \label{haro15lineas}
  \tiny
  \begin{tabular}{l  r rrr r}
  \noalign{\smallskip}
  \tableline
   \noalign{\smallskip}
Line  & $f(\lambda)$&  C  &  A  &  B  & D\\
\noalign{\smallskip}
 \tableline
\noalign{\smallskip}
 3705.04   He    I &   0.260 &       1.05:   &        \nodata   &        \nodata&        \nodata\\
 3728.00   [O  II] &   0.256 &   294$\pm$19   & 113$\pm$21   &        402$\pm$106 & 336$\pm$46\\
 3770.63    H    I &   0.249 &       0.68:   &  1.69$\pm$0.52   &        \nodata&        \nodata\\
 3797.90    H    I &   0.244 &       0.33:   &  3.10$\pm$1.00  &        \nodata&        \nodata\\
 3835.39    H    I &   0.237 &       1.69:   &  4.59$\pm$0.99   &        \nodata&        \nodata\\
 3868.75  [Ne III] &   0.230 &  14.2$\pm$2.8   &  48.6$\pm$8.4   &        \nodata& 39.4:\\
 3889.05    H    I &   0.226 &   7.6$\pm$3.0   &  15.5$\pm$2.8   &        \nodata&        \nodata\\
 3967.46  [Ne III] &   0.210 &        \nodata   &  29.6$\pm$4.8   &        \nodata&        \nodata\\
 3970.07    H    I &   0.210 &  19.6$\pm$1.7   &        \nodata   &        \nodata&        \nodata\\
 4026.21   He    I &   0.198 &        \nodata   &  1.05:   &        \nodata&        \nodata\\
 4068.60   [S  II] &   0.189 &       2.69:   &        \nodata   &        \nodata&        \nodata\\
 4101.74    H    I &   0.182 &  28.1$\pm$2.7   &  26.1$\pm$3.5   &        \nodata&26.1$\pm$7.0\\
 4340.47    H    I &   0.127 &  46.2$\pm$3.5   &  47.0$\pm$4.8   &        \nodata&46.7$\pm$9.9\\
 4363.21   [O III] &   0.121 &       1.22:   &  8.59$\pm$0.87   &        \nodata&3.9:\\
 4471.48   He    I &   0.095 &       3.56:   &  4.17$\pm$0.68   &        \nodata&        \nodata\\
 4658.10  [Fe III] &   0.050 &       1.75:   &       0.90:   &        \nodata&        \nodata\\
 4686.00   He   II &   0.043 &       0.49:   &  1.65$\pm$0.42   &        \nodata&        \nodata\\
 4711.37  [Ar  IV] &   0.037 &        \nodata   &  1.10:   &        \nodata&        \nodata\\
 4740.16  [Ar  IV] &   0.030 &        \nodata   &  0.73:   &        \nodata&        \nodata\\
 4754.83  [Fe III] &   0.026 &        \nodata   &       0.25:   &        \nodata&        \nodata\\
 4861.33    H    I &   0.000 & 100.0$\pm$6.7   & 100.0$\pm$5.9   &   100$\pm$39& 100$\pm$19\\
 4921.93   He    I &  -0.015 &        \nodata   &       0.55:   &        \nodata&        \nodata\\
 4958.91   [O III] &  -0.024 &  67.2$\pm$5.0   &   212$\pm$11   &    81$\pm$30& 132$\pm$19\\
 5006.84   [O III] &  -0.036 &   210$\pm$13   &   648$\pm$36   &   232$\pm$63& 383$\pm$44\\
 5015.68   He    I &  -0.038 &       1.02:   &        \nodata   &        \nodata&        \nodata\\
 5158.81  [Fe  II] &  -0.073 &       1.95:   &        \nodata   &        \nodata&        \nodata\\
 5197.90   [N   I] &  -0.082 &       2.76:   &        \nodata   &        \nodata&        \nodata\\
 5517.71  [Cl III] &  -0.154 &        \nodata   &       0.33:   &        \nodata&        \nodata\\
 5537.88  [Cl III] &  -0.158 &        \nodata   &       0.26:   &        \nodata&        \nodata\\
 5875.64   He    I &  -0.215 &  12.8$\pm$1.6   &  9.95$\pm$1.70   &      46.1:& 10.5$\pm$2.8\\
 6300.30   [O   I] &  -0.282 &   7.5$\pm$1.1   &  2.21$\pm$0.76   &        \nodata&7.7$\pm$3.1\\
 6312.10   [S III] &  -0.283 &       1.10:   &  1.16$\pm$0.31   &        \nodata&        \nodata\\
 6363.78   [O   I] &  -0.291 &       2.18:   &     0.65:   &        \nodata&        \nodata\\
 6548.03   [N  II] &  -0.318 &  20.1$\pm$1.8   &  2.82$\pm$0.71   &        \nodata&6.0:\\
 6562.82    H    I &  -0.320 &   288$\pm$19   &   202$\pm$46   &   284$\pm$75& 288$\pm$35\\
 6583.41   [N  II] &  -0.323 &  64.8$\pm$4.6   &  7.76$\pm$1.84   &      23.4:& 23.2$\pm$6.1\\
 6678.15   He    I &  -0.336 &       2.81:   &  2.55$\pm$0.74   &        \nodata& 3.9:\\
 6730.85   [S  II] &  -0.344 &  18.0$\pm$2.0   &  5.09$\pm$1.32   &      28.5:& 17.6$\pm$8.3 \\
\noalign{\smallskip}
 \tableline
 \noalign{\smallskip}
Aperture size (arcsec)       &   &    6$\times$1       &     8.4$\times$1     &   8$\times$1         & 4.4 $\times$1 \\
Distance$^b$ (arcsec)   &   &       0             &     13               &  23                  &  12  \\
$F$(\Hb)$^a$          &   &  23.25 $\pm$   1.08 &   23.42 $\pm$   0.89 &    0.52 $\pm$   0.10 & 1.32 $\pm$ 0.12 \\ 
$C$(\Hb)              &   &   0.11 $\pm$   0.03 &    0.33 $\pm$   0.03 &    0.06 $\pm$   0.03 & 0.37 $\pm$ 0.02\\
$W_{abs}$ (\AA)     &   &    2.4 $\pm$    0.4 &     1.3 $\pm$   0.3  &    0.5               & 2.2 $\pm$ 0.2\\            
\noalign{\smallskip}
$-W$(\Ha) (\AA)         &   &   75.2 $\pm$    5.0 &   423.6 $\pm$   22.5 &    43.9 $\pm$   10.0 & 48.8 $\pm$ 2.7\\ 
$-W$(\Hb) (\AA)         &   &   16.4 $\pm$    1.1 &    75.7 $\pm$    4.2 &     8.4 $\pm$    3.2 & 20.8 $\pm$ 1.3\\ 
$-W$(H$\gamma$) (\AA)   &   &    5.5 $\pm$    0.4 &    26.9 $\pm$    1.5 &     \nodata          &  7.5 $\pm$ 0.9  \\ 
$-W$([O III]) 5007 (\AA)&   &   29.4 $\pm$    1.8 &   462.7 $\pm$   23.1 &    20.2 $\pm$    5.1 & 77.9 $\pm$ 3.9 \\ 
\noalign{\smallskip}
 \tableline
  \end{tabular}
  \begin{flushleft}
  $^a$ In units of 10$^{-15}$ erg s$^{-1}$ cm$^{-2}$ and not corrected for extinction.\\
   $^b$ Relative distance with respect to the center of Haro~15.
  \end{flushleft}
\end{table*}

\begin{table*}[t!]
  \caption{\footnotesize{Physical conditions and chemical abundances of the ionized gas of the regions analyzed in Haro 15.}}
  \label{haro15abun}
  \tiny
  \centering
  \begin{tabular}{l  rrrr}
\noalign{\smallskip}
\tableline
\noalign{\smallskip}
Region                 &         C       &        A          &         B          &   D  \\
\noalign{\smallskip}
\tableline
\noalign{\smallskip}				      
$T_e$(O III) (K)        &    9500 $\pm$ 800$^a$&  12900 $\pm$ 700   &	11500 $\pm$ 1000$^a$& 11800 $\pm$ 800$^a$ \\
$T_e$(O II) (K)         &    9600 $\pm$ 600   &   12000 $\pm$ 500   &	11000 $\pm$ 700   & 11260 $\pm$ 600 \\
$n_e$ (cm$^{-3}$)       &     100             &     100             &	  100             & 100\\
\noalign{\smallskip}
12+log(O$^{+}$/H$^+$)   &    8.16 $\pm$ 0.11  &    7.35 $\pm$ 0.08  &	 8.04 $\pm$ 0.14  &	7.93 $\pm$ 0.13   \\
12+log(O$^{++}$/H$^+$)  &    7.94 $\pm$ 0.10  &    8.01 $\pm$ 0.06  &	 7.72 $\pm$ 0.13  &	7.90 $\pm$ 0.10   \\
12+log(O/H)             &    8.37 $\pm$ 0.10  &    8.10 $\pm$ 0.06  &	 8.21 $\pm$ 0.14  &	8.22 $\pm$ 0.11    \\
\noalign{\smallskip}
log(O$^{++}$/O$^+$)     & $-$0.23 $\pm$ 0.16  &    0.66 $\pm$ 0.10  & $-$0.32 $\pm$ 0.18  & $-$0.03 $\pm$ 0.15 \\ 
12+log(N$^+$/H$^+$)     &    7.13 $\pm$ 0.07  &    6.00 $\pm$ 0.06  &	 6.55 $\pm$ 0.19  &	6.47 $\pm$ 0.14   \\
12+log(N/H)             &    7.34 $\pm$ 0.10  &    6.75 $\pm$ 0.10  &	 6.72 $\pm$ 0.21  &	6.76 $\pm$ 0.15   \\
log(N/O)                & $-$1.03 $\pm$ 0.15  & $-$1.35 $\pm$ 0.11  & $-$1.49 $\pm$ 0.20  &$-$1.46 $\pm$ 0.16	   \\
\noalign{\smallskip}
12+log(S$^+$/H$^+$)     &    6.05 $\pm$ 0.10  &    5.26 $\pm$ 0.08  &	 6.08 $\pm$ 0.22  &	5.85 $\pm$ 0.21 \\
12+log(S$^{++}$/H$^+$)  &    6.52 $\pm$ 0.25  &    6.02 $\pm$ 0.13  &	 \nodata	      &	 \nodata  \\
12+log(S/H)             &    6.65 $\pm$ 0.20  &    6.20 $\pm$ 0.11  &	 \nodata	    &	  \nodata\\
log(S/O)                & $-$1.71 $\pm$ 0.18  & $-$1.89 $\pm$ 0.15  &	 \nodata	    &	  \nodata\\
\noalign{\smallskip}
12+log(Ne$^{++}$/H$^+$) &    7.29 $\pm$ 0.15  &    7.33 $\pm$ 0.10  &	 \nodata        &	7.14 $\pm$ 0.20      \\
12+log(Ne/H)            &    7.72 $\pm$ 0.15  &    7.42 $\pm$ 0.10  &	 \nodata	    &	7.46 $\pm$ 0.20 \\
log(Ne/O)               & $-$0.65 $\pm$ 0.18  & $-$0.68 $\pm$ 0.12  &    \nodata	    &$-$0.76 $\pm$ 0.18 \\
\noalign{\smallskip} 									  
12+log(Ar$^{+3}$/H$^+$) &   \nodata           &    4.92 $\pm$ 0.17  &	 \nodata        &	  \nodata    \\
12+log(Cl$^{++}$/H$^+$) &   \nodata           &    4.26 $\pm$ 0.28  &	 \nodata	    &	  \nodata\\
12+log(Fe$^{++}$/H$^+$) &   5.2:              &    5.5:             &	 \nodata	    &	   \nodata\\
12+log(Fe/H)            &   6.2:              &    6.5:             &	 \nodata	    &	   \nodata\\
log(Fe/O)               &   $-$2.2:           & $-$1.6:             &	 \nodata	    &	   \nodata\\
\noalign{\smallskip} 
12+log(He$^+$/H$^+$)    & 10.97 $\pm$ 0.05    &  10.88 $\pm$ 0.06   &  10.96:           &  10.96 $\pm$ 0.12\\
\noalign{\smallskip}
\tableline
\noalign{\smallskip}
[O/H]                   &   $-$0.29 & $-$0.56$\pm$0.11 & $-$0.45 & $-$0.44 \\
\noalign{\smallskip}
\tableline
  \end{tabular}
  \begin{flushleft}
  $^a$Estimated using empirical relations. \\
  $^b$[O/H]=log(O/H)-log(O/H)$_{\odot}$, using 12+log(O/H)$_{\odot}$ = 8.66$\pm$0.05 \citep*{ASP05}.
  \end{flushleft}
\end{table*}

\begin{table*}[t!]
\centering
  \caption{\footnotesize{Dereddened line intensity ratios with respect to $I$(\Hb)=100 for knots analyzed in Mkn 1199.}}
  \label{mkn1199lineas}
  \tiny
  \begin{tabular}{l  r  rrrrr}
  \noalign{\smallskip}
  \tableline
   \noalign{\smallskip}
Line   & $f(\lambda)$&  C   &  NE  &  A  & B  & D \\
\noalign{\smallskip}
 \tableline
\noalign{\smallskip}
 3728.00  [O  II]  &  0.256 & 124.6$\pm$7.8   &   254$\pm$20   &     204:	   &	       145: &  192$\pm$45\\   
 3835.39   H	I  &  0.237 &	    1.23:   &	     \nodata   &	\nodata    &	    \nodata &  \nodata\\
 3868.75 [Ne III]  &  0.230 &  1.39$\pm$0.46   &       6.64:   &	\nodata    &	    \nodata &  \nodata\\
 3889.05   H	I  &  0.226 &	9.2$\pm$1.9   &       9.40:   &        \nodata     &	    \nodata &  \nodata\\
 3967.46 [Ne III]  &  0.210 &	     \nodata   &       6.22:   &	\nodata    &	    \nodata &  \nodata\\
 3970.07   H	I  &  0.210 &  18.0$\pm$1.2   &        \nodata   &	  \nodata  &	    \nodata &  \nodata\\
 4101.74   H	I  &  0.182 &  27.0$\pm$1.9   &  26.0$\pm$5.9	&	 \nodata   &	    \nodata &  30.5$\pm$7.6\\
 4340.47   H	I  &  0.127 &  45.9$\pm$3.4   &  46.7$\pm$6.2	&      31:	   &	    \nodata &  47$\pm$18\\
 4471.48  He	I  &  0.095 &	    2.48:   &	    5.46:   &	     \nodata	   &	    \nodata &  \nodata\\
 4658.10 [Fe III]  &  0.050 &  3.44$\pm$0.68   &	\nodata   &	   \nodata &	    \nodata &  \nodata\\
 4686.00  He   II  &  0.043 &  0.24:   &	\nodata   &	   \nodata &	    \nodata &  \nodata\\
 4861.33   H	I  &  0.000 & 100.0$\pm$6.0   &   100$\pm$11   &     100:	   &	       100: & 100$\pm$27 \\
 4958.91  [O III]  & -0.024 &  10.7$\pm$1.4   &  60.0$\pm$8.4	&      23:	   &	    \nodata & 23.5: \\
 5006.84  [O III]  & -0.036 &  30.6$\pm$2.4   &   167$\pm$15   &      62:	   &	      85: & 60$\pm$20 \\
 5055.98  Si   II  & -0.048 &	    0.81:   &	    2.37:   &	     \nodata	   &	    \nodata &  \nodata\\
 5197.90  [N   I]  & -0.082 &  2.49$\pm$0.60   &	\nodata   &	   \nodata &	    \nodata &  \nodata\\
 5754.64  [N  II]  & -0.194 &	    0.61:   &	     \nodata   &	\nodata    &	    \nodata &  \nodata\\
 5875.64  He	I  & -0.215 &  9.46$\pm$0.93   &  12.7$\pm$3.1   &	  \nodata  &	    \nodata &  11.6:\\
 6300.30  [O   I]  & -0.282 &  3.57$\pm$0.80   &   6.6$\pm$2.4   &	  \nodata  &	    \nodata &  16.8:\\
 6312.10  [S III]  & -0.283 &	    0.14:   &	    1.19:   &	     \nodata	   &	    \nodata &  \nodata\\
 6363.78  [O   I]  & -0.291 &	    0.85:   &	     \nodata   &	\nodata    &	    \nodata &  \nodata\\
 6548.03  [N  II]  & -0.318 &  48.2$\pm$3.1   &  13.9$\pm$3.5	&      51:	   &	      51: &  57$\pm$14\\
 6562.82   H	I  & -0.320 &	293$\pm$17   &   290$\pm$25   &        294:        &	       296: &  298$\pm$65\\
 6583.41  [N  II]  & -0.323 & 149.8$\pm$8.9   &  41.2$\pm$5.6	&      142:        &           141: &  172$\pm$35\\
 6678.15  He	I  & -0.336 &  1.90$\pm$0.49   &       3.88:   &	\nodata    &	    \nodata &  \nodata\\
 6716.47  [S  II]  & -0.342 &  35.8$\pm$2.3   &  35.4$\pm$5.2	&      66:	   &	      74: &  74$\pm$19\\
 6730.85  [S  II]  & -0.344 &  32.1$\pm$6.4   &  25.5$\pm$4.4	&      40:	   &	      44: &  49$\pm$14\\
 7065.28  He	I  & -0.387 &  1.30$\pm$0.42   &	\nodata   &	   \nodata &	    \nodata &  \nodata\\
 7135.78 [Ar III]  & -0.396 &  1.91$\pm$0.56   &       6.29:   &	\nodata    &	    \nodata &  \nodata\\
 7318.39  [O  II]  & -0.418 &  0.87$\pm$0.32   &	\nodata   &	   \nodata &	    \nodata &  \nodata\\
 7329.66  [O  II]  & -0.420 &	    0.39:   &	     \nodata   &	\nodata    &	    \nodata &  \nodata\\
\noalign{\smallskip}
 \tableline
 \noalign{\smallskip}
Aperture size (arcsec)       &   &    10$\times$1      &     6$\times$1     &   8$\times$1         & 8$\times$1       & 5.6$\times$1\\
Distance$^b$ (arcsec)   &   &       0             &     26               &  18                  &  14                   & 8.4  \\
$F$(\Hb)$^a$          &        &  74.2 $\pm$   3.1 &    3.17 $\pm$   0.21 &    0.33 $\pm$   0.08 &    0.42 $\pm$   0.10  &    0.98 $\pm$   0.14 \\ 
$C$(\Hb)              &        &   0.30 $\pm$   0.03 &    0.16 $\pm$   0.03 &    0.17 $\pm$   0.04 &    0.44 $\pm$   0.06  &    0.27 $\pm$   0.04 \\
$W_{abs}$ (\AA)     &        &    1.8 $\pm$   0.4  &     0.6 $\pm$   0.3  &     1.7 $\pm$ 0.3    &    2                  &     2.5 $\pm$   0.3 \\ 
\noalign{\smallskip}
$-W$(\Ha) (\AA)     &        &  129.1 $\pm$    7.9 &     110 $\pm$    10  &     21 $\pm$    7    &     22 $\pm$    8     &    34.7 $\pm$    7.6 \\ 
$-W$(\Hb) (\AA)     &        &   21.4 $\pm$    1.3 &    20.2 $\pm$    2.3 &     5.0 $\pm$    2.6 &     5.2 $\pm$    2.4  &     8.3 $\pm$    2.3 \\ 
$-W$(H$\gamma$) (\AA)&       &    6.7 $\pm$    0.5 &     8.4 $\pm$    1.1 &     1.5 $\pm$    1.7 &     \nodata           &     2.1 $\pm$    0.8 \\ 
$-W$([O III]) 5007 (\AA)&    &    6.8 $\pm$    0.5 &    35.5 $\pm$    3.4 &     2.8 $\pm$    1.6 &     4.2 $\pm$    1.8  &     5.1 $\pm$    1.7 \\ 
\noalign{\smallskip}
 \tableline
  \end{tabular}
  \begin{flushleft}
   $^a$ In units of 10$^{-15}$ erg s$^{-1}$ cm$^{-2}$ and not corrected for extinction.\\
    $^b$ Relative distance with respect to the center of Mkn~1199.
  \end{flushleft}
\end{table*}

\begin{table*}[t!]
  \caption{\footnotesize{Physical conditions and chemical abundances of the ionized gas of the regions analyzed in Mkn 1199.}}
  \label{mkn1199abun}
  \tiny
  \centering
  \begin{tabular}{l  rrrrr}
\tableline
\noalign{\smallskip}
Region                  &         Center      &        NE$^a$           &         A$^a$           &   B$^a$                &  D$^a$ \\
\noalign{\smallskip}
\tableline
\noalign{\smallskip}				      
$T_e$(O III) (K)        &    5400 $\pm$  700  &    8450 $\pm$ 800   &	 6950 $\pm$ 800   &    6300 $\pm$ 800  &     6750 $\pm$ 800  \\
$T_e$(O II) (K)         &    6800 $\pm$  600$^b$  &    8900 $\pm$ 600   &	 7850 $\pm$ 600   &    7400 $\pm$ 600  &     7700 $\pm$ 600  \\
$n_e$ (cm$^{-3}$)       &    300  $\pm$  100  &     100   	    &	 100		  &     100	       &     100 	     \\
\noalign{\smallskip}
12+log(O$^{+}$/H$^+$)   &    8.59 $\pm$ 0.11  &    8.25 $\pm$ 0.12  &	 8.43 $\pm$ 0.20  &    8.43 $\pm$ 0.21 &     8.44 $\pm$ 0.16 \\
12+log(O$^{++}$/H$^+$)  &    8.24 $\pm$ 0.12  &    8.05 $\pm$ 0.11  &	 7.99 $\pm$ 0.23  &    8.32 $\pm$ 0.26 &     8.05 $\pm$ 0.19 \\
12+log(O/H)             &    8.75 $\pm$ 0.12  &    8.46 $\pm$ 0.13  &	 8.57 $\pm$ 0.21  &    8.68 $\pm$ 0.23 &     8.59 $\pm$ 0.17 \\
\noalign{\smallskip}
log(O$^{++}$/O$^+$)     & $-$0.36 $\pm$ 0.16  & $-$0.19 $\pm$ 0.09  & $-$0.44 $\pm$ 0.18  & $-$0.10 $\pm$ 0.32 &  $-$0.37 $\pm$ 0.22 \\ 
12+log(N$^+$/H$^+$)     &    7.98 $\pm$ 0.09  &    7.05 $\pm$ 0.10  &	 7.76 $\pm$ 0.14  &    7.85 $\pm$ 0.18 &     7.86 $\pm$ 0.11 \\
12+log(N/H)             &    8.14 $\pm$ 0.11  &    7.26 $\pm$ 0.13  &	 7.90 $\pm$ 0.18  &    8.10 $\pm$ 0.28 &     8.00 $\pm$ 0.15 \\
log(N/O)                & $-$0.62 $\pm$ 0.10  & $-$1.20 $\pm$ 0.11  & $-$0.67 $\pm$ 0.20  & $-$0.58 $\pm$ 0.30 &  $-$0.59 $\pm$ 0.17 \\
\noalign{\smallskip}
12+log(S$^+$/H$^+$)     &    6.70 $\pm$ 0.09  &    6.28 $\pm$ 0.09  &	 6.67 $\pm$ 0.16  &    6.80 $\pm$ 0.16 &     6.76 $\pm$ 0.13 \\
12+log(S$^{++}$/H$^+$)  &    7.05 $\pm$ 0.17  &    6.80 $\pm$ 0.22  & \nodata		  & \nodata	       &  \nodata	     \\
12+log(S/H)             &    7.22 $\pm$ 0.15  &    6.92 $\pm$ 0.18  & \nodata		  & \nodata	       &  \nodata	     \\
log(S/O)                & $-$1.54 $\pm$ 0.14  & $-$1.54 $\pm$ 0.17  & \nodata		  & \nodata	       &  \nodata	     \\
\noalign{\smallskip}
12+log(Ne$^{++}$/H$^+$) &    7.65 $\pm$ 0.17  &    7.40 $\pm$ 0.20  & \nodata		  & \nodata	       &  \nodata	     \\
12+log(Ne/H)            &    8.17 $\pm$ 0.19  &    7.81 $\pm$ 0.20  & \nodata		  & \nodata	       &  \nodata	     \\
log(Ne/O)               & $-$0.58 $\pm$ 0.17  & $-$0.65 $\pm$ 0.18  & \nodata		  & \nodata	       &  \nodata	     \\
\noalign{\smallskip} 	     	    	    	              	    			  		         		    
12+log(Ar$^{+3}$/H$^+$) &    6.07 $\pm$ 0.16  &    5.95 $\pm$ 0.22  & \nodata		  & \nodata	       &  \nodata	     \\
12+log(Fe$^{++}$/H$^+$) &    6.75 $\pm$ 0.13  & \nodata 	  & \nodata	 & \nodata	       &  \nodata	     \\
12+log(Fe/H)            &    6.89 $\pm$ 0.13  & \nodata 	  & \nodata	 & \nodata	       &  \nodata	     \\
log(Fe/O)               & $-$1.86 $\pm$ 0.26  & \nodata 	  & \nodata	 & \nodata	       &  \nodata	     \\
\noalign{\smallskip} 
12+log(He$^+$/H$^+$)    &   10.79 $\pm$ 0.07  &  10.9:        &\nodata  & \nodata & 10.9: \\
\noalign{\smallskip}
\tableline
\noalign{\smallskip}
[O/H]$^b$    &    +0.09$\pm$0.17  &  $-$0.20 &  $-$0.09   &	 +0.02   & $-$0.07\\
\noalign{\smallskip}
\tableline
  \end{tabular}
  \begin{flushleft}
   $^a$Electron temperatures estimated using empirical relations. \\
   $^b$Derived from [\ion{N}{ii}] and [\ion{O}{ii}] ratios, see \S3.4.1\\
  $^c$[O/H]=log(O/H)-log(O/H)$_{\odot}$, using 12+log(O/H)$_{\odot}$ = 8.66$\pm$0.05 \citep{ASP05}.
  \end{flushleft}
\end{table*}

\begin{table*}[t!]
\centering
  \caption{\footnotesize{Dereddened line intensity ratios with respect to $I$(\Hb)=100 for knots analyzed in Mkn 5. Region A was observed using three 
slit positions with a PA of 0$^{\circ}$ (INT-1), 349$^{\circ}$ = $-$11$^{\circ}$ (INT-2) and 354$^{\circ}$ = $-$6$^{\circ}$ (WHT).}}
  \label{mkn5lineas}
  \tiny
  \begin{tabular}{l  r  rrrr}
  \noalign{\smallskip}
  \tableline
   \noalign{\smallskip}
Line   & $f(\lambda)$&  A-INT-1   & A-INT-2  &  A-WHT  & B  \\
\noalign{\smallskip}
 \tableline
\noalign{\smallskip}

 3666.10   H	I &  0.267 &  1.99$\pm$0.71   &     \nodata   &	  \nodata &   \nodata\\
 3697.15   H	I &  0.262 &	   1.16:   &	    \nodata   &           \nodata &\nodata   \\
 3705.04  He	I &  0.260 &	   1.65:   &	    \nodata   &           \nodata & \nodata  \\
 3711.97   H	I &  0.259 &	    \nodata   &  2.03$\pm$0.76   &	  \nodata &  \nodata \\
 3728.00  [O  II] &  0.256 &   191$\pm$12   &	213$\pm$12   &            \nodata &  252: \\
 3750.15   H	I &  0.253 &  1.79$\pm$0.68   &        \nodata   &	  \nodata &  \nodata \\
 3770.63   H	I &  0.249 &	    \nodata   &  2.47$\pm$0.79   &	  \nodata &  \nodata \\
 3797.90   H	I &  0.244 &	    \nodata   &       1.54:   &           \nodata &  \nodata \\
 3819.61  He	I &  0.240 &	    \nodata   &       0.65:   &           \nodata &  \nodata \\
 3835.39   H	I &  0.237 &   7.4$\pm$1.7   &   5.6$\pm$1.3   &	  \nodata &   \nodata\\
 3868.75 [Ne III] &  0.230 &  23.1$\pm$3.3   &  31.0$\pm$2.4   &	  \nodata &  \nodata \\
 3889.05   H	I &  0.226 &  13.6$\pm$2.9   &  17.7$\pm$2.1   &	  \nodata &  \nodata \\
 3967.46 [Ne III] &  0.210 &  18.7$\pm$1.9   &  21.3$\pm$1.8   &	  \nodata &  \nodata \\
 4026.21  He	I &  0.198 &	    \nodata   &       1.55:   &           \nodata &  \nodata \\
 4068.60  [S  II] &  0.189 &  2.24$\pm$0.70   & 2.29$\pm$0.76   &	  \nodata &  \nodata \\
 4101.74   H	I &  0.182 &  26.0$\pm$2.8   &  26.8$\pm$2.1   &	  \nodata &  \nodata \\
 4168.97  He	I &  0.167 &	    \nodata   &       1.06:   &           \nodata &  \nodata \\
 4340.47   H	I &  0.127 &  47.0$\pm$3.2   &  47.0$\pm$3.1    &    47.0$\pm$2.8 &  \nodata \\
 4363.21  [O III] &  0.121 &  5.14$\pm$0.91   &  5.29$\pm$0.93   &  4.93$\pm$0.74 &  \nodata \\
 4387.93  He	I &  0.115 &	    \nodata   &       0.56:   &     1.32$\pm$0.40 &  \nodata \\
 4416.27 [Fe  II] &  0.109 &	    \nodata   &        \nodata   &	    0.38: &   \nodata\\
 4471.48  He	I &  0.095 &  4.43$\pm$0.85   &  3.96$\pm$0.84   &  4.16$\pm$0.60 &   \nodata\\
 4658.10 [Fe III] &  0.050 &	    \nodata   &       1.10:   &             0.72: &   \nodata\\
 4686.00  He   II &  0.043 &  0.92$\pm$0.19   &       1.01:   &             0.82: &   \nodata\\
 4711.37 [Ar  IV] &  0.037 &	    \nodata   &       0.59:   &           \nodata &   \nodata\\
 4713.14  He	I &  0.037 &	    \nodata   &        \nodata   &	    0.28: &  \nodata \\
 4740.16 [Ar  IV] &  0.030 &	    \nodata   &        \nodata   &	    0.31: &  \nodata \\
 4861.33   H	I &  0.000 & 100.0$\pm$6.3   & 100.0$\pm$5.7   &    100.0$\pm$5.2 & 100:  \\
 4921.93  He	I & -0.015 &	    \nodata   &       1.70:   &     0.95$\pm$0.36 &  \nodata \\
 4958.91  [O III] & -0.024 & 144.1$\pm$8.5   & 144.7$\pm$7.8   &    133.8$\pm$7.2 & 70.43:  \\
 4985.90 [Fe III] & -0.031 &  2.33$\pm$0.65   &       2.02:   &             2.20: &  \nodata \\
 5006.84  [O III] & -0.036 &   423$\pm$22   &	430$\pm$21   &         374$\pm$19 & 214$\pm$69  \\
 5015.68  He	I & -0.038 &	    \nodata   &        \nodata   &  2.13$\pm$0.62 &  \nodata \\
 5197.90  [N   I] & -0.082 &	    \nodata   &       0.46:   &           \nodata &  \nodata \\
 5875.64  He	I & -0.215 &  8.46$\pm$0.83   &  8.41$\pm$0.88   &   11.0$\pm$1.0 &  \nodata \\
 6300.30  [O   I] & -0.282 &  5.12$\pm$0.68   &  4.57$\pm$0.66   &  4.05$\pm$0.58 &  \nodata \\
 6312.10  [S III] & -0.283 &  2.31$\pm$0.55   &  1.92$\pm$0.43   &  1.79$\pm$0.47 &  \nodata \\
 6363.78  [O   I] & -0.291 &  1.90$\pm$0.51   &       1.52:   &     1.19$\pm$0.38 &  \nodata \\
 6548.03  [N  II] & -0.318 &  4.56$\pm$0.56   &  5.25$\pm$0.69   &  5.49$\pm$0.63 &  \nodata \\
 6562.82   H	I & -0.320 &   283$\pm$16   &	284$\pm$15   &         284$\pm$14 & 284$\pm$85\\
 6583.41  [N  II] & -0.323 &  14.7$\pm$1.2   &  14.0$\pm$1.1   &     14.9$\pm$1.1 & 10.13:  \\
 6678.15  He	I & -0.336 &  3.43$\pm$0.63   &  3.49$\pm$0.66   &  3.05$\pm$0.51 &  \nodata \\
 6716.47  [S  II] & -0.342 &  21.5$\pm$1.5   &  22.2$\pm$1.7   &     24.3$\pm$1.5 &  67.4: \\
 6730.85  [S  II] & -0.344 &  15.6$\pm$1.2   &  16.0$\pm$1.4   &     18.1$\pm$1.2 &  51.6:\\
 7065.28  He	I & -0.387 &  2.13$\pm$0.50   &  2.40$\pm$0.64   &  2.17$\pm$0.45 &  \nodata \\
 7135.78 [Ar III] & -0.396 &  8.49$\pm$0.61   &  8.56$\pm$0.97   &  7.70$\pm$0.52 &  \nodata \\
 7281.35  He	I & -0.414 &	    \nodata   &        \nodata   &	    0.41: &   \nodata\\
 7318.39  [O  II] & -0.418 &	    \nodata   &    2.87:         &  3.62$\pm$0.54 & \nodata \\
 7329.66  [O  II] & -0.420 &	    \nodata   &    2.57:         &  2.74$\pm$0.44 &  \nodata \\
 7751.10 [Ar III] & -0.467 &	    \nodata   &        \nodata   &  1.64$\pm$0.38 &  \nodata \\
\noalign{\smallskip}
 \tableline
 \noalign{\smallskip}
Aperture size (arcsec)       &   &    14.4$\times$1       &     16$\times$1     &   3.6$\times$1         & 6 $\times$1 \\
Distance$^b$ (arcsec)   &   &       0             &     0               &  0                  &  16  \\
$F$(\Hb)$^a$          &        &  17.66 $\pm$   0.73 &   18.13 $\pm$   0.69 &   10.83 $\pm$   0.40 &    0.35 $\pm$   0.08\\ 
$C$(\Hb)              &        &   0.36 $\pm$   0.02 &    0.17 $\pm$   0.02 &    0.03 $\pm$   0.02 &    0.30 $\pm$   0.06\\
$W_{abs}$ (\AA)     &        &    1.1 $\pm$    0.2 &     0.8 $\pm$   0.2  &    0                 &     1.5 $\pm$   0.5 \\ 
\noalign{\smallskip}
$-W$(\Ha) (\AA)     &        &      449 $\pm$	26 &       435 $\pm$   23 &	678 $\pm$     35 &     43 $\pm$    12\\ 
$-W$(\Hb) (\AA)     &        &       75 $\pm$	 5 &        80 $\pm$	5 &	135 $\pm$      7 &    10: \\ 
$-W$(H$\gamma$) (\AA)&       &       43 $\pm$	 3 &        34 $\pm$	2 &	 44 $\pm$      3 &     \nodata \\ 
$-W$([O III]) 5007 (\AA)&    &      320 $\pm$	17 &       360 $\pm$   18 &	530 $\pm$     28 &    33 $\pm$  11\\ 
\noalign{\smallskip}
 \tableline
  \end{tabular}
  \begin{flushleft}
   $^a$ In units of 10$^{-15}$ erg s$^{-1}$ cm$^{-2}$ and not corrected for extinction.\\
   $^b$ Relative distance with respect to the center of Mkn~5.
  \end{flushleft}
\end{table*}

\begin{table*}[t!]
  \caption{\footnotesize{Physical conditions and chemical abundances of the ionized gas of the regions analyzed in Mkn 5.}}
  \label{mkn5abun}
  \tiny
  \centering
  \begin{tabular}{l  rrrr}
\tableline
\noalign{\smallskip}
Region               &        A-INT-1       &       A-INT-2        &      A-WHT          &   B$^a$ \\
\noalign{\smallskip}
\tableline
\noalign{\smallskip}				      
$T_e$(O III) (K)        &   12400 $\pm$ 700   &   12450 $\pm$ 650   &	12700 $\pm$ 600   &   13250 $\pm$ 900 \\
$T_e$(O II) (K)         &   11700 $\pm$ 500   &   11700 $\pm$ 450   &	11900 $\pm$ 400   &   12300 $\pm$ 700 \\
$n_e$ (cm$^{-3}$)       &   $\leq$100  	      &   $\leq$100   	    &	$\leq$100  		  &   $\leq$110	       \\
\noalign{\smallskip}
12+log(O$^{+}$/H$^+$)   &    7.62 $\pm$ 0.09  &    7.67 $\pm$ 0.08  &	 7.71 $\pm$ 0.11  &    7.66 $\pm$ 0.19 \\
12+log(O$^{++}$/H$^+$)  &    7.88 $\pm$ 0.06  &    7.87 $\pm$ 0.06  &	 7.80 $\pm$ 0.06  &    7.49 $\pm$ 0.13 \\
12+log(O/H)             &    8.07 $\pm$ 0.07  &    8.08 $\pm$ 0.07  &	 8.06 $\pm$ 0.08  &    7.89 $\pm$ 0.17 \\
\noalign{\smallskip}
log(O$^{++}$/O$^+$)     &    0.25 $\pm$ 0.10  &    0.21 $\pm$ 0.09  &    0.09 $\pm$ 0.11  & $-$0.17 $\pm$ 0.18 \\
12+log(N$^+$/H$^+$)     &    6.27 $\pm$ 0.06  &    6.29 $\pm$ 0.05  &	 6.30 $\pm$ 0.05  &    6.08 $\pm$ 0.22 \\
12+log(N/H)             &    6.72 $\pm$ 0.09  &    6.71 $\pm$ 0.08  &	 6.65 $\pm$ 0.08  &    6.30 $\pm$ 0.22 \\
log(N/O)                & $-$1.35 $\pm$ 0.10  & $-$1.38 $\pm$ 0.10  & $-$1.41 $\pm$ 0.10  & $-$1.58 $\pm$ 0.20 \\
\noalign{\smallskip}
12+log(S$^+$/H$^+$)     &    5.78 $\pm$ 0.04  &    5.79 $\pm$ 0.04  &	 5.82 $\pm$ 0.04  &    6.24 $\pm$ 0.11 \\
12+log(S$^{++}$/H$^+$)  &    6.37 $\pm$ 0.13  &    6.28 $\pm$ 0.13  &	 6.23 $\pm$ 0.12  & \nodata	       \\
12+log(S/H)             &    6.51 $\pm$ 0.12  &    6.44 $\pm$ 0.12  &	 6.40 $\pm$ 0.12  & \nodata	       \\
log(S/O)                & $-$1.56 $\pm$ 0.13  & $-$1.64 $\pm$ 0.12  & $-$1.67 $\pm$ 0.13  & \nodata	       \\
\noalign{\smallskip}
12+log(Ne$^{++}$/H$^+$) &    7.02 $\pm$ 0.12  &    7.14 $\pm$ 0.10  & \nodata 	          & \nodata 	     \\
12+log(Ne/H)            &    7.22 $\pm$ 0.12  &    7.35 $\pm$ 0.11  & \nodata 	          & \nodata 	     \\
log(Ne/O)               & $-$0.85 $\pm$ 0.14  & $-$0.74 $\pm$ 0.13  & \nodata 	          & \nodata \\
\noalign{\smallskip} 
12+log(Ar$^{++}$/H$^+$)	&    5.69 $\pm$ 0.07  &    5.69 $\pm$ 0.09  &	 5.60 $\pm$ 0.09  & \nodata	       \\								  
12+log(Ar$^{+3}$/H$^+$) & \nodata             &    4.69 $\pm$ 0.25  &	 4.54 $\pm$ 0.21  & \nodata		\\
12+log(Ar/H)            & \nodata 	          &    5.79 $\pm$ 0.12  &	 5.73 $\pm$ 0.10  & \nodata	       \\
log(Ar/O)               & \nodata 	          & $-$2.29 $\pm$ 0.15  & $-$2.33 $\pm$ 0.12  & \nodata	       \\
\noalign{\smallskip} 
12+log(Fe$^{++}$/H$^+$) &    5.73 $\pm$ 0.13  &    5.66: 		    &    5.68:	          & \nodata		\\
12+log(Fe/H)            &    6.11 $\pm$ 0.16  &    6.00:    	    &    5.98:            & \nodata		\\
log(Fe/O)               & $-$1.96 $\pm$ 0.18  & $-$2.08:            & $-$2.08:            & \nodata     \\
\noalign{\smallskip}
12+log(He$^+$/H$^+$)    &   10.96 $\pm$ 0.06  &  10.91 $\pm$ 0.06  &  10.92 $\pm$ 0.05 & \nodata\\
\noalign{\smallskip}
\tableline
\noalign{\smallskip}
[O/H]$^b$                &  $-$0.59$\pm$0.12 & $-$0.58$\pm$0.12 & $-$0.60$\pm$0.13 & $-$0.77 \\ 
\noalign{\smallskip}
\tableline
  \end{tabular}
  \begin{flushleft}
  $^a$Electron temperatures estimated using empirical relations. \\
  $^b$[O/H]=log(O/H)-log(O/H)$_{\odot}$, using 12+log(O/H)$_{\odot}$ = 8.66$\pm$0.05 \citep{ASP05}.
  \end{flushleft}
\end{table*}

\begin{table*}[t!]
\centering
  \caption{\footnotesize{Dereddened line intensity ratios with respect to $I$(\Hb)=100 for regions analyzed in IRAS 08208+2816. The slit positions we 
used for each knot are: PA 345$^{\circ}$ for C, PA 355$^{\circ}$ for \#8 and \#10 and PA 10$^{\circ}$ for \#3 y \#5.}}
  \label{iras08208lineas}
  \tiny
  \begin{tabular}{l  r@{\hspace{5pt}}  r@{\hspace{8pt}}    r@{\hspace{8pt}}r@{\hspace{8pt}}r@{\hspace{8pt}}r@{\hspace{8pt}}  }
  \noalign{\smallskip}
  \tableline
   \noalign{\smallskip}
\noalign{\smallskip}
Line   & $f(\lambda)$&  C      &         \#3            &     \#5          &      \#8          &    \#10         \\
\noalign{\smallskip}
 \tableline
\noalign{\smallskip}
 3728.00  [O  II] &   0.256 & 146.8$\pm$9.8   &   279$\pm$20    &   251$\pm$24     &	   164$\pm$22 	  &   324$\pm$64    \\
 3750.15   H	I &   0.253 &  1.74$\pm$0.51  &	\nodata    &	    \nodata   &     \nodata 	  &	   \nodata  \\
 3770.63   H	I &   0.249 &  1.55$\pm$0.49  &	  \nodata  &	    \nodata   &     \nodata 	  &	   \nodata  \\
 3797.90   H	I &   0.244 &  2.86$\pm$0.60  &	 3.17:     &	    \nodata   &     \nodata  	  &	   \nodata  \\
 3835.39   H	I &   0.237 &  5.53$\pm$0.79  &	  \nodata  &	    \nodata         \nodata  	  &	   \nodata  \\
 3868.75 [Ne III] &   0.230 &  32.8$\pm$2.2   &  15.1$\pm$3.4   &  26.3$\pm$7.0    &       5.27:  	  &	   \nodata  \\
 3889.05   H	I &   0.226 &  15.1$\pm$1.7   &  13.5$\pm$3.2   &  12.9$\pm$5.0    &       6.36:  	  &	   \nodata  \\
 3967.46 [NeIII]H7&   0.210 &  22.1$\pm$1.9   &  15.9$\pm$2.9   &  17.0$\pm$5.3    &    22.0$\pm$5.7   &	26.23:      \\
 4026.21  He	I &   0.198 &	    0.92:     &	\nodata    &	    \nodata   &        \nodata    &	   \nodata  \\
 4068.60  [S  II] &   0.189 &  1.14$\pm$0.45  &	   \nodata &	    \nodata   &        \nodata    &	   \nodata  \\
 4101.74   H	I &   0.182 &  26.1$\pm$2.1   &  26.1$\pm$2.9   &  24.1$\pm$5.1    &  26.1$\pm$5.7  	  &	 26.23:     \\
 4340.47   H	I &   0.127 &  46.7$\pm$2.7   &  46.8$\pm$4.6   &  43.8$\pm$6.8    &  43.5$\pm$7.4  	  &    46$\pm$15    \\
 4363.21  [O III] &   0.121 &  3.12$\pm$0.48  &  1.66:          &	   2.26:      &	   \nodata  	  &	   \nodata  \\
 4471.48  He	I &   0.095 &  4.00$\pm$0.67  &   3.7$\pm$1.3   &	    \nodata   &        \nodata    &	   \nodata  \\
 4658.10 [Fe III] &   0.050 &  1.68$\pm$0.58  &       1.72:     &	    \nodata   &        \nodata    &	   \nodata  \\
 4861.33   H	I &   0.000 & 100.0$\pm$5.7   & 100.0$\pm$8.4   &   100$\pm$11     &	100$\pm$12   	  &   100$\pm$22    \\
 4958.91  [O III] &  -0.024 & 152.6$\pm$8.3   &  76.6$\pm$5.8   &   104$\pm$10     &  44.7$\pm$8.9  	  &    54$\pm$15    \\
 5006.84  [O III] &  -0.036 &	470$\pm$24    &   228$\pm$14    &   305$\pm$24     &	 89$\pm$13   	  &   151$\pm$30    \\
 5197.90  [N   I] &  -0.082 &  1.09$\pm$0.42  &	   \nodata &	   2.46:      &	    \nodata  	  &	   \nodata  \\
 5875.64  He	I &  -0.215 &  12.8$\pm$1.2   &  12.4$\pm$1.9   &  14.6$\pm$2.9    &  12.3$\pm$3.4  	  &	 11.28:     \\
 6300.30  [O   I] &  -0.282 &  4.74$\pm$0.62  &   5.6$\pm$1.1   &   9.8$\pm$2.3    &  10.1$\pm$3.1  	  &	  9.00:     \\
 6312.10  [S III] &  -0.283 &  1.46$\pm$0.42  &       1.13:     &	    \nodata   &        \nodata    &	   \nodata  \\
 6363.78  [O   I] &  -0.291 &  1.70$\pm$0.46  &	\nodata    &	   3.21:      &	   2.24:  	  &	   \nodata  \\
 6548.03  [N  II] &  -0.318 &  14.2$\pm$1.1   &  16.0$\pm$2.4   &  16.2$\pm$3.9    &  30.2$\pm$5.3  	  &	 14.95:     \\
 6562.82   H	I &  -0.320 &	285$\pm$15    &   283$\pm$20    &   281$\pm$25     &	288$\pm$33  	  &   286$\pm$51    \\
 6583.41  [N  II] &  -0.323 &  36.9$\pm$2.6   &  39.2$\pm$4.2   &  37.8$\pm$5.3    &  75.3$\pm$9.3  	  &    42$\pm$11    \\
 6678.15  He	I &  -0.336 &  3.26$\pm$0.63  &  3.04$\pm$0.93  &	    \nodata   &       2.93:  	  &	  3.65:     \\
 6716.47  [S  II] &  -0.342 &  22.0$\pm$1.5   &  40.2$\pm$4.2   &  52.9$\pm$6.8    &  66.4$\pm$9.6  	  &    50$\pm$13    \\
 6730.85  [S  II] &  -0.344 &  16.6$\pm$1.3   &  28.5$\pm$3.3   &  34.5$\pm$5.2    &  46.7$\pm$7.5  	  &    36$\pm$10    \\
 7065.28  He	I &  -0.387 &  2.42$\pm$0.57  &	   \nodata &	    \nodata   &      \nodata      &	   \nodata  \\
 7135.78 [Ar III] &  -0.396 &  6.37$\pm$0.78  &  11.8$\pm$3.0   &	    \nodata   &      \nodata  	  &	   \nodata  \\
\noalign{\smallskip}
 \tableline
 \noalign{\smallskip}
Aperture size (arcsec)     &        &   1.4$\times$1   &   2.8$\times$1   & 4.0$\times$1    &  6.4$\times$1     &   4.0$\times$1     \\
Distance (arcsec)$^b$&      &       -          &   12.6           &  6.8             &     8.8           &    16   \\
$F$(\Hb)$^a$          &       & 12.9  $\pm$ 0.5  & 5.97 $\pm$ 0.28 &  2.98 $\pm$ 0.19 &  4.05 $\pm$ 0.34  & 1.36 $\pm$ 0.18 \\ 
$C$(\Hb)              &       &  0.11 $\pm$ 0.02 & 0.47 $\pm$ 0.04 &  0.41 $\pm$ 0.04 &  0.17 $\pm$ 0.03  & 0.12 $\pm$ 0.02 \\
$W_{abs}$ (\AA)     &       &	3.2 $\pm$ 0.1  & 1.6  $\pm$ 0.2  &  1.4  $\pm$ 0.2  &   4.9 $\pm$ 0.1   & 1.9  $\pm$ 0.1  \\ 
\noalign{\smallskip}													
$-W$(\Ha) (\AA)     &       &   331 $\pm$ 18   &  202 $\pm$ 15   &   346 $\pm$   32 &    89 $\pm$ 10    &   98 $\pm$  18  \\    
$-W$(\Hb) (\AA)     &       &    80 $\pm$  5   &   56 $\pm$  5   &	62 $\pm$    7 &    24 $\pm$  3    &   17 $\pm$   4  \\    
$-W$(H$\gamma$) (\AA)&      &    30 $\pm$  2  &   20 $\pm$  2   &	27 $\pm$    4 &    10 $\pm$  2    &    9 $\pm$   3  \\   
$-W$([O III]) (\AA)&   &   370 $\pm$ 19  &    130 $\pm$  8   &   200 $\pm$   16 &    18 $\pm$  3    &   19 $\pm$   4  \\   
\noalign{\smallskip}
 \tableline
   \end{tabular}
  \begin{flushleft}		
     $^a$ In units of 10$^{-15}$ erg s$^{-1}$ cm$^{-2}$ and not corrected for extinction.\\
	 $^b$ Relative distance with respect to the center of IRAS 08208+2816.
   \end{flushleft}
\end{table*}

\begin{table*}[t!]
  \caption{\footnotesize{Physical conditions and chemical abundances of the ionized gas of the regions analyzed in IRAS 08208+2816.}}
  \label{iras08208abun}
  \tiny
  \centering
  \begin{tabular}{l@{\hspace{8pt}}  r@{\hspace{6pt}}     r@{\hspace{6pt}}r@{\hspace{6pt}} r@{\hspace{6pt}}r@{\hspace{6pt}}  } 
\tableline
\noalign{\smallskip}
Region               &      C              &   \#3$^a$           &   \#5$^a$           &   \#8$^a$	      &   \#10$^a$	   \\
\noalign{\smallskip}
\tableline
\noalign{\smallskip}				      
$T_e$(O III) (K)        &   10100 $\pm$ 700   &	 9400 $\pm$ 900   &    9600 $\pm$ 1000  &    6750 $\pm$ 1000  &    9500 $\pm$ 1000 \\
$T_e$(O II) (K)         &   10100 $\pm$ 500   &	 9600 $\pm$ 700   &    9700 $\pm$ 800	&    7700 $\pm$ 800   &    9650 $\pm$ 800  \\
$n_e$ (cm$^{-3}$)       &   $<$100            &  $<$100  	      &  $<$100       	&  $<$100	      &  $<$100	    \\
\noalign{\smallskip}
12+log(O$^{+}$/H$^+$)   &    7.77 $\pm$ 0.09  &	 8.15 $\pm$ 0.13  &    8.08 $\pm$ 0.15  &    8.39 $\pm$ 0.15  &    8.20 $\pm$ 0.18 \\
12+log(O$^{++}$/H$^+$)  &    8.20 $\pm$ 0.07  &	 8.01 $\pm$ 0.11  &    8.10 $\pm$ 0.12  &    8.28 $\pm$ 0.14  &    7.82 $\pm$ 0.17 \\
12+log(O/H)$^c$         &    8.33 $\pm$ 0.08  &	 8.24:            &    8.27:            &    \nodata          &    \nodata         \\
12+log(O/H)$^d$         &    8.41 $\pm$ 0.10  &	 8.39 $\pm$ 0.11  &    8.39 $\pm$ 0.13  &    8.64 $\pm$ 0.15  &    8.35 $\pm$ 0.17 \\
\noalign{\smallskip}
log(O$^{++}$/O$^+$)     &    0.43 $\pm$ 0.12  &	 0.15 $\pm$ 0.17  &    0.02 $\pm$ 0.19  & $-$0.10 $\pm$ 0.20  & $-$0.38 $\pm$ 0.22 \\
12+log(N$^+$/H$^+$)     &    6.88 $\pm$ 0.05  &	 6.98 $\pm$ 0.08  &    6.96 $\pm$ 0.10  &    7.54 $\pm$ 0.10  &    6.97 $\pm$ 0.14 \\
12+log(N/H)             &    7.44 $\pm$ 0.10  &	 7.22 $\pm$ 0.11  &    7.27 $\pm$ 0.13  &    7.80 $\pm$ 0.16  &    7.12 $\pm$ 0.17 \\
log(N/O)                & $-$0.89 $\pm$ 0.11  & $-$1.17 $\pm$ 0.13  & $-$1.12 $\pm$ 0.14  & $-$0.84 $\pm$ 0.15  & $-$1.22 $\pm$ 0.18 \\
\noalign{\smallskip}
12+log(S$^+$/H$^+$)     &    5.94 $\pm$ 0.05  &	 6.25 $\pm$ 0.08  &    6.33 $\pm$ 0.09  &    6.73 $\pm$ 0.12  &    6.34 $\pm$ 0.13 \\
12+log(S$^{++}$/H$^+$)  &    6.52 $\pm$ 0.16  &	 6.56 $\pm$ 0.28  &    \nodata  	&    \nodata	      & \nodata 	\\
12+log(S/H)             &    6.69 $\pm$ 0.13  &	 6.75 $\pm$ 0.23  &    \nodata  	&    \nodata	      & \nodata 	\\
log(S/O)                & $-$1.64 $\pm$ 0.16  & $-$1.64 $\pm$ 0.25  &    \nodata  	&    \nodata	      & \nodata 	\\
\noalign{\smallskip}
12+log(Ne$^{++}$/H$^+$) &    7.52 $\pm$ 0.11  &	 7.37 $\pm$ 0.19  &    7.53 $\pm$ 0.22  &    7.60 $\pm$ 0.25  &    \nodata	    \\
12+log(Ne/H)            &    7.66 $\pm$ 0.11  &	 7.75 $\pm$ 0.18  &    7.82 $\pm$ 0.22  &    7.96 $\pm$ 0.25  &    \nodata	    \\
log(Ne/O)               & $-$0.67 $\pm$ 0.13  & $-$0.64 $\pm$ 0.21 & $-$0.57 $\pm$ 0.25  & $-$0.68 $\pm$ 0.26  &    \nodata	    \\
\noalign{\smallskip}
12+log(Ar$^{+2}$/H$^+$) &    5.78 $\pm$ 0.08  &	 6.11 $\pm$ 0.15  &    \nodata  	&    \nodata	      &    \nodata	    \\
12+log(Ar/H)            &    5.86 $\pm$ 0.11  &	 5.92 $\pm$ 0.15  &    \nodata  	&    \nodata	      &    \nodata	    \\
log(Ar/O)               & $-$2.51 $\pm$ 0.15  & $-$2.47 $\pm$ 0.18  &    \nodata  	&    \nodata	      &    \nodata	    \\
\noalign{\smallskip} 
12+log(Fe$^{++}$/H$^+$) &    5.90 $\pm$ 0.15  &	 \nodata	  &    \nodata  	&    \nodata	      &    \nodata	   \\
12+log(Fe/H)            &    6.47 $\pm$ 0.15  &	 \nodata	  &    \nodata  	&    \nodata	      &    \nodata	   \\
log(Fe/O)               & $-$1.95 $\pm$ 0.17  &	 \nodata	  &    \nodata  	&    \nodata	      &    \nodata	   \\
\noalign{\smallskip} 
12+log(He$^+$/H$^+$)    &   10.97 $\pm$ 0.04  &	10.95 $\pm$ 0.07  & 11.03 $\pm$ 0.09 &   10.90 $\pm$ 0.12  &  10.94:     \\
\noalign{\smallskip}
\tableline
\noalign{\smallskip}
[O/H]$^b$               & $-$0.33 $\pm$ 0.13  & $-$0.27             & $-$0.27              & $-$0.02             & $-$0.31	    \\
\noalign{\smallskip}
\tableline
  \end{tabular}
  \begin{flushleft}
  $^a$Electron temperatures estimated using empirical relations. \\
  $^b$[O/H]=log(O/H)-log(O/H)$_{\odot}$, using 12+log(O/H)$_{\odot}$ = 8.66$\pm$0.05 \citep{ASP05}.\\
  $^c$Oxygen abundance computed using the direct method.\\ 
  $^d$Oxygen abundance computed using the empirical calibrations given by \citet{P01a,P01b}.
  \end{flushleft}
\end{table*}

\begin{table*}[t!]
\centering
  \caption{\footnotesize{Dereddened line intensity ratios with respect to $I$(\Hb)=100 for regions analyzed in POX 4, UM 420 and SBS 0926+606.}}
  \label{pox4lineas}
  \tiny
  \begin{tabular}{l  r@{\hspace{5pt}}  r@{\hspace{8pt}}    r@{\hspace{8pt}}r@{\hspace{8pt}}r@{\hspace{8pt}}r@{\hspace{8pt}}  }
  \noalign{\smallskip}
  \tableline
   \noalign{\smallskip}
\noalign{\smallskip}
Line  & $f(\lambda)$&  POX 4    &        POX 4 Comp            &     UM 420          &      SBS 0926+606A     &    SBS 0926+606B        \\
\noalign{\smallskip}
 \tableline
\noalign{\smallskip}

 3679.36   H	I &  0.265 &  0.30$\pm$0.07   &	\nodata          &	  \nodata &	   \nodata &	    \nodata\\
 3682.81   H	I &  0.264 &  0.40$\pm$0.08   &	\nodata          &	  \nodata &	   \nodata &	    \nodata\\
 3686.83   H	I &  0.263 &  0.74$\pm$0.10   &        \nodata   &	  \nodata &	   \nodata &	    \nodata\\
 3691.56   H	I &  0.263 &  0.92$\pm$0.11   &        \nodata   &	  \nodata &	   \nodata &	    \nodata\\
 3697.15   H	I &  0.262 &  1.08$\pm$0.12   &        \nodata   &	 0.53:    &	   \nodata &	    \nodata\\
 3703.86   H	I &  0.260 &  2.02$\pm$0.17   &        \nodata   &  0.93$\pm$0.36 &	   \nodata &	    \nodata\\
 3711.97   H	I &  0.259 &  1.86$\pm$0.16   &        \nodata   &  1.14$\pm$0.38 &	   \nodata &	    \nodata\\
 3721.83  [S III] &  0.257 &  3.68$\pm$0.24   &        \nodata   &  1.30$\pm$0.39 &	   \nodata &	    \nodata\\
 3726.03  [O  II] &  0.257 &  42.8$\pm$2.0    &   142$\pm$35     &  85.9$\pm$4.7  &	   \nodata &	    \nodata\\
 3728.82  [O  II] &  0.256 &  57.1$\pm$2.7    &   165$\pm$40     & 140.2$\pm$7.5  &	   \nodata &	    \nodata\\
 3734.17   H	I &  0.255 &  2.58$\pm$0.19   &        \nodata   &  2.19$\pm$0.45 &	   \nodata &	    \nodata\\
 3750.15   H	I &  0.253 &  2.88$\pm$0.21   &        \nodata   &  2.32$\pm$0.46 &	   \nodata &	    \nodata\\
 3770.63   H	I &  0.249 &  3.77$\pm$0.25   &        \nodata   &  2.41$\pm$0.47 &	   \nodata &	    \nodata\\
 3797.90   H	I &  0.244 &  5.30$\pm$0.37   &        \nodata   &  3.36$\pm$0.75 &	   \nodata &	    \nodata\\
 3819.61  He	I &  0.240 &  0.91$\pm$0.11   &        \nodata   &	 0.87:    &	   \nodata &	    \nodata\\
 3835.39   H	I &  0.237 &  7.18$\pm$0.40   &        \nodata   &  5.96$\pm$0.70 &	   \nodata &	    \nodata\\
 3868.75 [Ne III] &  0.230 &  51.7$\pm$2.3    &      35.6:       &  29.7$\pm$2.0  &	   \nodata &	    \nodata\\
 3889.05   H	I &  0.226 & 17.75$\pm$0.95   &        \nodata   &  16.8$\pm$1.5  &	   \nodata &	    \nodata\\
 3967.46 [Ne III] &  0.210 & 16.17$\pm$0.78   &        \nodata   &  6.45$\pm$0.72 &	   \nodata &	    \nodata\\
 3970.07   H	I &  0.210 & 15.36$\pm$0.71   &        \nodata   &  16.1$\pm$1.2  &	   \nodata &	    \nodata\\
 4009.22  He	I &  0.202 &  0.15$\pm$0.06   &	\nodata          &	  \nodata &	   \nodata &	    \nodata\\
 4026.21  He	I &  0.198 &  1.57$\pm$0.14   &        \nodata   &  1.55$\pm$0.40 &	   \nodata &	    \nodata\\
 4068.60  [S  II] &  0.189 &  0.92$\pm$0.11   &        \nodata   &  2.40$\pm$0.46 &	   \nodata &	    \nodata\\
 4076.35  [S  II] &  0.187 &  0.28$\pm$0.07   &	\nodata          &	 0.71:    &	   \nodata &	    \nodata\\
 4101.74   H	I &  0.182 &  26.0$\pm$1.2    &      \nodata	 &  26.1$\pm$2.1  &	   \nodata &	    \nodata\\
 4143.76  He	I &  0.172 &	   0.13:      &	    \nodata      &	  \nodata &	   \nodata &	    \nodata\\
 4168.97  He	I &  0.167 &	    \nodata   &        \nodata   &       \nodata  &  0.77$\pm$0.25 &	   \nodata\\
 4276.83 [Fe  II] &  0.142 &	   0.08:      &	    \nodata      &	  \nodata &	   \nodata &	    \nodata\\
 4287.40 [Fe  II] &  0.139 &	   0.11:      &	    \nodata      &	 0.42:    &	   \nodata &	    \nodata\\
 4340.47   H	I &  0.127 &  51.2$\pm$2.3    &    49$\pm$16     &  47.0$\pm$3.0  &	 47.5$\pm$2.8 &  47.4$\pm$9.1\\
 4363.21  [O III] &  0.121 & 11.89$\pm$0.61   &        \nodata   &  4.48$\pm$0.82 &  7.00$\pm$0.66 &	    \nodata\\
 4387.93  He	I &  0.115 &  0.42$\pm$0.08   &	\nodata          &	 0.21:    &	   \nodata &	    \nodata\\
 4413.78 [Fe III] &  0.109 &	   0.07:      &	    \nodata      &	  \nodata &	   \nodata &	    \nodata\\
 4471.48  He	I &  0.095 &  4.12$\pm$0.26   &        \nodata   &  3.07$\pm$0.50 &  3.69$\pm$0.41 &	    \nodata\\
 4562.60  Mg   I] &  0.073 &  0.19$\pm$0.06   &	\nodata          &	  \nodata &	     0.44: &	    \nodata\\
 4571.20  Mg   I] &  0.071 &	   0.13:      &	    \nodata      &	  \nodata &	   \nodata &	    \nodata\\
 4658.10 [Fe III] &  0.050 &  0.64$\pm$0.09   &	\nodata          &  1.31$\pm$0.46 & 1.20$\pm$0.40 &	    \nodata\\
 4686.00  He   II &  0.043 &  1.21$\pm$0.16   &        \nodata   &  1.04$\pm$0.28 &  0.72$\pm$0.21 &	    \nodata\\
 4701.53 [Fe III] &  0.039 &	   0.14:      &	    \nodata      &	  \nodata &	   \nodata &	    \nodata\\
 4711.37 [Ar  IV] &  0.037 &  1.86$\pm$0.15   &        \nodata   &	  \nodata &	     0.49: &	    \nodata\\
 4713.14  He	I &  0.037 &	    \nodata   &        \nodata   &	 0.45:    &  0.72$\pm$0.24 &	    \nodata\\
 4740.16 [Ar  IV] &  0.030 &  1.34$\pm$0.12   &        \nodata   &   \nodata      &	    0.49: &	   \nodata\\
 4754.83 [Fe III] &  0.026 &	    \nodata   &        \nodata   &    \nodata     &	    0.29: &	   \nodata\\
 4861.33   H	I &  0.000 & 100.0$\pm$4.3    &   100$\pm$19     & 100.0$\pm$5.5  &	100.0$\pm$4.9 &   100$\pm$14\\
 4881.00 [Fe III] & -0.005 &  0.15$\pm$0.06   &	\nodata          &	  \nodata &	     0.29: &	    \nodata\\
 4921.93  He	I & -0.015 &  0.49$\pm$0.10   &        \nodata   &	 0.22:    &  0.38: &	    \nodata\\
 4958.91  [O III] & -0.024 &   237$\pm$10     &	 93$\pm$19       & 106.8$\pm$6.0  &	 157.8$\pm$7.7 &   134$\pm$18\\
 4985.90 [Fe III] & -0.031 &  0.57$\pm$0.09   &	\nodata          &	 0.57:    &	   \nodata &	    \nodata\\
 5006.84  [O III] & -0.036 &   731$\pm$32     &	255$\pm$55       &   312$\pm$17   &	   \nodata &	    \nodata\\
 5015.68  He	I & -0.038 &  2.57$\pm$0.21   &        \nodata   &	  \nodata &	    \nodata &	     \nodata\\
 5041.03  Si   II & -0.044 &	   0.14:      &	    \nodata      &	  \nodata &	    \nodata &	     \nodata\\
 5047.74  He	I & -0.046 &  0.16$\pm$0.06   &	\nodata          &	  \nodata &	    \nodata &	     \nodata\\
 5197.90  [N   I] & -0.082 &  0.23$\pm$0.06   &	\nodata          &	  \nodata &	    \nodata &	     \nodata\\
 5200.26  [N   I] & -0.083 &  0.30$\pm$0.07   &	\nodata          &	  \nodata &	    \nodata &	     \nodata\\
 5270.40 [Fe III] & -0.100 &        \nodata   &        \nodata   &       0.48:    &          \nodata&        \nodata\\
 5517.71 [Cl III] & -0.154 &  0.31$\pm$0.07   &	\nodata          &	 0.25:    &	   \nodata &	    \nodata\\
 5537.88 [Cl III] & -0.158 &  0.23$\pm$0.08   &	\nodata          &	 0.26:    &	   \nodata &	    \nodata\\
 5754.64  [N  II] & -0.194 &	   0.08:      &	    \nodata      &	 0.59:    &	   \nodata &	    \nodata\\
 5875.64  He	I & -0.215 & 10.93$\pm$0.56   &        \nodata   & 10.51$\pm$0.79 & 10.77$\pm$0.85 &  12.4$\pm$4.0\\
 6300.30  [O   I] & -0.282 &  1.81$\pm$0.12   &        \nodata   &  7.49$\pm$0.49 &  3.13$\pm$0.30 &	   5.32:\\
 6312.10  [S III] & -0.283 &  1.60$\pm$0.12   &        \nodata   &  1.75$\pm$0.18 &  1.98$\pm$0.24 &	    \nodata\\
 6363.78  [O   I] & -0.291 &  0.61$\pm$0.09   &	\nodata          &  2.06$\pm$0.41 &  0.92$\pm$0.23 &	   2.24:\\
 6371.36  Si   II & -0.292 &	   0.05:      &	    \nodata      &	  \nodata &	   \nodata &	    \nodata\\
 6548.03  [N  II] & -0.318 &  1.58$\pm$0.11   &        \nodata   &  9.14$\pm$0.71 &  2.47$\pm$0.30 &	   4.51:\\
 6562.82   H	I & -0.320 &   285$\pm$12     &	282$\pm$50       &   281$\pm$14   &	    286$\pm$15 &   286$\pm$34\\
 6583.41  [N  II] & -0.323 &  4.20$\pm$0.23   &      11.7:       &  28.4$\pm$1.7  &	7.50$\pm$0.64 &  18.2$\pm$4.6\\
 6678.15  He	I & -0.336 &  2.93$\pm$0.20   &        \nodata   &	   \nodata &  2.72$\pm$0.34 &	    2.63:\\
 6716.47  [S  II] & -0.342 &  7.80$\pm$0.38   &      15.9:       &	   \nodata &	  16.8$\pm$1.0 &  42.8$\pm$7.0\\
 6730.85  [S  II] & -0.344 &  6.03$\pm$0.30   &      10.2:       &	   \nodata &	12.12$\pm$0.77 &  29.5$\pm$5.6\\
 7065.28  He	I & -0.387 &	    \nodata   &        \nodata   &	   \nodata &  2.45$\pm$0.32 &	     \nodata\\
 7135.78 [Ar III] & -0.396 &	    \nodata   &        \nodata   &	   \nodata &  6.54$\pm$0.45 &	    5.09:\\
 7281.35  He	I & -0.414 &	    \nodata   &        \nodata   &	  \nodata  &  0.61$\pm$0.20 &	     \nodata\\
 7318.39  [O  II] & -0.418 &	    \nodata   &        \nodata   &	  \nodata  &  2.11$\pm$0.23 &	    2.22:\\
 7329.66  [O  II] & -0.420 &	    \nodata   &        \nodata   &	  \nodata  &  1.71$\pm$0.22 &	    2.05:\\
 7751.10 [Ar III] & -0.467 &	    \nodata   &        \nodata   &	  \nodata  &  1.75$\pm$0.25 &	     \nodata\\
 \noalign{\smallskip}
 \tableline
 \noalign{\smallskip}
Aperture size ($\arcsec$)   &       &   7.2$\times$1      &   3.6$\times$1       &  3.6$\times$1     &   4.0$\times$1       &  5.6$\times$1       \\
Distance ($\arcsec$)$^a$&     &      --             &   20.4               &    --             &  --                  &  74.4               \\              
$F$(\Hb)$^b$          &       &   56.0 $\pm$   1.8  &    0.11 $\pm$   0.02 &  6.88 $\pm$ 0.27  &   16.20 $\pm$   0.58 &	 1.58 $\pm$   0.15 \\
$C$(\Hb)              &       &	0.08 $\pm$   0.01 &    0.06 $\pm$   0.03 &  0.09 $\pm$ 0.01  &    0.12 $\pm$   0.03 &	 0.18 $\pm$   0.04 \\
$W_{abs}$ (\AA)     &       &	 2.0 $\pm$    0.1 &     1.4 $\pm$   0.2  &   2.0 $\pm$ 0.1   &    0.7  $\pm$   0.1  &	  1.0 $\pm$   0.3 \\
\noalign{\smallskip}
$-W$(\Ha) (\AA)     &       &   1075 $\pm$   48   &     380 $\pm$   65   &  1076 $\pm$   55  &     613 $\pm$   33   &	   92 $\pm$	11 \\
$-W$(\Hb) (\AA)     &       &    200 $\pm$    9   &      12 $\pm$    4   &   169 $\pm$   10  &     125 $\pm$	6   &	   18 $\pm$	 3 \\
$-W$(H$\gamma$) (\AA)&     &     71 $\pm$    3   &       4  $\pm$    1   &    69 $\pm$    4  &      41 $\pm$	2   &	    9 $\pm$	 2 \\
$-W$([O III]) 5007 (\AA)&   &   1366 $\pm$   60   &      19 $\pm$    3   &   581 $\pm$   32  &       \nodata	    &	   \nodata	   \\
\noalign{\smallskip}
 \tableline
   \end{tabular}
  \begin{flushleft}		
	 $^a$ Relative distance of the knot with respect to the main region in the galaxy.\\
     $^b$ In units of 10$^{-15}$ erg s$^{-1}$ cm$^{-2}$ and not corrected for extinction.
   \end{flushleft}
\end{table*}

\begin{table*}[t!]
  \caption{\footnotesize{Physical conditions and chemical abundances of the ionized gas in POX 4, UM 420 and \mbox{SBS 0926+606.}}}
  \label{pox4abun}
  \tiny
  \centering
  \begin{tabular}{l  rrrrr}
\tableline
\noalign{\smallskip}
Object                 &      POX 4         &     POX 4 Comp$^a$   &	   UM 420 	   &   SBS 0926+606A	&   SBS 0926+606B$^a$  \\
\noalign{\smallskip}
\tableline
\noalign{\smallskip}				      		   		    
$T_e$(O III) (K)        &   14000 $\pm$ 500   &   12400 $\pm$ 1000   &	13200 $\pm$ 600    &   13600 $\pm$ 700  &      11500 $\pm$ 1000  \\
$T_e$(O II) (K)         &   12800 $\pm$ 400   &   11800 $\pm$ 600   &	12200 $\pm$ 500    &   12500 $\pm$ 500  &      11000 $\pm$ 800   \\
$n_e$ (cm$^{-3}$)       &     250 $\pm$  80   &    $<$100 	        &	  140 $\pm$  80    &  $<$100	        &  $<$100	      \\
\noalign{\smallskip}
12+log(O$^{+}$/H$^+$)   &    7.21 $\pm$ 0.04  &    7.85 $\pm$ 0.14  &	 7.63 $\pm$ 0.05   &	7.38 $\pm$ 0.10  &	  7.73 $\pm$ 0.16   \\
12+log(O$^{++}$/H$^+$)  &    7.96 $\pm$ 0.04  &    7.55 $\pm$ 0.11  &	 7.67 $\pm$ 0.05   &	7.80 $\pm$ 0.08  &	  7.04 $\pm$ 0.15  \\
12+log(O/H)             &    8.03 $\pm$ 0.04$^c$&  8.03 $\pm$ 0.13  &	 7.95 $\pm$ 0.05$^c$&	7.94 $\pm$ 0.08$^c$&  8.15 $\pm$ 0.16	\\
\noalign{\smallskip}
log(O$^{++}$/O$^+$)     &    0.74 $\pm$ 0.06  & $-$0.30 $\pm$ 0.22  &	 0.00 $\pm$ 0.08   &	0.42 $\pm$ 0.12  &    0.21 $\pm$ 0.14	\\
12+log(N$^+$/H$^+$)     &    5.68 $\pm$ 0.04  &    6.25 $\pm$ 0.18  &	 6.52 $\pm$ 0.05   &	5.93 $\pm$ 0.05  &	  6.39 $\pm$ 0.09   \\
12+log(N/H)             &    6.50 $\pm$ 0.06  &    6.43 $\pm$ 0.22  &	 6.84 $\pm$ 0.06   &	6.48 $\pm$ 0.10  &	  6.80 $\pm$ 0.13   \\
log(N/O)                & $-$1.54 $\pm$ 0.06  & $-$1.60 $\pm$ 0.20  & $-$1.11 $\pm$ 0.07   & $-$1.45 $\pm$ 0.09  & $-$1.35 $\pm$ 0.12	\\
\noalign{\smallskip}
12+log(S$^+$/H$^+$)     &    5.28 $\pm$ 0.03  &    5.62 $\pm$ 0.20  &	 5.61 $\pm$ 0.11   &	5.61 $\pm$ 0.04  &	  6.12 $\pm$ 0.11   \\
12+log(S$^{++}$/H$^+$)  &    6.03 $\pm$ 0.08  & \nodata             &	 6.16 $\pm$ 0.09   &	6.17 $\pm$ 0.11  & \nodata	      \\
12+log(S/H)             &    6.24 $\pm$ 0.07  & \nodata             &	 6.29 $\pm$ 0.10   &	6.34 $\pm$ 0.10  & \nodata	      \\
log(S/O)                & $-$1.80 $\pm$ 0.10  & \nodata             & $-$1.66 $\pm$ 0.13   & $-$1.60 $\pm$ 0.13  & \nodata	      \\
\noalign{\smallskip}
12+log(Ne$^{++}$/H$^+$) &    7.18 $\pm$ 0.06  &    6.97:            &	 6.96 $\pm$ 0.09   &	\nodata 	 & \nodata		  \\
12+log(Ne/H)            &    7.26 $\pm$ 0.06  &    7.40:            &	 7.24 $\pm$ 0.09   &	\nodata 	 & \nodata		 \\
log(Ne/O)               & $-$0.78 $\pm$ 0.10  & $-$0.60:            & $-$0.71 $\pm$ 0.13   &	\nodata 	 & \nodata		  \\
\noalign{\smallskip}
12+log(Ar$^{+2}$/H$^+$) & \nodata             & \nodata             & \nodata		   &	5.52 $\pm$ 0.07  &  5.54 $\pm$ 0.15  \\
12+log(Ar$^{+3}$/H$^+$) &    5.03 $\pm$ 0.07  & \nodata 	        &	\nodata        &	4.61 $\pm$ 0.18  & \nodata			\\
12+log(Ar/H)            & \nodata             & \nodata             & \nodata		   &	5.60 $\pm$ 0.11  & \nodata	\\
log(Ar/O)               & \nodata             & \nodata             & \nodata		   & $-$2.34 $\pm$ 0.13  & \nodata	\\
\noalign{\smallskip} 
12+log(Cl$^{++}$/H$^+$) &    3.83 $\pm$ 0.14  & \nodata 	        &	 4.18 $\pm$ 0.26   & \nodata		     & \nodata        \\
12+log(Fe$^{++}$/H$^+$) &    5.14 $\pm$ 0.10  &	\nodata             &	 5.52 $\pm$ 0.12&	5.47 $\pm$ 0.14  & \nodata	      \\
12+log(Fe/H)            &    5.86 $\pm$ 0.10  &	\nodata             &	 5.79 $\pm$ 0.13&	5.95 $\pm$ 0.15  & \nodata	      \\
log(Fe/O)               & $-$2.17 $\pm$ 0.11  & \nodata             & $-$2.16 $\pm$ 0.13& $-$1.99 $\pm$ 0.16  & \nodata	      \\
\noalign{\smallskip} 
12+log(He$^+$/H$^+$)    &   10.91 $\pm$ 0.03  & \nodata             &	10.88 $\pm$ 0.04  &   10.94 $\pm$ 0.04  & 11.0:\\
\noalign{\smallskip}
\tableline
\noalign{\smallskip}
[O/H]$^b$               &   $-$0.63 $\pm$ 0.09 & $-$0.63   &  $-$0.71$\pm$0.10   &   $-$0.72 $\pm$ 0.13   & $-$0.51 \\
\noalign{\smallskip}
\tableline
  \end{tabular}
  \begin{flushleft}
  $^a$Electron temperatures estimated using empirical relations. \\
  $^b$[O/H]=log(O/H)-log(O/H)$_{\odot}$, using 12+log(O/H)$_{\odot}$ = 8.66$\pm$0.05 \citep{ASP05}.\\
  $^c$Considering the existence of O$^{+3}$ because of the detection of \ion{He}{ii} $\lambda$4686, this value should be $\sim$0.01--0.02 dex higher.
  \end{flushleft}
\end{table*}

\begin{table*}[t!]
\centering
  \caption{\footnotesize{Dereddened line intensity ratios with respect to $I$(\Hb)=100 for regions analyzed in SBS 0948+532, SBS 1054+365 and SBS 
1211+540.}}
  \label{sbs1054lineas}
  \tiny
  \begin{tabular}{l  r@{\hspace{5pt}}  r@{\hspace{8pt}}    r@{\hspace{8pt}}r@{\hspace{8pt}}r@{\hspace{8pt}}  }
  \noalign{\smallskip}
  \tableline
   \noalign{\smallskip}
\noalign{\smallskip}
Line  & $f(\lambda)$&  SBS 0948+532    &        SBS 1054+365      &     SBS 1054+365 b      &     SBS 1211+540     \\
\noalign{\smallskip}
 \tableline
\noalign{\smallskip}

 3697.15    H	 I  &  0.262 &       0.48:   &        \nodata	&	   \nodata  &	     \nodata\\
 3703.86    H	 I  &  0.260 &  1.96$\pm$0.38	&	 \nodata   &	   \nodata  &	     \nodata\\
 3711.97    H	 I  &  0.259 &  1.25$\pm$0.32	&	 \nodata   &	   \nodata  &	     \nodata\\
 3721.83   [S III]  &  0.257 &  1.57$\pm$0.35	&	 \nodata   &	   \nodata  &	     \nodata\\
 3726.03   [O  II]  &  0.257 &  45.4$\pm$2.7   &	\nodata   &	   \nodata  &  33.0$\pm$3.8\\ 
 3728.00   [O  II]  &  0.256 &        \nodata	& 100.2$\pm$7.5   &	350$\pm$95  &	     \nodata\\
 3728.82   [O  II]  &  0.256 &  65.7$\pm$3.8   &	\nodata   &	   \nodata  &  44.8$\pm$4.6\\ 
 3734.17    H	 I  &  0.255 &  2.52$\pm$0.43	&	 \nodata   &	   \nodata  &	    2.61:\\   
 3750.15    H	 I  &  0.253 &  2.56$\pm$0.43	&	 \nodata   &	   \nodata  &	3.7$\pm$1.3\\ 
 3770.63    H	 I  &  0.249 &  2.88$\pm$0.45	&	1.53:	&	   \nodata  &	3.2$\pm$1.2\\ 
 3797.90    H	 I  &  0.244 &  5.11$\pm$0.71	&	2.85:	&	   \nodata  &	    3.99:\\   
 3819.61   He	 I  &  0.240 &  0.65$\pm$0.25	&	 \nodata   &	   \nodata  &	     \nodata\\
 3835.39    H	 I  &  0.237 &  7.74$\pm$0.77	&   6.8$\pm$1.3   &	   \nodata  &	5.9$\pm$1.5\\ 
 3868.75  [Ne III]  &  0.230 &  41.6$\pm$2.6   &  49.6$\pm$4.8   &	   \nodata  &  37.6$\pm$4.0\\ 
 3889.05    H	 I  &  0.226 &  19.7$\pm$1.5   &  19.8$\pm$2.8   &	   \nodata  &  19.9$\pm$3.0\\ 
 3967.46  [Ne III]  &  0.210 &  15.0$\pm$1.2   &  27.4$\pm$2.8   &	   \nodata  &  13.3$\pm$2.1\\ 
 3970.07    H	 I  &  0.210 &  16.2$\pm$1.2   &	\nodata   &	   \nodata  &  16.1$\pm$2.2\\ 
 4026.21   He	 I  &  0.198 &  1.45$\pm$0.32	&	 \nodata   &	   \nodata  &	     \nodata\\
 4068.60   [S  II]  &  0.189 &  1.53$\pm$0.33	&	2.05:	&	 \nodata    &	     \nodata\\
 4101.74    H	 I  &  0.182 &  26.2$\pm$1.7   &  26.4$\pm$3.0   &	  \nodata   &  26.2$\pm$3.0\\ 
 4340.47    H	 I  &  0.127 &  47.2$\pm$2.9   &  47.2$\pm$4.1   &	28.51:      &  47.3$\pm$4.1\\ 
 4363.21   [O III]  &  0.121 &  8.13$\pm$0.82	&   9.7$\pm$1.5   &	   \nodata  &  12.2$\pm$1.8\\ 
 4471.48   He	 I  &  0.095 &  3.81$\pm$0.49	&  3.31$\pm$0.81   &	   \nodata  &	4.6$\pm$1.3\\ 
 4658.10  [Fe III]  &  0.050 &       1.13:   &        \nodata	&	 \nodata    &	     \nodata\\
 4686.00   He	II  &  0.043 &  1.3$\pm$0.3	&  0.61$\pm$0.23   &	   \nodata  &	     \nodata\\
 4711.37  [Ar  IV]  &  0.037 &  0.85$\pm$0.25	&	1.18:	&	 \nodata    &	     \nodata\\
 4713.14   He	 I  &  0.037 &        \nodata	&	 \nodata   & \nodata	    &	    0.75:\\   
 4733.93  [Fe III]  &  0.031 &       0.18:   &        \nodata	&	 \nodata    &	     \nodata\\
 4740.16  [Ar  IV]  &  0.030 &  0.61$\pm$0.24   &        \nodata	&	 \nodata    &	    0.98:\\   
 4754.83  [Fe III]  &  0.026 &       0.34:   &        \nodata	&	 \nodata    &	     \nodata\\
 4861.33    H	 I  &  0.000 & 100.0$\pm$5.2   & 100.0$\pm$7.4   &   100$\pm$39     & 100.0$\pm$7.5\\ 
 4958.91   [O III]  & -0.024 & 189.5$\pm$9.2   &   210$\pm$13	&    75$\pm$23      &	163$\pm$10\\  
 4985.90  [Fe III]  & -0.031 &  1.49$\pm$0.30	&	 \nodata   &	   \nodata  &	     \nodata\\
 5006.84   [O III]  & -0.036 &   584$\pm$28   &   623$\pm$37   &   183$\pm$53	    &	481$\pm$29\\  
 5015.68   He	 I  & -0.038 &        \nodata	&	 \nodata   &   \nodata      &	    2.24:\\   
 5197.90   [N	I]  & -0.082 &        \nodata	&	1.03:	&	 \nodata    &	     \nodata\\
 5517.71  [Cl III]  & -0.154 &       0.39:   &        \nodata	&	 \nodata    &	     \nodata\\
 5537.88  [Cl III]  & -0.158 &       0.24:   &        \nodata	&	 \nodata    &	     \nodata\\
 5875.64   He	 I  & -0.215 & 10.76$\pm$0.81	&   8.7$\pm$1.6   &	   \nodata  &	     \nodata\\
 6300.30   [O	I]  & -0.282 &  3.00$\pm$0.31	&	1.12:	&	 \nodata    &	    2.36:\\   
 6312.10   [S III]  & -0.283 &  1.74$\pm$0.24	&	1.36:	&	 \nodata    &	2.8$\pm$1.1\\ 
 6363.78   [O	I]  & -0.291 &  1.07$\pm$0.23	&	0.65:	&	 \nodata    &	    0.57:\\   
 6548.03   [N  II]  & -0.318 &  2.45$\pm$0.30	&  2.02$\pm$0.74   &	  7.66:    &	    0.82:\\   
 6562.82    H	 I  & -0.320 &   278$\pm$14     &   277$\pm$17   &   279$\pm$74	    &	280$\pm$17\\  
 6583.41   [N  II]  & -0.323 &  6.21$\pm$0.60	&   5.6$\pm$1.0   &	 20.2:     &  2.24$\pm$0.89\\
 6678.15   He	 I  & -0.336 &  2.81$\pm$0.36	&  2.94$\pm$0.96   &	  7.40:    &	3.6$\pm$1.1\\ 
 6716.47   [S  II]  & -0.342 &        \nodata	&  10.1$\pm$1.2   &	    22.4:     &  5.65$\pm$0.88\\
 6730.85   [S  II]  & -0.344 &        \nodata	&  7.03$\pm$0.97   &	 19.6:    &  4.91$\pm$0.82\\
 7065.28   He	 I  & -0.387 &        \nodata	&  2.08$\pm$0.83   &	  \nodata   &	     \nodata\\
 7135.78  [Ar III]  & -0.396 &        \nodata	&   8.8$\pm$1.5   &	  \nodata   &	     \nodata\\									  									    
\noalign{\smallskip}
 \tableline
 \noalign{\smallskip}
Aperture size (arcsec)   &         &   3.6               &    6.4               &    5.8               &     3.6             \\
Distance$^a$ (arcsec)&       &   0                 &      0               &    17.8              &      0              \\
$F$(\Hb)$^b$          &         &   8.44 $\pm$   0.32 &	14.57 $\pm$   0.68 &	0.66 $\pm$   0.13 &    1.84 $\pm$   0.09 \\ 
$C$(\Hb)              &         &   0.35 $\pm$   0.03 &	 0.02 $\pm$   0.02 &	0.6 $\pm$   0.1   &    0.12 $\pm$   0.01 \\
$W_{abs}$ (\AA)     &         &    0.3 $\pm$	0.1 &	 0.8  $\pm$   0.1  &	0.3  $\pm$   0.1  &    1.3  $\pm$   0.1  \\ 
\noalign{\smallskip}
$-W$(\Ha) (\AA)     &         &  788   $\pm$   43   &	422   $\pm$   27   &	32   $\pm$    8   &   705   $\pm$   45   \\ 
$-W$(\Hb) (\AA)     &         &  213   $\pm$   11   &	 89   $\pm$    7   &	 8   $\pm$    3   &   135   $\pm$   10   \\ 
$-W$(H$\gamma$) (\AA)&        &   57   $\pm$	4   &	 43   $\pm$    4   &	 2   $\pm$    1   &    74   $\pm$    7   \\ 
$-W$([O III]) 5007 (\AA)&     &  689   $\pm$   34   &	567   $\pm$   35   &	12   $\pm$    3   &   618   $\pm$   38   \\ 
\noalign{\smallskip}
 \tableline
   \end{tabular}
  \begin{flushleft}		
	 $^a$ Relative distance of the knot with respect to the main region in the galaxy.\\
     $^b$ In units of 10$^{-15}$ erg s$^{-1}$ cm$^{-2}$ and not corrected for extinction.
   \end{flushleft}
\end{table*}

\begin{table*}[t!]
  \caption{\footnotesize{Physical conditions and chemical abundances of the ionized gas in SBS 0948+532, SBS 1054+365 and SBS 1211+540.}}
 \label{sbs1054abun}
  \tiny
  \centering
  \begin{tabular}{l  rrrr}
\tableline
\noalign{\smallskip}
Object                  &    SBS 0948+532     &     SBS 1054+365    &  SBS 1054+365b$^a$  &   SBS 1211+540     \\
\noalign{\smallskip}
\tableline
\noalign{\smallskip}				      
$T_e$(O III) (K)        &   13100 $\pm$ 600   &   13700 $\pm$ 900   &   11800 $\pm$ 1100  &   17100 $\pm$ 600  \\
$T_e$(O II) (K)         &   12200 $\pm$ 400   &   12600 $\pm$ 700   &   11300 $\pm$  900  &   15000 $\pm$ 400  \\
$N_e$ (cm$^{-3}$)       &     250 $\pm$  80   &  $<$100 	        &     300 $\pm$  200  &     320 $\pm$  50  \\
\noalign{\smallskip}
12+log(O$^{+}$/H$^+$)   &    7.33 $\pm$ 0.05  &    7.22 $\pm$ 0.10  &	 7.97 $\pm$ 0.18  &    6.88 $\pm$ 0.05 \\
12+log(O$^{++}$/H$^+$)  &    7.94 $\pm$ 0.05  &    7.92 $\pm$ 0.07  &	 7.62 $\pm$ 0.12  &    7.57 $\pm$ 0.04 \\
12+log(O/H)             &    8.03 $\pm$ 0.05  &    8.00 $\pm$ 0.07  &	 8.13 $\pm$ 0.16  &    7.65 $\pm$ 0.04 \\
\noalign{\smallskip}
log(O$^{++}$/O$^+$)     &    0.61 $\pm$ 0.08  &    0.70 $\pm$ 0.11  & $-$0.35 $\pm$ 0.20  &    0.69 $\pm$ 0.07 \\
12+log(N$^+$/H$^+$)     &    5.91 $\pm$ 0.05  &    5.81 $\pm$ 0.08  &	 6.49 $\pm$ 0.20  &    5.26 $\pm$ 0.12 \\
12+log(N/H)             &    6.61 $\pm$ 0.07  &    6.59 $\pm$ 0.09  &	 6.65 $\pm$ 0.21  &    6.03 $\pm$ 0.13 \\
log(N/O)                & $-$1.42 $\pm$ 0.08  & $-$1.41 $\pm$ 0.08  & $-$1.47 $\pm$ 0.20  & $-$1.62 $\pm$ 0.10 \\  
\noalign{\smallskip}
12+log(S$^+$/H$^+$)     &    5.43 $\pm$ 0.12  &    5.37 $\pm$ 0.07  &	 5.89 $\pm$ 0.16  &    5.04 $\pm$ 0.06 \\
12+log(S$^{++}$/H$^+$)  &    6.16 $\pm$ 0.11  &    5.99 $\pm$ 0.22  &	 \nodata	  &    6.02 $\pm$ 0.14 \\
12+log(S/H)             &    6.34 $\pm$ 0.11  &    6.21 $\pm$ 0.18  &	 \nodata	  &    6.18 $\pm$ 0.12 \\
log(S/O)                & $-$1.69 $\pm$ 0.14  & $-$1.79 $\pm$ 0.15  &	 \nodata	  & $-$1.47 $\pm$ 0.14 \\
\noalign{\smallskip}
12+log(Ne$^{++}$/H$^+$) &    7.21 $\pm$ 0.09  &    7.25 $\pm$ 0.09  &	 \nodata	  &    6.82 $\pm$ 0.08 \\
12+log(Ne/H)            &    7.30 $\pm$ 0.09  &    7.33 $\pm$ 0.12  &	 \nodata	  &    6.90 $\pm$ 0.08 \\
log(Ne/O)               & $-$0.73 $\pm$ 0.12  & $-$0.67 $\pm$ 0.11  &	 \nodata	  & $-$0.75 $\pm$ 0.10 \\
\noalign{\smallskip}
12+log(Ar$^{+2}$/H$^+$) &    \nodata	      &    5.62 $\pm$ 0.10  &	 \nodata	  &    \nodata         \\
12+log(Ar$^{+3}$/H$^+$) &    4.79 $\pm$ 0.15  &    4.90 $\pm$ 0.20  &	 \nodata	  &    4.77 $\pm$ 0.22 \\
12+log(Ar/H)            &    \nodata	      &    5.71 $\pm$ 0.17  &	 \nodata	  &    \nodata         \\
log(Ar/O)               &    \nodata	      & $-$2.29 $\pm$ 0.14  &	 \nodata	  &    \nodata         \\
\noalign{\smallskip} 
12+log(Cl$^{++}$/H$^+$) &    3.97 $\pm$ 0.18  &    \nodata	    &	 \nodata	  &    \nodata         \\
12+log(Fe$^{++}$/H$^+$) &    5.64 $\pm$ 0.09  &	   \nodata 	    &    \nodata	  &    \nodata	       \\
12+log(Fe/H)            &    6.25 $\pm$ 0.09  &	   \nodata 	    &    \nodata	  &    \nodata	       \\
log(Fe/O)               & $-$1.78 $\pm$ 0.10  &	   \nodata 	    &    \nodata      &    \nodata	       \\
\noalign{\smallskip} 
12+log(He$^+$/H$^+$)    &   10.88 $\pm$ 0.04  &   10.88 $\pm$ 0.07  &   11.30:          &    10.90 $\pm$ 0.15  \\
\noalign{\smallskip}
\tableline
\noalign{\smallskip}
[O/H]$^b$               & $-$0.63 $\pm$ 0.10  & $-$0.66 $\pm$ 0.12  & $-$0.53             & $-$1.01 $\pm$ 0.09 \\
\noalign{\smallskip}
\tableline
  \end{tabular}
  \begin{flushleft}
  $^a$Electron temperatures estimated using empirical relations. \\
  $^b$[O/H]=log(O/H)-log(O/H)$_{\odot}$, using 12+log(O/H)$_{\odot}$ = 8.66$\pm$0.05 \citep{ASP05}.
  \end{flushleft}
\end{table*}

\begin{table*}[t!]
\centering
  \caption{\footnotesize{Dereddened line intensity ratios with respect to $I$(\Hb)=100 for knots analyzed in SBS 1319+579 (regions A, B and C) and SBS 
1415+457 (regions C and A).}}
  \label{sbs1319lineas}
  \tiny
  \begin{tabular}{l  r@{\hspace{3pt}}  rrrrr}
  \noalign{\smallskip}
  \tableline
   \noalign{\smallskip}
                     &       &        \multicolumn{3}{c}{SBS 1319+579}  &       \multicolumn{2}{c}{SBS 1415+437} \\   
Line & $f(\lambda)$& A     &  B             &  C       &   C     &  A \\
\noalign{\smallskip}
 \tableline
\noalign{\smallskip}
 4340.47   H	I  &  0.127 &  47.2$\pm$2.6    &  47.4$\pm$4.6    &  47.4$\pm$3.1  &  47.5$\pm$2.8   &  47.4$\pm$3.8\\
 4363.21  [O III]  &  0.121 &  9.98$\pm$0.62   &   4.2$\pm$1.0    &  3.73$\pm$0.64 &  7.11$\pm$0.61   &   6.1$\pm$1.2\\
 4387.93  He	I  &  0.115 &  0.43$\pm$0.17   &	\nodata   &	  1.12:    &	    \nodata   &        \nodata\\
 4437.55  He	I  &  0.104 &	    0.20:      &	\nodata   &	   \nodata &	    \nodata   &        \nodata\\
 4471.48  He	I  &  0.095 &  3.85$\pm$0.36   &   4.4$\pm$1.6    &  3.88$\pm$0.68 &  3.86$\pm$0.49   &  2.80$\pm$0.66\\
 4658.10 [Fe III]  &  0.050 &	    \nodata    &	\nodata   &	  0.72:    &  1.05$\pm$0.32   &       0.99:\\
 4686.00 He   II   &  0.043 &       0.8:     &        \nodata   &        \nodata &  2.35$\pm$0.23   &        \nodata\\
 4711.37 [Ar  IV]  &  0.037 &  2.14$\pm$0.35   &	\nodata   &	   \nodata &	    \nodata  &        \nodata\\
 4713.14 He    I   &  0.037 &        \nodata   &        \nodata   &        \nodata &	   0.30:   &	    \nodata\\
 4740.16 [Ar  IV]  &  0.030 &  1.61$\pm$0.26   &	\nodata   &	   \nodata &	   0.56:   &	    \nodata\\
 4861.33   H	I  &  0.000 & 100.0$\pm$4.9    & 100.0$\pm$8.5    & 100.0$\pm$6.2  & 100.0$\pm$4.9   & 100.0$\pm$6.3\\
 4881.00  [Fe III] & -0.005 &        \nodata   &        \nodata   &        \nodata &	   0.21:   &	    \nodata\\
 4921.93  He	I  & -0.015 &  0.96$\pm$0.23   &	\nodata   &	  0.75:    &  1.41$\pm$0.33   &        \nodata\\
 4958.91  [O III]  & -0.024 &	232$\pm$12     &   132$\pm$10	  & 129.9$\pm$7.4  & 107.5$\pm$5.5   & 107.3$\pm$6.5\\
 4985.90 [Fe III]  & -0.031 &	    \nodata    &	\nodata   &	   \nodata &  1.35$\pm$0.33   &       1.19:\\
 5006.84  [O III]  & -0.036 &	     \nodata   &	\nodata   &	   \nodata &   301$\pm$14   &	286$\pm$16\\
 5015.68 He    I   & -0.038 &	     \nodata   & 	\nodata   &	  \nodata  &  1.52$\pm$0.34   &        \nodata\\
 5875.64  He	I  & -0.215 & 11.90$\pm$0.86   &  11.7$\pm$3.2    &  11.2$\pm$1.2  &  9.61$\pm$0.92   &   7.7$\pm$1.1\\
 6300.30  [O   I]  & -0.282 &  1.74$\pm$0.26   &       3.08:	  &  4.77$\pm$0.72 &  3.44$\pm$0.43   &  2.41$\pm$0.56\\
 6312.10  [S III]  & -0.283 &  1.56$\pm$0.29   &       1.14:	  &  1.80$\pm$0.52 &  1.34$\pm$0.31   &       0.98:\\
 6363.78  [O   I]  & -0.291 &  0.60$\pm$0.18   &       1.28:	  &  1.48$\pm$0.52 &  0.96$\pm$0.31   &       0.86:\\
 6548.03  [N  II]  & -0.318 &  1.44$\pm$0.27   &   3.8$\pm$1.5    &  5.50$\pm$0.72 &  1.13$\pm$0.35   &       1.07:\\
 6562.82   H	I  & -0.320 &	280$\pm$13     &   279$\pm$19	  &   281$\pm$15   &   274$\pm$13   &	278$\pm$17\\
 6583.41  [N  II]  & -0.323 &  4.08$\pm$0.42   &  13.1$\pm$2.4    &  14.8$\pm$1.3  &  4.30$\pm$0.45   &  3.49$\pm$0.98\\
 6678.15  He	I  & -0.336 &  2.97$\pm$0.34   &   2.7$\pm$1.1	  &  2.87$\pm$0.66 &  2.31$\pm$0.41   &  2.28$\pm$0.66\\
 6716.47  [S  II]  & -0.342 &  7.79$\pm$0.56   &  26.0$\pm$3.2    &  27.9$\pm$1.9  & 13.15$\pm$0.83   &   9.2$\pm$1.1\\
 6730.85  [S  II]  & -0.344 &  5.76$\pm$0.44   &  17.8$\pm$2.9    &  18.9$\pm$1.3  &  9.60$\pm$0.63   &  6.71$\pm$0.94\\
 7065.28  He	I  & -0.387 &  2.69$\pm$0.36   &	\nodata       &  2.09$\pm$0.54 &  1.49$\pm$0.34   &  2.09$\pm$0.66\\
 7135.78 [Ar III]  & -0.396 &  4.76$\pm$0.47   &   6.6$\pm$1.6    &  6.06$\pm$0.72 &  3.77$\pm$0.38   &  3.28$\pm$0.61\\
 7281.35  He	I  & -0.414 &	    0.62:      &	\nodata       &	   \nodata &	    \nodata   &        \nodata\\
 7318.39  [O  II]  & -0.418 &  1.30$\pm$0.23   &   3.1$\pm$1.0    &  2.69$\pm$0.33 &  2.05$\pm$0.33   &  1.56$\pm$0.62\\
 7329.66  [O  II]  & -0.420 &  1.17$\pm$0.24   &       2.01:	  &  1.94$\pm$0.27 &  1.73$\pm$0.35   &       1.35:\\
 7751.10 [Ar III]  & -0.467 &  1.55$\pm$0.25   &	\nodata   &	   \nodata &  0.96$\pm$0.31   &       0.94:\\
\noalign{\smallskip}
 \tableline
 \noalign{\smallskip}
Aperture size (arcsec)   &         &    6$\times$1     &     2.8$\times$1   &   5.6$\times$1       & 6$\times$1           & 3.4$\times$1\\
Distance (arcsec)$^a$&       &       -           &    10              &   29                 &  -                   & 6  \\
$F$(\Hb)$^b$          &         &  14.57 $\pm$ 0.53 &	 1.97 $\pm$   0.10 &	8.18 $\pm$   0.32 &   18.51 $\pm$   0.66 &    4.07 $\pm$   0.17 \\
$C$(\Hb)              &         &   0.03 $\pm$ 0.01 &	 0.11 $\pm$   0.03 &	0.02 $\pm$   0.02 &    0.01 $\pm$   0.02 &    0.16 $\pm$   0.03 \\
$W_{abs}$ (\AA)     &         &    0.0 $\pm$  0.1 &	  0.4 $\pm$   0.1  &	0.2  $\pm$   0.1  &    0.8  $\pm$   0.1  &    1.0  $\pm$   0.2 \\
\noalign{\smallskip}
$-W$(\Ha) (\AA)     &         & 1530 $\pm$   75 &   162 $\pm$	11 &   295 $\pm$   23 &  1300 $\pm$	65 &  1187 $\pm$   75 \\
$-W$(\Hb) (\AA)     &         &  285 $\pm$   14 &    42 $\pm$	 4 &	94 $\pm$    6 &   222 $\pm$	11 &   130 $\pm$    8 \\
$-W$(H$\gamma$) (\AA)&        &   84 $\pm$    5 &    15 $\pm$	 1 &	23 $\pm$    1 &    73 $\pm$	 4 &	58 $\pm$    5 \\
$-W$([O III]) 5007 (\AA)&     &      \nodata    &      \nodata     &    \nodata       &   542 $\pm$	26 &   574 $\pm$   32 \\
\noalign{\smallskip}
 \tableline
   \end{tabular}
  \begin{flushleft}		
	 $^a$ Relative distance of the knot with respect to the main region in the galaxy.\\
     $^b$ In units of 10$^{-15}$ erg s$^{-1}$ cm$^{-2}$ and not corrected for extinction.
   \end{flushleft}
\end{table*}

\begin{table*}[t!]
  \caption{\footnotesize{Physical conditions and chemical abundances of the ionized gas for the regions analyzed in SBS 1319+579 and \mbox{SBS 
1415+437.}}}
  \label{sbs1319abun}
  \tiny
  \centering
  \begin{tabular}{l@{\hspace{3pt}}  rrrrr}
\tableline
\noalign{\smallskip}
Object                  &   SBS 1319+579A     &     SBS 1319+579B   &   SBS 1319+ 579C    &   SBS 1415+ 437C     &   SBS 1415+ 437A \\
\noalign{\smallskip}
\tableline
\noalign{\smallskip}				      
$T_e$(O III) (K)        &   13400 $\pm$ 500   &   11900 $\pm$ 800   &   11500 $\pm$ 600   &   16400 $\pm$ 600  &   15500 $\pm$ 700  \\
$T_e$(O II) (K)         &   12400 $\pm$ 400   &   11300 $\pm$ 600   &   11050 $\pm$ 400   &   14500 $\pm$ 400  &   13850 $\pm$ 500  \\
$N_e$ (cm$^{-3}$)       &    $<$100           &  $<$100 	        &  $<$100		      &  $<$100	           &  $<$100	    \\
\noalign{\smallskip}
12+log(O$^{+}$/H$^+$)   &    7.22 $\pm$ 0.07  &    7.73 $\pm$ 0.12  &	 7.75 $\pm$ 0.07  &    7.07 $\pm$ 0.07 &    7.05 $\pm$ 0.10 \\
12+log(O$^{++}$/H$^+$)  &    7.98 $\pm$ 0.06  &    7.89 $\pm$ 0.10  &	 7.93 $\pm$ 0.08  &    7.42 $\pm$ 0.04 &    7.47 $\pm$ 0.05 \\
12+log(O/H)             &    8.05 $\pm$ 0.06  &    8.12 $\pm$ 0.10  &	 8.15 $\pm$ 0.07  &    7.58 $\pm$ 0.05$^b$& 7.61 $\pm$ 0.06 \\
\noalign{\smallskip}
log(O$^{++}$/O$^+$)     &    0.77 $\pm$ 0.12  &    0.16 $\pm$ 0.19  &    0.18 $\pm$ 0.13  &    0.35 $\pm$ 0.08 &    0.42 $\pm$ 0.14 \\
12+log(N$^+$/H$^+$)     &    5.69 $\pm$ 0.05  &    6.24 $\pm$ 0.08  &	 6.37 $\pm$ 0.05  &    5.50 $\pm$ 0.05 &    5.48 $\pm$ 0.10 \\
12+log(N/H)             &    6.52 $\pm$ 0.09  &    6.63 $\pm$ 0.12  &	 6.77 $\pm$ 0.07  &    6.01 $\pm$ 0.07 &    6.04 $\pm$ 0.11 \\
log(N/O)                & $-$1.53 $\pm$ 0.10  & $-$1.49 $\pm$ 0.12  & $-$1.38 $\pm$ 0.10  & $-$1.57 $\pm$ 0.08 & $-$1.57 $\pm$ 0.09 \\
\noalign{\smallskip}
12+log(S$^+$/H$^+$)     &    5.29 $\pm$ 0.03  &    5.88 $\pm$ 0.05  &	 5.93 $\pm$ 0.04  &    5.38 $\pm$ 0.03 &    5.26 $\pm$ 0.05 \\
12+log(S$^{++}$/H$^+$)  &    6.09 $\pm$ 0.12  &    6.13 $\pm$ 0.18  &	 6.39 $\pm$ 0.15  &    5.75 $\pm$ 0.11 &    5.68 $\pm$ 0.15 \\
12+log(S/H)             &    6.29 $\pm$ 0.11  &    6.36 $\pm$ 0.15  &	 6.55 $\pm$ 0.12  &    5.96 $\pm$ 0.08 &    5.89 $\pm$ 0.13 \\
log(S/O)                & $-$1.76 $\pm$ 0.10  & $-$1.76 $\pm$ 0.14  & $-$1.60 $\pm$ 0.11  & $-$1.62 $\pm$ 0.12 & $-$1.72 $\pm$ 0.14 \\
\noalign{\smallskip}
12+log(Ar$^{+2}$/H$^+$) &    5.44 $\pm$ 0.07  &    5.61 $\pm$ 0.12  &	 5.61 $\pm$ 0.09  &    5.13 $\pm$ 0.10 &    5.14 $\pm$ 0.14 \\
12+log(Ar$^{+3}$/H$^+$) &    5.19 $\pm$ 0.09  &    \nodata	        &	 \nodata	      &    4.56 $\pm$ 0.16 &    \nodata	    \\
12+log(Ar/H)            &    5.65 $\pm$ 0.08  &    \nodata	        &	 \nodata	      &    5.27 $\pm$ 0.12 &    \nodata	    \\
log(Ar/O)               & $-$2.41 $\pm$ 0.11  &    \nodata	        &	 \nodata	      & $-$2.31 $\pm$ 0.13 &    \nodata	    \\
\noalign{\smallskip} 
12+log(Fe$^{++}$/H$^+$) &      \nodata        &	   \nodata 	        &    5.46:            &    5.23 $\pm$ 0.12 &	5.24:      \\
12+log(Fe/H)            &      \nodata        &	   \nodata 	        &    5.80:            &    5.67 $\pm$ 0.12 &	5.72:      \\
log(Fe/O)               &      \nodata        &    \nodata 	        & $-$2.35:            & $-$1.91 $\pm$ 0.13 & $-$1.89:      \\
\noalign{\smallskip} 
12+log(He$^+$/H$^+$)    &   10.94 $\pm$ 0.04  &  10.94 $\pm$ 0.11   &  10.92 $\pm$ 0.05   &  10.75  $\pm$ 0.06 & 10.77 $\pm$ 0.07 \\
\noalign{\smallskip}
\tableline
\noalign{\smallskip}
[O/H]$^a$               & $-$0.61 $\pm$ 0.11  & $-$0.54 $\pm$ 0.15  & $-$0.51 $\pm$ 0.13  & $-$1.08 $\pm$ 0.10 & $-$1.05 $\pm$ 0.12 \\
\noalign{\smallskip}
\tableline
  \end{tabular}
  \begin{flushleft}
  $^a$[O/H]=log(O/H)-log(O/H)$_{\odot}$, using 12+log(O/H)$_{\odot}$ = 8.66$\pm$0.05 \citep{ASP05}.\\
  $^b$Considering the existence of O$^{+3}$ because of the detection of \ion{He}{ii} $\lambda$4686, this value should be $\sim$0.01--0.02 dex higher.
  \end{flushleft}
\end{table*}

\begin{table*}[t!]
\centering
  \caption{\footnotesize{Dereddened line intensity ratios with respect to $I$(\Hb)=100 for knots analyzed in III Zw 107  (regions A, B and C) and Tol 
9 (spectra obtained with \INTe\ and \NOTe).}}
  \label{iiizw107lineas}
  \tiny
  \begin{tabular}{l  r@{\hspace{10pt}}  rrrr r}
  \noalign{\smallskip}
  \tableline
   \noalign{\smallskip}
\noalign{\smallskip}
Line &   $f(\lambda)$ & III Zw 107 A &  III Zw 107 B &  III Zw 107 C  &     Tol 9 INT        &   Tol 9 NOT \\
\noalign{\smallskip}
 \tableline
\noalign{\smallskip}
 3554.42   He    I  &  0.283 &   \nodata	&	 \nodata    &  \nodata   &    3.8$\pm$1.4    & \nodata \\
 3728.00   [O  II]  &  0.256 &   213$\pm$12   &   306$\pm$23   &      20.32: &   142$\pm$10  & 177$\pm$27\\
 3770.63    H	 I  &  0.249 &       1.00:   &       0.54:   &        \nodata    & 	 0.56:   & \nodata \\
 3797.90    H	 I  &  0.244 &       2.37:   &        \nodata	&	\nodata  & 	 0.83:       & \nodata \\
 3835.39    H	 I  &  0.237 &  3.60$\pm$0.95	&	1.69:	&	5.02:    & 	 1.94:           & \nodata \\
 3868.75  [Ne III]  &  0.230 &  23.3$\pm$2.1   &  21.1$\pm$4.0   &	20.64:   &  10.9$\pm$2.0 & 12.08:\\
 3889.05    H	 I  &  0.226 &  12.6$\pm$1.7   &  11.1$\pm$4.2   &	 8.84:   &   6.9$\pm$2.1 & \nodata \\
 3967.46  [NeIII]H7 &  0.210 &  17.1$\pm$1.7   &  22.4$\pm$2.9   &	\nodata  &  21.6$\pm$1.9 & \nodata  \\
 4068.60   [S  II]  &  0.189 &  2.26$\pm$0.82	&	4.69:	&	\nodata  &   3.3$\pm$1.1     &  \nodata \\
 4101.74    H	 I  &  0.182 &  26.2$\pm$1.8  & 26.1$\pm$4.2 &  27.5$\pm$6.5 &  25.3$\pm$2.5 & 26.3$\pm$4.0 \\
 4340.47    H	 I  &  0.127 &  46.7$\pm$2.8  & 46.9$\pm$4.8 &  47.5$\pm$7.9 &  45.8$\pm$3.2 & 46.9$\pm$5.8 \\
 4363.21   [O III]  &  0.121 &  3.14$\pm$0.67	&	1.97:	&	 \nodata & 	 0.55:           & \nodata \\
 4471.48   He	 I  &  0.095 &  4.08$\pm$0.75	&	3.40:	&	 \nodata &  3.94$\pm$0.65    & 3.96:\\
 4658.10  [Fe III]  &  0.050 &       1.06:   &        \nodata	&	 \nodata & 	 1.09:       & \nodata \\
 4686.00   He	II  &  0.043 &  \nodata	&	 \nodata   &	\nodata  &	 0.33:       & \nodata \\
 4814.55  [Fe  II]  &  0.012 &   \nodata        &       \nodata   &     \nodata  & 	 0.85:   & \nodata \\  
 4861.33    H	 I  &  0.000 & 100.0$\pm$5.3   & 100.0$\pm$8.2 & 100$\pm$15  & 100.0$\pm$6.0 & 100$\pm$10\\
 4921.93   He	 I  & -0.015 &       0.42:   &        \nodata	&	 \nodata & 	  \nodata    & \nodata \\
 4958.91   [O III]  & -0.024 & 123.7$\pm$6.5 &  99.5$\pm$8.1  &  93$\pm$14  &  78.3$\pm$5.2  & 75.5$\pm$9.1\\
 5006.84   [O III]  & -0.036 &   375$\pm$18  &   293$\pm$19  &   257$\pm$32  &   236$\pm$13  & 225$\pm$21 \\
 5197.90   [N	I]  & -0.082 &       1.08:   &       2.07:   &        \nodata    & \nodata	 & \nodata \\
 5200.26   [N	I]  & -0.083 &       0.36:   &        \nodata	&	 \nodata & 	 2.63$\pm$0.78  & \nodata \\
 5517.71  [Cl III]  & -0.154 &       0.22:   &        \nodata	&	 \nodata & 	 0.57:       & \nodata \\
 5537.88  [Cl III]  & -0.158 &       0.20:   &        \nodata	&	 \nodata & 	 0.64:       & \nodata \\
 5754.64   [N  II]  & -0.198 & \nodata       &       \nodata    &   \nodata  &   0.41:       & \nodata \\
 5875.64   He	 I  & -0.215 &  12.5$\pm$1.1   &  13.1$\pm$2.3   &	12.94:   &  12.6$\pm$1.4 & 16.3$\pm$3.8 \\
 6300.30   [O	I]  & -0.282 &  4.56$\pm$0.61	&   6.1$\pm$2.0   &	  7.57:  &  7.78$\pm$0.71& 7.32$\pm$0.96\\
 6312.10   [S III]  & -0.283 &  1.02$\pm$0.23	&	 \nodata   &	\nodata  & 	 0.66:       & 0.66:\\
 6363.78   [O	I]  & -0.291 &  1.55$\pm$0.42	&	2.01:	&	3.58:    &  2.14$\pm$0.59    & 2.30$\pm$0.91\\
 6548.03   [N  II]  & -0.318 &  9.98$\pm$0.75	&  12.1$\pm$1.4   & 13.6$\pm$4.5 &  23.8$\pm$1.8 & 24.3$\pm$2.3 \\
 6562.82    H	 I  & -0.320 &   282$\pm$16   &   273$\pm$18   &   280$\pm$35    &   267$\pm$18  & 284$\pm$30\\
 6583.41   [N  II]  & -0.323 &  28.7$\pm$1.8   &  37.0$\pm$3.2   &  41.4$\pm$8.3 &  72.2$\pm$4.9 & 82.1$\pm$7.3\\
 6678.15   He	 I  & -0.336 &  3.17$\pm$0.50	&	4.01:	&	 \nodata &  3.16$\pm$0.64        & 4.1$\pm$1.3 \\
 6716.47   [S  II]  & -0.342 &  19.6$\pm$1.3   &  36.4$\pm$4.3   &  52.2$\pm$8.7 &  37.0$\pm$2.6 & 42.1$\pm$3.9 \\
 6730.85   [S  II]  & -0.344 &  15.9$\pm$1.1   &  27.5$\pm$3.6   &  35.6$\pm$6.9 &  29.8$\pm$2.2 & 35.7$\pm$3.4\\
 7065.28   He	 I  & -0.387 &  2.27$\pm$0.40	&	 \nodata   &	 \nodata &  1.93$\pm$0.51    & 2.41$\pm$0.94\\
 7135.78  [Ar III]  & -0.396 &  11.4$\pm$0.9	&   9.4$\pm$1.7   &	 \nodata &  10.5$\pm$1.0     & 10.8$\pm$1.5\\
 7318.39   [O  II]  & -0.418 &  2.12$\pm$0.58	&	 \nodata   &	 \nodata & 	 1.96:           &  1.57$\pm$0.43\\
 7329.66   [O  II]  & -0.420 &  2.48$\pm$0.66	&	 \nodata   &	 \nodata & 	 1.08:           &  1.30$\pm$0.34\\
 7751.10  [Ar III]  & -0.467 & \nodata          & \nodata      & \nodata     & \nodata           &  2.68$\pm$0.68 \\
\noalign{\smallskip}								  
 \tableline									  
 \noalign{\smallskip}
Aperture size (arcsec)   &        &    7.2$\times$1     &     5.6$\times$1   &   5.2$\times$1         & 6.4$\times$1     & 3.8$\times$1  \\
Distance (arcsec)$^a$&      &       -             &    7.2             &   12.4                 &  -               & -  \\
$F$(\Hb)$^b$          &        &   22.3 $\pm$   0.8  &    8.6 $\pm$   0.4 &    1.56 $\pm$   0.14 &   23.4 $\pm$  0.9  & 3.8$\times$0.3\\ 
$C$(\Hb)              &        &   0.68 $\pm$   0.04 &    0.15 $\pm$ 0.02 &    0.22 $\pm$   0.03 &   0.50 $\pm$  0.05 & 0.40 $\pm$ 0.05\\
$W_{abs}$ (\AA)     &        &    2.0 $\pm$    0.3 &    1.30 $\pm$   0.10 &    0.50 $\pm$   0.10 &    7.5 $\pm$ 0.8 &  6.2 $\pm$ 0.6 \\ 
\noalign{\smallskip}
$-W$(\Ha) (\AA)     &        &  306 $\pm$   18 &    76 $\pm$    5 &      30 $\pm$    4   &  178 $\pm$   12 &  186 $\pm$ 15 \\ 
$-W$(\Hb) (\AA)     &        &   44 $\pm$    3 &    15 $\pm$    2 &     4.7 $\pm$    0.7 &   33 $\pm$    2 &   17 $\pm$  3 \\ 
$-W$(H$\gamma$) (\AA)&       &   16.4 $\pm$    1.0 &     5.5 $\pm$    0.6 &     1.8 $\pm$    0.3 &   12.3 $\pm$    0.8 \\ 
$-W$([O III]) 5007 (\AA)&    &  172 $\pm$    8 &    41 $\pm$    3 &    11.6 $\pm$    1.5 &   77 $\pm$    4 &   32 $\pm$  4\\ 
\noalign{\smallskip}
 \tableline
   \end{tabular}
  \begin{flushleft}		
	 $^a$ Relative distance of the knot with respect to the main region in the galaxy.\\
     $^b$ In units of 10$^{-15}$ erg s$^{-1}$ cm$^{-2}$ and not corrected for extinction.
   \end{flushleft}
\end{table*}

\begin{table*}[t!]
  \caption{\footnotesize{Physical conditions and chemical abundances of the ionized gas for the regions analyzed in III Zw 107 \mbox{and Tol 9.}}}
  \label{iiizw107abun}
  \tiny
  \centering
  \begin{tabular}{l@{\hspace{3pt}}  rrrr r}
\tableline
\noalign{\smallskip}
Object                  &   III Zw 107 A     &  III Zw 107 B$^a$  &   III Zw 107 C$^a$    &  Tol 9  INT     &   Tol 9 NOT    \\
\noalign{\smallskip}
\tableline
\noalign{\smallskip}				      
$T_e$(O III) (K)        &   10900 $\pm$ 900   &   10400 $\pm$ 1000  &   10350 $\pm$ 1000  &    7600 $\pm$ 1000  & 7850 $\pm$ 1000\\
$T_e$(O II) (K)         &   10500 $\pm$ 800   &   10300 $\pm$ 800   &   10250 $\pm$ 800   &    8300 $\pm$ 700  & 8500 $\pm$ 800 \\
$N_e$ (cm$^{-3}$)       &    200  $\pm$  60   &  $<$100 	        &  $<$100		      &     180 $\pm$  60  &  260 $\pm$  80 \\
\noalign{\smallskip}
12+log(O$^{+}$/H$^+$)   &    7.87 $\pm$ 0.11  &    8.05 $\pm$ 0.14  &	 8.08 $\pm$ 0.15  &    8.15 $\pm$ 0.18 &   8.21 $\pm$ 0.19 \\
12+log(O$^{++}$/H$^+$)  &    7.99 $\pm$ 0.08  &    7.96 $\pm$ 0.10  &	 7.92 $\pm$ 0.11  &    8.38 $\pm$ 0.14 &   8.29 $\pm$ 0.14\\
12+log(O/H)             &    8.23 $\pm$ 0.09  &    8.31 $\pm$ 0.12  &	 8.31 $\pm$ 0.13  &    8.58 $\pm$ 0.15 &   8.55 $\pm$ 0.16\\
\noalign{\smallskip}
log(O$^{++}$/O$^+$)     &    0.12 $\pm$ 0.14  & $-$0.09 $\pm$ 0.18  & $-$0.15 $\pm$ 0.21  &    0.22 $\pm$ 0.18 &   0.09 $\pm$ 0.20 \\
12+log(N$^+$/H$^+$)     &    6.70 $\pm$ 0.06  &    6.82 $\pm$ 0.08  &	 6.88 $\pm$ 0.11  &    7.37 $\pm$ 0.08 &   7.37 $\pm$ 0.09\\
12+log(N/H)             &    7.07 $\pm$ 0.08  &    7.08 $\pm$ 0.10  &	 7.11 $\pm$ 0.12  &    7.80 $\pm$ 0.14 &   7.72 $\pm$ 0.14\\
log(N/O)                & $-$1.16 $\pm$ 0.10  & $-$1.23 $\pm$ 0.15  & $-$1.20 $\pm$ 0.16  & $-$0.78 $\pm$ 0.15 &$-$0.84 $\pm$ 0.17 \\
\noalign{\smallskip}
12+log(S$^+$/H$^+$)     &    5.87 $\pm$ 0.06  &    6.14 $\pm$ 0.08  &	 6.28 $\pm$ 0.09  &    6.41 $\pm$ 0.08 &   6.46 $\pm$ 0.08\\
12+log(S$^{++}$/H$^+$)  &    6.23 $\pm$ 0.16  &    \nodata          &	  \nodata         &    6.78 $\pm$ 0.24 &   6.70 $\pm$ 0.25 \\
12+log(S/H)             &    6.42 $\pm$ 0.13  &    \nodata          &	  \nodata         &    6.97 $\pm$ 0.20 &   6.92 $\pm$ 0.21\\
log(S/O)                & $-$1.82 $\pm$ 0.15  &    \nodata          &     \nodata         & $-$1.61 $\pm$ 0.17 &$-$1.63 $\pm$ 0.18 \\
\noalign{\smallskip}
12+log(Ne$^{++}$/H$^+$) &    7.26 $\pm$ 0.13  &    7.30 $\pm$ 0.16  &	 7.30 $\pm$ 0.19  &    7.64 $\pm$ 0.18 &   7.60 $\pm$ 0.23\\
12+log(Ne/H)            &    7.51 $\pm$ 0.13  &    7.65 $\pm$ 0.16  &	 7.68 $\pm$ 0.20  &    7.84 $\pm$ 0.18 &   7.86 $\pm$ 0.23\\
log(Ne/O)               & $-$0.73 $\pm$ 0.15  & $-$0.66 $\pm$ 0.20  & $-$0.62 $\pm$ 0.20  & $-$0.74 $\pm$ 0.18 &$-$0.69 $\pm$ 0.22 \\
\noalign{\smallskip}
12+log(Ar$^{+2}$/H$^+$) &    5.94 $\pm$ 0.08  &    5.87 $\pm$ 0.14  &	 \nodata          &    6.30 $\pm$ 0.13 &   6.28 $\pm$ 0.15\\
12+log(Ar/H)            &    5.77 $\pm$ 0.09  &    5.69 $\pm$ 0.14  &	 \nodata	      &    6.13 $\pm$ 0.13 &   6.12 $\pm$ 0.15\\
log(Ar/O)               & $-$2.46 $\pm$ 0.13  & $-$2.52 $\pm$ 0.18  &	 \nodata	      & $-$2.45 $\pm$ 0.20 &$-$2.44 $\pm$ 0.22\\
\noalign{\smallskip}
12+log(Cl$^{++}$/H$^+$) &    4.32:            &    \nodata          &    \nodata          &    5.31:           &  \nodata  \\
12+log(Fe$^{++}$/H$^+$) &    5.61:            &	   \nodata 	        &    \nodata          &    6.14:           &  \nodata    \\
12+log(Fe/H)            &    5.92:            &	   \nodata 	        &    \nodata          &    6.51:           &  \nodata \\
log(Fe/O)               & $-$2.31:            &    \nodata 	        &    \nodata          & $-$2.07:           &  \nodata \\
\noalign{\smallskip} 
12+log(He$^+$/H$^+$)    &   10.94 $\pm$ 0.05  &  10.99 $\pm$ 0.08   &  10.99:             &  10.93 $\pm$ 0.06  &  11.04 $\pm$ 0.12  \\
\noalign{\smallskip}
\tableline
\noalign{\smallskip}
[O/H]$^b$               & $-$0.43 $\pm$ 0.14  & $-$0.35             & $-$0.35            & $-$0.08 $\pm$ 0.20 & $-$0.11 $\pm$ 0.21\\
\noalign{\smallskip}
\tableline
  \end{tabular}
  \begin{flushleft}
  $^a$Electron temperatures estimated using empirical relations. \\
  $^b$[O/H]=log(O/H)-log(O/H)$_{\odot}$, using 12+log(O/H)$_{\odot}$ = 8.66$\pm$0.05 \citep{ASP05}.
  \end{flushleft}
\end{table*}

\begin{table*}[t!]
\centering
  \caption{\footnotesize{Dereddened line intensity ratios with respect to $I$(\Hb)=100 for knots analyzed in Tol 1457-262 (regions A, B and C) and Arp 
252 (galaxy A, ESO 566-8, and galaxy B, ESO 566-7).}}
  \label{tol1457lineas}
  \tiny
  \begin{tabular}{l  r@{\hspace{10pt}}  r@{\hspace{8pt}}r@{\hspace{8pt}}r@{\hspace{8pt}}r@{\hspace{8pt}}r@{\hspace{8pt}} }
  \noalign{\smallskip}
  \tableline
   \noalign{\smallskip}
\noalign{\smallskip}
Line & $f(\lambda)$  & Tol 1457-262A    &  Tol 1457-262B  &  Tol 1457-262C &   ESO 566-8      &  ESO 566-7    \\
\noalign{\smallskip}
 \tableline
\noalign{\smallskip}
 3728.00   [O  II] &   0.256 &   224$\pm$17     &   187$\pm$16    &   270$\pm$25  &   256$\pm$18     &   280$\pm$51   \\
 3797.90    H	 I &   0.244 &  4.46$\pm$0.81	&      \nodata    &	 \nodata      &	   \nodata       &        \nodata \\
 3835.39    H	 I &   0.237 &  4.29$\pm$0.79	&  11.8$\pm$2.1   &	\nodata       &   6.0$\pm$1.3    &        \nodata \\
 3868.75  [Ne III] &   0.230 &  30.4$\pm$8.9    &  27.1$\pm$9.1	  &   23$\pm$11   &	  4.77:          &        \nodata \\
 3889.05    H	 I &   0.226 &  12.0$\pm$3.0    &  27.3$\pm$5.6   &  23.1$\pm$5.5 &  10.2$\pm$2.9    &        \nodata \\
 3967.46 [NeIII]H7 &   0.210 &  21.9$\pm$1.8    &  25.3$\pm$3.5   &  23.3$\pm$4.5 &  14.8$\pm$2.1    &        \nodata \\
 4068.60   [S  II] &   0.189 &        \nodata   &   \nodata       &   \nodata     &	  1.38:          &        \nodata \\
 4101.74    H	 I &   0.182 &  26.0$\pm$2.3    &  31.0$\pm$4.6   &  27.0$\pm$5.0 &  25.9$\pm$3.0    &        \nodata \\
 4243.97  [Fe  II] &   0.149 &        \nodata   &   \nodata       &   \nodata     &	  1.12:          &        \nodata \\
 4340.47    H	 I &   0.127 &  46.7$\pm$2.8    &  50.8$\pm$5.6   &  48.7$\pm$6.7 &  46.7$\pm$4.2    &    46$\pm$10   \\
 4363.21   [O III] &   0.121 &  8.68$\pm$0.77	&  10.6$\pm$3.1  &	6.8$\pm$2.3     &	   \nodata       &         \nodata \\
 4471.48   He	 I &   0.095 &  4.10$\pm$0.66	&   4.3$\pm$1.0   &   4.9$\pm$1.5 &  5.0$\pm$1.4     &        \nodata \\
 4658.10  [Fe III] &   0.050 &       0.85:      &  1.12$\pm$0.43  &	\nodata       &	 1.09:           &        \nodata \\
 4686.00   He	II &   0.043 &   1.9$\pm$0.4  &  2.9$\pm$0.3  &	 \nodata      &  0.74$\pm$0.25   &        \nodata \\
 4711.37  [Ar  IV] &   0.037 &        \nodata   &  \nodata        &     \nodata   &  1.25:           &        \nodata \\
 4861.33    H	 I &   0.000 & 100.0$\pm$5.5    & 100.0$\pm$9.0   &   100$\pm$10  & 100.0$\pm$7.1    &   100$\pm$16   \\
 4958.91   [O III] &  -0.024 &   203$\pm$10     &   190$\pm$12    &   163$\pm$11  &  56.9$\pm$4.3    &  25.9$\pm$5.4  \\
 5006.84   [O III] &  -0.036 &   560$\pm$27     &   522$\pm$29    &   455$\pm$28  &   204$\pm$12     &    75$\pm$11   \\
 5041.03   Si   II &  -0.044 &        \nodata	&       \nodata   &      \nodata  &	  0.48:          &        \nodata \\
 5055.98   Si   II &  -0.048 &        \nodata	&       \nodata   &      \nodata  &	  0.33:          &        \nodata \\
 5197.90   [N   I] &  -0.082 &        \nodata	&       \nodata   &      \nodata  &  2.96$\pm$0.79   &        \nodata \\
 5270.40  [Fe III] &  -0.100 &        \nodata	&       \nodata   &      \nodata  &	  0.50:          &        \nodata \\
 5754.64   [N  II] &  -0.194 &        \nodata	&       \nodata   &      \nodata  &  1.23$\pm$0.32   &        \nodata \\
 5875.64   He	 I &  -0.215 &  12.6$\pm$2.8	&  11.3$\pm$2.5   & 13.4$\pm$2.4  &  14.8$\pm$4.4    &      16.25:    \\
 6300.30   [O	I] &  -0.282 &  2.88$\pm$0.33	&  2.23$\pm$0.61  &   4.0$\pm$1.1 &  4.45$\pm$0.54   &       4.42:    \\
 6312.10   [S III] &  -0.283 &  1.34$\pm$0.24	&  1.76$\pm$0.56  &	  \nodata     &	  0.23:          &        \nodata \\
 6363.78   [O	I] &  -0.291 &  0.95$\pm$0.23	&	0.96:	      &	1.00:         &  1.64$\pm$0.32   &       3.15:    \\
 6548.03   [N  II] &  -0.318 &  3.13$\pm$0.35	&  2.00$\pm$0.61  &   3.8$\pm$1.3 &  42.4$\pm$3.2    &  40.3$\pm$7.2  \\
 6562.82    H	 I &  -0.320 &   285$\pm$15     &   286$\pm$17    &   279$\pm$18  &   286$\pm$17     &   287$\pm$35   \\
 6583.41   [N  II] &  -0.323 &  9.32$\pm$0.70	&   5.7$\pm$1.2   &  10.4$\pm$2.2 & 118.3$\pm$7.7    &   121$\pm$15   \\
 6678.15   He	 I &  -0.336 &  3.51$\pm$0.42	&  3.26$\pm$0.94  &   3.6$\pm$1.3 &  5.03$\pm$0.90   &        \nodata \\
 6716.47   [S  II] &  -0.342 &  17.1$\pm$1.0    &  17.6$\pm$2.3   &  26.3$\pm$3.3 &  36.7$\pm$2.7    &    89$\pm$12   \\
 6730.85   [S  II] &  -0.344 & 13.81$\pm$0.87	&  13.3$\pm$1.9   &  21.2$\pm$2.7 &  32.0$\pm$2.4    &    66$\pm$10   \\
 7065.28   He	 I &  -0.387 &  3.61$\pm$0.47	&   3.1$\pm$1.1   &	   3.12:      &  3.47$\pm$0.82   &        \nodata \\
 7135.78  [Ar III] &  -0.396 & 12.39$\pm$0.86	&  11.6$\pm$1.7   &  12.0$\pm$2.5 &   7.7$\pm$1.6    &       14.9:    \\
 7155.14  [Fe  II] &  -0.399 &       0.38:      &        \nodata  &	 \nodata      &	   \nodata       &        \nodata \\
 7281.35   He	 I &  -0.414 &  0.83$\pm$0.21	&	 \nodata      &	  \nodata     &	   \nodata       &        \nodata \\
 7318.39   [O  II] &  -0.418 &  3.79$\pm$0.32	&  3.46$\pm$0.75  &   4.7$\pm$1.3 &  2.98$\pm$0.67   &        \nodata \\
 7329.66   [O  II] &  -0.420 &  3.02$\pm$0.26	&  2.98$\pm$0.70  &   3.4$\pm$1.3 &  2.25$\pm$0.60   &        \nodata \\
 7751.10  [Ar III] &  -0.467 &  2.64$\pm$0.33	&	2.80:	      &	 \nodata      &  1.93$\pm$0.52   &        \nodata \\
\noalign{\smallskip}
 \tableline
 \noalign{\smallskip}
Aperture size (arcsec)   &        &   3.8$\times$1   &  1.5$\times$1  &   2.25$\times$1  & 3.2$\times$1    &  3.0$\times$1   \\
Distance (arcsec)$^a$&      &       -          &    9.7         &   5.5            &  -              &     49.7       \\
$F$(\Hb)$^b$          &        & 222  $\pm$ 8     & 47.7 $\pm$  2.3 & 33.6 $\pm$  1.7 &  111 $\pm$ 5    &  10.2 $\pm$ 0.8 \\
$C$(\Hb)              &        & 0.57 $\pm$ 0.03  & 0.00 $\pm$ 0.05 & 0.15 $\pm$ 0.02 & 0.49 $\pm$ 0.03 &  0.23 $\pm$ 0.05 \\
$W_{abs}$ (\AA)     &        &  1.4 $\pm$ 0.2   & 0.0  $\pm$  0.1 &  0.7 $\pm$ 0.1  & 1.3  $\pm$ 0.1  &   2.7 $\pm$ 0.2  \\
\noalign{\smallskip}
$-W$(\Ha) (\AA)     &        &  603 $\pm$   32  &  390 $\pm$   28 &  342 $\pm$   23 &	 472 $\pm$ 29 &    79 $\pm$  10 \\
$-W$(\Hb) (\AA)     &        &  101 $\pm$    6  &   82 $\pm$	7 &   92 $\pm$    9 &	  95 $\pm$  7 &    13 $\pm$   2 \\
$-W$(H$\gamma$) (\AA)&       &   31 $\pm$    2  &   24 $\pm$	3 &   30 $\pm$    4 &	  38 $\pm$  3 &     4 $\pm$   2 \\
$-W$([O III]) 5007 (\AA)&    &  560 $\pm$   27  &  430 $\pm$   25 &  411 $\pm$   26 &	 197 $\pm$ 12 &    10 $\pm$   3 \\
\noalign{\smallskip}
 \tableline
   \end{tabular}
  \begin{flushleft}		
	 $^a$ Relative distance of the knot with respect to the main region in the galaxy.\\
     $^b$ In units of 10$^{-15}$ erg s$^{-1}$ cm$^{-2}$ and not corrected for extinction.
   \end{flushleft}
\end{table*}

\begin{table*}[t!]
  \caption{\footnotesize{Physical conditions and chemical abundances of the ionized gas for the regions analyzed in Tol 1457-262 and Arp 252  (galaxy 
A, ESO 566-8, and galaxy B, ESO 566-7).}}
  \label{tol1457abun}
  \tiny
  \centering
  \begin{tabular}{l  rrrrr}
\tableline
\noalign{\smallskip}
Object                 &     Tol 1457-262A   &    Tol 1457-262B    &  Tol 1457-262C      &   ESO 566-8         &  ESO 566-7$^a$  \\
\noalign{\smallskip}
\tableline
\noalign{\smallskip}				      
$T_e$(O III) (K)        &   14000 $\pm$ 700   &   15200 $\pm$ 900  &   13400 $\pm$ 1100  &    8700 $\pm$ 900   &    7900 $\pm$ 1000 \\
$T_e$(O II) (K)         &   12500 $\pm$ 600   &   14200 $\pm$ 700   &   12400 $\pm$ 1000  &    9100 $\pm$ 800   &    8500 $\pm$ 900  \\
$N_e$ (cm$^{-3}$)       &     200 $\pm$  80   &  $<$100 	        &     200 $\pm$ 100	  &     300 $\pm$ 100   &     100 $\pm$ 150  \\
\noalign{\smallskip}
12+log(O$^{+}$/H$^+$)   &    7.59 $\pm$ 0.10  &    7.32 $\pm$ 0.10  &	 7.69 $\pm$ 0.15  &    8.23 $\pm$ 0.13  &    8.39 $\pm$ 0.18 \\
12+log(O$^{++}$/H$^+$)  &    7.87 $\pm$ 0.05  &    7.74 $\pm$ 0.06  &	 7.82 $\pm$ 0.09  &    8.04 $\pm$ 0.10  &    7.82 $\pm$ 0.15 \\
12+log(O/H)             &    8.05 $\pm$ 0.07$^c$&  7.88 $\pm$ 0.07  &    8.06 $\pm$ 0.11  &    8.46 $\pm$ 0.11$^c$&  8.50 $\pm$ 0.16 \\
\noalign{\smallskip}
log(O$^{++}$/O$^+$)     &    0.27 $\pm$ 0.11  &    0.43 $\pm$ 0.11  &	 0.14 $\pm$ 0.16  & $-$0.19 $\pm$ 0.17  & $-$0.57 $\pm$ 0.22  \\
12+log(N$^+$/H$^+$)     &    6.02 $\pm$ 0.06  &    5.70 $\pm$ 0.09  &	 6.10 $\pm$ 0.10  &    7.49 $\pm$ 0.06  &    7.57 $\pm$ 0.10 \\
12+log(N/H)             &    6.48 $\pm$ 0.09  &    6.27 $\pm$ 0.12  &	 6.47 $\pm$ 0.15  &    7.71 $\pm$ 0.08  &    7.67 $\pm$ 0.11 \\
log(N/O)                & $-$1.57 $\pm$ 0.11  & $-$1.61 $\pm$ 0.12  & $-$1.59 $\pm$ 0.16  & $-$0.76 $\pm$ 0.12  & $-$0.82 $\pm$ 0.16 \\
\noalign{\smallskip}
12+log(S$^+$/H$^+$)     &    5.65 $\pm$ 0.05  &    5.53 $\pm$ 0.06  &	 5.84 $\pm$ 0.09  &    6.32 $\pm$ 0.07  &    6.74 $\pm$ 0.12 \\
12+log(S$^{++}$/H$^+$)  &    5.95 $\pm$ 0.14  &    5.95 $\pm$ 0.16  &	 \nodata	      &    \nodata	        &    \nodata	     \\
12+log(S/H)             &    6.18 $\pm$ 0.10  &    6.16 $\pm$ 0.15  &	 \nodata	      &    \nodata	        &    \nodata	     \\
log(S/O)                & $-$1.88 $\pm$ 0.13  & $-$1.72 $\pm$ 0.18  &	 \nodata	      &    \nodata	        &    \nodata	     \\
\noalign{\smallskip}
12+log(Ne$^{++}$/H$^+$) &    6.99 $\pm$ 0.15  &    6.87 $\pm$ 0.18  &    6.98 $\pm$ 0.20  &    7.48 $\pm$ 0.16  &    \nodata	     \\
12+log(Ne/H)            &    7.17 $\pm$ 0.15  &    7.00 $\pm$ 0.18  &    7.22 $\pm$ 0.20  &    7.90 $\pm$ 0.16  &    \nodata	     \\
log(Ne/O)               & $-$0.88 $\pm$ 0.18  & $-$0.88 $\pm$ 0.20  & $-$0.84 $\pm$ 0.22  & $-$0.56 $\pm$ 0.19  &    \nodata	     \\
\noalign{\smallskip}
12+log(Ar$^{+2}$/H$^+$) &    5.73 $\pm$ 0.08  &    5.66 $\pm$ 0.14  &    5.77 $\pm$ 0.17  &    6.01 $\pm$ 0.14  &    6.41 $\pm$ 0.18 \\
12+log(Ar$^{+3}$/H$^+$) &    \nodata          &    \nodata 	        &    \nodata	      &    5.48 $\pm$ 0.20  &    \nodata      \\
12+log(Ar/H)            &    5.55 $\pm$ 0.08  &    5.44 $\pm$ 0.14  &	 5.61 $\pm$ 0.17  &    6.30 $\pm$ 0.17  &	 6.00 $\pm$ 0.20     \\
log(Ar/O)               & $-$2.50 $\pm$ 0.13  & $-$2.44	$\pm$ 0.18  & $-$2.45 $\pm$ 0.20  & $-$2.17 $\pm$ 0.19  & $-$2.49 $\pm$ 0.25	      \\
\noalign{\smallskip} 
12+log(Fe$^{++}$/H$^+$) &    5.43:            &	   5.62 $\pm$ 0.19  &	 \nodata	      &    5.80:            &    \nodata	     \\
12+log(Fe/H)            &    5.81:            &	   5.98 $\pm$ 0.19  &	 \nodata	      &    6.00:            &    \nodata	     \\
log(Fe/O)               & $-$2.24:            & $-$1.90 $\pm$ 0.22  &	 \nodata	      & $-$2.46:            &    \nodata	     \\
\noalign{\smallskip} 
12+log(He$^+$/H$^+$)    &   10.99 $\pm$ 0.06  &   10.95 $\pm$ 0.10  &  10.99  $\pm$ 0.08  &  11.09 $\pm$ 0.07   & 	11.06:	     \\
\noalign{\smallskip}
\tableline
\noalign{\smallskip}
[O/H]$^b$               & $-$0.61 $\pm$ 0.12  & $-$0.78 $\pm$ 0.12  & $-$0.60 $\pm$ 0.16  & $-$0.20 $\pm$ 0.16  & $-$0.16            \\
\noalign{\smallskip}
\tableline
  \end{tabular}
  \begin{flushleft}
  $^a$Electron temperatures estimated using empirical relations. \\
  $^b$[O/H]=log(O/H)-log(O/H)$_{\odot}$, using 12+log(O/H)$_{\odot}$ = 8.66$\pm$0.05 \citep{ASP05}.\\
  $^c$Considering the existence of O$^{+3}$ because of the detection of \ion{He}{ii} $\lambda$4686, this value should be $\sim$0.01 dex higher.
  \end{flushleft}
\end{table*}


\end{document}